\documentclass{article}
\usepackage{epubtk}    
\usepackage{booktabs}  
\usepackage{graphicx}  
\usepackage{amsmath}
\usepackage{amssymb}
\usepackage{latexsym}

\showlistoftablesfalse

\newcommand{\scri}{\ensuremath{\mathcal{J}^+}}

\DeclareMathOperator{\p}{\ensuremath{\partial}}

\usepackage[]{amsmath}
\usepackage[]{dsfont}
\usepackage[]{amsfonts}
\usepackage[]{amssymb}
\usepackage[]{comment}
\usepackage[]{color} 
\usepackage[]{empheq}       
\usepackage[]{mathrsfs}
\usepackage[]{units}
\usepackage[]{hyperref}
\usepackage{hyperref}
\hypersetup{
  colorlinks=true,        
  linkcolor=blue,         
  citecolor=cyan,         
}

\newcommand{\cf}{cf.~}
\newcommand{\eg}{e.g.,~}
\newcommand{\ie}{i.e.,~}
\newcommand{\bs}[1]{\boldsymbol{#1}}
\newcommand{\htt}{h^{^{\rm TT}}}
\newcommand{\bItf}{\hbox{$\hbox{${\boldsymbol{I}}$}\kern-.60em 
        \raise.4ex \hbox{$-$}$}}
\newcommand{\Itf}{\hbox{$\hbox{${\boldsymbol{I}}$}\kern-.60em 
        \raise.4ex \hbox{$-$}$}}
\newcommand{\m}{\mu}
\newcommand{\n}{\nu}
\newcommand{\tu}{{^{^{(3)}}}\!}

\begin{document}

\title{Extraction of Gravitational Waves in Numerical Relativity}

\author{\epubtkAuthorData{Nigel T.\ Bishop}
{Department of Mathematics, Rhodes University, Grahamstown 6140, South Africa}
{n.bishop@ru.ac.za}
{http://www.ru.ac.za/mathematics/people/staff/nigelbishop}
\and
\epubtkAuthorData{Luciano Rezzolla}
{Institute for Theoretical Physics, 60438 Frankfurt am Main, Germany\\ 
Frankfurt Institute for Advanced Studies, 60438 Frankfurt am Main, Germany}
{rezzolla@itp.uni-frankfurt.de}
{http://astro.uni-frankfurt.de/rezzolla}}

\maketitle

\begin{abstract}
A numerical-relativity calculation yields in general a solution of the
Einstein equations including also a radiative part, which is in practice
computed in a region of finite extent. Since gravitational radiation is
properly defined only at null infinity and in an appropriate coordinate
system, the accurate estimation of the emitted gravitational waves
represents an old and non-trivial problem in numerical relativity. A
number of methods have been developed over the years to ``extract'' the
radiative part of the solution from a numerical simulation and these
include: quadrupole formulas, gauge-invariant metric perturbations, Weyl
scalars, and characteristic extraction. We review and discuss each
method, in terms of both its theoretical background as well as its
implementation. Finally, we provide a brief comparison of the various
methods in terms of their inherent advantages and disadvantages.
\end{abstract}

\epubtkKeywords{Numerical relativity, Gravitational waves}

\newpage
\tableofcontents

\newpage

\section{Introduction}
\label{s-intro}

With the commissioning of the second generation of laser interferometric
gravitational-wave detectors, and the recent detection of
gravitational waves \cite{Abbott2016a}, there is
considerable interest in gravitational-wave astronomy. This is a huge
field, covering the diverse topics of: detector hardware construction and
design; data analysis; astrophysical source modeling; approximate methods
for gravitational-wave calculation; and, when the weak field approach is
not valid, numerical relativity.

Numerical relativity is concerned with the construction of a numerical
solution to the Einstein equations, so obtaining an approximate
description of a spacetime, and is reviewed, for example, in the textbooks
\cite{Alcubierre:2008, Bona2009, Baumgarte2010a, Gourgoulhon2012, 
Rezzolla_book:2013}. The physics in the simulation may be only gravity,
as is the case of a binary black hole scenario, but it may also include
matter fields and / or electromagnetic fields. Thus numerical relativity
may be included in the modeling of a wide range of astrophysical
processes. Often (but not always), an important desired outcome of the
modeling process will be a prediction of the emitted gravitational
waves. However, obtaining an accurate estimate of gravitational waves
from the variables evolved in the simulation is normally a rather
complicated process. The key difficulty is that gravitational waves are
unambiguously defined only at future null infinity ($\scri$), whereas in
practice the domain of numerical simulations is a region of finite extent
using a ``3+1'' foliation of the spacetime. This is true for most of the
numerical codes, but there are also notable exceptions. Indeed, there
have been attempts towards the construction of codes that include both
null infinity and the central dynamic region in the domain, but they have
not been successful in the general case. These attempts include the
hyperboloidal method \cite{Frauendiener04}, Cauchy Characteristic
Matching \cite{Winicour05}, and a characteristic
code \cite{Bishop97b}. The only successful application to an
astrophysical problem has been to axisymmetric core collapse using a
characteristic code \cite{Siebel03}.

In the linearized approximation, where gravitational fields are weak and
velocities are small, it is straightforward to derive a relationship
between the matter dynamics and the emission of gravitational waves, the
well-known quadrupole formula. This can be traced back to work by
Einstein \cite{Einstein1916,Einstein1918} shortly after the publication
of general relativity. The method is widely used to estimate
gravitational-wave production in many astrophysical processes. However,
the strongest gravitational-wave signals come from highly compact systems
with large velocities, that is from processes where the linearized
assumptions do not apply. And of course, it is an event producing a
powerful signal that is most likely to be found in gravitational-wave
detector data. Thus it is important to be able to calculate
gravitational-wave emission accurately for processes such as black hole
or neutron star inspiral and merger, stellar core collapse, etc. Such
problems cannot be solved analytically and instead are modeled by
numerical relativity, as described in the previous paragraph, to compute
the gravitational field near the source. The procedure of using this data
to measure the gravitational radiation far from the source is called
\emph{``extraction''} of gravitational waves from the numerical solution.

In addition to the quadrupole formula and full numerical relativity,
there are a number of other approaches to calculating gravitational-wave
emission from astrophysical sources. These techniques are not discussed
here and are reviewed elsewhere. They include post-Newtonian
methods \cite{Blanchet06}, effective one-body methods \cite{Damour2016},
and self-force methods \cite{Poisson2011}. Another approach, now
no-longer pursued, is the so-called ``Lazarus approach'', that combined
analytical and numerical techniques \cite{Baker00b, Baker:2001sf,
  Baker:2002qf}.

In this article we will review a number of different extraction methods:
\emph{(a)} Quadrupole formula and its variations (section~\ref{sec:qf}); 
\emph{(b)}  methods using the Newman--Penrose scalar $\psi_4$ evaluated on a
worldtube ($\Gamma$) (section~\ref{s-NP}); \emph{(c)} Cauchy Perturbative
methods, using data on $\Gamma$ to construct an approximation to a
perturbative solution on a known curved background (sections~\ref{s-CPA}
and \ref{ae} \cite{Abrahams88b, Abrahams90}); and \emph{(d)}
Characteristic extraction, using data on $\Gamma$ as inner boundary data
for a characteristic code to find the waveform at $\scri$
(sections~\ref{s-charac} and \ref{s-CE}). The description of the methods
is fairly complete, with derivations given from first principles and in
some detail. In cases (c) and (d), the theory involved is quite lengthy,
so we also provide implementation summaries for the reader who is more
interested in applying, rather than fully understanding, a particular
method, see sections~\ref{s-CPIS} and
\ref{s-characIS}.

In addition, this review provides background material on gravitational
waves (section~\ref{s-IGW}), on the ``3+1'' formalism for evolving the
Einstein equations (section~\ref{s-3+1}), and on the characteristic
formalism with particular reference to its use in estimating
gravitational radiation (section~\ref{s-charac}). The review concludes
with a comparison of the various methods for extracting gravitational
waves (section~\ref{s-comp}). This review uses many different symbols,
and their use and meaning is summarized in
Appendix~\ref{a-not}. Spin-weighted, and other, spherical harmonics are
discussed in Appendix~\ref{a-sYlm}, and various computer algebra scripts
and numerical codes are given in Appendix~\ref{a-codesscripts}.

\medskip

Throughout, we will use a spacelike signature $(-,+,+,+)$ and a system of
geometrised units in which $G = c = 1$, although when needed we will also
indicate the speed of light, $c$, explicitly. We will indicate with a
boldface any tensor, \eg $\bs{V}$ and with the standard arrow any
three-dimensional vector or operator, \eg $\vec{\bs{v}}$ and
$\vec{\nabla}$. Four-dimensional covariant and partial derivatives will
be indicated in general with $\nabla_{\mu}$ and $\partial_{\mu}$, but
other symbols may be introduced for less common definitions, or when we
want to aid the comparison with classical Newtonian expressions. Within
the standard convention of a summation of repeated indices, Greek letters
will be taken to run from 0 to 3, while Latin indices run from 1 to
3.

\medskip
We note that some of the material in this review has already appeared in
books or other review articles. In particular, we have abundantly used
parts of the text from the book \emph{``Relativistic Hydrodynamics''}, by
L.\ Rezzolla and O.\ Zanotti (Oxford University Press,
2013) \cite{Rezzolla_book:2013}, from the review
article \emph{``Gauge-invariant non-spherical metric perturbations of
Schwarzschild black-hole spacetimes''}, by A.\ Nagar and
L.\ Rezzolla \cite{Nagar05}, as well as adaptations of the text from the
article \emph{``Cauchy-characteristic matching''}, by N.\ T.\ Bishop,
R.\ Isaacson, R.\ G{\'o}mez, L.\ Lehner, B.\ Szil{\'a}gyi and
J.\ Winicour \cite{Bishop98a}.

\newpage
\section{A Quick Review of Gravitational Waves}
\label{s-IGW}
\subsection{Linearized Einstein equations}

When considering the Einstein equations
\begin{equation}
\label{einstein_equations:1}
G_{\mu\nu}  = 
R_{\mu\nu}-\frac{1}{2}R\, g_{\mu\nu} =
8\pi T_{\mu\nu} \,,
\end{equation}
as a set of second-order partial differential equations it is not easy to
predict that there exist solutions behaving as waves. Indeed, the concept
of gravitational waves as solutions of the Einstein equations written as
linear and homogeneous wave equations is valid only under some rather
idealised assumptions, such as a vacuum and asymptotically flat
spacetime, a linearised regime for the gravitational fields and suitable
gauges. If these assumptions are removed, the definition of gravitational
waves becomes much more difficult, although still possible. It should be
noted, however, that in this respect gravitational waves are not
peculiar. Any wave-like phenomenon, in fact, can be described in terms of
homogeneous wave equations only under simplified assumptions, such as
those requiring a uniform ``background'' for the fields propagating as
waves.

These considerations suggest that the search for wave-like solutions to
the Einstein equations should be made in a spacetime with very modest
curvature and with a line element which is that of flat spacetime but for
small deviations of nonzero curvature, \ie
\begin{equation}
\label{eq:pertmetric}
g_{\mu \nu} = \eta_{\mu \nu} + h_{\mu \nu} + 
\mathcal{O}\left((h_{\m \n})^2\right)\,,
\end{equation}
where the linearised regime is guaranteed by the fact that $|h_{\mu \nu}|
\ll 1$. Before writing the linearised version of the Einstein equations
(\ref{einstein_equations:1}) it is necessary to derive the linearised
expression for the Christoffel symbols. In a Cartesian coordinate basis
(such as the one we will assume hereafter), we recall that the general
expression for the affine connection is given by 
\begin{equation}
\label{eq:Gammas}
\Gamma _{\beta \gamma }^{\alpha }=\frac{1}{2}g^{\alpha \delta }
\left(
\partial_{\gamma} g_{\delta\beta} + \partial_{\beta} g_{\delta \gamma} -
\partial_{\delta} g_{\beta \gamma}
\right) \,.
\end{equation}
where the partial derivatives are readily calculated as
\begin{equation}
\partial_{\beta}g_{\n \alpha} = \partial_{\beta}\eta_{\n \alpha} + 
\partial_{\beta}h_{\n \alpha} 
= \partial_{\beta}h_{\n \alpha} \,.
\end{equation}
As a result, the linearised Christoffel symbols become
\begin{eqnarray}
\label{lin_gammas}
\Gamma^{\m}_{\ \alpha\beta} = 
\frac{1}{2} 
\eta^{\m \n}(
\partial_{\beta}h_{\n \alpha} + 
\partial_{\alpha}h_{\n \beta} - 
\partial_{\nu}h_{\alpha\beta}) =
\frac{1}{2}
(
 \partial_{\beta}h^{\mu}_{\ \ \alpha} + 
 \partial_{\alpha}h^{\mu}_{\ \ \beta} - 
 \partial^{\mu}h_{\alpha \beta}) \,.
\end{eqnarray}
Note that the operation of lowering and raising the indices in expression
(\ref{lin_gammas}) is not made through the metric tensors $g_{\m \n}$ and
$g^{\m \n}$ but, rather, through the spacetime metric tensors $\eta_{\m
  \n}$ and $\eta^{\m \n}$. This is just the consequence of the linearised
approximation and, despite this, the spacetime is really curved!

Once the linearised Christoffel symbols have been computed, it is
possible to derive the linearised expression for the Ricci tensor which
takes the form
\begin{equation}
\label{lin_rt}
R_{\m \n} = 
\partial_{\alpha}\Gamma^{\alpha}_{\ \m \n} - 
\partial_{\nu}\Gamma^{\alpha}_{\ \m \alpha}
= \frac{1}{2}
(
\partial_{\alpha}\partial_{\nu} h^{\ \;\alpha}_{\m} + 
\partial_{\alpha}\partial_{\mu} h^{\ \;\alpha}_{\n} -
\partial_{\alpha}\partial^{\alpha} h_{\m \n} - 
\partial_{\m}\partial_{\nu} h
) \,,
\end{equation}
where
\begin{equation}
h \coloneqq h^{\alpha}_{\ \; \alpha}  = \eta^{\m \alpha} h_{\m \alpha} 
\end{equation}
is the trace of the metric perturbations. The resulting Ricci scalar is
then given by
\begin{equation}
\label{lin_rs}
R \coloneqq g^{\m \n} R_{\m \n}  \simeq \eta^{\m \n} R_{\m \n} \,.
\end{equation}
Making use of (\ref{lin_rt}) and (\ref{lin_rs}), it is possible to
rewrite the Einstein equations (\ref{einstein_equations:1}) in a
linearised form as
\begin{equation}
\label{efe2}
\partial^{\alpha}\partial_{\nu}   h_{\m \alpha} + 
\partial^{\alpha}\partial_{\mu}   h_{\n \alpha} - 
\partial_{\alpha}\partial^{\alpha} h_{\m \n} - 
\partial_{\mu}\partial_{\nu}h - 
\eta_{\m \n}(
\partial^{\alpha}\partial^{\beta}h_{\alpha\beta} - 
\partial^{\alpha}\partial_{\alpha}h)
	= 16 \pi T_{\m \n} \,.
\end{equation}

Although linearised, the Einstein equations (\ref{efe2}) do not yet seem
to suggest a wave-like behaviour. A good step in the direction of
unveiling this behaviour can be made if we introduce a more compact
notation, which makes use of ``trace-free'' tensors defined as
\begin{equation}
\label{eq:barop}
{\bar h}_{\m \n} \coloneqq h_{\m \n} - \frac{1}{2} \eta_{\m \n} h \,,
\end{equation}
where the ``bar-operator'' in \eqref{eq:barop} can be applied to any
symmetric tensor so that, for instance, \hbox{${\bar R}_{\m \n} = G_{\m
    \n}$}, and also iteratively, \ie ${\bar {\bar h}}_{\m \n} = h_{\m
  \n}$.\footnote{Note that the ``bar'' operator can in principle be
  applied also to the trace so that ${\bar h}=-h$.} Using this notation,
the linearised Einstein equations (\ref{efe2}) take the more compact form
\begin{equation}
\label{efe3}
-\partial^{\alpha}\partial_{\alpha}{\bar h}_{\m \n} -
\eta_{\m \n} \,
\partial^{\alpha}\partial^{\beta}{\bar h}_{\alpha\beta} + 
\partial^{\alpha}\partial_{\mu}{\bar h}_{\n \alpha}
	= 16 \pi T_{\m \n} \,,
\end{equation}
where the first term on the left-hand side of \eqref{efe3} can be easily
recognised as the \emph{Dalambertian} (or wave) operator, \ie
$\partial_{\alpha}\partial^{\alpha}{\bar h}_{\m \n} = \square {\bar
  h}_{\m \n}$. At this stage, we can exploit the gauge freedom inherent
in general relativity (see also below for an extended discussion) to
recast Eqs. (\ref{efe3}) in a more convenient form. More specifically, we
exploit this gauge freedom by choosing the metric perturbations $h_{\m
  \n}$ so as to eliminate the terms in (\ref{efe3}) that spoil the
wave-like structure. Most notably, the coordinates can be
selected so that the metric perturbations satisfy
\begin{equation}
\label{lg}
\partial_{\alpha} {\bar h}^{\m \alpha} = 0 \,.
\end{equation}
Making use of the gauge (\ref{lg}), which is also known as the
\emph{Lorenz} (or Hilbert) \emph{gauge},
the linearised field equations take the form
\begin{equation}
\label{efe4}
\square\,{\bar h}_{\m \n} = - 16 \pi T_{\m \n} \,,
\end{equation}
that, in vacuum reduce to the desired result
\begin{equation}
\label{efe5}
\square\,{\bar h}_{\m \n} = 0 \,.
\end{equation}
Equations \eqref{efe5} show that, in the Lorenz gauge and in vacuum, the
metric perturbations propagate as waves distorting flat spacetime.

The simplest solution to the linearised Einstein equations (\ref{efe5})
is that of a plane wave of the type
\begin{equation}
\label{sol_efe5}
{\bar h}_{\m \n} = A_{\m \n} 
	\exp (i \kappa_{\alpha}x^{\alpha}) \,,
\end{equation}
where of course we are interested only in the real part
of \eqref{sol_efe5}, with $\boldsymbol{A}$ being the \emph{amplitude
  tensor}.
Substitution of the ansatz~\eqref{sol_efe5} into Eq.~\eqref{efe5} implies that
$\kappa^{\alpha}\kappa_{\alpha}=0$ so that $\boldsymbol{\kappa}$ is a
null four-vector.
In such a solution, the plane wave (\ref{sol_efe5}) travels in the
spatial direction $\vec{\boldsymbol{k}} =
(\kappa_x,\kappa_y,\kappa_z)/\kappa^0$ with frequency $\omega \coloneqq
\kappa^0 = (\kappa^j\kappa_j)^{1/2}$. The next step is to substitute the
ansatz~\eqref{sol_efe5} into the Lorenz gauge condition Eq.~\eqref{lg},
yielding $A_{\m \n}\kappa^{\m}=0$ so that $\boldsymbol{A}$ and
$\boldsymbol{\kappa}$ are orthogonal. Consequently, the amplitude tensor
$\boldsymbol{A}$, which in principle has $16-6=10$ independent
components, satisfies four conditions. Thus the imposition of the Lorenz
gauge reduces the independent components of $\boldsymbol{A}$ to six. We
now investigate how to reduce the number of independent components to
match the number of dynamical degrees of freedom of general relativity,
\ie two.

While a Lorenz gauge has been imposed [\cf
Eq. (\ref{lg})], this does not completely fix the coordinate system of
a linearised theory. A residual ambiguity, in fact, is preserved through
arbitrary \emph{gauge changes}, \ie through infinitesimal coordinate
transformations that are 
consistent with the gauge that
has been selected. The freedom to make such a transformation follows from
a foundation of general relativity, the principle of general covariance.
To better appreciate this matter, consider an infinitesimal
coordinate transformation in terms of a small but otherwise arbitrary
displacement four-vector $\boldsymbol{\xi}$
\begin{eqnarray}
\label{coord_trans}
x^{\alpha'} = x^{\alpha} + \xi^{\alpha} \,. 
\end{eqnarray}
Applying this transformation to the linearised metric
(\ref{eq:pertmetric}) generates a ``new'' metric tensor that, to the
lowest order, is
\begin{equation}
\label{metric'}
g^{\rm new}_{\mu'\nu'} = \eta_{\mu \nu} + h^{\rm old}_{\mu \nu} 
	- \partial_{\nu}\xi_{\mu} - \partial_{\mu}\xi_{\nu}	\,,
\end{equation}
so that the ``new'' and ``old'' perturbations are related by the
following expression
\begin{equation}
\label{hmn'}
h^{\rm new}_{\mu'\nu'} = h^{\rm old}_{\mu \nu} 
	- \partial_{\nu}\xi_{\mu} - \partial_{\mu}\xi_{\nu} \,,
\end{equation}
or, alternatively, by 
\begin{equation}
\label{hbmn'}
{\bar h}^{\rm new}_{\mu'\nu'} = {\bar h}^{\rm old}_{\mu \nu} -
	\partial_{\nu}\xi_{\mu} - \partial_{\mu}\xi_{\nu} +
	\eta_{\mu \nu}\, \partial_{\alpha}\xi^{\alpha}\,.
\end{equation}
Requiring now that the new coordinates satisfy the condition (\ref{lg})
of the Lorenz gauge $\partial^{\alpha} {\bar h}^{\rm new}_{\m \alpha} =
0$, forces the displacement vector to be solution of the homogeneous wave
equation
\begin{equation}
\label{we_xi}
\partial_{\beta}\partial^{\beta}\xi^{\alpha} = 0 \,. 
\end{equation}
As a result, the plane-wave vector with components
\begin{equation}
\label{xi_alpha}
\xi^{\alpha} \coloneqq -i C^{\alpha} {\rm exp}(i \kappa_{\beta}x ^{\beta}) 
\end{equation}
generates, through the \emph{four} arbitrary constants $C^{\alpha}$, a
gauge transformation that changes arbitrarily \emph{four} components of
$\boldsymbol{A}$ in addition to those coming from the condition
$\boldsymbol{A} \cdot \boldsymbol{\kappa}=0$. Effectively, therefore,
$A_{\mu \nu}$ has only $10-4-4=2$ linearly independent components,
corresponding to the number of degrees of freedom in general
relativity \cite{MTW1973}. 

Note that these considerations are not unique to general relativity and
similar arguments can also be made in classical electrodynamics, where
the Maxwell equations are invariant under transformations of the vector
potentials of the type $A_{\mu} \rightarrow A_{\mu'} = A_{\mu}
+ \partial_{\mu}\Psi$, where $\Psi$ is an arbitrary scalar function, so
that the corresponding electromagnetic tensor is $F^{\rm new}_{\m' \n'}
= \partial_{\nu'} A_{\m'} - \partial_{\mu'} A_{\n'} = F^{\rm
old}_{\m' \n'}$. Similarly, in a linearised theory of general relativity,
the gauge transformation (\ref{hmn'}) will preserve the components of the
Riemann tensor, \ie $R^{\rm new}_{\alpha \beta \m \nu} = R^{\rm
old}_{\alpha \beta \m \nu} + \mathcal{O}(R^2)$.

To summarise, it is convenient to constrain the components of the
amplitude tensor through the following conditions:
\begin{description}

\item {\it (a)}: \emph{orthogonality condition}: four components of the
  amplitude tensor can be specified since the Lorenz gauge implies that
  $\boldsymbol{A}$ and
  $\boldsymbol{\kappa}$ are 
  orthogonal, \ie $A_{\m \n}
  \kappa^{\n} = 0$.
  
\item {\it (b)}: \emph{choice of observer}: three components of the
  amplitude tensor can be eliminated after selecting the infinitesimal
  displacement vector $\xi^{\m} = iC^{\m}
  \exp(i\kappa_{\alpha}x^{\alpha})$ so that $A^{\m\n}u_{\m} =0$ for some
  chosen four-velocity vector $\boldsymbol{u}$. This means that the
  coordinates are chosen so that for an observer with four-velocity
  $u^{\m}$ the gravitational wave has an effect only in spatial
  directions\footnote{Note that the orthogonality condition fixes three
    and not four components since one further constraint needs to be
    satisfied, \ie $\kappa^{\m} A_{\m \n} u^{\n} = 0$.}.


\item {\it (c)}: 
      \emph{traceless condition}:
  one final component of the amplitude tensor can be eliminated after
  selecting the infinitesimal displacement vector $\xi^{\m} =
  iC^{\m} \exp(i\kappa_{\alpha}x^{\alpha})$ so that $A^{\m}_{\ \,\m} =
  0$.
\end{description}
Conditions {\it (a), (b)} and {\it (c)} define the so-called
\emph{transverse--traceless} (TT) \emph{gauge},
which represents a most convenient gauge for the analysis of
gravitational waves. 
To appreciate the significance of these conditions, consider them
implemented in a reference frame which is globally at rest,
\ie with four-velocity $u^{\alpha} = (1,0,0,0)$, where the amplitude
tensor must satisfy:
\begin{description}
\item {\it (a):}
\begin{equation}
\label{(b)}
A_{\m \n} \kappa^{\n} = 0 \qquad \Longleftrightarrow 
	\qquad \partial^j h_{ij} = 0 \,,
\end{equation}
\ie the spatial components of $h_{\m \n}$ are \emph{divergence-free}.

\item {\it (b):}
\begin{equation}
\label{(a)}
A_{\m \n} u^{\n} = 0 \qquad \Longleftrightarrow 
	\qquad h_{\m t} = 0 \,,
\end{equation}
\ie only the spatial components of $h_{\m \n}$ are \emph{nonzero}, hence
the \textit{transverse} character of the TT gauge.

\item {\it (c):}
\begin{equation}
\label{(c)}
A^{\m}_{\ \, \m} = 0 \qquad \Longleftrightarrow 
	\qquad h=h^{j}_{\ j} = 0 \,,
\end{equation}
\ie the spatial components of $h_{\m \n}$ are \emph{trace free} hence the
\textit{trace-free} character of the TT gauge. Because of this, and only
in this gauge, ${\bar h}_{\m \n} = h_{\m \n}$
\end{description}

\subsection{Making sense of the TT gauge}	

As introduced so far, the TT gauge might appear rather abstract and not
particularly interesting. Quite the opposite, the TT gauge introduces a
number of important advantages and simplifications in the study of
gravitational waves. The most important of these is that, in this gauge,
the only nonzero components of the Riemann tensor are
\begin{equation}
R_{j0k0}=R_{0j0k} = -R_{j00k} = -R_{0jk0}\,.
\end{equation}
However, since
\begin{equation}
\label{rj0k0}
R_{j0k0}=-\frac{1}{2} \partial^2_t \htt_{jk} \,,
\end{equation}
the use of the TT gauge indicates that a travelling gravitational wave
with periodic time behaviour $\htt_{jk} \propto \exp (i \omega t)$ can be
associated to a local oscillation of the spacetime, \ie
\begin{equation}
\partial^2_t \htt_{jk}
 \sim -\omega^2 \exp (i \omega t) \sim R_{j0k0}\,, 
	\qquad {\rm and} \qquad \ 
	R_{j0k0} = \frac{1}{2} \omega^{2} \htt_{jk} \,.
\end{equation}

	To better appreciate the effects of the propagation of a
gravitational wave, it is useful to consider the separation between two
neighbouring particles $A$ and $B$ on a geodesic motion and how this
separation changes in the presence of an incident gravitational wave (see
Fig.~\ref{fig0}). For this purpose, let us introduce a coordinate
system $x^{\hat \alpha}$ in the neighbourhood of particle $A$ so that along
the worldline of the particle $A$ the line element will have the form
\begin{equation}
ds^2 = -d\tau^2 + \delta_{\hat i \hat j}\, dx^{\hat i} dx^{\hat j} + {\cal
	O}(|x^{\hat j}|^2) dx^{\hat \alpha} dx^{\hat \beta} \,.
\end{equation}
The arrival of a gravitational wave will perturb the geodesic motion of
the two particles and produce a nonzero contribution to the
geodesic-deviation equation. We recall that the changes in the separation
four-vector $\boldsymbol{{\it n}}$ between two geodesic trajectories with
tangent four-vector $\boldsymbol{{\it u}}$ are expressed through the
geodesic-deviation equation (see Fig. \ref{fig0})
%
\begin{equation}
\label{gde_full}
\frac{D^2 n^{\alpha}}{D \tau^2} = 
          u^{\gamma}\nabla_{\gamma}\,\left(u^{\beta}\nabla_{\beta} 
          n^{\alpha}\right) = 
	-R^{\alpha}_{\ \beta \delta \gamma} u^{\beta} u^{\delta} n^{\gamma}\,,
\end{equation}
where the operator
\begin{equation}
\frac{D}{D \tau}\coloneqq u^{\alpha}  \nabla_{\alpha}\,,
\end{equation}
is the covariant time derivative along the worldline (in this case a
geodesic) of a particle.

\epubtkImage{}{%
\begin{figure}
\begin{center}
\includegraphics[width=0.65\columnwidth,angle=0]{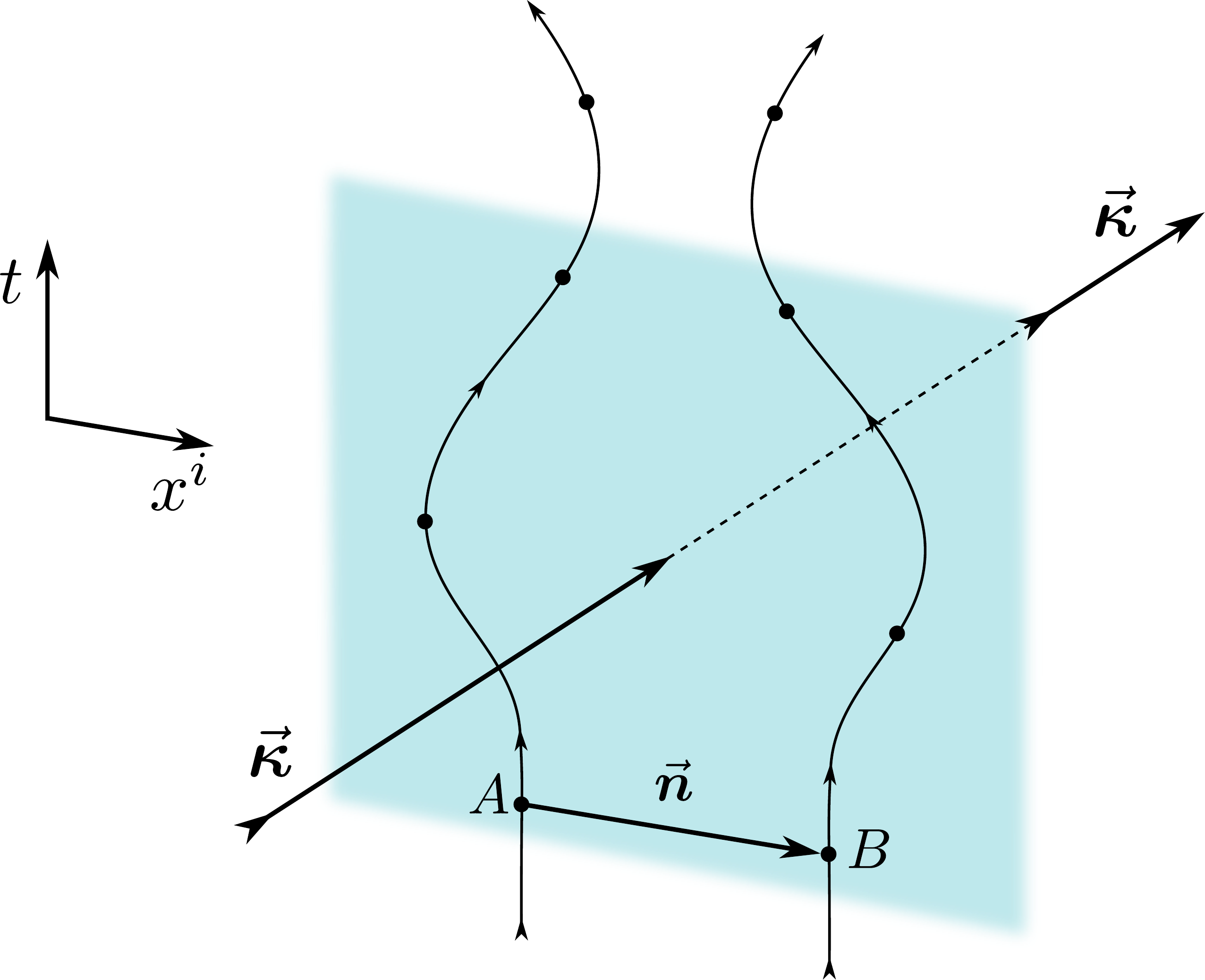}
\caption{Schematic diagram for the changes in the separation vector
  $\vec{\boldsymbol{n}}$ between two particles $A$ and $B$ moving along
  geodesic trajectories produced by the interaction with a gravitational
  wave propagating along the direction $\vec{\boldsymbol{\kappa}}$.
\label{fig0}}
\end{center}
\end{figure}}

Indicating now with $n^{\hat j}_{_{\rm B}} \coloneqq x^{\hat j}_{_{\rm B}} -
x^{\hat j}_{_{\rm A}} = x^{\hat j}_{_{\rm B}}$ the components of the
separation three-vector in the positions of the two particles, the
geodesic-deviation equation (\ref{gde_full}) can be written as
\begin{equation}
\label{gde1}
\frac{D^2 x^{\hat j}_{_{\rm B}}}{D \tau^2} =
	-R^{\hat j}_{\ 0 \hat k 0} x^{\hat k}_{_{\rm B}} \,. 
\end{equation}
A first simplification to these equations comes from the fact that around
the particle $A$, the affine connections vanish (\ie $\Gamma^{\hat
j}_{{\hat \alpha} {\hat \beta}}=0$) and the covariant derivative in
(\ref{gde1}) can be replaced by an ordinary total derivative.
Furthermore, because in the TT gauge the coordinate system
$x^{\hat \alpha}$ moves together with the particle $A$, the proper and
the coordinate time coincide at first order in the metric perturbation
[\ie $\tau=t + {\cal O}((h^{^{\rm TT}}_{\mu \nu})^2)$]. As a result,
equation (\ref{gde1}) effectively becomes
\begin{equation}
\label{gde2}
\frac{d^2 x^{\hat j}_{_{\rm B}}}{d t^2} =
	\frac{1}{2}\left(\frac{\partial^2 \htt_{{\hat j} {\hat k}}}
	{\partial t^2} \right) x^{\hat k}_{_{\rm B}} \,,
\end{equation}
and has solution
\begin{equation}
\label{gde_sol}
x^{\hat j}_{_{\rm B}}(t) = x^{\hat k}_{_{\rm B}}(0) \left[
	\delta_{{\hat j} {\hat k}} + \frac{1}{2} 
	\htt_{{\hat j} {\hat k}}(t)\right] \,.
\end{equation}
Equation (\ref{gde_sol}) has a straightforward interpretation and
indicates that, in the reference frame comoving with $A$, the particle
$B$ is seen oscillating with an amplitude proportional to $\htt_{{\hat j}
  {\hat k}}$.

Note that because these are transverse waves, they will produce a local
deformation of the spacetime only in the plane orthogonal to their
direction of propagation. As a result, if the two particles lay along the
direction of propagation (\ie if $\vec{\boldsymbol{n}} \parallel
\vec{\boldsymbol{\kappa}}$), then $\htt_{{\hat j} {\hat k}} x^{\hat
  j}_{_{\rm B}}(0) \propto \htt_{{\hat j} {\hat k}} \kappa^{\hat
  j}_{_{\rm B}}(0) = 0$ and no oscillation will be recorded by $A$
    [cf. equation (\ref{(b)})]

\epubtkImage{}{%
\begin{figure}
\begin{center}
\includegraphics[width=0.45\columnwidth,angle=0]{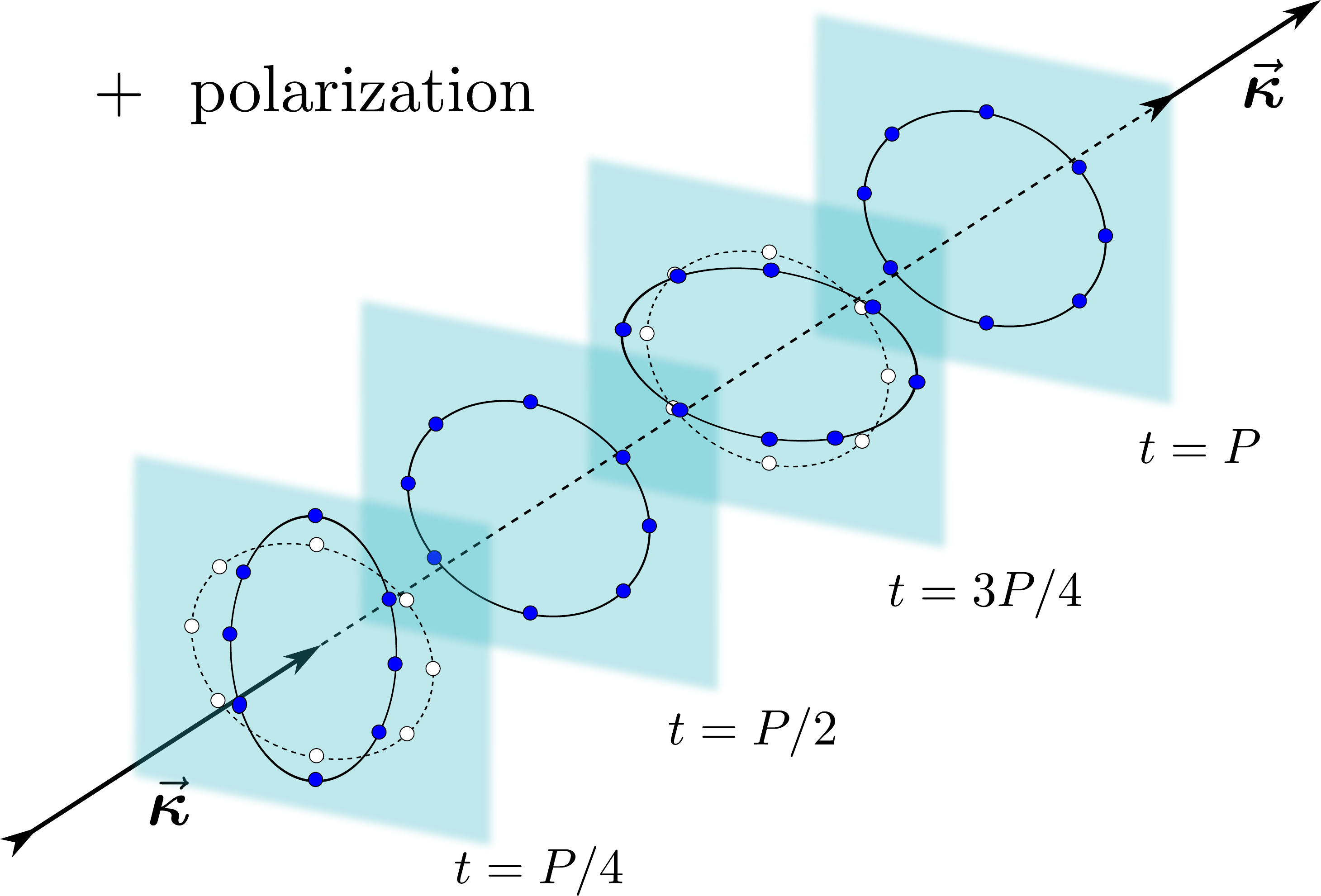}
\hskip 0.5cm
\includegraphics[width=0.45\columnwidth,angle=0]{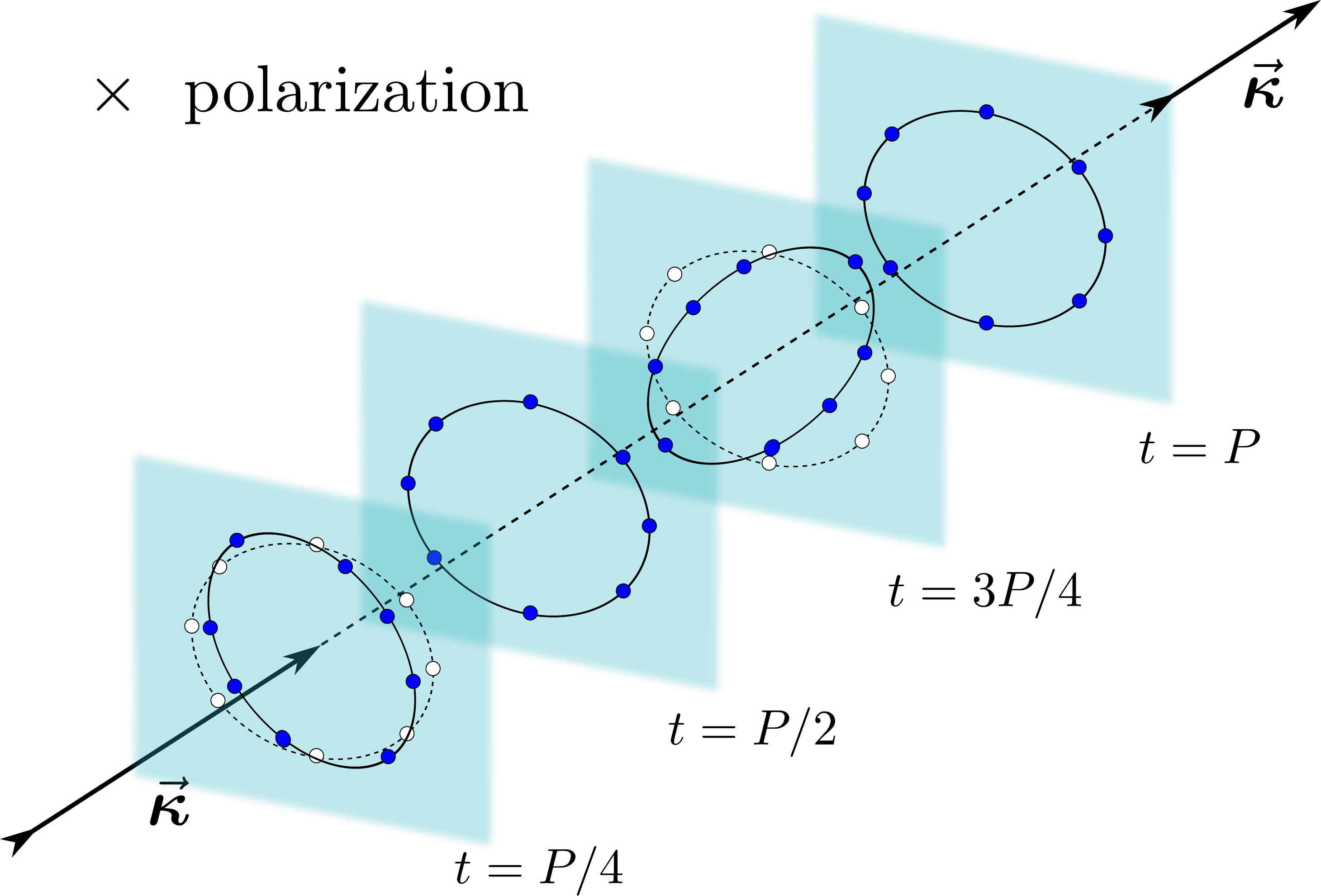}
\vskip 1.0cm
\caption{Schematic deformations produced on a ring of freely-falling
  particles by gravitational waves that are linear polarized in the
  ``$+$'' (``plus'') and ``$\times$'' (``cross'') modes. The continuous
  lines and the dark filled dots show the positions of the particles at
  different times, while the dashed lines and the open dots show the
  unperturbed positions.\label{fig1}}
\end{center}
\end{figure}}

%
	Let us now consider a concrete example and in particular a planar
gravitational wave propagating in the positive $z$-direction. In this
case
\begin{eqnarray}
\htt_{xx} &=& - \htt_{yy} = {\Re  \left\{ A_{+} 
	\exp[-i\omega(t-z)]\right\} } \,,
	\\ \nonumber \\
\htt_{xy} &=& \htt_{yx} = {\Re \left\{ A_{\times}
	\exp[-i\omega(t-z)]\right\}} \	, 
\end{eqnarray}
where $A_{+}$, $A_{\times}$ represent the two independent modes of
polarization, and the symbol $\Re$ refers to the real part. As in
classical electromagnetism, in fact, it is possible to decompose a
gravitational wave in two \emph{linearly} polarized plane waves or in two
\emph{circularly} polarized ones. In the first case, and for a
gravitational wave propagating in the $z$-direction, the polarization
{\sl tensors} $+$ (``plus'') and $\times$ (``cross'') are defined as
\begin{eqnarray}
\boldsymbol{e}_{+} &\coloneqq& \vec{\boldsymbol{e}}_{x} \otimes 
	\vec{\boldsymbol{e}}_{x} - \vec{\boldsymbol{e}}_{y} 
	\otimes \vec{\boldsymbol{e}}_{y} \,,
\\ \nonumber \\
\boldsymbol{e}_{\times} &\coloneqq& \vec{\boldsymbol{e}}_{x} \otimes 
	\vec{\boldsymbol{e}}_{x} + \vec{\boldsymbol{e}}_{y} 
	\otimes \vec{\boldsymbol{e}}_{y} \,.
\end{eqnarray}

	The deformations that are associated with these two modes of
linear polarization are shown in Fig.~\ref{fig1} where the positions of
a ring of freely-falling particles are schematically represented at
different fractions of an oscillation period. Note that the two linear
polarization modes are simply rotated of $\pi/4$.

\epubtkImage{}{%
\begin{figure}
\begin{center}
\includegraphics[width=0.45\columnwidth,angle=0]{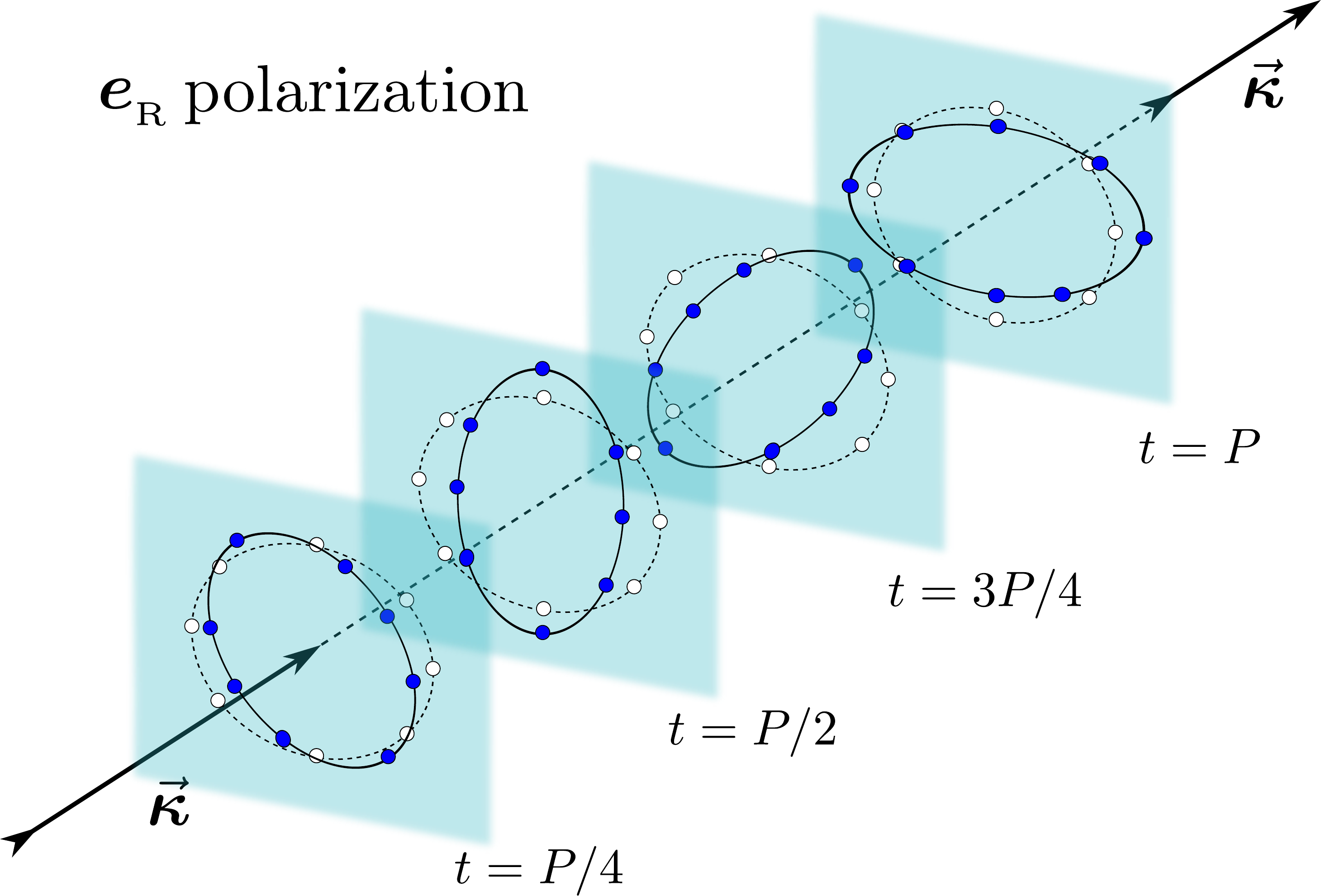}
\hskip 0.5cm
\includegraphics[width=0.45\columnwidth,angle=0]{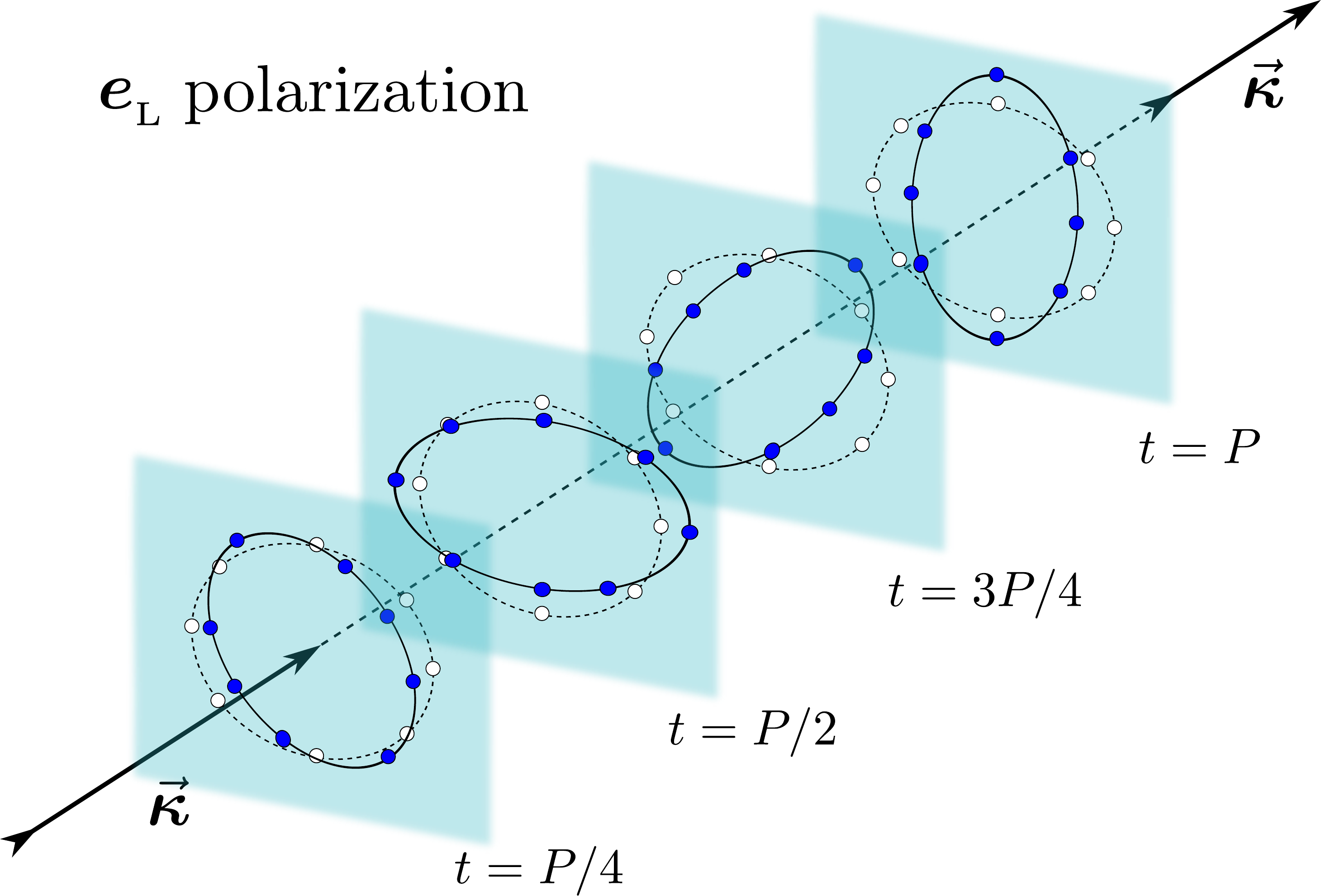}
\vskip 1.0cm
\caption{Schematic deformations produced on a ring of freely-falling
  particles by gravitational waves that are circularly polarized in the
  $R$ (clockwise) and $L$ (counter-clockwise) modes. The continuous lines
  and the dark filled dots show the positions of the particles at
  different times, while the dashed lines and the open dots show the
  unperturbed positions. \label{fig2}}
\end{center}
\end{figure}}

	In a similar way, it is possible to define two {\sl tensors}
describing the two states of circular polarization and indicate with
$\boldsymbol{e}_{_{\rm R}}$ the circular polarization that rotates clockwise
(see Fig.~\ref{fig2})
\begin{equation}
\boldsymbol{e}_{_{\rm R}} \coloneqq \frac{\boldsymbol{e}_{+} + i 
	\boldsymbol{e}_{\times}}{\sqrt 2} \,,
\end{equation}
and with $\boldsymbol{e}_{_{\rm L}}$ the circular polarization that rotates
counter-clockwise (see Fig.~\ref{fig2})
\begin{equation}
\boldsymbol{e}_{_{\rm L}} \coloneqq \frac{\boldsymbol{e}_{+} - i
	\boldsymbol{e}_{\times}}{\sqrt 2} \,.
\end{equation}
The deformations that are associated to these two modes of circular
polarization are shown in Fig.~\ref{fig2}

\subsection{The quadrupole formula}
\label{sec:qf}

The quadrupole formula and its domain of applicability were mentioned in
Sec.~\ref{s-intro}, and some examples of its use in a numerical
simulation are presented in Sec.~\ref{s-comp}. In practice, the
quadrupole formula represents a low-velocity, weak-field approximation to
measure the gravitational-wave emission within a purely Newtonian
description of gravity\footnote{Of course no gravitational waves are
  present in Newton's theory of gravity and the formula merely estimates
  the time variations of the quadrupole moment of a given distribution of
  matter.}. In practice, the formula is employed in those numerical
simulations that either treat gravity in an approximate manner (\eg via a
post-Newtonian approximation or a conformally flat metric) or that,
although in full general relativity, have computational domains that are
too small for an accurate calculation of the radiative emission.

In what follows we briefly discuss the amounts of energy carried by
gravitational waves and provide simple expressions to estimate the
gravitational radiation luminosity of potential sources. Although the
estimates made here come from analogies with electromagnetism, they
provide a reasonable approximation to more accurate expressions from
which they differ for factors of a few. Note also that while obtaining
such a level of accuracy requires only a small effort, reaching the
accuracy required of a template to be used in the realistic detection of
gravitational waves is far more difficult and often imposes the use of
numerical relativity calculations on modern supercomputers.

In classical electrodynamics, the energy emitted per unit time by an
oscillating electric dipole $d = qx$, with $q$ the electrical charges and
$x$ their separation, is easily estimated to be
\begin{equation}
\label{ed}
L_{\rm electric\ dip.} \coloneqq \frac{(\rm energy\ emitted)}{(\rm unit\ time)}
	=\frac{2}{3}q^2 (\ddot{x}\,)^2 =\frac{2}{3}(\ddot{d}\,)^2 \,,
\end{equation}
where the number of ``dots'' counts the order of the total time
derivative. Equally simple is to calculate the corresponding luminosity
in gravitational waves produced by an oscillating mass dipole. In the
case of a system of $N$ point-like particles of mass $m_{_{A}}$
($A=1,2,\ldots,N$), in fact, the total mass dipole and its first time
derivative are
\begin{equation}
\vec{\boldsymbol{d}} \coloneqq \sum^N_{A=1} 
m_{_{A}} \vec{\boldsymbol{x}}_{_{A}} \,,
\end{equation}
and
\begin{equation}
\dot{\vec{\boldsymbol{d}}} \coloneqq 
\sum^N_{A=1} m_{_{A}} \dot{\vec{\boldsymbol{x}}}_{_{A}} = 
	\vec{\boldsymbol{p}}\,,
\end{equation}
respectively. However, the requirement that the system conserves its
total linear momentum
\begin{equation}
\ddot{\vec{\boldsymbol{d}}} \coloneqq \dot{\vec{\boldsymbol{p}}}=0\,, 
\end{equation}
forces to conclude that $L_{\rm mass\ dip.}=0$, \ie that there is no
mass-dipole radiation in general relativity (This is equivalent to the
impossibility of having electromagnetic radiation from an
electric monopole oscillating in time.). Next, consider the electromagnetic energy emission
produced by an oscillating electric quadrupole. In classical
electrodynamics, this energy loss is given by
\begin{equation}
\label{eq}
L_{\rm electric\ quad.} \coloneqq 
	\frac{1}{20}({\dddot Q})^2 =\frac{1}{20}
	({\dddot Q_{jk}}{\dddot Q_{jk}}) \,,
\end{equation}
where	
\begin{equation}
Q_{jk} \coloneqq \sum^N_{A=1} q_{_{A}} \left[ (x_{_{A}})_j (x_{_{A}})_k - 
	\frac{1}{3}\delta_{jk} (x_{_{A}})_i (x_{_{A}})^i\right]\,,
\end{equation}
is the electric quadrupole for a distribution of $N$ charges $(q_1, q_2,
\ldots, q_N)$.

In close analogy with expression (\ref{eq}), the energy loss per unit
time due to an oscillating mass quadrupole is calculated to be
\begin{equation}
\label{mq}
L_{\rm mass\ quad.} \coloneqq 
	\frac{1}{5}
	\langle\; {(\,\dddot\bItf\,)^2 \; }\rangle =
	\frac{1}{5}
	\langle {\dddot \bItf_{jk}}{\dddot \bItf_{jk}} \rangle \,,
\end{equation}
where $\Itf_{jk}$ is the trace-less mass quadrupole (or ``reduced'' mass
quadrupole), defined as
\begin{eqnarray}
\label{itf}
\Itf_{jk} &\coloneqq& \sum^N_{A=1} m_{_{A}} \left[ (x_{_{A}})_j (x_{_{A}})_k - 
	\frac{1}{3}\delta_{jk} (x_{_{A}})_i (x_{_{A}})^i\right] 
\nonumber \\
	&=& 
	\int \rho \left(x_j x_k - \frac{1}{3} 
	\delta_{jk} x_i x^i\right) dV\,,
\end{eqnarray}
and the brackets $\langle ~ \rangle$ indicate a time average [Clearly,
the second expression in (\ref{itf}) refers to a continuous distribution
of particles with rest-mass density $\rho$.].

A crude estimate of the third time derivative of the mass quadrupole of
the system is given by
\begin{equation}
\label{rough}
{\dddot \Itf_{jk}} \sim \frac{(\rm mass\ of\ the\ system\ in\ motion)
	\times (size\ of\ the\ system)^2}{(\rm timescale)^3} 
	\sim \frac{M R^2}{\tau^3} \sim 
	\frac{M \langle v^2 \rangle}{\tau} \,,
\end{equation}
where $\langle v \rangle$ is the mean internal velocity. Stated
differently, 
\begin{equation}
{\dddot \Itf_{jk}} \sim L_{\rm int} \,,
\end{equation}
where $L_{\rm int}$ is the power of the system flowing from one part
of the system to the other. 

As a result, the gravitational-wave luminosity in the quadrupole
approximation can be calculated to be (we here restore the explicit use
of the gravitational constant and of the speed of light)
\begin{equation}
\label{crude}
L_{\rm mass-quad} \sim
	\left(\frac{G}{c^5}\right)
	\left(\frac{M \langle v^2 \rangle}{\tau} \right)^2\ \sim
	\left(\frac{G^4}{c^5}\right)
	\left( \frac{M}{R} \right)^5 \sim
	\left(\frac{c^5}{G}\right)
	\left(\frac{R_{_{\rm S}}}{R}\right)^2
	\left(\frac{\langle v^2\rangle}{c^2} \right)^3 
\,.	
\end{equation}
The second equality has been derived using the virial theorem for which
the kinetic energy is of the same order of the potential one, \ie $M
\langle v^2 \rangle \sim G M^2/R$, and assuming that the oscillation
timescale is inversely proportional to the mean stellar density, \ie
$\tau \sim (1/G\langle \rho \rangle)^{1/2} \sim
(R^3/GM)^{1/2}$. Similarly, the third equality expresses the luminosity
in terms of dimensionless quantities such as the size of the source
relative to the Schwarzschild radius $R_{_{\rm S}} = 2GM/c^2$, and the
source speed in units of the speed of light. Note that the quantity
${c^5}{G}$ has indeed the units of a luminosity, \ie ${\rm
  erg\ s^{-1}\ =\ cm^2\ g\ s^{-3}}$ in cgs units.

Although extremely simplified, expressions (\ref{rough}) and
(\ref{crude}) contain the two most important pieces of information about
the generation of gravitational waves. The first one is that the
conversion of any type of energy into gravitational waves is, in general,
not efficient. To see this it is necessary to bear in mind that
expression (\ref{mq}) is in geometrized units and that the conversion to
conventional units, say cgs units, requires dividing (\ref{mq}) by the
very large factor $c^5/G \simeq 3.63 \times 10^{59}\ {\rm
  erg\ s}^{-1}$. The second one is contained in the last expression in
Eq. (\ref{crude}) and that highlights how the gravitational-wave
luminosity can also be extremely large. There are in fact astrophysical
situations, such as those right before the merger of a binary system of
compact objects, in which $\sqrt{\langle v^2\rangle} \sim 0.1\,c$ and $R
\sim 10\,R_{_{\rm S}}$, so that $L_{\rm mass-quad} \sim 10^{51}\ {\rm
  erg\ s}^{-1} \sim 10^{18}\, L_{\odot}$, that is, $10^{18}$ times the
luminosity of the Sun; this is surely an impressive release of energy.

\subsubsection{Extensions of the quadrupole formula}
\label{sec:iqf}

Although valid only in the low-velocity, weak-field limit, the
quadrupole-formula approximation has been used extensively in the
past and still finds use in several simulations, ranging from
stellar-core collapse (see, \eg \cite{Zwerger97} for some initial
application) to binary neutron-star mergers (see, \eg \cite{Oechslin02}
for some initial application). In many of the simulations carried out to
study stellar collapse, one makes the additional assumption that the
system remains axisymmetric and the presence of an azimuthal Killing
vector has two important consequences. Firstly, the gravitational waves
produced in this case will carry away energy but not angular momentum,
which is a conserved quantity in this spacetime. Secondly, the
gravitational waves produced will have a single polarization state, so
that the transverse traceless gravitational field is completely
determined in terms of its only nonzero transverse and traceless (TT)
independent component. Following \cite{Zwerger97} and considering for
simplicity an axisymmetric system, it is useful to express the
gravitational strain $h^{^{\rm TT}}(t)$ observed at a distance $R$ from
the source in terms of the quadrupole wave amplitude $A_{20}$
\cite{Zanotti03}
\begin{equation}
\label{htt}
h^{^{\rm TT}}(t) = F_+ \left(\frac{1}{8}
	  \sqrt{\frac{15}{\pi}}\right) \frac{A_{20}(t-R)}{R} \,,
\end{equation}
where $F_+=F_+(R,\theta,\phi)$ is the detector's beam pattern function
and depends on the orientation of the source with respect to the
observer. As customary in these calculations, we will assume it to be
optimal, \ie $F_+=1$. The $\ell=2, m=0$ wave amplitude $A_{20}$ in
Eq. (\ref{htt}) is simply the second time derivative of the reduced mass
quadrupole moment in axisymmetry and can effectively be calculated
without taking time derivatives numerically, which are instead replaced
by spatial derivatives of evolved quantities after exploiting the
continuity and the Euler equations
\cite{Finn_L:1990,Blanchet90,Rezzolla:1999sz}.  The result in a spherical
coordinate system is
\begin{eqnarray}
\label{stress}
&&A_{20} \coloneqq \frac{d^2 I_{[ax]}}{dt^2} =
	k \!\int\!\! \rho \biggl[v_r v^r (3 z^2 \!-\! 1) \!+\!
	v_\theta v^\theta (2 \!-\! 3 z^2) \!-\!
	v_\phi v^\phi  \nonumber \\
	& & \qquad \qquad 
        \qquad -6 z \sqrt{(v^r v_r)(v_\theta v^\theta)
	(1 \!-\! z^2)} \biggl. - r \frac{\partial \Phi}{\partial r}
	(3 z^2 \!-\! 1) + 
	\, 3 z\frac{\partial \Phi}{\partial \theta}
	\sqrt{1 \!-\! z^2}\biggr] r^2 dr dz \,,
\end{eqnarray}
where $z\coloneqq \cos\theta$, $k = 16 \pi^{3/2} / \sqrt{15} $, $\Phi$ is
the Newtonian gravitational potential, and $I_{[ax]}$ is the
appropriate component  of the Newtonian reduced mass-quadrupole moment
in axisymmetry
\begin{equation}
\label{mass_quad}
I_{[ax]}\coloneqq \int \rho \left(\frac{3}{2}z^2 - \frac{1}{2}\right)r^4 dr dz \,.
\end{equation}

Of course it is possible to consider more generic conditions and derive
expressions for the strain coming from more realistic sources, such as a
an astrophysical system with equatorial symmetry. In this case, focussing
on the lowest $\ell=2$ moments, the relevant multipolar components for
the strain are \cite{Baiotti:2008nf}
\begin{align}
\label{eq:multipoles_20}
h^{20} &=\dfrac{1}{r}\sqrt{\dfrac{24\pi}{5}}\left(\ddot{\mathcal{I}}_{zz}
- \dfrac{1}{3}{\rm Tr}{(\ddot{\mathcal{I}})}\right)\,, \\
\label{eq:multipoles_21}
h^{21}&=-\dfrac{\i}{r}\sqrt{\dfrac{128\pi}{45}}\left(\ddot{\mathcal{J}}_{xz} 
- i \ddot{\mathcal{J}}_{yz}\right)\,,\\
\label{eq:multipoles_22}
h^{22} &=\dfrac{1}{r}\sqrt{\dfrac{4\pi}{5}}\left(\ddot{\mathcal{I}}_{xx}
- 2 i\ddot{\mathcal{I}}_{xy}-\ddot{\mathcal{I}}_{yy}\right)\,,
\end{align}
where 
\begin{align}
\label{gen_mass_quad}
\mathcal{I}_{ij} &= \int d^3 x \,\rho\, x_i x_j\,, \\
\label{gen_curr_quad}
\mathcal{J}_{ij} &= \int
d^3x \,\rho\,\epsilon_{abi}\,x_j x_a\, v^b\,,
\end{align}
are the more general Newtonian mass and mass-current quadrupoles.

Expressions \eqref{stress}--\eqref{gen_curr_quad} are strictly
Newtonian. Yet, these expression are often implemented in numerical codes
that are either fully general relativistic or exploit some level of
general-relativistic approximation. More seriously, these expressions
completely ignore considerations that emerge in a relativistic context,
such as the significance of the coordinate chosen for their calculation.
As a way to resolve these inconsistencies, improvements to these
expressions have been made to increase the accuracy of the computed
gravitational-wave emission. For instance, for calculations on known
spacetime metrics, the gravitational potential in expression
\eqref{stress} is often approximated with expressions derived from the
metric, \eg as $\Phi = (1 - g_{rr})/2$ \cite{Zanotti03}, which is correct
to the first Post-Newtonian (PN) order. Improvements to the mass
quadrupole \eqref{mass_quad} inspired by a similar spirit have been
computed in Ref. \cite{Blanchet90}, and further refined and tested in
Refs. \cite{Shibata04,Nagar:2005fz,Cerda05,Pazos2006,Baiotti06b,Dimmelmeier07,Corvino:2010}.

A systematic comparison among the different expressions of the quadrupole
formulas developed over the years was carried out in
\cite{Baiotti:2008nf}, where a generalization of the mass-quadrupole
formula \eqref{gen_mass_quad} was introduced. In essence, following
previous work in \cite{Nagar:2005fz}, Ref. \cite{Baiotti:2008nf}
introduced a ``generalized'' mass-quadrupole moment of the form
\begin{equation}
    \label{eq:quadrupole}
    \mathcal{I}_{ij}[\varrho] \coloneqq \int d^3 x \varrho x_i x_j\,,
\end{equation}
where the generalized rest-mass density $\varrho$ was assumed to take a
number of possible expressions, namely,
\begin{align}
\label{gqf_1}
& \varrho\coloneqq \rho\,,\\
\label{gqf_2}
& \varrho\coloneqq \alpha^2\sqrt{\gamma}T^{00}\,,\\
\label{gqf_3}
& \varrho\coloneqq \sqrt{\gamma}W\rho\,,\\
\label{gqf_4}
& \varrho\coloneqq u^0\rho=\frac{W}{\alpha}\rho\,,
\end{align}
Clearly, the first option corresponds to the ``standard'' quadrupole
formula, but, as remarked in Ref. \cite{Baiotti:2008nf} none of the
alternative quadrupole formulas obtained using these generalized
quadrupole moments should be considered better than the others, at least
mathematically. None of them is gauge invariant and indeed they yield
different results depending on the underlining choice made for the
coordinates. Yet, the comparison is meaningful in that these expressions
were and still are in use in many numerical codes, and it is therefore
useful to determine which expression is effectively closer to the fully
general-relativistic one.

Making use of a fully general-relativistic measurement of the
gravitational-wave emission from a neutron star oscillating nonradially
as a result of an initial pressure perturbation,
Ref. \cite{Baiotti:2008nf} concluded that the various quadrupole formulas
are comparable and give a very good approximation to the phasing of the
gravitational-wave signals. At the same time, they also suffer from
systematic over-estimate [expression \eqref{gqf_2}] or under-estimates of
the gravitational-wave amplitude [expressions \eqref{gqf_1}, and
  \eqref{gqf_3}--\eqref{gqf_4}]. In all cases, however, the relative
difference in amplitude was of $50\%$ at most, which is probably
acceptable given that these formulas are usually employed in complex
astrophysical calculations in which the systematic errors coming from the
microphysical modelling are often much larger.

\newpage
\section{Basic Numerical Approaches}
\label{s-3+1}
\subsection{The 3+1 decomposition of spacetime}
\label{3p1splitting}

At the heart of Einstein's theory of general relativity is the
equivalence among all coordinates, so that the distinction of spatial and
time coordinates is more an organisational matter than a requirement of
the theory. Despite this ``covariant view'', however, our experience, and
the laws of physics on sufficiently large scales, do suggest that a
distinction of the time coordinate from the spatial ones is the most
natural one in describing physical processes. Furthermore, while not
strictly necessary, such a distinction of time and space is the simplest
way to exploit a large literature on the numerical solution of hyperbolic
partial differential equations as those of relativistic hydrodynamics.
In a generic spacetime, analytic solutions to the Einstein equations are not
known, and a numerical approach is often the only way to obtain an estimate
of the solution.

Following this principle, a decomposition of spacetime into ``space'' and
``time'' was already proposed in the 1960s within a Hamiltonian
formulation of general relativity and later as an aid to the numerical
solution of the Einstein equations in vacuum. The basic idea is rather
simple and consists in ``foliating'' spacetime in terms of a set of
non-intersecting spacelike hypersurfaces $\Sigma \coloneqq \Sigma(t)$,
each of which is parameterised by a constant value of the coordinate
$t$. In this way, the three spatial coordinates are split from the one
temporal coordinate and the resulting construction is called the
\emph{3+1 decomposition} of spacetime \cite{MTW1973}.

Given one such constant-time hypersurface, $\Sigma_t$, belonging to the
\emph{foliation} $\Sigma$, we can introduce a timelike four-vector
$\bs{n}$ normal to the hypersurface at each event in the spacetime and
such that its dual one-form $\bs{\Omega} \coloneqq \bs{\nabla} t$ is
parallel to the gradient of the coordinate $t$, \ie
\begin{equation}
\label{eq:n_mu}
n_\mu=A \Omega_{\mu} = A \nabla_\mu t  \,,
\end{equation}
with $n_{\mu} = \{A,0,0,0 \}$ and $A$ a constant to be determined. If we
now require that the four-vector $\bs{n}$ defines an observer and thus
that it measures the corresponding four-velocity, then from the
normalisation condition on timelike four-vectors, $n^\mu n_\mu=-1$, we
find that
\begin{equation}
\label{eq:nmu_nnu}
n^\mu n_\mu=g^{\mu\nu}n_\mu n_\nu = 
g^{tt}A^2=-\frac{1}{\alpha^2}A^2=-1 \,,
\end{equation}
where we have defined $\alpha^2 \coloneqq - 1/g^{tt}$. From the last
equality in expression \eqref{eq:nmu_nnu} it follows that $A=\pm\alpha$
and we will select $A=-\alpha$, such that the associated vector field
$n^\mu$ is future directed. The quantity $\alpha$ is commonly referred to
as the \emph{lapse} function, it measures the rate of change of the
coordinate time along the vector $n^\mu$ (see Fig.~\ref{fig:3p1}), and
will be a building block of the metric in a 3+1 decomposition [\cf
  Eq. \eqref{eq:ds2_3p1}].

The specification of the normal vector $\bs{n}$ allows us to define the
metric associated to each hypersurface, \ie
\begin{align}
&\gamma_{\mu \nu} \coloneqq g_{\mu \nu} + n_{\mu} n_{\nu}\,,
&\gamma^{\mu \nu} \coloneqq g^{\mu \nu} + n^{\mu} n^{\nu}\,,
\end{align}
where $\gamma^{0\mu}=0$, $\gamma_{ij} = g_{ij}$, but in general
$\gamma^{ij} \neq g^{ij}$. Also note that $\gamma^{ik} \gamma_{kj}=
\delta^i_j$, that is, $\gamma^{ij}$ and $\gamma_{ij}$ are the inverse of
each other,
and so can be used for
raising and lowering the indices of purely spatial vectors and tensors
(that is, defined on the hypersurface $\Sigma_t$).

The tensors $\bs{n}$ and $\bs{\gamma}$ provide us with two useful tools
to decompose any four-dimensional tensor into a purely spatial part
(hence contained in the hypersurface $\Sigma_t$) and a purely timelike
part (hence orthogonal to $\Sigma_t$ and aligned with $\bs{n}$). Not
surprisingly, the spatial part is readily obtained after contracting with
the \emph{spatial projection operator} (or \emph{spatial projection
  tensor})
\begin{equation}
\label{eq:space_proj}
\gamma^{\mu}_{\ \,\nu} \coloneqq g^{\mu \alpha} \gamma_{\alpha \nu} =
g^{\mu}_{\ \,\nu} + n^{\mu} n_{\nu} = \delta^{\mu}_{\ \,\nu} + n^{\mu} n_{\nu}\,,
\end{equation}
while the timelike part is obtained after contracting with the \emph{time
  projection operator} (or \emph{time projection tensor})
\begin{equation}
\label{eq:time_proj}
N^{\mu}_{\ \,\nu} \coloneqq - n^{\mu} n_{\nu}\,,
\end{equation}
and where the two projectors are obviously orthogonal, \ie
\begin{equation}
\gamma^{\alpha}_{\ \,\mu} N^{\mu}_{\ \nu} = 0\,.
\end{equation}

We can now  introduce a new vector, $\bs{t}$, along which to carry
out the time evolutions and that is dual to the surface one-form
$\bs{\Omega}$. Such a vector is just the time-coordinate basis vector and
is defined as the linear superposition of a purely temporal part
(parallel to $\bs{n}$) and of a purely spatial one (orthogonal to
$\bs{n}$), namely
\begin{equation}
\label{eq:tpo_7}
\bs{t} = \bs{e}_t = \partial_t \coloneqq \alpha \bs{n} + \bs{\beta}\,.
\end{equation}
The purely spatial vector $\bs{\beta}$ [\ie $\beta^{\mu} = (0,\beta^i)$]
is usually referred to as the \emph{shift vector} and will be another
building block of the metric in a 3+1 decomposition [\cf
  Eq. \eqref{eq:ds2_3p1}]. The decomposition of the vector $\bs{t}$ into a
timelike component $n\bs{\alpha}$ and a spatial component $\bs{\beta}$ is
shown in Fig.~\ref{fig:3p1}. 

\epubtkImage{}{%
\begin{figure}
\begin{center}
\includegraphics[angle=0,width=11.0cm]{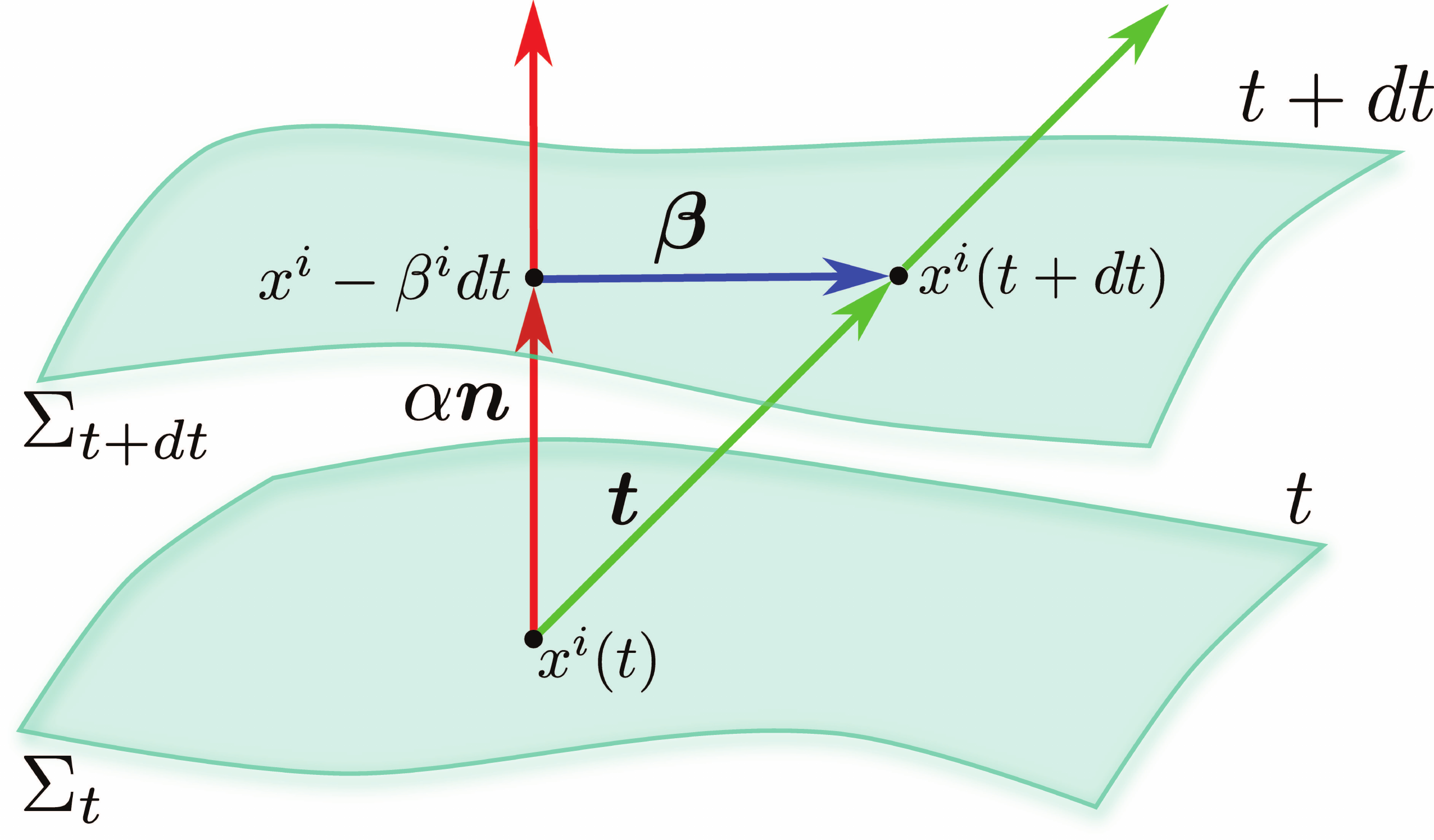}
\caption{Schematic representation of the 3+1 decomposition of spacetime
  with hypersurfaces of constant time coordinate $\Sigma_t$ and
  $\Sigma_{t+dt}$ foliating the spacetime. The four-vector $\bs{t}$
  represents the direction of evolution of the time coordinate $t$ and
  can be split into a timelike component $\alpha\bs{n}$, where $\bs{n}$
  is a timelike unit normal to the hypersurface, and into a spacelike
  component, represented by the spacelike four-vector $\bs{\beta}$. The
  function $\alpha$ is the ``lapse'' and measures the proper time between
  adjacent hypersurfaces, while the components of the ``shift'' vector
  $\beta^{i}$ measure the change of coordinates from one hypersurface to
  the subsequent one.}
\label{fig:3p1}
\end{center}
\end{figure}}

Because $\bs{t}$ is a coordinate basis vector, the integral curves of
$t^{\mu}$ are naturally parameterised by the time coordinate. As a
result, all infinitesimal vectors $t^{\mu}$ originating
at a given point $x_0^i$
on one
hypersurface $\Sigma_t$ would end up on the hypersurface
$\Sigma_{t+dt}$ at a point whose coordinates are also $x_0^i$. 
This condition is not guaranteed for translations along
$\Omega_{\mu}$ unless $\beta^\mu=0$ since $t^{\mu}t_{\mu} = g_{tt}= -\alpha^2 +
\beta^{\mu}\beta_{\mu}$, and as illustrated in Fig.~\ref{fig:3p1}.

In summary, the components of $\bs{n}$ are given by
\begin{align}
\label{eq:eulerian_observer}
&  n_\mu= \left(-\alpha,0,0,0 \right)\,, 
&& n^\mu=\frac{1}{\alpha}\left(1,-\beta^i \right)\,,
\end{align}
and we are now ready to deduce that the lapse function and the shift
vector can be employed to express the generic \emph{line element} in a
3+1 decomposition as
\begin{equation}
\label{eq:ds2_3p1}
ds^{2} = -(\alpha^{2}-\beta_{i}\beta^{i}) dt^{2}+ 
2 \beta_{i} dx^{i} dt + \gamma_{ij} dx^{i}dx^{j} \,.
\end{equation}
Expression \eqref{eq:ds2_3p1} clearly emphasises that when
$\beta^i=0=dx^i$, the lapse measures the proper time, $d\tau^2 = -ds^2$,
between two adjacent hypersurfaces, \ie
\begin{equation}
\label{eq:lapse_pt}
d\tau^{2} = \alpha^{2}(t,x^j) dt^{2}  \,,
\end{equation}
while the shift vector measures the change of coordinates of a point that
is moved along $\bs{n}$ from the hypersurface $\Sigma_t$ to the
hypersurface $\Sigma_{t+dt}$, \ie
\begin{equation}
x^i_{t+dt}=x^i_{t}-\beta^i (t,x^j) dt \,.
\end{equation}
Similarly, the covariant and contravariant components of the
metric \eqref{eq:ds2_3p1} can be written explicitly as
\begin{align}
&g_{\mu\nu} = \left(
\begin{array}{cc}
-\alpha^2 + \beta_i \beta^i ~~&~~ \beta_i   \\
~~&~~\\
 \beta_i            ~~&~~ \gamma_{ij} \\
\end{array} 
\right) \,,
&&g^{\mu\nu} = 
\left(
\begin{array}{cc}
-1/\alpha^2      ~~&~~ \beta^i/\alpha^2   \\
~~&~~\\
 \beta^i/\alpha^2   ~~&~~ \gamma^{ij} - \beta^i \beta^j/ \alpha^2 \\
\end{array} \right) \,,
\label{eq:ADM}
\end{align}
from which it is easy to obtain an important identity which will be used
extensively hereafter, \ie
\begin{equation}
\label{eq:sqrtg}
\sqrt{-g} = \alpha \sqrt{\gamma}\,,
\end{equation}
where $g \coloneqq \det(g_{\mu \nu})$ and $\gamma \coloneqq \det(\gamma_{ij})$.

When defining the unit timelike normal $\bs{n}$ in
Eq. \eqref{eq:nmu_nnu}, we have mentioned that it can be associated to
the four-velocity of a special class of observers, which are referred to
as \emph{normal} or \emph{Eulerian observers}. Although this
denomination is somewhat confusing, since such observers are not at rest
with respect to infinity but have a coordinate velocity \hbox{$dx^i/dt =
  n^i =-\beta^i/\alpha$}, we will adopt this traditional nomenclature
also in the following and thus take an ``Eulerian observer'' as one with
four-velocity given by (\ref{eq:eulerian_observer}).

When considering a fluid with four-velocity $\bs{u}$, the spatial
four-velocity $\bs{v}$ measured by an Eulerian observer will be given by
the ratio between the projection of $\bs{u}$ in the space orthogonal to
$\bs{n}$, \ie $\gamma^i_{\ \,\mu} u^{\mu} = u^i$, and the \emph{Lorentz
  factor} $W$ of $\bs{u}$ as measured by $\bs{n}$ \cite{deFelice90a}
\begin{equation}
-n_{\mu} u^{\mu} = \alpha u^t = W\,.
\end{equation}
As a result, the spatial four-velocity of a fluid as measured by an
Eulerian observer will be given by
\begin{align}
\label{projection_of_u_1}
\bs{v} \coloneqq 
\frac{\bs{\gamma} \cdot \bs{u}}{-\bs{n}\cdot\bs{u}}\,.
\end{align}
Using now the normalisation condition $u^{\mu}u_{\mu}=-1$,
we obtain
\begin{align}
\label{eq:LorFact}
& \alpha u^t = -\bs{n}\cdot\bs{u} =
\frac{1}{\sqrt{1 - v^iv_i}} = W\,,
&& u_t = W(-\alpha+\beta_i v^i) \,, 
\end{align}
so that the components of $\bs{v}$ can be written as
\begin{align}
\label{projection_of_u_4}
& v^i = \frac{u^i}{W} + \frac{\beta^i}{\alpha} = 
\frac{1}{\alpha}\left(\frac{u^i}{u^t} + \beta^i\right)\,,
&&v_i = \frac{u_i}{W} = \frac{u_i}{\alpha u^t} \,,
\end{align}
where in the last equality we have exploited the fact that
$\gamma_{ij}u^j = u_i - \beta_i W/\alpha$.

\subsection{The ADM formalism: 3+1 decomposition of the Einstein equations}
\label{s:ADM}

The 3+1 decomposition introduced in Section~\ref{3p1splitting} can be
used not only to decompose tensors, but also equations and, in
particular, the Einstein equations, which are then cast into an
initial-value form suitable to be solved numerically. A 3+1 decomposition
of the Einstein equations was presented by Arnowitt, Deser and
Misner \cite{Arnowitt62}, but it is really the reformulation suggested by
York \cite{York79} that represents what is now widely known as
the \emph{ADM formulation} [see, \eg \cite{Alcubierre:2008}
and \cite{Gourgoulhon2012} for a detailed and historical discussion]. As
we will see in detail later on, in this formulation the Einstein
equations are written in terms of purely spatial tensors that can be
integrated forward in time once some constraints are satisfied initially.

Here, we only outline the ADM formalism, and refer to the literature for
the derivation and justification. Further, it is important to note that
the ADM formulation is, nowadays, not used in practice because it is only
weakly hyperbolic. However, the variables used in the
ADM method, in particular the three-metric and the extrinsic
curvature, are what will be needed later for
gravitational-wave extraction, and are easily obtained from the output of
other evolution methods [see discussion in Sections \ref{ae}
and \ref{s-CE}].

Instead of the ADM formalism, modern simulations mainly formulate the
Einstein equations using: the BSSNOK method \cite{Nakamura87, Shibata95,
  Baumgarte99}; the CCZ4 formulation \cite{Alic:2011a}, which was
developed from the Z4 method \cite{Bona:2003fj, Bona:2003qn,
  Bona-and-Palenzuela-Luque-2005:numrel-book} (see also
\cite{Bernuzzi:2009ex} for the so-called Z4c formulation and
\cite{Alic2013} for some comparisons); or the generalized harmonic
method \cite{Pretorius:2004jg} (see also \cite{Baumgarte2010b,
  Rezzolla_book:2013} for more details).

We start by noting that once a 3+1
decomposition is introduced as discussed in Section~\ref{3p1splitting},
it is then possible to define the \emph{three-dimensional covariant
derivative} $D_i$. Formally, this is done by
projecting the standard covariant derivative onto the
space orthogonal to $n^\mu$, and the result is a covariant derivative
defined with respect to the connection coefficients
\begin{equation}
\label{eq:tpo_3}
\tu\Gamma _{j k }^{i }=\frac{1}{2}
\gamma^{i \ell }\left(
\partial_{j} \gamma_{k \ell}+ \partial_{k} \gamma_{\ell j}-
\partial_{\ell} \gamma_{j k}\right) \,,
\end{equation}
where we will use the upper left index $\tu$ to mark a purely spatial
quantity that needs to be distinguished from its spacetime
counterpart.\footnote{An alternative notation is to mark with an upper left
  index $^{(4)}$ the four-dimensional tensors and to leave unmarked the
  three-dimensional ones \cite{Baumgarte2010,Gourgoulhon2012}.}
Similarly, the three-dimensional Riemann tensor $\tu
R^{i}_{\ \,j k\ell}$ associated with $\bs{\gamma}$ has an explicit expression given by
\begin{equation}
\label{riemann_tensor_3D}
\tu R^i_{\ \,jk\ell} = 
\partial_{k}\!\! \tu \Gamma^i_{j\ell}-
\partial_{\ell}\!\! \tu \Gamma^i_{jk}+
\tu \Gamma^i_{m k} \tu \Gamma^m_{j\ell}-
\tu \Gamma^i_{m\ell} \tu \Gamma^m_{jk}\,.
\end{equation}
In a similar manner, the three-dimensional contractions of the
three-dimensional Riemann tensor, \ie the three-dimensional Ricci tensor
and the three-dimensional Ricci scalar, are defined respectively as their
four-dimensional counterparts, \ie
\begin{align}
\label{Ricci_tensor_3D}
& \tu R_{ij} \coloneqq \tu R^k_{\ \,ik j} \,,
&&\tu R \coloneqq \tu R^k_{\ \,k}\,.
\end{align}

The information present in $R^\mu_{\ \,\nu\kappa\sigma}$ and missing in
$\tu R^i_{\ \,jk\ell}$ can be found in another symmetric
tensor, the \emph{extrinsic curvature} $K_{ij}$, which is purely
spatial. Loosely speaking, the extrinsic
curvature provides a measure of how the three-dimensional hypersurface
$\Sigma_t$ is curved with respect to the four-dimensional spacetime.
For our purposes, it is convenient to define the extrinsic curvature
as (but note that other definitions, which can be shown to be equivalent,
are common)
\begin{equation}
\label{eq:K_ij}
K_{ij}=-\frac{1}{2}\mathscr{L}_{\bs{n}}\gamma_{ij}\,,
\end{equation}
where $\mathscr{L}_{\bs{n}}$ is the Lie derivative relative to the normal
vector field $\bs{n}$. Expression \eqref{eq:K_ij} provides a simple
interpretation of the extrinsic curvature $K_{ij}$ as the rate of change
of the three-metric $\gamma_{ij}$ as measured by an Eulerian
observer. Using properties of the Lie derivative, it follows that
\begin{equation}
\partial_t \gamma_{ij} =
-2\alpha K_{ij} + D_i\beta_j + D_j\beta_i \,.
\label{eq:ADM_evol1b}
\end{equation}
Note that Eq. \eqref{eq:ADM_evol1b} is a
geometrical result and is independent of the Einstein equations.

The next step is to note further purely geometric relations,
how the spacetime curvature is related to the intrinsic and extrinsic
curvatures of the hypersurface $\Sigma_t$. These formulas are known as the
Gauss--Codazzi equations and the Codazzi--Mainardi equations. They are
\begin{align}
\label{eq:GC}
\gamma^{\mu}_{\ \,i}\,
\gamma^{\nu}_{\ \,j}\,
\gamma^{\rho}_{\ \,k}\,
\gamma^{\sigma}_{\ \,\ell}\, 
R_{\mu\nu\rho\sigma} &= 
\tu R_{ijk\ell} + 
K_{ik}K_{j\ell} -
K_{i\ell}K_{jk}\,, \\
\label{eq:CM}
\gamma^{\rho}_{\ \,j}
\gamma^{\mu}_{\ \,i}
\gamma^{\nu}_{\ \,\ell}
n^{\sigma} R_{\rho\mu\nu\sigma} &= D_{i} K_{j\ell} - D_{j}
K_{i\ell}\,,\\
\label{eq:2sp_2t}
\gamma^{\alpha}_{\ \,i}
\gamma^{\beta}_{\ \,j}
n^{\delta} n^{\lambda}
R_{\alpha\delta\beta\lambda} &= 
\mathscr{L}_{\bs{n}} K_{ij} 
- \frac{1}{\alpha}D_{i} D_{j} \alpha  
+ K^{k}_{\ \,j} \,K_{ik}\,.
\end{align}
We now have enough identities to rewrite the Einstein equations in a 3+1
decomposition. After contraction, we can use the Einstein equations to
replace the spacetime Ricci tensor with terms involving the stress-energy
tensor, and then after further manipulation the final result is:
\begin{equation}
\begin{split}
\label{eq:ADM_evol2}
\partial_t K_{ij} = &-D_i D_j \alpha +
\beta^k\partial_k K_{ij} + K_{ik}\partial_j \beta^k +
K_{kj}\partial_i\beta^k  
\\
&
+ \alpha\left(\tu R_{ij}+K K_{ij} - 2 K_{ik}K^{k}_{\ \,j}\right)
+ 4\pi\alpha\left[\gamma_{ij}\left(S-E)-2S_{ij}\right)\right]
\,, 
\end{split}
\end{equation}
\begin{equation}
\label{einstein_constraints_1}
\tu R + K^2 - K_{ij}K^{ij}=16\pi E \,,
\end{equation}
\begin{equation}
\label{einstein_constraints_2}
D_j(K^{ij}-\gamma^{ij}K) = 8\pi S^i \,.
\end{equation}
The following definitions have been made for the \emph{``matter''}
quantities
\begin{equation}
\label{eq:mat_1}
S_{\mu\nu} \coloneqq  
\gamma^{\alpha}_{\ \,\mu} \, \gamma^{\beta}_{\ \,\n} T_{\alpha\beta}\,, \qquad
S_{\mu} \coloneqq -\gamma^{\alpha}_{\ \,\mu}\, n^\beta T_{\alpha\beta}\,,\qquad
S \coloneqq  S^\mu_{\ \,\mu} \,, \;\;
E \coloneqq n^\alpha \, n^\beta T_{\alpha\beta} \,,
\end{equation}
that is, for contractions of the energy--momentum tensor that would
obviously be zero in vacuum spacetimes.

Overall, the six equations \eqref{eq:ADM_evol2}, together with the six
equations \eqref{eq:ADM_evol1b} represent the time-evolving part of the
ADM equations and prescribe how the three-metric and the extrinsic
curvature change from one hypersurface to the following one. In contrast,
Eqs. \eqref{einstein_constraints_1} and \eqref{einstein_constraints_2}
are constraints that need to be satisfied on each hypersurface. This
distinction into evolution equations and constraint equations is not
unique to the ADM formulation and is indeed present also in classical
electromagnetism.  Just as in electrodynamics the divergence of the
magnetic field remains zero if the field is divergence-free at the
initial time, so the constraint equations \eqref{einstein_constraints_1}
and \eqref{einstein_constraints_2}, by virtue of the Bianchi identities
\cite{Alcubierre:2008, Bona2009, Baumgarte2010a, Gourgoulhon2012,
  Rezzolla_book:2013}, will remain satisfied during the evolution if they
are satisfied initially \cite{Frittelli97a}. Of course, this concept is
strictly true in the continuum limit, while numerically the situation is
rather different. However, that issue is not pursued here.

Two remarks should be made before concluding this section. The first one
is about the gauge quantities, namely, the lapse function $\alpha$ and
the shift vector $\beta^i$. Since they represent the four degrees of
freedom of general relativity, they are not specified by the equations
discussed above and indeed they can be prescribed arbitrarily, although
in practice great care must be taken in deciding which prescription is
the most useful. The second comment is about the mathematical properties
of the time-evolution ADM equations \eqref{eq:ADM_evol2} and
\eqref{eq:ADM_evol1b}. The analysis of these properties can be found, for
instance, in \cite{Reula98a} or in \cite{Frittelli:2000uj}, and reveals
that such a system is only weakly hyperbolic with zero eigenvalues and,
as such, not necessarily well-posed. The weak-hyperbolicity of the ADM
equations explains why, while an historical cornerstone in the 3+1
formulation of the Einstein equations, they are rarely used in practice
and have met only limited successes in multidimensional calculations
\cite{Cook97a_shortlist,Abrahams97a_shortlist}. At the same time, the
weak hyperbolicity of the ADM equations and the difficulty in obtaining
stable evolutions, has motivated, and still motivates, the search for
alternative formulations.

\subsection{Gravitational waves from $\psi_4$ on a finite worldtube(s)}
\label{s-NP}

The Newman--Penrose scalars are scalar quantities defined as
contractions between the Weyl, or conformal, tensor
\begin{equation}
C_{\alpha\beta\mu\nu}=R_{\alpha\beta\mu\nu}-g_{\alpha[\mu}R_{\nu]\beta}+g_{\beta[\mu}R_{\nu]\alpha}
+\frac{g_{\alpha[\mu}g_{\nu]\beta}R}{3}\,,
\end{equation}
(in four dimensions), and an orthonormal null tetrad
$\ell^\alpha,n_{_{[NP]}}^\alpha,m^\alpha,\bar{m}^\alpha$ \cite{Newman62a}.
The null tetrad is constructed from an orthonormal tetrad, and we use the
notation $n_{_{[NP]}}^\alpha$, rather than the usual $n^\alpha$, because
$n_{_{[NP]}}^\alpha$ is obtained in terms of the hypersurface normal
$n^\alpha$. Supposing that the spatial coordinates $(x,y,z)$ are
approximately Cartesian, then spherical polar coordinates are defined
using Eqs.~\eqref{e-C2sp} and \eqref{e-C2spi}. However, in general these
coordinates are not exactly spherical polar, and in particular the radial
coordinate is not a surface area coordinate (for which the 2-surface
$r=t=$ constant must have area $4\pi r^2$). We reserve the notation $r$
for a surface area radial coordinate, so the radial coordinate just
constructed will be denoted by $s$. Then the outward-pointing radial unit
normal $\bs{e}_{s}$ is
\begin{equation}
\left(\bs{e}_{s}\right)^i=\frac{\gamma^{ij} s_j}{\sqrt{\gamma^{ij} s_i s_j}}
\qquad\mbox{where} \quad s_j=\nabla_j \sqrt{x^2+y^2+z^2}.
\label{e-sxyz}
\end{equation}
An orthonormal basis $(\bs{e}_{s},\bs{e}_{\theta},\bs{e}_{\phi})$ of
$\Sigma_t$ is obtained by Gram-Schmidt orthogonalization, and is extended
to be an orthonormal tetrad of the spacetime by incuding the hypersurface
normal $\bs{n}$. Then the orthonormal null tetrad is
\begin{equation}
\bs{\ell}=\frac{1}{\sqrt{2}}\left(\bs{n}+\bs{e}_{s}\right)\,,\qquad
\bs{n}_{_{[NP]}}=\frac{1}{\sqrt{2}}\left(\bs{n}-\bs{e}_{s}\right)\,,\qquad
\bs{m}=\frac{1}{\sqrt{2}}\left(\bs{e}_{\theta}+i\bs{e}_{\phi}\right)\,.
\end{equation}
(The reader should be aware that some authors use different conventions,
\eg without a factor of $\sqrt{2}$, leading to different forms for
various equations). For example, in Minkowski spacetime there is no
distinction between $s$ and $r$, and in spherical polar coordinates
$(t,r,\theta,\phi)$
\begin{equation}
\ell^\alpha=\left(\frac{1}{\sqrt{2}},\frac{1}{\sqrt{2}},0,0\right)\,,\qquad
n_{_{[NP]}}^\alpha=\left(\frac{1}{\sqrt{2}},-\frac{1}{\sqrt{2}},0,0\right)\,,\qquad
m^\alpha=\left(0,0,\frac{1}{r\sqrt{2}},\frac{i}{r\sqrt{2}\sin\theta}\right)\,.
\label{e-nt}
\end{equation}
The null tetrad satisfies the orthonormality conditions
\begin{equation}
0=\ell^\alpha\ell_\alpha=n_{_{[NP]}}^\alpha n_{_{[NP]}\alpha}=m^\alpha m_\alpha=m^\alpha n_\alpha=m^\alpha \ell_\alpha,\;
\ell^\alpha n_{_{[NP]}\alpha} =-1,\; m^\alpha \bar{m}_\alpha=1.
\label{e-nt_ortho}
\end{equation}

\epubtkImage{}{%
\begin{figure}
\begin{center}
  \includegraphics[width=0.49\linewidth]{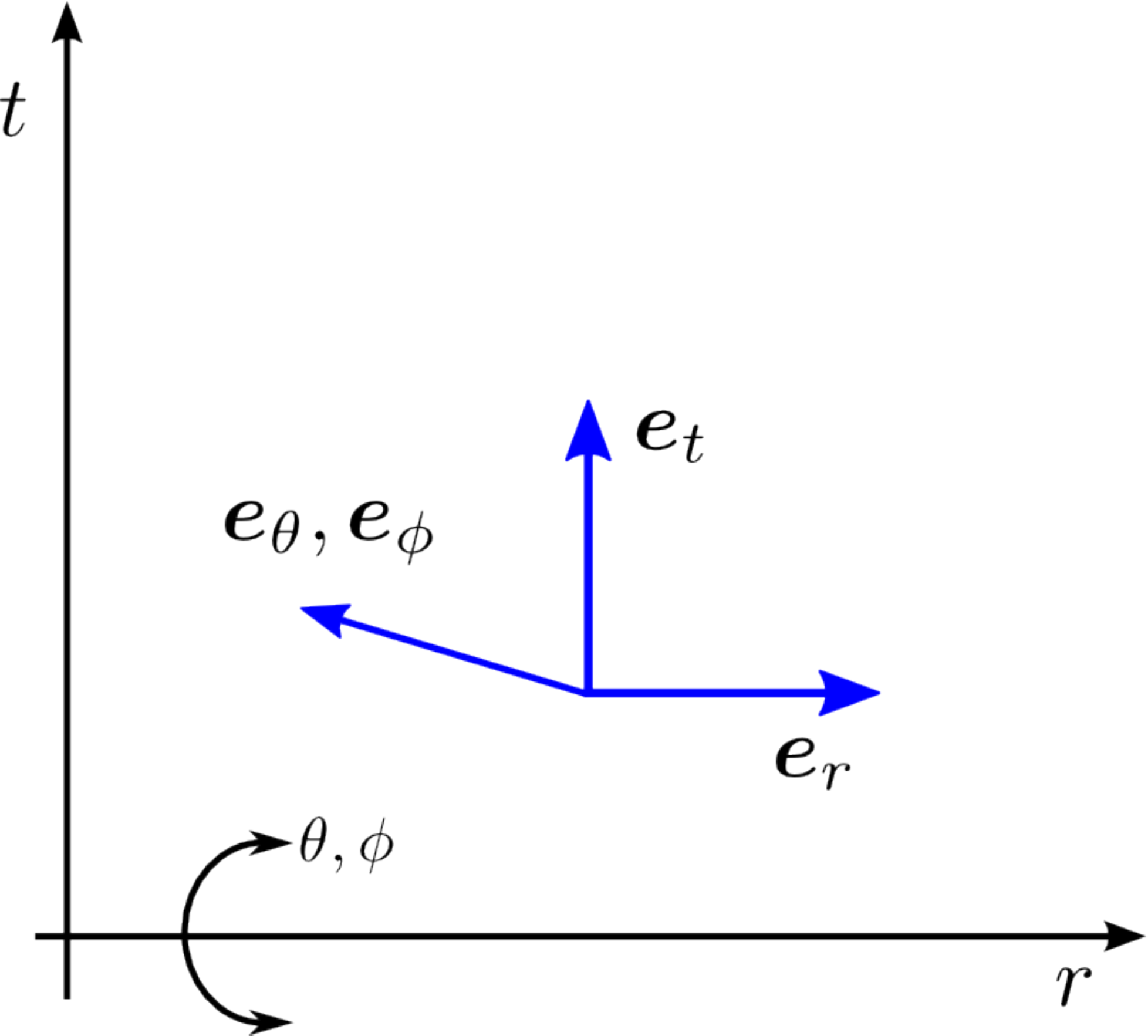}
  \includegraphics[width=0.49\linewidth]{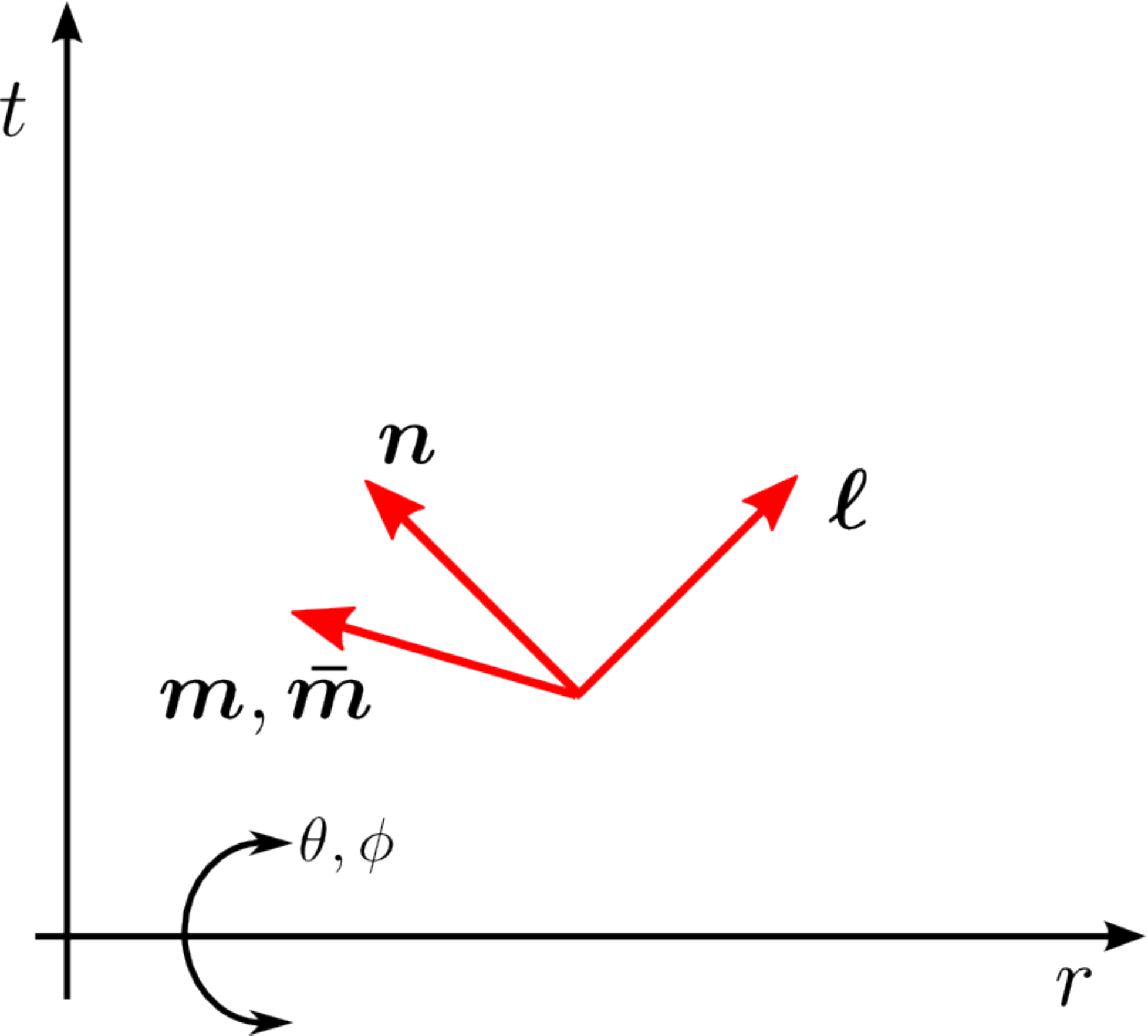}
\caption{Schematic representation of an orthonormal tetrad and a null
  tetrad in Minkowski spacetime in spherical polar $(t,r,\theta,\phi)$
  coordinates. The left panel shows the orthonormal tetrad
  $(\bs{e}_t,\bs{e}_r,\bs{e}_{\theta}, \bs{e}_{\phi})$, and the right
  panel illustrates the null tetrad $(\bs{\ell},\bs{n},
  \bs{m},\bar{\bs{m}})$. Both $(\bs{e}_{\theta},\bs{e}_{\phi})$ and
  $(\bs{m},\bar{\bs{m}})$ constitute a basis of the $(\theta,\phi)$
  subspace; and both $(\bs{e}_t,\bs{e}_r)$ and $(\bs{\ell},\bs{n})$
  constitute a basis of the $(t,r)$ subspace.}
\label{f-tetrad}
\end{center}
\end{figure}}

The Newman--Penrose, or Weyl, scalars \cite{Newman62a} are defined as
\begin{align}
\psi_0=&\,-C_{\alpha\beta\mu\nu}\ell^\alpha m^\beta\ell^\mu m^\nu \,, \\
\psi_1=&\,-C_{\alpha\beta\mu\nu}\ell^\alpha n_{_{[NP]}}^\beta\ell^\mu m^\nu \,, \\
\psi_2=&\,-C_{\alpha\beta\mu\nu}\ell^\alpha m^\beta\bar{m}^\mu n_{_{[NP]}}^\nu \,, \\
\psi_3=&\,-C_{\alpha\beta\mu\nu}\ell^\alpha n_{_{[NP]}}^\beta\bar{m}^\mu n_{_{[NP]}}^\nu \,, \\
\psi_4=&\,-C_{\alpha\beta\mu\nu}n_{_{[NP]}}^\alpha \bar{m}^\beta n_{_{[NP]}}^\mu \bar{m}^\nu.
\label{e-psi01224}
\end{align}
For our purposes, the most important of these quantities is $\psi_4$
since in the asymptotic limit it completely describes the outgoing
gravitational radiation field: far from a source, a gravitational wave is
locally plane and $\psi_4$ is directly related to the metric perturbation
in the TT gauge
\begin{equation}
\psi_4={\partial^2_t}\left(h_+ -ih_\times\right)\,.
\label{e-psi4h}
\end{equation}
In an asymptotically flat spacetime using appropriate coordinates (these issues are
discussed more formally in Sec.~\ref{s-charac}), the peeling theorem \cite{Penrose65,
Geroch77,Hinder:2011} shows that $\psi_4$ falls off as $r^{-1}$, and more generally
that $\psi_n$ falls off as $r^{n-5}$. Thus, gravitational waves are normally
described not by $\psi_4$ but by $r\psi_4$ which should be evaluated in the limit
as $r\rightarrow\infty$ (which in practice may mean evaluated at as large
a value of $r$ as is feasible). Often $\lim_{r\rightarrow\infty}r\psi_4$
is denoted by $\psi_4^0(t,\theta,\phi)$, but that notation will not be
used in this section. These issues are discussed further in
Sec.~\ref{s-charac}, but for now we will regard gravitational waves, and
specifically $r\psi_4$, as properly defined only in a spacetime whose
metric can be written in a form that tends to the Minkowski metric, and
for which the appropriate definition of the null tetrad is one that tends
to the form Eq.~\eqref{e-nt}, as $r\rightarrow \infty$.

Equation~\eqref{e-psi01224} for $\psi_4$ involves spacetime, rather than
hypersurface, quantities, and this is not convenient in a ``3+1''
simulation. However, the expression for $\psi_4$ can be manipulated into
a form involving hypersurface quantities only \cite{Gunnarsen95} (There
is also a derivation in the text-book \cite{Alcubierre:2008}, but note
the sign difference in the definition of $\psi_4$ used there):
\begin{equation}
\psi_4=(-R_{ij}-KK_{ij}+K_{ik}K^k_j+i\epsilon_i^{k\ell}\nabla_k K_{\ell j}) \bar{m}^i\bar{m}^j.
\label{e-psi4ADM}
\end{equation}
The proof is not given here, but in summary is based on using an
arbitrary timelike vector, in this case the hypersurface normal $\bs{n}$,
to decompose the Weyl tensor into its ``electric'' and ``magnetic''
parts.

The above procedures lead to an estimate $\psi_4$, but results are rarely
reported in this form. Instead, $\psi_4$ is decomposed into spin-weighted
spherical harmonics (see Appendix\ref{s-sYlm}),
\begin{equation}
\psi_4=\sum_{\ell\ge 2,|m|\le\ell}\psi_4^{\ell\,m}\,{}_{-2}Y^{\ell\,m}\,
\qquad \mbox{where}\qquad
\psi_4^{\ell\, m}=\int_{S^2} \psi_4\, \;{}_{-2}\bar{Y}^{\ell\,m}\, d\Omega\,,
\label{e-psi04lm}
\end{equation}
and the $r\psi_4^{\ell\, m}$ are evaluated and reported. Although,
normally, the dominant part of a gravitational-wave signal is in the
lowest modes with $\ell = 2$, the other modes are important to
gravitational-wave data analysis, recoil calculations, etc.

\subsubsection{Extracting gravitational waves using $\psi_4$ on a finite
worldtube}
\label{s-psi4WT}

``3+1'' numerical simulations are restricted to a finite domain, so it is
not normally possible to calculate exactly a quantity given by an
asymptotic formula (but see Secs.~\ref{s-charac} and \ref{s-CE}).
 A simple estimate of $r\psi_4$ can be
  obtained by constructing coordinates $(s,\theta,\phi)$ and an angular
  null tetrad vector $\bs{m}$ as discussed at the beginning of
  Sec.~\ref{s-NP}. Then $r\psi_4$ can be evaluated using
  Eq.~\eqref{e-psi4ADM} on a worldtube $s=$ constant, and the estimate is
  $r\psi_4=s\psi_4$ or alternatively $r\psi_4=\psi_4 \sqrt{A/4\pi}$ where
  $A$ is the area of the worldtube at time $t$. This approach was first
  used in \cite{Smarr77}, and subsequently in, for
  example, \cite{Pollney:2007ss,Pfeiffer:2007yz,Scheel:2008rj}. This
  method does not give a unique answer, and there are many variations in
  the details of its implementation. However, the various estimates
  obtained for $r\psi_4$ should differ by no more than ${\mathcal
    O}(r^{-1})$.

The quantity $\psi_4$ has no free indices and so tensorially is a scalar,
but its value does depend on the choice of tetrad. \textit{However}, it
may be shown that $\psi_4$ is first-order tetrad-invariant if the tetrad
is a small perturbation about a natural tetrad of the Kerr
spacetime. This result was shown by
Teukolsky \cite{Teukolsky72,Teukolsky73}; see also \cite{Chandrasekhar78,
  Campanelli98c}. Briefly, the reasoning is as follows. The Kinnersley
null tetrad is an exact null tetrad field in the Kerr
geometry \cite{Kinnersley:1969}. It has the required asymptotic limit,
and the vectors $\ell^\alpha$, $n_{_{[NP]}}^\alpha$ are generators of
outgoing and ingoing radial null geodesics respectively. In the Kerr
geometry $C^{\rm [Kerr]}_{\alpha\beta\mu\nu}\ne 0$, but using the
Kinnersley tetrad all $\psi_n$ are zero except $\psi_2$. Thus, to
first-order, $\psi_4$ is evaluated using the perturbed Weyl tensor and
the background tetrad; provided terms of the form $C^{\rm
  [Kerr]}_{\alpha\beta\mu\nu}n_{_{[NP]}}^\alpha \bar{m}^\beta
n_{_{[NP]}}^\mu \bar{m}^\nu$, where three of the tetrad vectors take
background values and only one is perturbed, are ignorable. Allowing for
those $\psi_n$ that are zero, and using the symmetry properties of the
Weyl tensor, all such terms vanish. This implies that the ambiguity in
the choice of tetrad is of limited importance because it is a
second-order effect; see also \cite{Campanelli98a,Campanelli98b}. These
ideas have been used to develop analytic methods for estimating
$\psi_4$ \cite{Campanelli98c,Baker00c,Baker00a,Baker:2001nu}. Further,
the Kinnersley tetrad is the staring point for a numerical extraction
procedure.

In practice the spacetime being
  evolved is not Kerr, but in many cases at least far from the source it
  should be Kerr plus a small perturbation, and in the far future it
  should tend to Kerr. Thus an idea for an appropriate tetrad for use on
  a finite worldtube is to construct an approximation to the Kinnersley
  form, now known as the quasi-Kinnersley null tetrad
   \cite{Beetle:2004wu,Nerozzi:2004wv}. The quasi-Kinnersley tetrad has
  the property that as the spacetime tends to Kerr, then the
  quasi-Kinnersley tetrad tends to the Kinnersley tetrad. The method was
  used in a number of applications in the mid
  2000s \cite{Nerozzi:2005hz,Campanelli:2005ia,Fiske05,Nerozzi:2007ai}.
  
Despite the mathematical attraction of the quasi-Kinnersley approach,
nowadays the extrapolation method which assumes the simpler Schwarzschild
background (see next Section) is preferred since, at a practical level,
and as discussed in Sec.~\ref{s-comp}, extrapolation can give highly
accurate results.  Modern simulations typically extract on a worldtube at
between $100M$ to $1000M$ where the correction due to the background
being Kerr rather than Schwarzschild is negligible. More precisely, an
invariant measure of curvature is the square root of the Kretschmann
scalar, which for the Kerr geometry \cite{Henry2000} takes the asymptotic
form
\begin{equation}
\sqrt{R^{\rm [Kerr]}_{\alpha\beta\mu\nu}\,R_{\rm
    [Kerr]}^{\alpha\beta\mu\nu}}=4\sqrt{3}\frac{M}{r^3}\left(1-\frac{21a^2\cos^2\theta}{2r^2}+{\mathcal
  O}\left(\frac{a^4}{r^4}\right)\right)\,,
\end{equation}
where $a\coloneqq J/M^2$. The curvature is already small in the
Schwarzschild ($a=0$) case, and the effect of ignoring $a$ is a small
relative error of order $a^2/r^2$.
  
\subsubsection{Extracting gravitational waves using $\psi_4$ in practice: The extrapolation method}
\label{s-psi4Extrap}

\epubtkImage{}{%
\begin{figure}[htb]
  \centering \includegraphics[width=9cm]{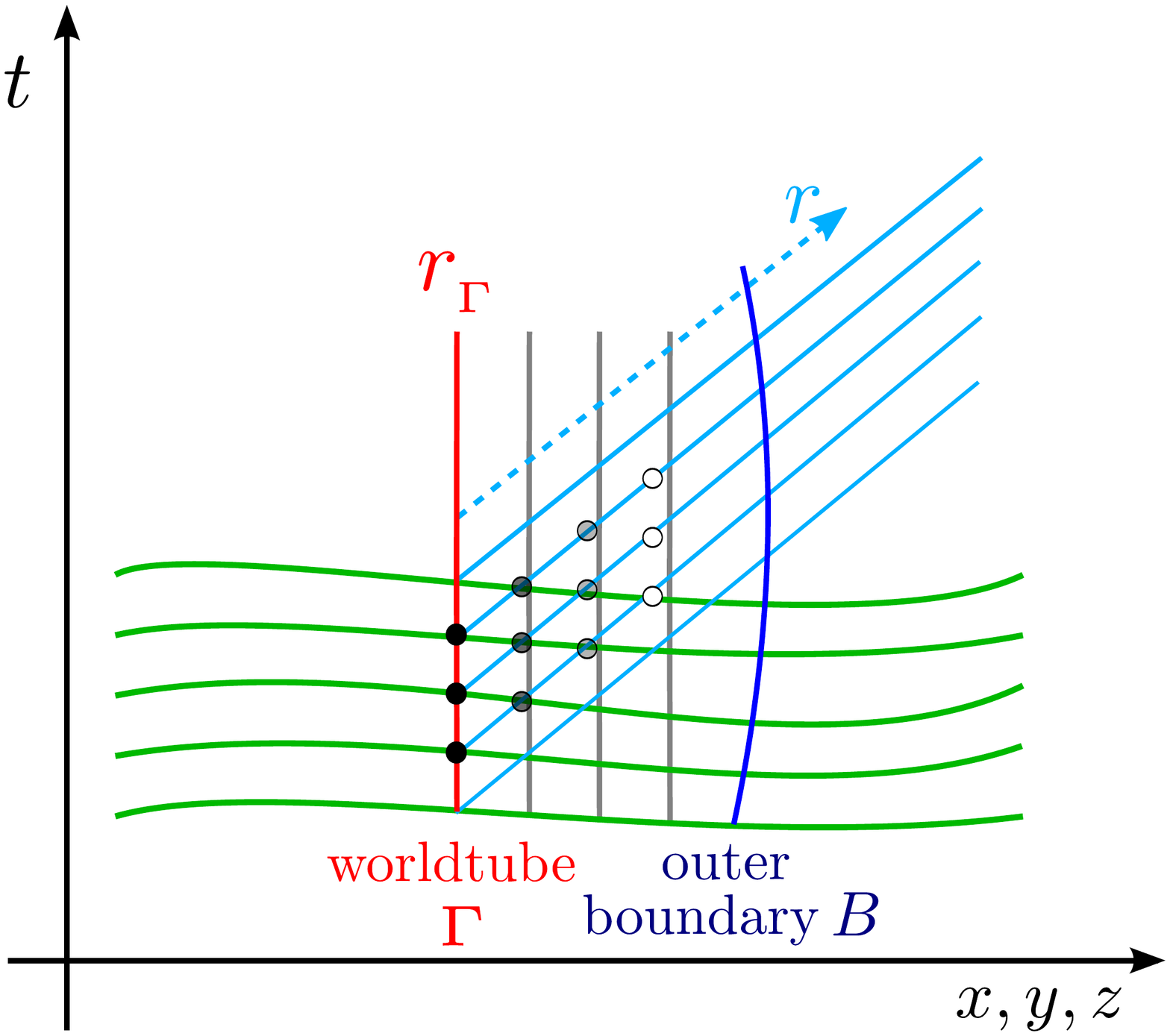}
    \caption{Schematic illustration of $\psi_4$ extrapolation. The
    Cauchy evolution is shown with {{green slices}},
    with an outer boundary in {{blue}} subject to a
    boundary condition that excludes incoming gravitational waves. The
    {{light blue}} lines are approximations to outgoing
    null slices. $\psi_4$ is evaluated where the Cauchy slices meet the
    innermost {{worldtube $r_\Gamma$}}; and also at fixed
    values of $r> r_\Gamma$ on each Cauchy slice, and then interpolated
    onto the black dots shown on the null slices. Values of $\psi_4$ at
    the black dots on a given null slice are then extrapolated to
    $r\rightarrow\infty$. }
    \label{f-psi4_logical}
\end{figure}}

The method most commonly used at present is an adaptation of a simple
estimate on a finite worldtube, and has become known as the
extrapolation method. A
schematic illustration of the method is given in
Fig.~\ref{f-psi4_logical}. A preliminary version of extrapolation was
used in 2005 \cite{Baker05a}. However, the method, as used at present,
was developed in 2009 by two different
groups \cite{Pollney2009,Boyle:2009vi}, and a recent description is given
by \cite{Taylor:2013}; see also \cite{Pollney2011b}. The essential idea
is that $\psi_4$ is estimated on worldtubes at a number of different
radii, and then the data is fitted to a polynomial of form
$\psi_4=\sum_{n=1}^N A_n/r^n$ so that $\lim_{r\rightarrow\infty}r\psi_4$
is approximated by $A_1$. However, there are some subtleties that
complicate the procedure a little. The expected polynomial form of
$\psi_4$ is applicable only on an outgoing null cone; and further $r$
should be a surface area coordinate (although often requiring this
property is not important). We assume that the data available is
$\psi_4^{\ell m}(t,s)$ obtained by decomposing $\psi_4$ into spherical
harmonic components on a spherical surface of fixed coordinate radius $s$
at a given coordinate time $t$. Then the first step in extrapolation is
to obtain $\psi_4^{\ell m}(t_{*},r)$, where $t_*$ is a retarted time
coordinate specified in Eq.~\eqref{e-t*} below, and where
\begin{equation}
r=r(t,s)=\sqrt{\frac{A}{4\pi}}\,,
\end{equation}
with $A$ the area of the coordinate 2-surface $t=~$constant,
$s=~$constant. Because the spacetime is dynamic, it would be a
complicated process to construct $t_*$ exactly. Instead, it is assumed
that extraction is performed in a region of spacetime in which the
geometry is approximately Schwarzschild with $(t,s)$ approximately
standard Schwarzschild coordinates. Then
\begin{equation}
t_{*}(t,s)=\int^{t}_0 \frac{\sqrt{-g^{ss}(t^\prime,s)/g^{tt}(t^\prime,s)}}{1-2 M / r(t^\prime,s)}
dt^\prime
-r(t,s)-2 M \ln \left(\frac{r(t,s)}{2M}-1\right)\,.
\label{e-t*}
\end{equation}
In Eq.~\eqref{e-t*}, $M$ is an estimate of the initial mass of the
system, usually the ADM mass, and $g^{tt},g^{ss}$ are averaged over the
2-sphere $t^\prime=~$constant, $s_j=~$constant. It is straightforward to
check that if $(t,s)$ are exactly Schwarzschild coordinates, then $t_*$
is null.

In this way we obtain, for fixed $t_*$, $\psi_4^{\ell m}$ at a number of
different extraction radii; that is, we have data of the form
$\psi_4^{\ell m}(t_*,r_k)$, $k=1,\cdots,K$\footnote{The values of $r_k$
  may vary with $t_*$ since the extraction spheres are constructed to be
  of constant coordinate radius $s$.}. In practice, the real and
imaginary parts of $\psi_4^{\ell m}$ may vary rapidly, and it has been
found to be smoother to fit the data to the amplitude and
phase\footnote{In the case of non-oscillatory modes, usually with $m=0$,
  fitting to the real and imaginary parts of $\psi_4^{\ell m}$ is
  preferred.}. For each spherical harmonic component two data-fitting
problems are solved
\begin{eqnarray}
|\psi_4^{\ell m}(t_*,r_k)| &\mbox{to}& \sum_{n=1}^N \frac{A_n(t_*)}{r^n}\,, \nonumber \\
\arg(\psi_4^{\ell m}(t_*,r_k)) &\mbox{to}& \sum_{n=0}^N \frac{\phi_n(t_*)}{r^n}\,,
\end{eqnarray}
using the least-squares method, and where the $A_n, \phi_n$ are all
real. Note that $\phi(t_*,r)$ must be continuous in $r$, and so for
certain values of $r_k$ it may be necessary to add $\pm 2\pi$ to
$\arg(\psi_4^{\ell m}(t_*,r_k))$. Then the estimate is
\begin{equation}
\lim_{r\rightarrow\infty}r\psi_4^{\ell m}= A_1 \exp(i\phi_0)\,.
\end{equation}

The remaining issue is the specification of $N$, and of the extraction
spheres $r_k$ (or more precisely, since extraction is performed on
spheres of specified coordinate radius, of the $s_k$). The key factors
are the innermost and outermost extraction spheres, \ie the values of
$r_1$ and $r_{_K}$, and of course the requirement that
$K>N+1$. Essentially, the extrapolation process uses data over the
interval $1/r\in [1/r_{_K},1/r_1]$ to construct an estimate at
$1/r=0$. Polynomial extrapolation can be unreliable, or even divergent,
as $N$ is increased; it can also be unreliable when the distance from the
closest data point is larger than the size of the interval over which the
data is fitted. As a result of this latter condition, it is normal to
require $r_{_K}>2 r_1$. On the other hand, increasing $r_{_K}$ increases
the computational cost of a simulation, and decreasing $r_1$ could mean
that a higher order polynomial is needed for accurate modelling of the
data at that point. A compromise is needed between these conflicting
factors. Values commonly used are that $N$ is between 3 and 5, $r_1$ is
normally of order $100M$, and $r_{_K}$ is $300 M$ with values as large as
$1000 M$ reported. Typically, $K$ is about 8, with the $1/r_k$ evenly
distributed over the interval $[1/r_{_K},1/r_1]$.

If the desired output of a computation is a waveform (to be used, say, in
the analysis of LIGO detector data), then $\psi_4$ needs to be translated
into its wave strain components $(h_+,h_\times)$. From Eq.~(\ref{e-psi4h}),
\begin{equation}
h_+^{\ell m}(t) - i h_\times^{\ell m}(t) = 
\int^t \left(\int^{t^\prime} \psi_4^{\ell m}(t^{\prime\prime}) dt^{\prime\prime}\right)
dt^\prime + A^{\ell m} t + B^{\ell m},
\end{equation}
where the constants of integration $A^{\ell m}, B^{\ell m}$ need to be
fixed by the imposition of some physical condition, for example that the
strain should tend to zero towards the end of the computation. While this
procedure is simple and straightforward, in practice it has been observed
that the double time integration may lead to a reduction in accuracy, and
in particular may introduce nonlinear drifts into the waveform. (The
presence of a linear drift is easily corrected by means of an adjustment
to the integration constants $A^{\ell m}, B^{\ell m}$). It was shown in
\cite{Reisswig:2011} that the cause of the problem is that $\psi_4^{\ell
  m}$ includes random noise, and this can lead to noticeable drifts after
a double integration. The usual procedure to control the effect is via a
transform to the Fourier domain. The process to construct the wave strain
from $\psi_4^{\ell m}$, without any correction for drift, is
\begin{align}
\tilde{\psi}_4^{\ell m}(\omega)=&{\mathcal F}[\psi_4^{\ell m}(t)] \,, \\
\tilde{h}^{\ell m}_+(\omega) - i \tilde{h}^{\ell m}_\times (\omega) =&
            -\frac{\tilde{\psi}_4^{\ell m}(\omega)}{\omega^2} \,, \\
h_+^{\ell m}(t) - i h_\times^{\ell m}(t)=&
             {\mathcal F}^{-1}[\tilde{h}^{\ell m}_+(\omega) - i \tilde{h}^{\ell m}_\times (\omega)]\,,
\label{e-Fpsi4}
\end{align}
where ${\mathcal F}$ is the Fourier transform operator, $\omega$ denotes
frequency in the Fourier domain, and ~$\tilde{ }$~ denotes a Fourier
transformed function.  The division by $\omega^2$ in the second line of
Eq. (\ref{e-Fpsi4}) is clearly potentially problematic for small
$\omega$, and an obvious strategy is to apply a filter to modify this
equation. A number of such filters have been proposed, based on reducing
those frequency components that are lower than $\omega_0$ -- the lowest
frequency expected, on physical grounds, in the waveform. The simplest
choice is a step function \cite{Campanelli:2008nk,Aylott:2009ya}, but it
has the drawback that it leads to Gibbs phenomena. To suppress this
effect, a smooth transition is needed near $\omega_0$ and various filters
have been investigated \cite{Santamaria2010,McKechan2010,Reisswig:2011}.
A particularly simple choice of filter, yet effective in many cases
\cite{Reisswig:2011}, is
\begin{align}
\tilde{h}^{\ell m}_+(\omega) - i \tilde{h}^{\ell m}_\times (\omega) =&
            -\frac{\tilde{\psi}_4^{\ell m}(\omega)}{\omega^2}  & (\omega\ge\omega_0)\,, \\
            =&-\frac{\tilde{\psi}_4^{\ell m}(\omega)}{\omega_0^2}  & (\omega<\omega_0)\,.
\end{align}

\subsubsection{Energy, momentum and angular momentum in the waves}
\label{s-psi4EMA}
Starting from the mass loss result of Bondi {\it et al.} \cite{Bondi62},
the theory of energy and momentum radiated as gravitational waves was
further developed in the 1960s \cite{Penrose:1963,Penrose65,
  Tamburino:1966, Winicour68, Isaacson68} and subsequently
\cite{Geroch77, Thorne80, Geroch1981}.  Formulas for the radiated angular
momentum were presented in Refs. \cite{Campanelli99, Lousto:2007mh} based
on earlier work by Winicour \cite{Winicour80}; formulas were also
obtained in \cite{Ruiz2007m, Ruiz:2008multipole} using the Isaacson
effective stress-energy tensor of gravitational waves \cite{Isaacson68}.

The result is formulas that express the energy, momentum and angular
momentum content of the gravitational radiation in terms of $\psi_4$.
Strictly, all the quantities should be evaluated in the limit as
$r\rightarrow \infty$ and using an appropriate null tetrad. The energy
equation is
\begin{equation}
\frac{d E}{d t}=\frac{1}{16\pi}\oint
\left|\int_{-\infty}^t r\psi_4 dt^\prime\right|^2 d\Omega\,.
\end{equation}
The linear momentum equations are
\begin{equation}
\frac{d P_i}{d t}=\frac{1}{16\pi}\oint
\hat{r}_i\left|\int_{-\infty}^t r\psi_4 dt^\prime\right|^2 d\Omega\,,
\end{equation}
where $\hat{r}_i$ is a unit radial vector. If the angular coordinate
system being used is spherical polars, then from Eq.~(\ref{e-C2sp})
$\hat{r}_i=(\sin\theta \cos\phi,\sin\theta\sin\phi,\cos\theta)$, whereas
if the coordinates are stereographic $\hat{r}_i$ would by given by
Eqs.~(\ref{e-C2N}) and (\ref{e-C2S}).  The angular momentum equations are
\begin{equation}
\frac{d J_i}{d t}=-\frac{1}{16\pi}\Re\left[\oint
\left(\int_{-\infty}^t r\bar{\psi}_4 dt^\prime\right)
\hat{J}_i\left(\int_{-\infty}^t \int_{-\infty}^{t^\prime}r\psi_4 dt^\prime dt^{\prime\prime}
\right) d\Omega\right]\,,
\end{equation}
where the $\hat{J}_i$ are operators given, in spherical polar coordinates, by
\begin{align}
\hat{J}_x =&\, -\sin\phi\partial_\theta-\cos\phi(\cot\theta\partial_\phi-is\csc\theta)\,, \\
\hat{J}_y =&\, \cos\phi\partial_\theta-\sin\phi(\cot\theta\partial_\phi-is\csc\theta)\,, \\
\hat{J}_z =&\, \partial_\phi\,.
\end{align}

In practice, the above formulas are rarely used directly, and instead
$\psi_4$ is first decomposed into spin-weighted spherical harmonics using
Eq.~\eqref{e-psi04lm}.  Then the energy equation is
\begin{equation}
\frac{d E}{d t}=\frac{1}{16\pi}\sum_{\ell\ge 2,|m|\le\ell}
\left|\int_{-\infty}^t r\psi_4^{\ell\,m} dt^\prime\right|^2\,.
\end{equation}
The momentum flux leaving the system is
\begin{align}
\frac{d P_x+iP_y}{d t}=&\,\frac{r^2}{8\pi}\sum_{\ell\ge 2,|m|\le\ell}
\int_{-\infty}^t \psi_4^{\ell\,m} dt^\prime
\int_{-\infty}^t \left(a_{\ell\,m}\bar{\psi}_4^{\ell,m+1} 
+b_{\ell,-m}\bar{\psi}_4^{\ell-1,m+1}
-b_{\ell+1,m+1}\bar{\psi}_4^{\ell+1,m+1}
\right)dt^\prime\,,  \\
\frac{d P_z}{d t}=&\,\frac{r^2}{16\pi}\sum_{\ell\ge 2,|m|\le\ell}
\int_{-\infty}^t \psi_4^{\ell\,m} dt^\prime
\int_{-\infty}^t \left(c_{\ell\,m}\bar{\psi}_4^{\ell\,m} 
+d_{\ell\,m}\bar{\psi}_4^{\ell-1,m}
+d_{\ell+1,m}\bar{\psi}_4^{\ell+1,m}
\right)dt^\prime\,,
\end{align}
where
\begin{align}
a_{\ell\,m}=&\,\frac{\sqrt{(\ell-m)(\ell+m+1)}}{\ell(\ell+1)}\,,&
\qquad
b_{\ell\,m}=&\, \frac{1}{2\ell}\sqrt{\frac{(\ell-2)(\ell+2)(\ell+m)(\ell+m-1)}{(2\ell-1)(2\ell+1)}}\,,&
\nonumber \\
c_{\ell\,m}=&\,\frac{2m}{\ell(\ell+1)}\,,&
\qquad
d_{\ell\,m}=&\, \frac{1}{\ell}\sqrt{\frac{(\ell-2)(\ell+2)(\ell-m)(\ell+m)}{(2\ell-1)(2\ell+1)}}\,.&
\end{align}
The angular momentum equations become
\begin{align}
\frac{d J_x}{d t}=&\,-\frac{ir^2}{32\pi}\Im\left(
\sum_{\ell\ge 2,|m|\le\ell}
\int_{-\infty}^t \int_{-\infty}^{t^\prime} \psi_4^{\ell\,m} dt^{\prime\prime} dt^\prime
\int_{-\infty}^t \left(f_{\ell\,m}\bar{\psi}_4^{\ell,m+1} 
+f_{\ell,-m}\bar{\psi}_4^{\ell,m-1}
\right)dt^\prime \right)\,, \\
\frac{d J_y}{d t}=&\,-\frac{r^2}{32\pi}\Re\left(
\sum_{\ell\ge 2,|m|\le\ell}
\int_{-\infty}^t \int_{-\infty}^{t^\prime} \psi_4^{\ell\,m} dt^{\prime\prime} dt^\prime
\int_{-\infty}^t \left(f_{\ell\,m}\bar{\psi}_4^{\ell,m+1} 
-f_{\ell,-m}\bar{\psi}_4^{\ell,m-1}
\right)dt^\prime \right)\,, \\
\frac{d J_z}{d t}=&\,-\frac{ir^2}{16\pi}\Im\left(
\sum_{\ell\ge 2,|m|\le\ell}
m\int_{-\infty}^t \int_{-\infty}^{t^\prime} \psi_4^{\ell\,m} dt^{\prime\prime} dt^\prime
\int_{-\infty}^t \bar{\psi}_4^{\ell\,m} 
dt^\prime \right)\,,
\end{align}
where the symbol $\Im$ refers to the imaginary part and 
\begin{equation}
f_{\ell\,m} \coloneqq \sqrt{\ell(\ell+1)-m(m+1)}\,.
\end{equation}

\newpage
\section{Gravitational Waves in the Cauchy-Perturbative Approach}
\label{s-CPA}

Black-hole perturbation theory has been fundamental not only for
understanding the stability and oscillations properties of black hole
spacetimes \cite{Regge57}, but also as an essential tool for clarifying
the dynamics that accompanies the process of black hole formation as a
result of gravitational collapse \cite{Price72,Price72b}. As one example
among the many possible, the use of perturbation theory has led to the
discovery that Schwarzschild black holes are characterised by decaying
modes of oscillation that depend on the black hole mass only, \ie
the black hole quasi-normal modes \cite{Vishveshwara70,Vishveshwara70b,press71,Chandrasekhar75}. Similarly, black-hole perturbation theory
and the identification of a power-law decay in the late-time dynamics of
generic black-hole perturbations has led to important theorems, such as
the ``no hair'' theorem, underlining the basic black-hole property of
removing all perturbations so that ``all that can be radiated away is
radiated away'' \cite{Price72,Price72b,MTW1973}.

The foundations of non-spherical metric perturbations of Schwarzschild
black holes date back to the work in 1957 of Regge and
Wheeler \cite{Regge57}, who first addressed the linear stability of the
Schwarzschild solution. A number of investigations, both gauge-invariant
and not, then followed in the 70's, when many different approaches were
proposed and some of the most important results about the physics of
perturbed spherical and rotating black holes
established \cite{Price72,Price72b,Vishveshwara70,Vishveshwara70b,Chandrasekhar75,Zerilli70,Zerilli70a,Moncrief74,Cunningham78,Cunningham79,Teukolsky72,Teukolsky73}. Building
on these studies, which defined most of the mathematical apparatus behind
generic perturbations of black holes, a number of applications have been
performed to study, for instance, the evolutions of perturbations on a
collapsing background
spacetime \cite{Gerlach79a,Gerlach79b,Gerlach80,Karlovini02,Seidel87a,Seidel88,Seidel90,Seidel91b}. Furthermore,
the gauge-invariant and coordinate independent formalism for
perturbations of spherically symmetric spectimes developed in the 70's by
Gerlach and Sengupta in \cite{Gerlach79a,Gerlach79b,Gerlach80}, has been
recently extended to higher-dimensional spacetimes with a maximally
symmetric subspace in Refs. \cite{Kodama00,Kodama03a,Kodama03b,Kodama04},
for the study of perturbations in brane-world models.

Also nowadays, when numerical relativity calculations allow to evolve the
Einstein equations in the absence of symmetries and in fully nonlinear
regimes, black hole perturbative techniques represent important
tools\footnote{All of our discussion hereafter will deal with
perturbative analyses in the time domain. However, a hybrid approach is
also possible in which the perturbation equations are solved in the
frequency domain. In this case, the source terms are given by
time-dependent perturbations created, for instance, by the motion of
matter and computed by fully nonlinear three-dimensional
codes \cite{Ferrari:2006mk}.}. Schwarzschild perturbation theory, for
instance, has been useful in studying the late-time behaviour of the
coalescence of compact binaries in a numerical simulation after the
apparent horizon has formed \cite{Price94a,Abrahams94c,Abrahams95d}. In
addition, methods have been developed that match a fully numerical and
three-dimensional Cauchy solution of Einstein's equations on spacelike
hypersurfaces with a perturbative solution in a region where the
components of three-metric (or of the extrinsic curvature) can be treated
as linear perturbations of a Schwarzschild black hole (this is usually
referred to as the
\emph{``Cauchy-Perturbative Matching''}) \cite{Abrahams97a,Rupright98,Camarda99,Allen98a,Rezzolla99a,Lousto:2010n,Nakano2015}. This method, in turn, allows to
``extract'' the gravitational waves generated by the simulation, evolve
them out to the wave-zone where they assume their asymptotic form, and
ultimately provide outer boundary conditions for the numerical evolution.

This section intends to review the mathematical aspects of the metric
perturbations of a Schwarzschild black hole, especially in its
gauge-invariant formulations. Special care is paid to ``filter'' those
technical details that may obscure the important results and provide the
reader with a set of expressions that can be readily used for the
calculation of the odd and even-parity perturbations of a Schwarzschild
spacetime in the presence of generic matter-sources. Also, an effort is
made to ``steer'' the reader through the numerous conventions and
notations that have accompanied the development of the formalism over the
years. Finally, as mentioned in the Introduction, a lot of the material
presented here has already appeared in the Topical Review by Nagar and
Rezzolla \cite{Nagar05}.

\subsection{Gauge-invariant metric perturbations}
\label{gi_intro}

It is useful to recall that even if the coordinate system of the
background spacetime has been fixed, the coordinate freedom of general
relativity introduces a problem when linear perturbations are added. In
particular, it is not possible to distinguish an infinitesimal
``physical'' perturbation from one produced as a result of an
infinitesimal coordinate transformation (or gauge-transformation). This
difficulty, however, can be removed either by explicitly fixing a gauge
(see, \eg \cite{Regge57,Price72,Price72b,Vishveshwara70,Vishveshwara70b,Zerilli70,Zerilli70a}),
or by introducing linearly gauge--invariant perturbations (as initially
suggested by Moncrief \cite{Moncrief74} and subsequently adopted in
several
applications \cite{Cunningham78,Cunningham79,Seidel87a,Seidel88,Seidel90,Seidel91b}).

More specifically, given a tensor field $\boldsymbol{X}$
and its
infinitesimal perturbation $\delta \boldsymbol{X}$, an infinitesimal
coordinate transformation $x^{\mu}\rightarrow x^{\mu'} \coloneqq
x^{\mu}+\xi^{\mu}$ with $\xi^{\mu}\ll 1$ will yield a new tensor field
\begin{equation}
\delta \boldsymbol{X}\rightarrow 
\delta \boldsymbol{X}'=\delta \boldsymbol{X}+ 
{\cal L}_{\boldsymbol{\xi}} \boldsymbol{X}\,,
\end{equation}
where ${\cal L}_{\boldsymbol{\xi}}$ is the Lie derivative along
$\boldsymbol{\xi}$.
We will then consider $\delta \boldsymbol{X}$ to be gauge-invariant if
and only if ${\cal L}_{\boldsymbol{\xi}}\boldsymbol{X}=0$, \ie if $\delta
\boldsymbol{X}'=\delta \boldsymbol{X}$. In particular, since
gravitational waves are metric perturbations, we will consider the case
that $\boldsymbol{X}$ is the background metric
$\boldsymbol{g{\!\!\!\!\;\raisebox{-0.1ex}{$^{^0}$}}}$, and then metric
perturbations are gauge invariant if and only if ${\cal
  L}_{\boldsymbol{\xi}}\boldsymbol{g{\!\!\!\!\;\raisebox{-0.1ex}{$^{^0}$}}}=0$.

	Stated differently, the possibility of building gauge--invariant
metric perturbations relies on the existence of symmetries of the
background metric. In the case of a general spherically symmetric
background spacetime (\ie one allowing for a time dependence) and
which has been decomposed in multipoles (see Sect.~\ref{sec:multipole}),
the construction of gauge-invariant quantities is possible for multipoles
of order $\ell \geq 2$ only \cite{Gerlach79a, Gerlach79b,
Martin-Garcia98, Gundlach00b}. In practice, the advantage in the use of
gauge-invariant quantities is that they are naturally related to scalar
observables and, for what is relevant here, to the energy and momentum of
gravitational waves. At the same time, this choice guarantees that
possible gauge-dependent contributions are excluded by construction.

Of course, this procedure is possible if and only if the background
metric has the proper symmetries under infinitesimal coordinates
transformation; in turn, a gauge-invariant formulation of the Einstein
equations for the perturbations of a general spacetime is not
possible. Nevertheless, since any asymptotically flat spacetime can in
general be matched to a Schwarzschild one at sufficiently large
distances, a gauge-invariant formulation can be an effective tool to
extract physical information about the gravitational waves generated in
a numerically evolved, asymptotically flat spacetime \cite{Abrahams97a,Rupright98,Camarda99,Allen98a,Rezzolla99a} (see also
Sect.~\ref{s-CPIS} for additional implementational details). The
following section is dedicated to a review of the mathematical techniques
to obtain gauge-invariant perturbations of a the Schwarzschild metric.

\subsection{Multipolar expansion of metric perturbations}
\label{sec:multipole}

	Given a spherically symmetric Schwarzschild solution with metric
${\boldsymbol g{\!\!\!\!\; \raisebox{-0.1ex}{$^{^0}$}}}$ and line element
\begin{equation}
ds^2 \coloneqq g{\!\!\!\!\; \raisebox{-0.1ex}{$^{^0}$}}_{\mu\nu}
	dx^{\mu}dx^{\nu}
	=-e^{2a}dt^2+e^{2b}dr^2+r^2\left(d\theta^2+\sin^2\theta
	d\phi^2\right)\,, 
\end{equation}
where $e^{2a}=e^{-2b}=\left(1-2M/r\right)$, we generically consider small
non-spherical perturbations $h_{\mu \nu}$ such that the new perturbed
metric is
\begin{equation}
\label{gmunu}
g_{\mu\nu} \coloneqq  g{\!\!\!\!\; \raisebox{-0.1ex}{$^{^0}$}}_{\mu\nu} + 
	h_{\mu\nu}\,,
\end{equation}
where $|h_{\mu \nu}|/|g{\!\!\!\!\; \raisebox{-0.1ex}{$^{^0}$}}_{\mu\nu}|
\ll 1$. Although we have chosen to employ Schwarzschild coordinates to
facilitate the comparison with much of the previous literature, this is
not the only possible choice, nor the best one. Indeed, it is possible to
formulate the perturbations equations independently of the choice of
coordinates as discussed in Ref. \cite{Martel:2005ir}, or in
horizon-penetrating coordinates when the perturbations are in
vacuum \cite{Sarbach:2001qq}. Mostly to remain with the spirit of a
review and because most of the results have historically been derived in
these coordinates, we will hereafter continue to use Schwarzschild
coordinates although the reader should bear in mind that this is not the
optimal choice.

Because the background manifold ${\cal M}$ is spatially spherically
symmetric, it can be written as the product ${\cal M}={\sf M}^2\times
{\sf S}^2$, where ${\sf M}^2$ is a Lorentzian 2-dimensional manifold of
coordinates $(t,r)$ and ${\sf S}^2$ is the 2-sphere of unit radius and
coordinates $(\theta, \phi)$. As a result, the perturbations can be split
\emph{``ab initio''} in a part confined to ${\sf M}^2$ and in a part
confined on the 2-sphere ${\sf S}^2$ of metric
$\boldsymbol{\gamma}$. Exploiting this, we can expand the metric
perturbations $\boldsymbol{h}$ in multipoles referred to as ``odd'' or
``even-parity'' according to their transformation properties under
parity. In particular, are \emph{odd} (or \emph{axial}) multipoles those
that transform as $(-1)^{\ell+1}$, under a parity transformation
$(\theta, \phi) \to (\pi-\theta, \pi +\phi)$, while are \emph{even} (or
\emph{polar}) those multipoles that transform as $(-1)^{\ell}$. As a
result, the metric perturbations can be written as
\begin{equation}
\label{hmunu}
h_{\mu\nu}= 
	\sum_{\ell,m}\left[\left(h_{\mu\nu}^{{\ell m}}\right)^{({\rm o})} 
	+\left(h_{\mu\nu}^{\ell m}\right)^{({\rm e})}\right]\,,
\end{equation}
where $\sum_{\ell,m} \coloneqq \sum_{\ell=2}^{\infty} \;
\sum_{m=-\ell}^{\ell}$, and the upper indices $^{({\rm o})}$ and $^{({\rm
    e})}$ distinguish odd and even-parity objects, respectively. Adapting
now a notation inspired by that of Gerlach and Sengupta \cite{Gerlach79a,Gerlach79b,Gerlach80} and recently revived by Gundlach and
Mart\'in-Garc\'ia \cite{Martin-Garcia98,Gundlach00b,Gundlach00c}, we
use lower-case indices $a,\,b\, \ldots=0,1$ to label the coordinates of
${\sf M}^2$ and upper-case indices $C,\,D\ldots=2,3$ to label the
coordinates of ${\sf S}^2$. Using this notation, the scalar spherical
harmonics $Y^{{\ell m}}$ are then simply defined as
\begin{equation}
\label{def:harmonics}
\gamma^{_{CD}}\nabla_{_D}\nabla_{_C}Y^{{\ell m}}= -\Lambda Y^{{\ell m}}\,,
\end{equation}
where $\nabla_{_C}$ indicates the covariant derivative with respect to
the metric $\boldsymbol{\gamma}\coloneqq \mathrm{diag}(1,\sin^2\theta)$
of ${\sf S}^2$, and where
\begin{equation}
\Lambda \coloneqq \ell(\ell+1)\,. 
\end{equation}

It is now convenient to express the odd and even-parity metric functions
in (\ref{gmunu}) in terms of tensor spherical harmonics. To do this we
introduce the axial vector $S^{{\ell m}}_{_C}$ defined as
\begin{equation}
S_{_C}^{{\ell m}}
\coloneqq \epsilon_{_{CD}}\gamma^{_{DE}}\nabla_{_E} Y^{{\ell m}}\,, 
\end{equation}
where $\epsilon_{_{CD}}$ is the volume form on ${\sf S}^2$ as defined by the
condition $\epsilon_{_{CD}}\epsilon^{_{CE}}=\gamma_{_D}^{_{\;\;E}}$ and such that
$\nabla_{_C}\epsilon_{_{AB}}=0$. In this way, each odd-parity metric function
in (\ref{gmunu}) can be written as
\begin{eqnarray}
\label{odd:metric}
\left(h_{\mu\nu}^{{\ell m}}\right)^{({\rm o})}= \left(\begin{array}{c|c} 0 &
	h_{a}^{({\rm o})}S_{_C}^{{\ell m}} \\ 
	\hline  \\
	h_{a}^{({\rm o})}S_{_C}^{{\ell m}} & h
	\nabla_{_{(D}} S_{_C)}^{{\ell m}}\\
\end{array}
\right)\,,\\ \nonumber
\end{eqnarray}
where $h,h_{a}^{({\rm o})}$ are functions of $(t,r)$ only, and
where we have omitted the indices $\ell, m$ on $h,h_{a}^{({\rm o})}$for clarity.

	Proceeding in a similar manner, each even-parity metric function
can be decomposed in tensor spherical harmonics as
\begin{eqnarray}
\label{even:metric}
 \qquad \left(h_{\mu\nu}^{{\ell m}}\right)^{(\mathrm{e})}
	=\left(\begin{array}{cc|cc} e^{2a} H_0Y^{{\ell m}} & 
	H_1Y^{{\ell m}} & \qquad
	h_{a}^{({\rm e})}\nabla_{_C}Y^{{\ell m}} \\ 
	H_1Y^{{\ell m}} & e^{2b} H_2Y^{{\ell m}} & \\ 
\hline & &\\ 
	h_{a}^{({\rm e})}\nabla_{_C} Y^{{\ell m}} & &
	r^2\left(KY^{{\ell m}}\gamma_{_{CD}}+
	G \nabla_{_D}\nabla_{_C} Y^{{\ell m}}\right)\\
\end{array}\right)\,,\\
\nonumber
\end{eqnarray}
where $H_0,\,H_1,\,H_2,\, h_0^{({\rm e})},\,h_1^{({\rm e})}\, K$, and
$G$ (with the indices $\ell, m$ omitted for clarity) are the coefficients
of the even-parity expansion, are also functions of $(t,r)$ only.

Note that we have used the Regge--Wheeler set of tensor harmonics to
decompose the even-parity part of the metric in
multipoles \cite{Regge57}. Despite this being a popular choice in the
literature, it is not the most convenient one since the tensor harmonics
in this set are not linearly independent. An orthonormal set is instead
given by the Zerilli-Mathews tensor harmonics \cite{Zerilli70b,Mathews62}
and the transformation from one basis to the other is given by defining
the tensor $Z_{_{CD}}^{{\ell m}}$ confined on the 2-sphere ${\sf S}^2$
(see also Appendix \ref{app:vtsh})
\begin{equation}
Z_{_{CD}}^{{\ell m}} \coloneqq \nabla_{_C} \nabla_{_D} Y^{{\ell m}}+
	\frac{\Lambda}{2}\gamma_{_{CD}}Y^{{\ell m}}\,,
\end{equation}
and then replacing in equation~(\ref{even:metric}) the second covariant
derivative of the spherical harmonics $\nabla_{_C}\nabla_{_D}Y^{{\ell
    m}}$ with $Z^{{\ell m}}_{_{CD}}$. This transformation has to be taken
into account, for instance, when developing gauge-invariant procedures
for extracting the gravitational-wave content from numerically generated
spacetimes which are ``almost'' Schwarzschild
spacetimes \cite{Abrahams92a,Anninos93b,Anninos94b,Anninos95g,Abrahams95b,Abrahams95c}.

Besides vacuum tensor perturbations, the background Schwarzschild
spacetime can be modified if non-vacuum tensor perturbations are present
and have a nonzero mass-energy, but much smaller than that of the black
hole.  In this case, the generic stress-energy tensor $t_{\mu\nu}$
describing the matter-sources can be similarly decomposed in odd and
even-parity parts
\begin{equation}
t_{\mu\nu}=\sum_{\ell,m}\left[\left(t^{{\ell m}}_{\mu\nu}\right)^{({\rm
	o})}+\left(t^{{\ell m}}_{\mu\nu}\right)^{({\rm e})}\right]
	\,,
\end{equation}
that are naturally gauge-invariant since the background is the vacuum
Schwarzschild spacetime and are given explicitely by
\begin{eqnarray}
\label{source:odd}
\left(t^{{\ell m}}_{\mu\nu}\right)^{({\rm o})}= \left(
\begin{array}{cc}
	0 & L_{a}^{{\ell m}}S_{_C}^{{\ell m}} \\ \\ 
	L_{a}^{{\ell m}}S_{_C}^{{\ell m}} &
	L^{{\ell m}} \nabla_{_{(D}}S_{_{C)}}^{{\ell m}}\\
\end{array}
\right)\,,\\ \nonumber
\end{eqnarray}
for the odd-parity part and by
\begin{eqnarray}
\label{source:even}
\left(t^{{\ell m}}_{\mu\nu}\right)^{({\rm e})}= 
	\left(\begin{array}{c|c}
	T_{ab}^{{\ell m}}Y^{{\ell m}} & 
	T_{a}^{{\ell m}}\nabla_{_C} Y^{{\ell m}}\\ 
	& \\ \hline & \\ 
	T_{a}^{{\ell m}}\nabla_{_C}
	Y^{{\ell m}} & r^2T_3^{{\ell m}}Y^{{\ell m}}\gamma_{_{CD}}+
	T_2^{{\ell m}}Z_{_{CD}}^{{\ell m}}\\
\end{array}
\right) \,, \\ \nonumber
\end{eqnarray}
for the even-parity one. Note that we have now used the Zerilli-Matthews
set of harmonics for the expansion, that the ten coefficients
$L_{a}^{{\ell m}},\, L^{{\ell m}},\, T_{ab}^{{\ell m}},\, T_{a}^{{\ell
m}},\, T_2^{{\ell m}},\, T_3^{{\ell m}}$ are gauge-invariant, and that
explicit expressions for them will be presented in the following sections.

Let us now consider the Einstein field equations that, in the static
vacuum background, take the simple form
\begin{equation}
\label{efe_u}
R{\!\!\!\!\; \raisebox{0.3ex}{$^{^0}$}}_{\mu \nu} = 0 \,,
\end{equation}
where ${\boldsymbol R}{\!\!\!\!\!\; \raisebox{0.3ex}{$^{^0}$}}$ is the
Ricci tensor built from the background metric ${\boldsymbol
g}{\!\!\!\!\; \raisebox{-0.1ex}{$^{^0}$}}$. At first order in the
perturbations, the field equations reduce to
\begin{equation}
\label{efe_p}
R_{\mu\nu}-\frac{1}{2}g{\!\!\!\!\; \raisebox{-0.1ex}{$^{^0}$}}_{\mu\nu}R
=8\pi t_{\mu\nu} \,,
\end{equation}
where $\boldsymbol{R}$ is now the Ricci tensor built from the metric
perturbations $\boldsymbol{h}$. 

Note that while a generic perturbation will be a mixture of odd and
even-parity contributions, we will exploit the linearity of the approach
to handle them separately and simplify the treatment. In the following
two sections we will discuss the form the Einstein equations
(\ref{efe_p}) assume in response to purely odd and even-parity
perturbations over a Schwarzschild background. In particular, we will
show how the three odd-parity coefficients of the expansion in harmonics
of the metric, \ie $h_{a}^{({\rm o})},\,h$, and the seven even-parity
ones, \ie $H_0,\,H_1,\,H_2,\, h_0^{({\rm e})},\,h_1^{({\rm e})}\, K,\,
G$, can be combined to give two gauge-invariant master equations, named
respectively after Regge and Wheeler \cite{Regge57} and
Zerilli \cite{Zerilli70b}, each of which is a wave-like equation in a
scattering potential\footnote{These results were originally obtained by
Regge and Wheeler \cite{Regge57} and by
Zerilli \cite{Zerilli70,Zerilli70a} in a specific gauge (\ie the
Regge--Wheeler gauge). Subsequently, the work of Moncrief showed how to
reformulate the problem in a gauge-invariant form by deriving the
equations from a suitable variational principle \cite{Moncrief74}.}.

Although our attention is here focussed on the $radiative$ degrees of
freedom of the perturbations (\ie those with $\ell\ge 2$) because
of their obvious application to the modelling of sources of gravitational
waves, a comment should be made also on lower-order multipoles. In
particular, it is worth remarking that the monopole component of the
metric for a vacuum perturbation (\ie with $\ell = 0$) is only of
even-parity type and represents a variation in the mass-parameter of the
Schwarzschild solution. On the other hand, the dipole component of the
even-parity metric for a vacuum perturbation (\ie with $\ell=1$) is
of pure-gauge type and it can be removed by means of a suitable gauge
transformation \cite{Zerilli70a}. This is not the case for a dipolar
odd-parity metric perturbation, which can instead be associated to the
introduction of angular momentum onto the background metric.

\subsection{Gauge-invariant odd-parity perturbations}
\label{opp}

Before discussing the derivation of the odd-parity equation, we should
make a choice for the odd-parity master function. Unfortunately, this
choice has not been unique over the years and different authors have made
different choices, making comparisons between different approaches less
straightforward. Here, we will make a choice which highlights the
relation with the gravitational-wave amplitude measured by a distant
observer and, in particular, we construct the gauge-invariant combination
of multipoles \cite{Gerlach79a,Martin-Garcia98,Harada03a}
\begin{equation}
\label{def:kA}
k_{a} \coloneqq h_{a}-\nabla_{\!a} h +2h\frac{\nabla_{\!a}r}{r}\,,
\end{equation}
where, we recall, $\nabla_{\!a}$ represents the covariant derivative with
respect to the connection of the submanifold ${\sf M}^2$. If
$\epsilon_{ab}$ is the antisymmetric volume form on ${\sf M}^2$, then the
function
\begin{equation}
\label{phi:GM}
\Phi^{({\rm o})}(t,r) \coloneqq
	r^3\epsilon^{ab}\nabla_{\!b}\left(\frac{k_{a}}{r^2}\right)
	= r\left[\partial_{t}h_{1}^{({\rm o})}-
	r^2\partial_r\left(\frac{h_0^{({\rm o})}}{r^2}\right)\right]\,,
\end{equation}
is gauge-invariant and will be our choice for the Regge--Wheeler master
function \cite{Gerlach79a,Martin-Garcia98,Gundlach00b,Harada03a}.

A slight variation of the master function (\ref{phi:GM}) has been
introduced by Cunningham, Price and Moncrief \cite{Cunningham78} in terms
of the function $\widetilde{\psi} \coloneqq \Lambda \Phi^{({\rm o})}$ and
this has been used so extensively in the literature \cite{Seidel87a,Seidel88,Seidel91b} that it is now commonly referred to as the
Cunningham-Price-Moncrief (CPM) convention. We partly follow this
suggestion and introduce a new master function for the odd-parity
perturbations defined as
\begin{equation}
\label{CPM_n}
\Psi^{({\rm o})} \coloneqq \frac{1}{\Lambda-2} \,\Phi^{({\rm o})}\,.
\end{equation}

	With the choice (\ref{CPM_n}), the Einstein field equations
(\ref{efe_p}) with odd-parity perturbations lead to the inhomogeneous
\emph{``Regge--Wheeler''} equation
\begin{equation}
\label{rw_eq}
\partial^2_{t} \Psi^{({\rm o})} - \partial^2_{r_*} \Psi^{({\rm o})} + 	
	V_{\ell}^{({\rm o})}\Psi^{({\rm o})}=S^{({\rm o})}\,,
\end{equation}
where 
\begin{equation}
r_*\coloneqq r+2M\ln \left( \frac{r}{2M}-1\right) \,,
\end{equation}
is the ``tortoise coordinate'' \cite{MTW1973} and $V_{\ell}^{({\rm o})}$ is
the odd-parity potential, defined as
\begin{equation}
\label{odd:potential}
V_{\ell}^{({\rm o})} \coloneqq \left(1-\frac{2M}{r}\right)
	\left(\frac{\Lambda}{r^2}- \frac{6M}{r^3} \right)\,.
\end{equation}
The right-hand side of Eq. (\ref{rw_eq}) represents the generic
odd-parity ``matter-source'' and is given by
\begin{equation}
\label{odd:source}
 S^{({\rm o})} \coloneqq 
\frac{16\pi r}{\Lambda-2} \, e^{2a}\epsilon^{bc}\nabla_{\!c}L_{b} = 
\frac{16\pi r}{\Lambda-2} \left[\left(1-\frac{2M}{r}\right)
\partial_{t}L_{1}^{{\ell m}} - \partial_{r_*}
	L_{0}^{{\ell m}}\right]\,, 
\end{equation}
with the components of the odd-parity matter-source vector defined as
\begin{equation}
L_{a}^{{\ell m}} \coloneqq \frac{1}{\Lambda}\int
	\frac{d\Omega}{\sin\theta}\left({i}\,m\,
	t_{{a2}}Y^{*}_{{\ell m}}+
	t_{{a3}}\,\partial_{\theta}Y_{\ell m}^{*}\right)\,, \qquad a=0,1\,,
\end{equation}
and where $d\Omega=\sin\theta d\phi d\theta$ is the surface element on
the 2-sphere ${\sf S}^2$. 


Another choice for the gauge-invariant odd-parity master variable is
possible and indeed was originally proposed by
Moncrief \cite{Moncrief74}. This function, which hereafter we will refer
to as the odd-parity Moncrief function, is defined as
\begin{equation}
\label{mfop}
Q^{(\rm o)}\coloneqq 
	g{\!\!\!\!\; \raisebox{-0.1ex}{$^{^0}$}}^{ab}k_{b} 
        \frac{\nabla_{\!a} r}{r} =
	\frac{1}{r}\left(1-\frac{2M}{r}\right)
	\left[h_{1}^{({\rm o})}+\frac{r^2}{2} \partial_r
	\left(\frac{h_2}{r^2}\right)\right]\,,
\end{equation}
where the first expression is coordinate
independent \cite{Martel:2005ir}, while the second one is specialized to
Schwarzschild coordinates with $h_2 = - 2 h$ \cite{Moncrief74}. In the
Regge--Wheeler gauge, \ie for $h_2=h=0$, the definition (\ref{mfop})
coincides with the variable used by Regge and
Wheeler \cite{Regge57}. Historically, the choice of (\ref{mfop}) as
master variable has been the most common in the literature to describe
odd-parity perturbations of a Schwarzschild spacetime and we will refer
to it as ``Regge--Wheeler'' (RW) convention. It should be noted that while
(\ref{mfop}) is a solution of the Regge--Wheeler equation, the
corresponding source term differs from expression (\ref{odd:source}). A
general expression of the source in the RW convention can be found in
Ref. \cite{Martel:2005ir,Martel04} together with its specification for a
point-particle (see also Refs. \cite{Andrade99,Tominaga99,Ferrari00}).

The two master functions $Q^{(\rm o)}$ and $\Psi^{(\rm o)}$ are
intimately related through the variational formalism employed by Moncrief
in Ref. \cite{Moncrief74}, and through the explicit
expression \cite{Martel:2005ir}
\begin{equation}
\label{Q_vs_Psi}
	\partial_t\Psi^{(\rm o)}=-Q^{(\rm o)} +
	\frac{16\pi}{\Lambda-2}\frac{r}{e^{2b}} L_1^{\ell m} \,.
\end{equation} 
Note that Eq. (\ref{Q_vs_Psi}) highlights an important difference between
the two master functions which is not just a dimensional one (\ie
$\Psi^{({\rm o})}$ has the dimensions of a length, while $Q^{({\rm o})}$
is dimensionless) and this will have consequences on the asymptotic
expressions for the gravitational waveforms when these are expressed in
one or in the other convention. A detailed discussion of this will be
made in Sects.~\ref{sec:h_from_op}, \ref{a_general} and
\ref{e_am_losses}.

\subsection{Gauge-invariant even-parity perturbations}
\label{epp}

Also in the case of even-parity perturbations, it is possible to express
the evolution of the even-parity perturbations in terms of a wave-like
equation in a scattering potential [\cf the Regge--Wheeler equation
(\ref{rw_eq})]. In particular, following Moncrief \cite{Moncrief74}, we
define the gauge-invariant functions
\begin{eqnarray}
\label{kappa1}
\kappa_1 & \coloneqq & K+\frac{1}{e^{2b}}\left(r\partial_r G-
	\frac{2}{r}h_1^{({\rm e})}\right)\,,\\
\label{kappa2}
\kappa_2 & \coloneqq &\frac{1}{2}\left[e^{2b}H_2-
	e^b \partial_r \left(r e^b K\right)\right]\,,
\label{def:q1}
\end{eqnarray}
and where their linear combination
\begin{equation}
\label{q1}
q_1  \coloneqq r\Lambda\kappa_1 + \frac{4r}{e^{4b}}\kappa_2 \,.
\end{equation}
is also a gauge-invariant function. Strictly related to expression
(\ref{q1}) is the gauge-invariant function most frequently used in the
literature \cite{Seidel90,Gundlach00c,Martel04,Lousto97a,Ruoff01a,Ruoff01b,Martel:2001yf,Nagar04,Poisson04b}
\begin{equation}
\label{psi_e}
{\Psi}^{({\rm e})} \coloneqq \frac{r
	q_1}{\Lambda\left[r\left(\Lambda-2\right)+6M\right]} \,,
\end{equation}
which is also the solution of the inhomogeneous even-parity master
equation or \emph{``Zerilli''} equation
\begin{equation}
\label{zerilli_eq}
\partial^2_{t}{\Psi}^{({\rm e})} - 
	\partial^2_{r_*}{\Psi}^{({\rm e})}+
	V_{\ell}^{({\rm e})}{\Psi}^{({\rm e})}=S^{({\rm e})} \,,
\end{equation}
and, again, is a wave-like equation in the scattering Zerilli
potential \cite{Zerilli70}
\begin{equation}
 V_{\ell}^{({\rm e})} \coloneqq
	\left(1-\frac{2M}{r}\right)\frac{\Lambda(\Lambda-2)^2r^3
	+6(\Lambda-2)^2Mr^2+36(\Lambda-2)M^2r+72M^3}
	{r^3\left[(\Lambda-2)r+6M\right]^2}\,.
\end{equation}
The even-parity matter-source has a rather extended expression given
by \cite{Martel04,Nagar04,Nagar:2005fz}
\begin{eqnarray}
\label{source:polar}
S^{({\rm e})}=-\frac{8\pi}{\Lambda\left[(\Lambda-2)r+6M\right]}
\Bigg\{
\frac{\Lambda\Big(6r^3-16Mr^2\Big)-r^3\Lambda^2-8r^3+68Mr^2-108M^2r}
     {(\Lambda-2)r+6M} T_{00}^{{\ell m}}
\nonumber\\
\hskip 4.25cm
        +\frac{1}{e^{4b}}\Big[2Mr+r^2(\Lambda-4)\Big]T_{11}^{{\ell m}}+
        2r^3 \partial_{r_*} T_{00}^{{\ell m}}
        -\frac{2r^3}{e^{4b}} 
        \partial_{r_*} T_{11}^{{\ell m}}
        \nonumber\\ 
\hskip 4.25cm
        +\frac{4\Lambda r}{e^{4b}}T_{1}^{{\ell m}}
        +\frac{1}{e^{2b}}\left[2\Lambda\left(1-\frac{3M}{r}\right)-
          \Lambda^2\right]T_{2}^{{\ell m}}+
        \frac{4r^2}{e^{4b}}T_3^{{\ell m}}\Bigg\}
        \ .
        \nonumber \\ 
\end{eqnarray}
Note that the expressions of the even-parity vector and tensor
spherical-harmonics for the matter-source needed in (\ref{source:polar})
can be obtained from the orthogonality properties of the harmonics and
are
\begin{eqnarray}
 && T_{a}^{{\ell m}}=\frac{1}{\Lambda}\int d\Omega 
	\left[t_{a 2} \partial_{\theta}(\bar{Y}^{\ell m}) -
	t_{a 3} \frac{ i m \bar{Y}^{\ell m}}{\sin^2\theta} \right]\,,
	\hskip 2.7cm a = 0,1\,,\\
 && T_{2}^{{\ell m}}=\frac{2}{\Lambda(\Lambda-2)} \int
	d\Omega\bigg[t_{22} \frac{\bar{W}^{\ell m}}{2}  + 
	t_{23} \frac{2\bar{X}^{{\ell m}}}{\sin\theta} 
	+t_{33} \bigg(\frac{\Lambda \bar{Y}^{{\ell m}}}{2} 
	-\frac{m^2 \bar{Y}^{{\ell m}}}{\sin^2\theta} +
	\cot\theta \partial_{\theta}\bar{Y}^{\ell m}\bigg)
	\bigg]\,, 
	\nonumber\\ && ~\\
 && T_{3}^{{\ell m}} =\frac{1}{2r^2}\int d\Omega 
	\left(t_{22}+t_{33}\frac{1}{\sin^2\theta}\right)\bar{Y}^{\ell m}\,, \\
 && T_{ab}^{{\ell m}}=\int d\Omega\,t_{ab}\, \bar{Y}^{\ell m}\,,\;
	\hskip 6.0cm a,b = 0,1 \,,
\end{eqnarray}
where the angular functions $W^{\ell m}(\theta, \phi)$ and $X^{\ell
m}(\theta, \phi)$ are defined as \cite{Regge57}
\begin{align}
\label{Wlm}
 W^{{\ell m}}
	&\coloneqq\frac{\nabla_{(\phi}S_{\theta)}^{\ell m}}{\sin\theta}=
\partial^2_{\theta}Y^{{\ell m}}-\cot\theta\,\partial_{\theta}
	Y^{{\ell m}}-\frac{1}{\sin^2\theta}\partial^2_{\phi}
	Y^{{\ell m}}\,,\\ 
\label{Xlm}
 X^{{\ell m}}
	&\coloneqq-\sin\theta\left(\nabla_{\theta}S_{\theta}^{\ell m}-
	\frac{\nabla_\phi
	S_{\phi}^{\ell m}}{\sin^2\theta}\right)= 2\left(\partial^2_{\theta\phi}
	Y^{{\ell m}}-\cot\theta\partial_{\phi}Y^{{\ell m}}\right)\,,
\end{align}
and where the overbar stands for complex conjugation. 

We should note that even-parity functions can be found in the literature
under different notations. A particularly common choice is that proposed
by Moncrief in \cite{Moncrief74} for the even parity gauge-invariant
master function $Q^{({\rm e})}$ which is related to Zerilli function
(\ref{psi_e}) simply as $Q^{({\rm e})} = \Lambda \Psi^{({\rm e})}$, while
other authors use instead a master function defined as $Z\coloneqq
2{\Psi}^{({\rm e})}$ \cite{Martel04,Lousto97a}. Another even-parity
function can be introduced in terms of two new gauge-invariant metric
functions $k$ and $\chi$ are defined as \cite{Gundlach00b,Gundlach00c}
\begin{align}
k = \kappa_1 =&
	K+\frac{1}{e^{2b}}\left[r \partial_r G-
	\frac{2}{r}h_1^{({\rm e})}\right]\,, &
\\
\chi + k =& H_2 -\frac{2}{e^{2b}}\partial_rh_1^{(\rm e)}-
	\frac{2M}{r^2}h_1^{(\rm e)} 
        + \frac{1}{e^{2b}}\partial_r
        \left(r^2\partial_rG\right)+M\partial_r G\,, &
\end{align}
and such that
\begin{equation}
\kappa_2 = \frac{1}{2}e^{2b}\left(\chi-r\partial_r k+\frac{M}{r}e^{2b} k\right)\,.
\end{equation}
In this case, the Zerilli function (\ref{psi_e}) can be equivalently
defined as \cite{Gundlach00c}
\begin{equation}
\label{psi_e_gmg}
{\Psi}^{({\rm e})}\coloneqq \frac{2r^2}
	{\Lambda\left[(\Lambda-2)r+6M\right]e^{2b}}
	\left[\chi+\left(\frac{\Lambda}{2}+\frac{M}{r}\right)e^{2b} 
	k -r \partial_{r} k\right]\,.
\end{equation}

Finally, the homogeneous odd and even-parity master equations
(\ref{rw_eq}) and (\ref{zerilli_eq}) can be transformed into each other
by means of differential operations \cite{Chandrasekhar83}, and that they
are connected to the master equation that Bardeen and Press have derived
via the Newman--Penrose formalism \cite{Bardeen73a}.

\newpage
\section{Numerical Implementations of the Cauchy-Perturbative Approach}
\label{ae}

In the previous Chapter we have reviewed the derivation of the equations
describing the evolution of perturbations of nonrotating black holes
induced, for instance, by a nonzero stress-energy tensor. These
perturbations have been assumed to be generic in nature, needing to
satisfy only the condition of having a mass-energy much smaller than that
of the black hole. The solution of these equations with suitable initial
conditions completely specifies the reaction of the black hole to the
perturbations and this is essentially represented by the emission of
gravitational waves.

As mentioned in section \ref{gi_intro}, the importance of the
gauge-invariant variables used so far is that they are directly related
to the amplitude and energy of the gravitational-wave signal measured at
large distances. The purpose of this Chapter is to review the steps
necessary to obtain the relations between the master functions for the
odd and even-parity perturbations and the ``plus'' and ``cross''
polarisation amplitudes $h_+, h_\times$ of a gravitational wave in the TT
gauge. In practice, and following the guidelines tracked in
Refs. \cite{Cunningham78,Cunningham79}, we will derive an expression for
the perturbation metric $\boldsymbol{h}$ equivalent to that obtained in
the standard TT gauge on a Minkowski spacetime and relate it to the odd
and even-parity master functions $\Psi^{({\rm o})}$ and $\Psi^{({\rm
e})}$.

To obtain this result a number of conditions need to be met. First, we
need to evaluate each multipole of the decomposed metric perturbations in
the tetrad $\boldsymbol{e}_{\hat{\nu}}$ of stationary observers in the
background Schwarzschild spacetime, \ie $h_{\hat{\mu}\hat{\nu}} =
\boldsymbol{e}^{\mu}_{\hat{\mu}} \boldsymbol{e}^{\nu}_{\hat{\nu}}
h_{\mu\nu}$, where $\boldsymbol{e}$ is diagonal with components
\begin{equation}
\boldsymbol{e}_{\hat{\mu}}^{\mu}\coloneqq 
\left (e^b, e^{-b}, r^{-1}, (r \sin\theta)^{-1}\right)\,, 
\end{equation}
and where the indices ${\hat \mu}$ refer to the locally ``flat''
coordinates. Second, all of the quantities need to be evaluated far away
from the source (\ie in the ``wave-zone'') and in the so-called
\emph{radiation gauge}. In practice, this amounts to requiring that that
components $h_{\hat{\theta}\hat{\theta}}$, $h_{\hat{\phi}\hat{\phi}}$ and
$h_{\hat{\theta}\hat{\phi}}$ are functions of the type $f(t-r)/r$ ({\it
  \ie} they are outgoing spherical waves), while all the other components
have a more rapid decay of ${\cal O}(1/r^2)$. Finally, we need to impose
the condition that the metric is traceless modulo higher order terms,
\ie $h_{\hat\theta\hat\theta} + h_{\hat\phi\hat\phi}= 0 + {\cal
  O}(1/r^2)$.

In the following sections we will discuss the asymptotic expressions from
odd- and even-parity perturbations, and how to implement the
Cauchy-perturbative approach to extract gravitational-wave information
within a standard numerical-relativity code.

\subsection{Asymptotic expressions from odd-parity perturbations}
\label{sec:h_from_op}

We first consider odd-parity perturbations and recall that from the
radiation-gauge conditions and since for large $r$ the metric asymptotes
that of a flat spacetime, \ie $e^{b}\sim e^{-b}\sim 1$, we have
\begin{eqnarray}
\label{label:h0}
h_{\hat{\theta}\hat{t}}^{({\rm o})}& =
\displaystyle
	\frac{h_0^{({\rm o})}}{r}e^{b}S_{\theta}\sim 
	\frac{h_0^{({\rm o})}}{r}\sim
	{\cal O}\left(\frac{1}{r^2}\right)
	\;\longrightarrow h_0^{({\rm o})}\sim
	{\cal O}\left(\frac{1}{r}\right)\,,\\
\label{label:h1}
h_{\hat{\theta}\hat{r}}^{({\rm o})}& =
\displaystyle
	\frac{h_1^{({\rm o})}}{r}e^{b}S_{\theta}\sim 
	\frac{h_1^{({\rm o})}}{r}\sim {\cal O}\left(\frac{1}{r^2}\right)
	\;\longrightarrow h_1^{({\rm o})}\sim 
	{\cal O}\left(\frac{1}{r}\right)\,,
\end{eqnarray}
where the $\ell, m$ indices have been omitted for clarity. Similarly,
since $h_{\hat{\theta}\hat{\theta}}^{({\rm o})} = 2h r^{-2}
\nabla_{\theta} S_{\theta}\sim {\cal O}(1/r)$, we can deduce that $h\sim
      {\cal O}(r)$, so that the only components of the metric having
      wave-like properties at large $r$ are
\begin{eqnarray}
\label{odd:hplus}
h_{+}^{({\rm o})} &\coloneqq
	&\frac{1}{2}\left(h_{\hat{\theta}\hat{\theta}}^{({\rm
	o})}-h_{\hat{\phi}\hat{\phi}}^{({\rm o})}\right)
	=\frac{h}{r^2}\left(\nabla_{\theta}S_{\theta}-
	\frac{\nabla_{\phi}S_{\phi}}{\sin^2\theta}\right)+
	{\cal O}\left(\frac{1}{r^2}\right)\,, \\
\label{odd:hcross}
h_{\times}^{({\rm o})}&\coloneqq & h_{\hat{\theta}
	\hat{\phi}}^{({\rm o})}=\frac{h}{r^2}
	\frac{\nabla_{(\phi}S_{\theta)}}{\sin\theta}+
	{\cal O}\left(\frac{1}{r^2}\right)\,.
\end{eqnarray}
Note that since $h$ has the dimensions of a length squared, both $h_{+}$
and $h_{\times}$ are dimensionless. Next, we need to relate the
perturbation $h$ to the odd-parity master function $\Psi^{({\rm o})}$. To
do so, we follow the procedure outlined in Ref. \cite{Cunningham78}, and
note that (cf.\  Eq.~(III-20)\footnote{We recall that in the
notation of Ref. \cite{Cunningham78} $\widetilde{\psi}
= \Lambda(\Lambda-2)\Psi^{(\rm o)}$, and the multipoles
in \cite{Cunningham78} are related to ours as $\widetilde{h}_2 = h_2 =
-2h$, $\widetilde{h}_{0} = h^{(\rm o)}_{0}$ and $\widetilde{h}_{1} =
h^{(\rm o)}_{1}$})
\begin{equation}
\label{eq:III-20}
  \partial_t h =\left(1-\frac{2M}{r}\right)\partial_r\left(r\Psi^{(\rm o)}\right)
  +h_0^{(\rm o)}\,.
\end{equation}
Equation~(\ref{eq:III-20}) represents one of the Hamilton equations as
derived by Moncrief in a Hamiltonian formulation of perturbation
equations \cite{Moncrief74}. The radiation-gauge conditions on $h$ and
$h_0^{(\rm o)}$ imply that $\Psi^{({\rm o})}\sim {\cal O}(1)$, \ie in the
wave-zone $\Psi^{({\rm o})}$ has the dimensions of a length, behaves as
an outgoing-wave, but it does not depend explicitly on $r$.

Exploiting now the outgoing-wave behaviour of $h$ at large distances we
can write
\begin{equation}
\label{h_outgoing}
\partial_t h=-\partial_r h+{\cal O}\left(\frac{1}{r}\right)\,,
\end{equation}
so that asymptotically Eq.~(\ref{eq:III-20}) simply becomes
\begin{equation}
\partial_r h=-\partial_r\left(r\Psi^{(\rm o)}\right)+{\cal O}\left(\frac{1}{r}\right)\,,
\end{equation}
and its integration yields
\begin{equation}
\frac{h}{r}\sim -\Psi^{({\rm o})}+{\cal O}\left(\frac{1}{r}\right)\,.
\end{equation}
As a result, the ``$+$'' and ``$\times$'' polarisation amplitudes of the
gravitational wave can be calculated from
Eqs. (\ref{odd:hcross})--(\ref{odd:hplus}) as
\begin{align}
\label{odd:plus_b}
h_{+}^{({\rm o})} &
	=-\frac{1}{r}\Psi^{({\rm o})}\;
	\left(\nabla_\theta S_{\theta}-
	\frac{\nabla_\phi S_{\phi}}{\sin^2\theta}\right)+
	{\cal O}\left(\frac{1}{r^2}\right)\,, \\
\label{odd:cross_b}
h_{\times}^{({\rm o})}&
	=-\frac{1}{r}
	\Psi^{({\rm o})}\;
	\frac{\nabla_{(\phi}S_{\theta)}}
	{\sin\theta}+{\cal O}\left(\frac{1}{r^2}\right)\,.
\end{align}

Expressions (\ref{odd:plus_b}) and (\ref{odd:cross_b}) can be written in
a compact form using the $s=-2$ spin-weighted spherical harmonics (see
also Appendix \ref{app:vtsh})
\begin{equation}
\label{eq:spin_harmonic}
_{_{-2}}Y^{{\ell m}}(\theta,\phi)\coloneqq\sqrt{\frac{(\ell-2)!}{(\ell+2)!}}
	\left(W^{{\ell m}}-{i}\;
	\frac{X^{{\ell m}}}{\sin\theta}\right)\,,
\end{equation}
so that expressions (\ref{odd:plus_b}) and (\ref{odd:cross_b}) can be
combined into a single complex expression given by
\begin{equation}
\label{h+_hx_Psio}
\left(h^{({\rm o})}_+ -
	{i}h^{({\rm o})}_{\times}\right)_{{\ell m}}=
	\frac{{i}}{r}\;\sqrt{\frac{(\ell+2)!}{(\ell-2)!}}\;
	\Psi^{({\rm o})}_{{\ell m}}\;_{_{-2}}Y^{{\ell m}}(\theta,\phi) 
	+ {\cal O}\left(\frac{1}{r^2}\right) \,,
\end{equation}
where, for clarity, we have explicitly restored the multipole indices
$\ell, m$.

\subsubsection{The master function $Q^{({\rm o})}$}
As discussed in Sec.~\ref{opp}, the odd-parity metric perturbations
are sometimes expressed in terms of the odd-parity Moncrief function
$Q^{({\rm o})}$ [cf.\ Eq. (\ref{mfop})]; indeed it is not unusual to
find in the literature the gravitational-wave amplitudes expressed in
terms of this quantity. However, great care must be paid to the
asymptotic relation between the master function $Q^{({\rm o})}$ and the
gravitational-wave amplitudes and, indeed, this is sometimes a source of
confusion \cite{Kawamura04,kawamura03a}. To clarify this point, we recall
that the derivation of the asymptotic relation between $Q^{({\rm o})}$
and $h$ proceeds in a way similar to the one discussed above. In the
radiation gauge and at large distances from the black hole, we can use
relation (\ref{h_outgoing}) in the definition (\ref{mfop}) with
$h_2=-2h$, so that
\begin{equation}
\label{Q(o)_vs_h}
Q^{({\rm o})}\sim \frac{1}{r} \partial_t h + 
	{\cal O}\left(\frac{1}{r}\right)\,,
\end{equation}
which is also a dimensionless quantity. Since $h \sim {\cal O}(r)$, the
function $Q^{({\rm o})}$ does not depend on $r$ at leading order and
Eq. (\ref{Q(o)_vs_h}) can be integrated to give
\begin{equation}
\label{ht_over_r}
\frac{h(t)}{r}\sim \int_{-\infty}^{t} Q^{({\rm o})}(t') dt' 
	+ {\cal O}\left(\frac{1}{r}\right) + \mathrm{const.}\,,
\end{equation}
where the integration constant can be defined as 
\begin{equation}
\mathrm{const.}\coloneqq \lim_{t\rightarrow - \infty} \frac{h(t,r)}{r}\sim
	{\cal O}(1)\,,
\end{equation}
and it can be set to zero in the case of asymptotically flat
metric perturbations ($h=0$) at earlier times. Combining now expressions
(\ref{Wlm}), (\ref{Xlm}) and (\ref{ht_over_r}), the gravitational-wave
amplitudes in the two polarisations and with the new master function read
\begin{equation}
\label{h+_hx_Qo}
 \left(h^{({\rm o})}_+
	-{i}h^{({\rm o})}_{\times}\right)_{{\ell m}}=
	-\frac{{i}}{r}\sqrt{\frac{(\ell+2)!}{(\ell-2)!}}
	\left(\int_{-\infty}^{t} Q^{({\rm o})}_{{\ell m}}(t')dt'
	\right)\;_{_{-2}}Y^{{\ell m}}(\theta,\phi)+
	{\cal O}\left(\frac{1}{r^2}\right)\,.
\end{equation}

Note that in expressions (\ref{h+_hx_Psio}) and (\ref{h+_hx_Qo}) the
quantities $\Psi^{({\rm o})}$ and $Q^{({\rm o})}$ are both solutions of
the Regge--Wheeler equation (\ref{rw_eq}), but they yield two different
asymptotic expressions for the gravitational-wave amplitudes. This
difference, which is consistent with Eq.~(\ref{Q_vs_Psi}) when evaluated
in a an asymptotic region of the spacetime where $L_1^{\ell m}=0$, is
subtle but important and, as mentioned above, it has led to some
inconsistencies in the literature both for the determination of the
asymptotic gravitational-wave amplitudes and for the energy losses. This
will be further discussed in Sects.~\ref{a_general} and
\ref{e_am_losses}.

\subsection{Asymptotic expressions from even-parity perturbations}
\label{sec:h_from_ep}

	A calculation conceptually analogous to the one carried out in
Sect.~\ref{sec:h_from_op} leads to the relation between the
gravitational-wave amplitude and the even-parity master function. In
particular, after projecting Eq.~(\ref{even:metric}) along the stationary
tetrad, the asymptotic wave amplitudes in the two polarisation states are
\begin{align}
 h_{+}^{({\rm e})}&=\frac{1}{2}
	\left(h_{\hat{\theta}\hat{\theta}}^{({\rm e})}-
	h_{\hat{\phi}\hat{\phi}}^{({\rm e})}\right)=
	\frac{G}{2}\left(\nabla_\theta\nabla_\theta
	Y^{{\ell m}}-\frac{\nabla_\phi\nabla_\phi
	Y^{{\ell m}}}{\sin^2\theta}\right)=\frac{G}{2}\;W^{{\ell m}}\,,\\ 
 h_{\times}^{({\rm e})}&=h_{\hat{\theta}\hat{\phi}}^{({\rm
	e})}=G\,
	\frac{\nabla_\theta\nabla_\phi Y^{{\ell m}}}{\sin\theta}
	=\frac{G}{2}\;\frac{\;\;\;X^{{\ell m}}}{\sin\theta}\,,
\end{align}
so that we essentially need to relate the metric perturbation $G$ with the
even-parity function ${\Psi}^{({\rm e})}$. Firstly, it is easy to realize
that the even-parity metric projected onto the tetrad,
$h_{\hat\mu\hat\nu}^{({\rm e})}$, is such that 
\begin{equation}
H_2\sim {\cal O}\left(\frac{1}{r^2}\right)\,,  \qquad {\rm and} \qquad 
h_{1}^{({\rm e})} \sim {\cal O}\left(\frac{1}{r}\right) \,, 
\end{equation}
so that the terms proportional to these multipoles are of higher order
for large $r$ and can be neglected. Furthermore, from the transverse
traceless condition
\begin{equation}
h_{\hat{\theta}\hat{\theta}}^{({\rm e})}+
	h_{\hat{\phi}\hat{\phi}}^{({\rm e})}= 0 + 
	{\cal O}\left(\frac{1}{r^2}\right) \,,
\end{equation}
we obtain an asymptotic relation between the gauge-invariant functions
$K$ and $G$
\begin{equation}
\label{K_vs_G_1}
 2KY^{{\ell m}}+G\left(\nabla_\theta\nabla_\theta
	Y^{{\ell m}}+\frac{\nabla_\phi\nabla_\phi 
	Y^{{\ell m}}}{\sin^2\theta}\right)
	=\left(2K-G\Lambda\right)Y^{{\ell m}}\sim 
	{\cal O}\left(\frac{1}{r^2}\right)\,,
\end{equation}
where we have used the definition (\ref{def:harmonics}) to derive the
right-hand side of expression (\ref{K_vs_G_1}). As a result, the
asymptotic relation between the two components of the even-parity part of
the perturbation metric simply reads
\begin{equation}
K\sim\frac{\Lambda}{2}\,G+{\cal O}\left(\frac{1}{r^2}\right)\,.
\end{equation}
Using now the definitions (\ref{kappa1})--(\ref{kappa2}), we have that
asymptotically
\begin{align}
\kappa_1 \sim &\, \frac{\Lambda}{2}\,G + r \partial_r G+
	{\cal O}\left(\frac{1}{r^2}\right)\\ 
\kappa_2 \sim &\,
	-\frac{1}{2}\left(K+r \partial_r K\right)\sim
	-\frac{\Lambda}{4}\left(G+r \partial_r G\right)+
	{\cal O}\left(\frac{1}{r^2}\right)\,,
\end{align}
and their linear combination~(\ref{def:q1}) becomes
\begin{equation}
q_1\sim \frac{rG}{2}\;\Lambda\left(\Lambda-2\right)+ 
	{\cal O}\left(\frac{1}{r}\right)\,.
\end{equation}

	Finally, the asymptotic gauge-invariant even-parity master
function reads
\begin{equation}
{\Psi}^{({\rm e})}\sim \frac{r q_1}{\Lambda\left[r(\Lambda-2)+
	6M\right]}\sim \frac{1}{2}\,rG +{\cal O}\left(\frac{1}{r}\right)\,,
\end{equation}
so that, modulo higher-order terms, the even-parity gravitational-wave
amplitudes measured by a distant observer can be written in the compact
form
\begin{equation}
\left(h^{({\rm e})}_{+}-{i}h^{({\rm e})}_{\times}\right)_{{\ell m}}=
	\frac{1}{r}\sqrt{\frac{(\ell-2)!}{(\ell+2)!}} 
	{\Psi}^{({\rm e})}_{{\ell m}}\;_{_{-2}}Y^{{\ell m}}(\theta,\phi)+
	{\cal O}\left(\frac{1}{r^2}\right)\,.
\end{equation}

\subsection{Asymptotic general expressions}
\label{a_general}

It is often convenient to combine the expressions for the asymptotic
gravitational-wave amplitudes related to odd and even-parity
perturbations into the single expression
\begin{equation}
\label{hp_hc}
h_{+}-{i}h_{\times}=\frac{1}{r}\sum_{\ell,m}
	\sqrt{\frac{(\ell+2)!}{(\ell-2)!}}
	\left({\Psi}^{({\rm e})}_{{\ell m}}+
	{i}{\Psi}_{{\ell m}}^{({\rm o})}\right)
	\;_{_{-2}}Y^{{\ell m}}(\theta,\phi)+
	{\cal O}\left(\frac{1}{r^2}\right)\,,
\end{equation}
or, equivalently 
\begin{eqnarray}
\label{hp_hc_mf}
 h_+-{i}h_{\times}=\frac{1}{r}\sum_{\ell,m}
	\sqrt{\frac{(\ell+2)!}{(\ell-2)!}}
	\left({\Psi}^{({\rm e})}_{{\ell m}}-{i}
	\int_{-\infty}^tQ^{({\rm o})}_{{\ell m}}(t')dt'
	\right)\,_{_{-2}}Y^{{\ell m}}(\theta,\phi)
	+ {\cal O}\left(\frac{1}{r^2}\right)\,,
\nonumber \\
\end{eqnarray}
where we have defined $h_{+}\coloneqq h^{({\rm o})}_{+} + h^{({\rm e})}_{+}$
and $h_{\times}\coloneqq h^{({\rm o})}_{\times} + h^{({\rm
e})}_{\times}$. Note that $X^{{\ell 0}}=0$ for any value of $\ell$, so
that in the case of axisymmetry the gravitational-wave signal is
proportional to $W^{\ell 0}$ only.

It is also useful to underline that while expression (\ref{hp_hc})
resembles the corresponding expression (10) of Ref. \cite{Kawamura04}, it
is indeed different. Firstly, because in Ref. \cite{Kawamura04} the
Moncrief function is adopted for the odd-parity part of the perturbations
and hence, modulo a normalisation factor, the function $\Psi^{({\rm o})}$
appearing there corresponds to our function $Q^{({\rm o})}$ [cf.\
expression (\ref{mfop})]. Secondly, because with this choice for the
odd-parity perturbations a time derivative is needed in the asymptotic
expression for the gravitational-wave amplitudes [cf.\  the
discussion in the derivation of Eq. (\ref{h+_hx_Qo})]. As a result,
expression (10) of Ref. \cite{Kawamura04} (which is also missing the
distinction between the real and imaginary parts) should really be
replaced by expression (\ref{hp_hc_mf}). A similar use of the Moncrief
function for the odd-parity part is present also in
Refs. \cite{Shibata:2003ga,Shibata03d,Shibata:2004kb}, where it is
employed to calculate the gravitational-wave content of numerically
simulated spacetimes.

\subsection{Energy and angular momentum losses}
\label{e_am_losses}

Using the expressions derived in the previous sections we can now
estimate the energy and angular momentum losses due to gravitational
waves propagating outwards to spatial infinity. More specifically, this
can be done by using expression (\ref{hp_hc}) and the definition of
Isaacson's stress-energy pseudo-tensor $\tau_{\mu \nu}$ for the
gravitational-wave field ${\boldsymbol h}$ propagating in the curved
background field ${\boldsymbol g}{\!\!\!\!\;\raisebox{-0.1ex}{$^{^0}$}}$
and in a Lorentz gauge \cite{Isaacson68,Landau-Lifshitz2}
\begin{equation}
\label{gw_set_c}
\tau_{\mu \nu} \coloneqq \frac{1}{32 \pi}
	\left\langle 
	\nabla_{\mu} {h}_{\alpha \beta} 
	\nabla_{ \nu} {h}^{\alpha \beta}
	\right\rangle \,,
\end{equation}
where the brackets $\langle \ldots \rangle$ refer to a spatial average
over a length much larger than the typical gravitational
wavelength \cite{Isaacson68,MTW1973}. The averaged expression
(\ref{gw_set_c}) is gauge-invariant \cite{Isaacson68} and holds in the
``limit of high frequency'' (or \emph{short-wave} approximation), {\it
  \ie}  whenever the wavelength of the gravitational-wave field is small
when compared to the local radius of curvature of the background
geometry. In practice, gravitational radiation from isolated systems is
of high frequency whenever it is far enough away from its source.

Expression (\ref{gw_set_c}) accounts for the amount of energy and
momentum carried by the gravitational wave over a certain region of
spacetime, but since we are interested in the energy flux as measured by
an inertial observer, we need to project the pseudo-tensor on the
observer's locally orthonormal tetrad, where it becomes
\begin{equation}
\label{gw_set}
\tau_{\hat\mu \hat\nu} \coloneqq \frac{1}{32 \pi}
	\left\langle 
	\partial_{\hat\mu} {\bar h}_{\hat\alpha \hat\beta} 
	\partial_{\hat\nu} {\bar h}^{\hat\alpha \hat\beta}
	\right\rangle \,,
\end{equation}
with $\bar{h}_{\hat\mu\hat\nu} \coloneqq h_{\hat\mu\hat\nu} - \frac{1}{2}h
\eta_{\hat\mu\hat\nu}$ and $h$ being now the trace of
$h_{\hat\mu\hat\nu}$. As a result, the energy per unit time and angle
carried by the gravitational waves and measured by a stationary observer
at large distance is given by
\begin{equation}
\label{dedtdom}
\frac{d^2E}{dtd\Omega}=\frac{r^2}{16\pi}
	\left[ \left(\frac{d {h}_{\hat{\theta}\hat{\phi}}}{dt}\right)^2+
	\frac{1}{4}\left(\frac{d h_{\hat{\theta}\hat{\theta}}}{dt}-
	\frac{d h_{\hat{\theta}\hat{\phi}}}{dt}\right)^2\right]=
	\frac{r^2}{16\pi}\left(\left|\frac{d {h}_+}{dt}\right|^2
	+\left|\frac{d h_{\times}}{dt}\right|^2\right) \,, 
\end{equation}
where the total derivative is made with respect to the asymptotic
observer's time. Integrating (\ref{dedtdom}) over the solid angle, the
total power emitted in gravitational waves is then given by
\begin{eqnarray}
\label{power}
&& \frac{dE}{dt} =  
	\frac{1}{16\pi}\sum_{\ell,m}\frac{(\ell+2)!}{(\ell-2)!}
	\left(\left|\frac{d {\Psi}^{({\rm e})}_{{\ell m}}}{dt}\right|^2
	+\left|\frac{d {\Psi}^{({\rm o})}_{{\ell m}}}{dt}\right|^2\right) 
	\,, \\
\label{power:CPM}
&& \hskip 0.6cm	
	= \frac{1}{16\pi}\sum_{\ell,m} \left(
	\Lambda(\Lambda-2)
	\left|\frac{d {\Psi}^{({\rm e})}_{{\ell m}}}{dt}\right|^2
	+\frac{\Lambda}{\Lambda-2}
	\left|\frac{d {\Phi}^{({\rm o})}_{{\ell m}}}{dt}
	\right|^2\right)\,,
\end{eqnarray}
where expression (\ref{power:CPM}) was first presented in
Refs. \cite{Cunningham78,Cunningham79}.

Note that as discussed at the end of section~\ref{a_general}, these
expressions need to be suitably modified when the energy losses are
expressed in terms of the odd-parity Moncrief function $Q^{({\rm o})}$,
in which case the energy-loss rate needs to be modified as
\begin{equation}
\frac{dE}{dt} = \frac{1}{16\pi}\sum_{\ell,m}\frac{(\ell+2)!}{(\ell-2)!}
	\left(\left|\frac{d {\Psi}^{({\rm e})}_{{\ell m}}}{dt}\right|^2+ 
	\left|Q_{{\ell m}}^{({\rm o})}\right|^2\right) \,.
\end{equation}
Similarly, the angular momentum flux carried away in the form of
gravitational waves can also be calculated in terms of the
energy-momentum tensor (\ref{gw_set}). In particular, using spherical
coordinates and assuming that the rotation is parametrised by the angle
$\phi$, we have
\begin{equation}
\label{dJdtdom}
 \frac{d^2J}{dtd\Omega}=\frac{r^2}{32\pi}
	\left\langle \partial_{\phi}\bar{h}_{\hat{\mu}\hat{\nu}}
	\partial_r\bar{h}^{\hat{\mu}\hat{\nu}}\right\rangle 
	= -\frac{r^2}{16\pi}
	\left\langle \partial_rh_{\hat\theta\hat\theta}
 	\partial_\phi h_{\hat\theta\hat\theta}
	+\partial_rh_{\hat\theta\hat\phi}\partial_{\phi}
	h_{\hat\theta\hat\phi}\right\rangle \,.
\end{equation}
Since the metric components in the radiation-gauge behave like outgoing
spherical waves and since $h_{\hat\theta\hat\theta}=h_+$ and
$h_{\hat\theta\hat\phi}=h_{\times}$, the angular momentum carried away in
the form of gravitational waves (\ref{dJdtdom}) is then expressed as
\begin{equation}
\frac{d^2J}{dtd\Omega}=-\frac{r^2}{16\pi}\left(\partial_th_+\partial_\phi 
	\bar{h}_+ + \partial_t h_{\times}\partial_{\phi}\bar{h}_{\times}\right)\,,
\end{equation}
Proceeding in a way similar to the one followed in the calculation of the
emitted power, the total angular momentum lost per unit time to
gravitational wave reads \cite{Martel:2005ir}
\begin{equation}
\label{dJdt_CPM}
\frac{dJ}{dt}=\frac{1}{16\pi}\sum_{\ell,m} i m
	\frac{(\ell+2)!}{(\ell-2)!}
	\left[\frac{d\Psi^{(\rm e)}_{\ell m}}{dt}
	\left(\bar{\Psi}_{\ell m}^{({\rm e})}\right)+
	\frac{d\Psi^{(\rm o)}_{\ell m}}{dt}
	\left(\bar{\Psi}_{\ell m}^{({\rm o})}\right)\right]\,,
\end{equation}
or, using the Moncrief master function (\ref{mfop}) for the odd-parity
perturbations \cite{Poisson04b} 
\begin{equation}
\label{dJdt_RWM}
\frac{dJ}{dt}=\frac{1}{16\pi}\sum_{\ell, m} i m
	\frac{(\ell+2)!}{(\ell -2)!}
	\left[\frac{d\Psi^{(\rm e)}_{\ell m}}{dt}
	\left(\bar{\Psi}_{\ell m}^{({\rm e})}\right)+
	Q_{\ell m}^{(\rm o)}\int_{-\infty}^{t}
	\left(\bar{Q}_{\ell m}^{(\rm o)}\right)(t')dt'\right]\,.
\end{equation}

	To conclude, we report the expression for the energy spectrum
${dE}/{d\omega}$, which is readily calculated from Eq. (\ref{power})
after performing the Fourier transform of the odd and even-parity master
functions, \ie
\begin{equation}
\label{dedom}
\frac{dE}{d\omega}=\frac{1}{16\pi^2}\sum_{\ell,m}
	\frac{(\ell+2)!}{(\ell-2)!}\;\, \omega^2\left(
	\left|\widetilde{\Psi}^{({\rm e})}_{{\ell m}}\right|^2
	+\left|\widetilde{\Psi}^{({\rm o})}_{{\ell m}}\right|^2\right)\,,
\end{equation}
where we have indicated with $\widetilde{f}(\omega,r)$ the Fourier
transform of the timeseries $f(t,r)$. Similarly, when using the
odd-parity Moncrief function one obtains
\begin{equation}
\label{en_spectrum}
\frac{dE}{d\omega}=\frac{1}{16\pi^2}\sum_{\ell,m}
	\frac{(\ell+2)!}{(\ell-2)!}\;\,
	\left(\omega^2\left|\widetilde{\Psi}^{({\rm e})}_{{\ell m}}\right|^2
	+\left|\widetilde{Q}^{({\rm o})}_{{\ell m}}
	\right|^2\right) \,.
\end{equation}

\subsection{A commonly used convention}
\label{sec:RW}

A rather popular choice for the gauge-invariant master functions has
found successful application in the extraction of the gravitational-wave
content of numerically simulated spacetimes \cite{Abrahams95b,
Abrahams95c, Abrahams97a, Rupright98, Rezzolla99a}. For instance, the
convention discussed below has been implemented in the {\tt Cactus}
computational toolkit \cite{Camarda99,Allen98a}, a diffused and freely
available infrastructure for the numerical solution of the Einstein
equations \cite{Allen99a,CactuswebLR}. Numerous tests and applications of
this implementation have been performed over the years and we refer the
reader to Refs. \cite{Camarda99,Allen98a,Font02c,Baiotti04b} for examples
both in vacuum and non-vacuum spacetimes.

The reference work for this convention in the one by Abrahams and
Price \cite{Abrahams95b,Abrahams95c}, although a similar approach for the
even-parity part of the perturbations was also adopted in previous
works \cite{Abrahams92a,Anninos95g}. We first note that the coefficients
$c_0$, $c_1$ and $c_2$ introduced in Refs. \cite{Abrahams95b,Abrahams95c}
are related simply to the multipolar coefficients of the odd-parity part
introduced in section~\ref{sec:multipole}. More specifically, considering
that $c_2 = - 2h = h_2$, $c_0 = h_0^{({\rm o})}$, and $c_1 = h_1^{({\rm
o})}$, it is then easy to realise that the odd and even-parity master
functions $Q^{\times}_{{\ell m}}$ and $Q^{+}_{{\ell m}}$ defined in
Refs. \cite{Abrahams95b,Abrahams95c} are related to the master functions
discussed so far through the simple algebraic expressions
\begin{eqnarray}
\label{qc}
&& Q^{\times}_{{\ell m}} \coloneqq \sqrt{\frac{2(\ell+2)!}{(\ell-2)!}}\,
	Q^{({\rm o})}_{{\ell m}}\,, 
\\ 
\label{qp}
&& Q^{+}_{{\ell m}} \coloneqq \sqrt{\frac{2(\ell+2)!}{(\ell-2)!}}\,
	{\Psi}^{({\rm e})}_{{\ell m}}\,,
\end{eqnarray}
so that the asymptotic expression for the gravitational-wave amplitudes
in the two polarisations are given by
\begin{eqnarray}
\label{hp_ap}
 h_+=\frac{1}{\sqrt{2}r}\sum_{\ell,m}
	\sqrt{\frac{(\ell-2)!}{(\ell+2)!}}\,
	\left[Q^+_{{\ell m}}W^{{\ell m}}-\left(\int_{-\infty}^{t}
	Q^{\times}_{{\ell m}}(t')dt'\right)\frac{\;\;\;
	X^{{\ell m}}}{\sin\theta}\right]
	+ {\cal O}\left(\frac{1}{r^2}\right)\,, \\
\label{hc_ap}
 h_{\times}=\frac{1}{\sqrt{2}r}\sum_{\ell,m}
	\sqrt{\frac{(\ell-2)!}{(\ell+2)!}}\,
	\left[Q^+_{{\ell m}}\frac{\;\;\;X^{{\ell m}}}{\sin\theta}+
	\left(\int_{-\infty}^{t}Q^{\times}_{{\ell m}}(t')dt'\right)
	W^{{\ell m}}\right]
	+ {\cal O}\left(\frac{1}{r^2}\right)\,.
\end{eqnarray}
Similarly, expressions (\ref{hp_ap}) and (\ref{hc_ap}) can be combined
into a single one
\begin{equation}
\label{eq:wave_gi}
 h_{+}-{i}h_{\times}=\frac{1}{\sqrt{2}r}\sum_{\ell,m}\left(
	Q^{+}_{{\ell m}} - {i}\int_{-\infty}^{t}
	Q^{\times}_{{\ell m}}(t')dt'\right)
	\;_{_{-2}}Y^{{\ell m}}(\theta,\phi)
	+ {\cal O}\left(\frac{1}{r^2}\right)\,,
\end{equation}
which closely resembles expression (\ref{hp_hc_mf}) and that in its
compactness highlights the advantage of the normalisation
(\ref{qc})--(\ref{qp}). We should remark that the notation in
Eq.~(\ref{eq:wave_gi}) could be misleading as it seems to suggest that
$h_\times$ is always of odd-parity and $h_+$ is always of
even-parity. Indeed this is not true in general and in the absence of
axisymmetry, \ie when $m\neq0$, both $h_{\times}$ and $h_+$ are a
superposition of odd and even parity modes. It is only for axisymmetric
systems, for which only $m=0$ modes are present, that $Q^\times_{\ell m}$
and $Q^+_{\ell m}$ are \emph{real} numbers, that $h_+$ is \emph{only}
even-parity and $h_\times$ is \emph{only} odd-parity.

Also very compact is the expression for the emitted power that, with this
convention, simply reads
\begin{equation}
\label{p_ap}
\frac{dE}{dt}=\frac{1}{32\pi}\sum_{\ell,m}\left(\left|
	\frac{d {Q}^{+}_{{\ell m}}}{dt}\right|^2+
	\left|Q^{\times}_{{\ell m}}\right|^2\right)\,.
\end{equation}

Finally, the flux of linear momentum emitted in gravitational waves in
the $i$-direction can be computed from the Isaacson's energy-momentum
tensor and can be written in terms of the two polarization amplitudes
as \cite{Favata:2004wz}
\begin{equation}
  \label{eq:linmom_flux}
  {\cal F}_i\coloneqq \dot{P_i}=\dfrac{r^2}{16\pi}\int d\Omega\;
   n_i\left(\dot{h}_+^2+\dot{h}_{\times}^2\right) \,,
\end{equation}
where $n_i=x_i/r$ is the unit radial vector that points from the source
to the observer. The calculation of this flux in terms of $Q^+_{\ell m}$
and $Q^\times_{\ell m}$ can be computed after inserting
Eq.~(\ref{eq:wave_gi}) in Eq.~(\ref{eq:linmom_flux}), decomposing $n_i$
in spherical harmonics and performing the angular integral. This
procedure goes along the lines discussed by Thorne in
Ref. \cite{Thorne80b}, where all the relevant formulae are essentially
available [\textit{cf.} Eq.~(4.20) there, but see also
Refs. \cite{Sopuerta:2006wj,Pollney:2007ss}], so that we only need to
adapt them to our notation. More specifically, in Ref. \cite{Thorne80b}
the even-parity (or \emph{electric}) multipoles are indicated with
$I_{\ell m}$ and the odd-parity (or \emph{magnetic}) ones with $S_{\ell
m}$, and are related to our notation by
\begin{align}
^{(\ell)}I_{\ell m}  \coloneqq &\, Q^+_{\ell m} \,, \\
^{(\ell+1)}S_{\ell m} \coloneqq &\, Q_{\ell m}^\times\,,
\end{align}
where $^{(\ell)}f_{\ell m}\coloneqq d^{\ell} f_{\ell m}/dt^{\ell}$. From
the property $(Q^{+,\times}_{\ell m})^* = (-1)^mQ_{\ell\,-m}^{+,\times}$,
where the asterisk indicates complex conjugation, we rewrite Eq.~(4.20)
of Ref. \cite{Thorne80b} in a more compact form. Following
Ref. \cite{Damour-Gopakumar-2006} where the lowest multipolar
contribution was explicitly computed in this way, it is convenient to
combine the components of the linear momentum flux in the equatorial
plane in a complex number as ${\cal F}_x+{\rm i}{\cal F}_y$. The
multipolar expansion of the flux vector can be written
as \cite{Pollney:2007ss}
\begin{align}
\label{eq:recoil}
{\cal F}_x+ i {\cal F}_y &=
\sum_{\ell=2}^{\infty}\sum_{m=0}^{\ell}\delta_m\left({\cal
  F}_{x}^{\ell m} +  i {\cal F}_y^{\ell m}\right) \,, \\
{\cal F}_z &= \sum_{\ell=2}^{\infty}\sum_{m=0}^{\ell}\delta_m {\cal
  F}_{z}^{\ell m} \,,
\end{align}
where $\delta_m=1$ if $m\neq0$ and $\delta_m=1/2$ if $m=0$. A more
extended representation in terms of the various multipoles reads
\begin{align}
    \label{eq:gi}
{\cal F}_x^{\ell m}+ i {\cal F}_y^{\ell m} \coloneqq 
  \dfrac{(-1)^m}{16\pi\ell(\ell +1)}\Bigg\{& -2 i \bigg[a_{\ell
  m}^+ \dot{Q}^+_{\ell-m}Q_{\ell\,m-1}^\times +a_{\ell
  m}^-\dot{Q}^+_{\ell m} Q^\times_{\ell\;-(m+1)}\bigg] + \nonumber \\
 & \hskip -1.0cm \sqrt{\dfrac{\ell^2(\ell-1)(\ell+3)}{(2\ell+1)(2\ell+3)}}
   \bigg[ 
    b_{\ell m}^- \left(\dot{Q}^+_{\ell\;-m}\dot{Q}^+_{\ell+1\;m-1} +
    Q_{\ell\;-m}^\times\dot{Q}_{\ell +1\;m-1}^\times \right) + \nonumber \\
    & \hskip 2.0cm b_{\ell m}^+\left(\dot{Q}^+_{\ell m}\dot{Q}^+_{\ell
        +1\;-(m+1)}+Q_{\ell
        m}^\times\dot{Q}^\times_{\ell+1\;-(m+1)}\right)\bigg]
  \Bigg\} \,, \\
\label{eq:kick_z}
{\cal F}^{\ell m}_z 
 \coloneqq \dfrac{(-1)^m}{8\pi\ell(\ell+1)}
\bigg\{ & 2m\; \Im \left[\dot{Q}_{\ell\,-m}^+Q_{\ell
    m}^{\times}\right]+ \nonumber \\
& c_{\ell m}\sqrt{\dfrac{\ell^2(\ell-1)(\ell+3)}{(2\ell+1)(2\ell+3)}}
\Re \left[\dot{Q}_{\ell\,-m}^+Q^+_{\ell+1\,m}+
Q^\times_{\ell\,-m}\dot{Q}^\times_{\ell+1\,m}\right]\bigg\} \,,
\end{align}
where
\begin{eqnarray}
&&a_{\ell m}^{\pm}\coloneqq\sqrt{(\ell \pm m)(\ell \mp m+1)} \,, \\
&&b_{\ell m}^{\pm}\coloneqq\sqrt{(\ell \pm m+1)(\ell \pm  m+2}) \,, \\
&&c_{\ell m}\coloneqq\sqrt{(\ell - m+1)(\ell - m+1}) \,.
\end{eqnarray}
Note that here both ${\cal F}_x^{\ell m}$ and ${\cal F}_y^{\ell m}$
are \emph{real} numbers and are obtained as the real and imaginary part
of the right-hand side of Eq.~(\ref{eq:gi}).

\subsection{Implementation summary}
\label{s-CPIS}

All of the material presented in the previous sections about the
gauge-invariant description of the perturbations of a Schwarzschild black
hole has laid the ground for the actual implementation of the
Cauchy-perturbative extraction method in numerical-relativity
calculations. We recall that the goal of the Cauchy-perturbative method
is that of replacing, at least in parts of the three-dimensional
numerical domain, the solution of the full nonlinear Einstein's equations
with the solution of a set of simpler linear equations that can be
integrated to high accuracy with minimal computational cost. In turn,
this provides an unexpensive evolution of the radiative degrees of
freedom, the extraction of the gravitational-wave information, and, if
needed, the imposition of boundary conditions via the reconstruction of
the relevant quantities at the edge of the three-dimensional
computational domain.

In order to do this, it is necessary to determine the region of spacetime
where a perturbative approach can be applied. In general, the
three-dimensional numerical grid (indicated as \texttt{\textbf{N}} in
\hbox{Fig. \ref{f-pcem}}) will comprise an isolated region of spacetime
where the gravitational fields are strong and highly dynamical. In this
region, indicated as ${\cal S}$ in \hbox{Fig. \ref{f-pcem}}, the full
nonlinear Einstein equations must be solved. Outside of ${\cal S}$,
however, in what we will refer to as the perturbative region ${\cal P}$,
a perturbative approach is not only possible but highly advantageous.
Anywhere in the portion of ${\cal P}$ covered by \texttt{\textbf{N}} we
can place a two-dimensional surface, indicated as $\Gamma$ in
\hbox{Fig. \ref{f-pcem}}, which will serve as the surface joining
numerically the highly dynamical strong-field region ${\cal S}$ and the
perturbative one ${\cal P}$. In practice, it is easier to choose this
surface to be a 2-sphere of radius $r_{_\Gamma}$, where $r_{_\Gamma}$ can
either be the local coordinate radius, the corresponding Schwarzschild
radial coordinate, or some more sophisticated radial coordinate deduced
from the local values of the metric (\cf discussion in
Sec.~\ref{sec:multipole})\footnote{Note that in principle the gauge
  invariant quantities are independent of radius [\cf
    Eq.~\eqref{eq:wave_gi}]. In practice, however, their amplitudes may
  reach the correct asymptotic value only at sufficiently large
  distances. For this reason the extraction is in practice performed at
  different extraction radii and the amplitudes compared for convergence
  to an asymptotic value \cite{Rupright98,Rezzolla99a}.}. It is
important to emphasize that the 2-sphere $\Gamma$ need not be in a region
of spacetime where the gravitational fields are weak or the curvature is
small. In contrast to approaches which matched Einstein's equations onto
a Minkowski background \cite{Abrahams88b,Abrahams90}, the matching is
here made on a Schwarzschild background, so that the only requirement is
that the spacetime outside of ${\cal S}$ approaches a Schwarzschild
one. Of course, even in the case of a binary black-hole merger, it will
be possible to find a region of spacetime, sufficiently distant from the
black holes, where this requirement is met to the desired
precision \cite{Price94a,Abrahams94c,Abrahams95b,Abrahams95c,Abrahams95d}.

\epubtkImage{}{%
\begin{figure}[htb]
  \centering \includegraphics[width=13cm]{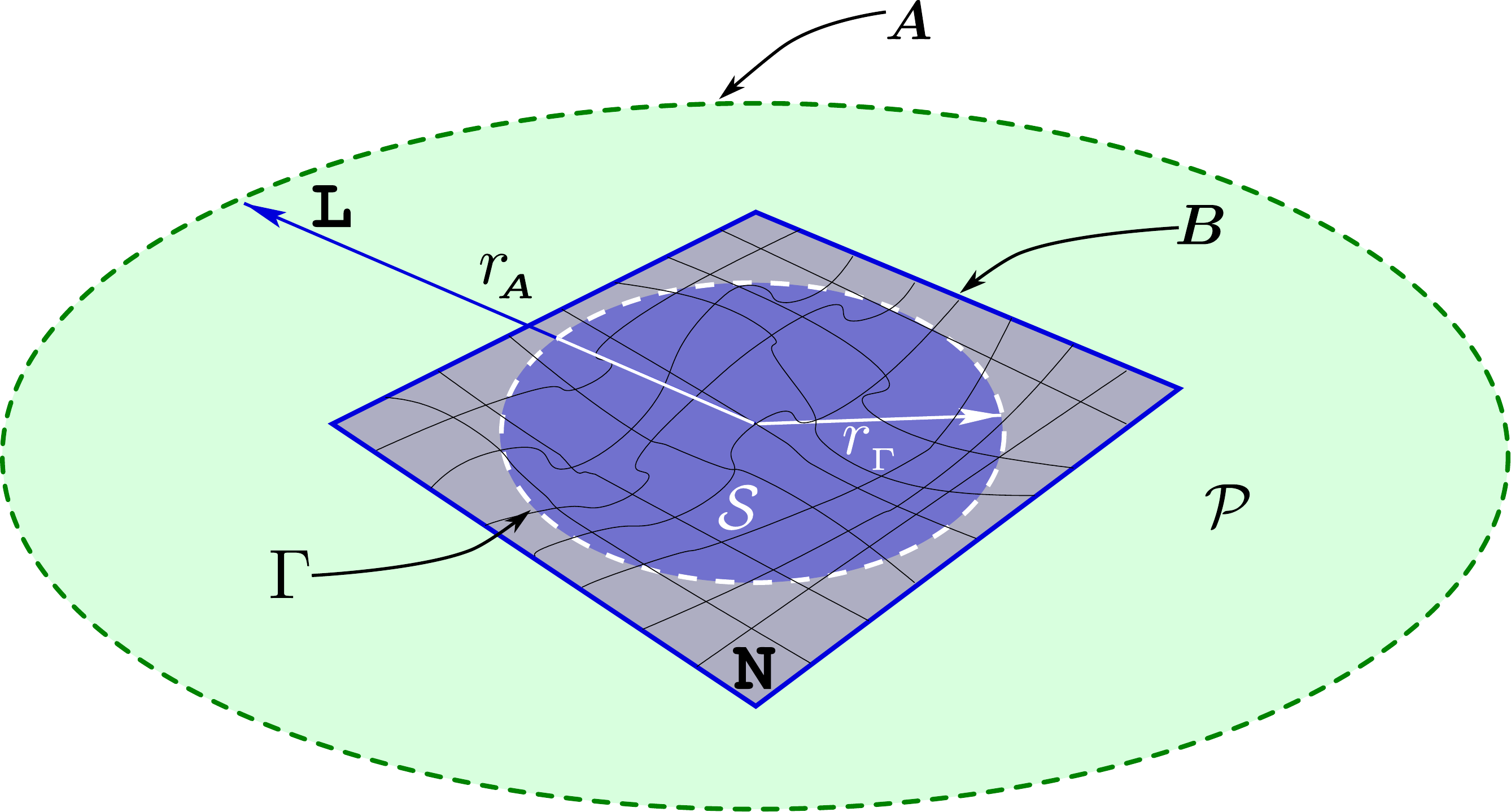} 
  \caption{Schematic illustration of a perturbative Cauchy extraction on
    a single spacelike hypersurface. Schematic picture of the
    Cauchy-perturbative matching procedure for a spacelike slice of
    spacetime (one spatial dimension has been
    suppressed). \texttt{\textbf{N}} is the three-dimensional numerical
    grid on the spacelike hypersurface $\Sigma_t$ in which the full
    Einstein equations are solved, and $\boldsymbol{B}$ its
    two-dimensional outer boundary. The interior (dark shaded) region
    ${\cal S}$ shows the strong-field highly dynamical region of
    spacetime fully covered by \texttt{\textbf{N}}. ${\cal P}$ is the
    region of spacetime where a perturbative solution can be performed
    and extends from the 2-sphere $\Gamma$ (of radius $r_{_\Gamma}$) to
    the 2-sphere $\boldsymbol{A}$ (of radius $r_{_A}$) located in the
    asymptotically flat region of spacetime. ${\cal P}$ is covered
    entirely by a one-dimensional grid \texttt{\textbf{L}} and partially
    by the three-dimensional grid \texttt{\textbf{N}}.}  \label{f-pcem}
\end{figure}}

In a practical implementation of the Cauchy-perturbative
approach \cite{Rupright98,Rezzolla99a}, a numerical code provides the
solution to the full nonlinear Einstein equations everywhere in the
three-dimensional grid \texttt{\textbf{N}} except at its outer boundary
surface $\boldsymbol{B}$. At the extraction 2-sphere $\Gamma$, a
different code (\ie the perturbative module) ``extracts'' the
gravitational wave information and transforms it into a set of multipole
amplitudes which are chosen to depend only on the radial and time
coordinates of the background Schwarzschild
metric \cite{Rupright98,Rezzolla99a}.

In this way, two of the three spatial dimensions of the problem are
suppressed and the propagation of gravitational waves on a curved
background is reduced to a one-dimensional problem. During each timestep,
information about the gravitational field read-off at $\Gamma$ is
propagated by the perturbative module out to the 2-sphere
$\boldsymbol{A}$ in the asymptotic flat region of spacetime. This is done
by solving a set of coupled one-dimensional linear differential equations
(one for each of the multipoles extracted at $\Gamma$) on the
one-dimensional grid \texttt{\textbf{L}} covering the perturbative region
${\cal P}$ and ranging between $r_{_\Gamma}$ and $r_{_A} \gg
r_{_\Gamma}$. From a computational point of view, this represents an
enormous advantage: with a few straightforward transformations, the
computationally expensive three-dimensional evolution of the
gravitational waves via the nonlinear Einstein equations is replaced with
a set of one-dimensional linear equations that can be integrated to high
accuracy with minimal computational cost. Although linear, these
equations account for all of the effects of wave propagation in a curved
spacetime and, in particular, automatically incorporate the effects of
backscatter off the curvature.

Note that as a result of this construction, (and as shown in
\hbox{Fig. \ref{f-pcem}}), the perturbative region ${\cal P}$ is entirely
covered by a one-dimensional grid {\tt L} and only partially by a
three-dimensional grid in the complement to ${\cal S}$
in \texttt{\textbf{N}}. The overlap between these two grids is
essential. In fact, the knowledge of the solution on ${\cal P}$ allows
the perturbative approach to provide boundary conditions at the outer
boundary surface $\boldsymbol{B}$ and, if useful, Dirichlet data on every
gridpoint of \texttt{\textbf{N}} outside the strong region ${\cal
S}$. This is also illustrated in Fig.~\ref{f-pce}, which represents a
one-dimensional cut of Fig. \ref{f-pcem}, and highlights the difference
between the asymptotic values of the gravitational waves extracted at the
boundary $A$ of the one-dimensional grid (filled blue circles) with and
the boundary values that can be instead specified (\ie ``injected'') on
the outer boundary surface $B$ of the three-dimensional grid.

The freedom to specify boundary data on a 2-surface of arbitrary shape as
well as on a whole three-dimensional region of \texttt{\textbf{N}}
represents an important advantage of the perturbative approach over
similar approaches to the problem of gravitational-wave extraction and
imposition of boundary conditions.

In what follows we briefly review the main steps necessary for the
numerical implementation of the Cauchy-perturbative approach in a
numerical-relativity code solving the Einstein equations in a $3+1$ split
of spacetime. This approach, which follows closely the discussion made in
Refs. \cite{Rupright98,Rezzolla99a}, basically consists of three steps:
{\sl (1)} \emph{extraction} of the independent multipole amplitudes on
$\Gamma$; {\sl (2)} \emph{evolution} of the radial wave equations
(\ref{oddwave})--(\ref{evenwave2}) on \texttt{\textbf{L}} out to the
distant wave zone; {\sl (3)} \emph{reconstruction} of $K_{ij}$ and
$\partial_t K_{ij}$ at specified gridpoints at the outer boundary
of \texttt{\textbf{N}}. We next discuss in detail each of these steps.

\epubtkImage{}{%
\begin{figure}[htb]
  \centering 
  \includegraphics[width=9cm]{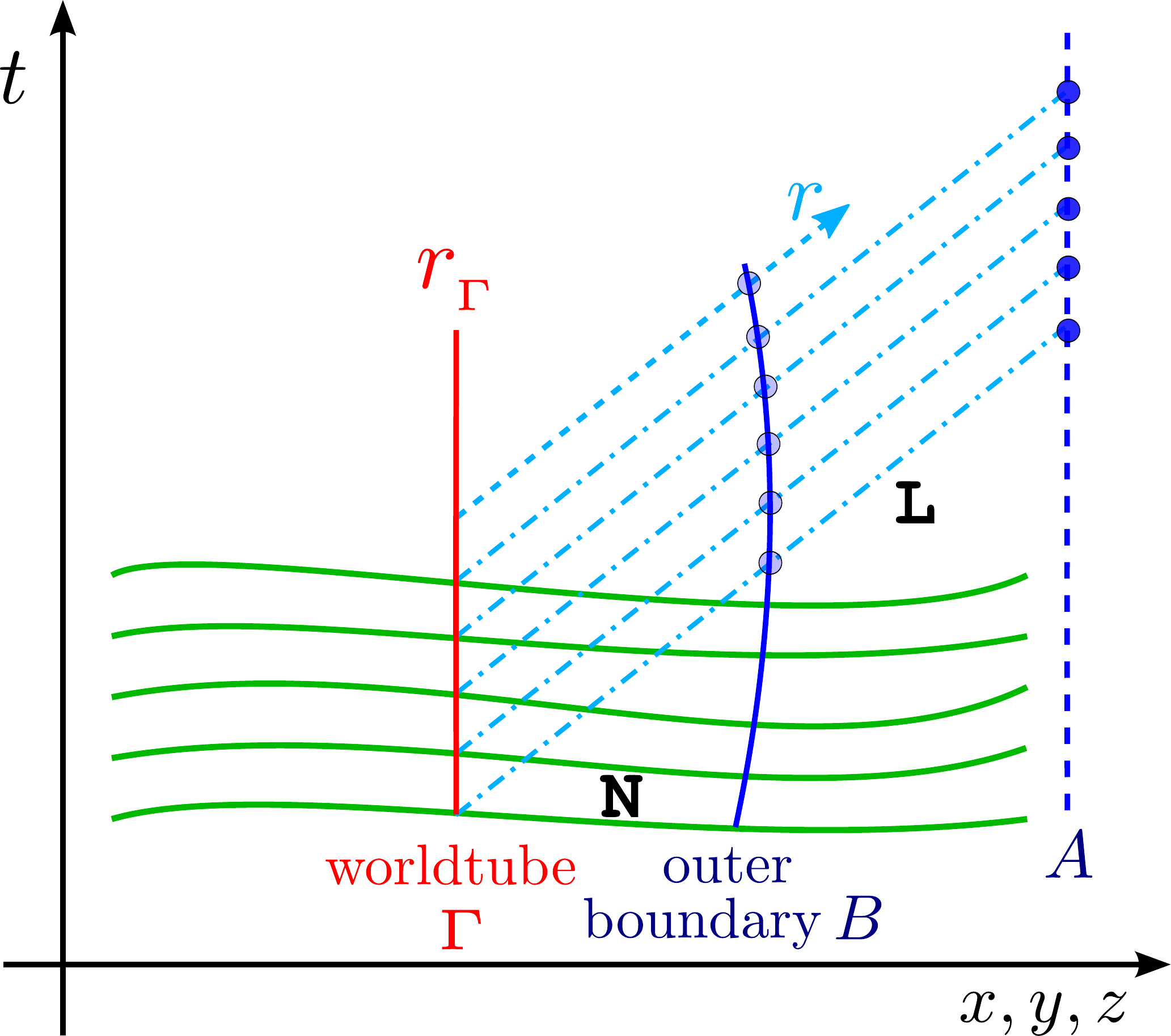} 
  \caption{Schematic illustration of a perturbative Cauchy
    extraction. The Cauchy evolution is shown with {{green slices}},
    comprising hypersurfaces $\Sigma_t$ on each of which is constructed a
    three-dimensional grid {\texttt{\textbf{N}}.}  The outer-boundary
    surface $B$ of the three-dimensional grid is shown in {{dark blue}},
    and is subject to a boundary condition that excludes incoming
    gravitational waves. Data from the Cauchy evolution on the
    {{worldtube $\Gamma$}} supplies boundary data to the perturbative
    equations, whose solution leads to the gravitational waves on the
    asymptotic boundary $A$. Note the difference between the asymptotic
    values of the gravitational waves extracted at $A$ (filled blue
    circles) with the boundary values that can instead be injected on
    $B$.}  \label{f-pce}
\end{figure}}

\subsubsection{Perturbative expansion}
\label{subsec:pertexpans}

The first step is to linearize the Einstein equations around a static
Schwarzschild background by separating the gravitational quantities of
interest into background (denoted by a tilde) and perturbed parts: the
three-metric $\gamma_{i j} = \widetilde{\gamma}_{i j} + h_{i j}$, the
extrinsic curvature $K_{i j} = \widetilde{K}_{i j} + \kappa_{i j}$, the
lapse $N = \widetilde{N} + \alpha$, and the shift vector $\beta^i =
\widetilde{\beta}^i + v^i$. Note that the large majority of modern
numerical-relativity codes implement the BSSNOK \cite{Nakamura87,
Shibata95, Baumgarte99} or the CCZ4 \cite{Alic:2011a} formulation of the
Einstein equations. As mentioned in Section \ref{s:ADM}, in these
formulations, the extrinsic curvature tensor is not evolved directly, but
rather a traceless tensor extrinsic curvature tensor related to a
conformal decomposition of the three-metric \cite{Alcubierre:2008,
Bona2009, Baumgarte2010a, Gourgoulhon2012, Rezzolla_book:2013}. Of
course, also in these formulations it is possible to reconstruct the
physically related extrinsic curvature tensor $K_{ij}$ and we will
therefore continue to make use of $K_{ij}$ hereafter.

In Schwarzschild coordinates $(t, r,
\theta, \phi)$, the background quantities are given by
\begin{align}
\label{eq:background}
\widetilde{N} =& \left(1 - \frac{2M}{r}\right)^{1/2} \,, &        \\
\widetilde{g}_{i j} dx^i dx^j = \, & \widetilde{N}^{-2} dr^2 +  
		r^2 (d\theta^2 + \sin^2\theta d\phi^2) \,, & \\
\widetilde{\beta}^i = \,& 0 \,, &                          	   \\
\widetilde{K}_{i j} = \,& 0 \,,  &                          
\end{align}
while the perturbed quantities have arbitrary angular dependence. The
background quantities satisfy the dynamical equations
$\partial_t \widetilde{\gamma}_{i j} = 0$, $\partial_t \widetilde{N} =
0$, and thus remain constant for all time. The perturbed quantities, on
the other hand, obey the following evolution equations

\begin{align}
\label{eq:h_dot} 
\partial_t h_{i j} =&\, -2 \widetilde{N} \kappa_{i j} + 
		2 \widetilde{\nabla}_{(i} v_{j)} \,,     \\
\label{eq:alpha_dot}
\partial_t \alpha =&\, v^i \widetilde{\nabla}_i \widetilde{N} -
		\widetilde{N}^2 \kappa \,,               \\
\label{eq:kappawave}
\widetilde N^{-1}\partial_t^2\kappa_{ij} -
	\widetilde N\widetilde\nabla^k\widetilde\nabla_k\kappa_{ij}
        =&\, - 4 \widetilde\nabla_{(i}\kappa^k_{\ j)} \widetilde\nabla_k 
	\widetilde N + \widetilde N^{-1} \kappa_{ij} \widetilde\nabla^k 
        \widetilde N \widetilde\nabla_k \widetilde N
        + 3 \widetilde\nabla^k \widetilde N \widetilde\nabla_k 
        \kappa_{ij}
        \nonumber\\
        &+ \kappa_{ij} \widetilde\nabla^k \widetilde\nabla_k \widetilde{N}
        - 2 \kappa^k_{\;(i} \widetilde\nabla_{j)} \widetilde\nabla_k 
	\widetilde{N}
        - 2 \widetilde{N}^{-1} \kappa^k_{\;(i} \widetilde\nabla_{j)} 
        \widetilde{N} \widetilde\nabla_k \widetilde{N}
        + 2 \kappa \widetilde\nabla_i \widetilde\nabla_j \widetilde{N}  
        \nonumber\\
        &+ 4 \partial_{(i} \kappa \partial_{j)} \widetilde{N}
        + 2 \widetilde{N}^{-1} \kappa \widetilde\nabla_i \widetilde{N} 
        \widetilde\nabla_j \widetilde{N}
        - 2 \widetilde{N} \widetilde{R}_{k(i} \kappa^k_{\ j)}
        - 2 \widetilde{N} \widetilde{R}_{kijm} \kappa^{km} \,,
\end{align}
where $\kappa \coloneqq \kappa^i_{\ i}$ and, as mentioned above, the
tilde denotes a spatial quantity defined in terms of the background
metric, $\widetilde{\gamma}_{i j}$. Note that the wave equation for
$\kappa_{i j}$ involves only the background lapse and curvature.

Next, it is possible to simplify the evolution equation
(\ref{eq:kappawave}) by separating out the angular dependence, thus
reducing it to a set of one-dimensional equations. This is accomplished
by expanding the extrinsic curvature in Regge--Wheeler tensor spherical
harmonics \cite{Regge57} and substituting this expansion into
(\ref{eq:kappawave}). Using the notation of Moncrief \cite{Moncrief74} we
express the expansion as
\begin{align}
\kappa_{i j} = &\, a_\times (t,r) (\hat e_1)_{i j} + 
              r b_\times (t,r) (\hat e_2)_{i j} + 
 	     \widetilde N^{-2} a_+ (t,r) (\hat f_2)_{i j} + 
                   r b_+ (t,r) (\hat f_1)_{i j} + 
        \nonumber\\
	        & r^2 c_+ (t,r) (\hat f_3)_{i j} +
		r^2 d_+ (t,r) (\hat f_4)_{i j} \,,
\label{eq:kappa_expand}
\end{align}
where $(\hat e_1)_{i j},\cdots,(\hat f_4)_{i j}$ are the Regge--Wheeler
harmonics, which are functions of $(\theta,\phi)$ and have suppressed
angular indices $(\ell,m)$ for each mode. Explicit expressions for these
tensors are given in Appendix \ref{app:rwh}.

The odd-parity multipoles ($a_\times$ and $b_\times$) and the even-parity
multipoles ($a_+$, $b_+$, $c_+$, and $d_+$) also have suppressed indices
for each angular mode and there is an implicit sum over all modes in
(\ref{eq:kappa_expand}). The six multipole amplitudes correspond to the
six components of $\kappa_{i j}$. However, using the linearized momentum
constraints

\begin{equation}
\widetilde{\nabla}_j (\kappa^j_{\ i} - \delta^j_{\ i} \kappa) = 0 \,,
\label{eq:MomCon}
\end{equation}
we reduce the number of independent components of $\kappa_{i j}$ to
three. An important relation is also obtained through the wave equation
for $\kappa$, whose multipole expansion is simply given by $\kappa =
h(t,r) Y_{_{\ell m}}$. Using this expansion, in conjunction with the
momentum constraints (\ref{eq:MomCon}), we derive a set of radial
constraint equations which relate the dependent amplitudes
$(b_\times)_{_{\ell m}}$, $(b_+)_{_{\ell m}}$, $(c_+)_{_{\ell m}}$ and
$(d_+)_{_{\ell m}}$ to the three independent amplitudes
$(a_\times)_{_{\ell m}}$, $(a_+)_{_{\ell m}}$, $(h)_{_{\ell m}}$
\begin{align}
\label{eq:Constraints}
(b_\times)_{_{\ell m}} =&\, -\frac{1}{(\ell+2)(\ell-1)}
		[(1+3\widetilde N^2) + 
		2\widetilde N^2 r\;\partial_r]\; 
		(a_\times)_{_{\ell m}} \,, \\
(b_+)_{_{\ell m}} =&\, \frac{1}{\ell(\ell+1)} [(3+r\partial_r)\; 
		(a_+)_{_{\ell m}} - (1+r\partial_r)\; 
		(h)_{_{\ell m}}] \,,        \\
(c_+)_{_{\ell m}} =&\, \frac{1}{2(\ell+2)(\ell-1)} 
		\{2(1-\ell-\ell^2)\; (a_+)_{_{\ell m}} - 
		2\; (h)_{_{\ell m}} +
		\ell(\ell+1) [(1+5\widetilde N^2) + 
		2\widetilde N^2 r\; \partial_r]\; (b_+)_{_{\ell m}}\} \,, \\
(d_+)_{_{\ell m}} =&\, \frac{1}{\ell(\ell+1)}[(a_+)_{_{\ell m}} 
		+ 2 (c_+)_{_{\ell m}} - (h)_{_{\ell m}}] \,,
\end{align}
for each $(\ell,m)$ mode.

\subsubsection{Extraction}
\label{subsec:angdecomp}

Taking the extraction 2-sphere $\Gamma$ as the surface joining the
evolution of the highly dynamical, strong field region (dark shaded area
of Fig. \ref{f-pcem}) and the perturbative regions (light shaded areas),
at each timestep, $K_{i j}$ and $\partial_t K_{i j}$ are computed
on \texttt{\textbf{N}} as a solution to Einstein's equations. Assuming
that \texttt{\textbf{N}} uses topologically Cartesian
coordinates\footnote{This is a standard choice in modern
numerical-relativity codes but there are no restrictions on the choice of
the coordinate system}, the Cartesian components of these tensors are
then transformed into their equivalents in a spherical coordinate basis
and their traces are computed using the inverse background metric,
\ie $H = \widetilde{\gamma}^{i j} K_{i j}$, $\partial_t H
= \widetilde{\gamma}^{i j}
\partial_t K_{i j}$. From the spherical components of $K_{i j}$ and
$\partial_t K_{i j}$, the independent multipole amplitudes for each
$(\ell,m)$ mode are then derived by an integration over the 2-sphere:
\begin{align}
\label{int_1}
(a_\times)_{_{\ell m}} =&\, 
	\frac{1}{\ell(\ell+1)} \int \frac{1}{\sin\theta}
	\left[ K_{r \phi} \, \partial_\theta - K_{r \theta} 
	\, \partial_\phi \right] \, 
	{Y^*_{_{\ell m}}} \, d\Omega \,,\\
\label{int_2}
(a_+)_{_{\ell m}} =&\, \int \widetilde{N}^2 \, K_{r r} 
		\, Y^*_{\ell m} \, d\Omega \,,\\
\label{int_3}
(h)_{_{\ell m}} =&\, \int \, H \, Y^*_{\ell m} d\Omega \,.
\end{align}
Their time derivatives are computed similarly. Rather than performing the
integrations (\ref{int_1})--(\ref{int_3}) using spherical polar
coordinates, it is useful to cover $\Gamma$ with two stereographic
coordinate ``patches''. These are uniformly spaced two-dimensional grids
onto which the values of $K_{i j}$ and $\partial_t K_{i j}$ are
interpolated using either a three-linear or a three-cubic polynomial
interpolation scheme. As a result of this transformation, the integrals
over the 2-sphere in (\ref{int_1})--(\ref{int_3}) are computed avoiding
polar singularities (see discussion in Appendix \ref{sec:stereo}).

\subsubsection{Perturbative evolution}
\label{sec:procimplement}

Substituting (\ref{eq:kappa_expand}) into (\ref{eq:kappawave}) and using
the constraint equations (\ref{eq:Constraints}), we obtain a set of
linearized radial wave equations for each independent amplitude. For each
$(\ell,m)$ mode we have one odd-parity equation

\begin{align}
\Biggl\{ \partial^2_t - \widetilde{N}^4 \partial^2_r - 
	\frac{2}{r}\widetilde{N}^2 \partial_r 
        - \frac{2 M}{r^3} \left(1 - \frac{3 M}{2 r} \right) + 
	\widetilde{N}^2 \left[ \frac{\ell(\ell+1)}{r^2} - 
	\frac{6 M}{r^3} \right] \Biggr\} 
        (a_\times)_{_{\ell m}} = 0 \,,
\label{oddwave}
\end{align}
and two coupled even-parity equations,
\begin{align}
	\Biggl[ \partial^2_t - \widetilde{N}^4 
	\partial^2_r - \frac{6}{r}\widetilde{N}^4 \partial_r
	+\widetilde{N}^2 \frac{\ell(\ell+1)}{r^2} - \frac{6}{r^2} + &\,
	\frac{14M}{r^3}-\frac{3M^2}{r^4}
	\Biggr] (a_+)_{_{\ell m}} +  
	\nonumber\\ 
	\Biggl[\frac{4}{r} \widetilde{N}^2 \left(1 -\frac{3M}{r}\right) 
	\partial_r + &\, \frac{2}{r^2} 
	\left(1 - \frac{M}{r} - \frac{3M^2}{r^2}\right) 
	\Biggr] (h)_{_{\ell m}} = 0 \,, 
\label{evenwave1} 
\end{align}	
\begin{align}	
	\Biggl[ \partial^2_t - \widetilde{N}^4 \partial^2_r - 
	\frac{2}{r}\widetilde{N}^2 \partial_r
	+ \widetilde{N}^2 \frac{\ell(\ell+1)}{r^2} 
	+ \frac{2 M}{r^3} -
	\frac{7 M^2}{r^4} \Biggr] (h)_{_{\ell m}} 
	- \frac{2 M}{r^3} \left(3 - \frac{7 M}{r}\right) 
	(a_+)_{_{\ell m}} = 0 \,.
\label{evenwave2}
\end{align}
These equations are related to the standard Regge--Wheeler and Zerilli
equations \cite{Regge57,Zerilli70}. 

Once the multipole amplitudes, $(a_\times)_{_{\ell m}}$, $(a_+)_{_{\ell
m}}$, $(h)_{_{\ell m}}$ and their time derivatives are computed on
$\Gamma$ in the timeslice $t=t_0$, they are imposed as inner boundary
conditions on the one-dimensional grid. Using a suitably accurate
integration scheme, the radial wave equations
(\ref{oddwave})--(\ref{evenwave2}) can be evolved for each $(\ell, m)$
mode forward to the next timeslice at $t=t_1$. The outer boundary of the
one-dimensional grid is always placed at a distance large enough that
background field and near-zone effects are unimportant, and a radial
Sommerfeld condition for the wave equations
(\ref{oddwave})--(\ref{evenwave2}) can be imposed there. The evolution
equations for $h_{i j}$ [Eq. \eqref{eq:h_dot}] and $\alpha$
[Eq. \eqref{eq:alpha_dot}] can also be integrated using the data for
$K_{i j}$ computed in this region. Note also that because $h_{i j}$ and
$\alpha$ evolve along the coordinate time axis, these equations need only
be integrated in the region in which their values are desired, not over
the whole region \texttt{\textbf{L}}.

Of course, the initial data on \texttt{\textbf{L}} must be consistent
with the initial data on \texttt{\textbf{N}}, and this can be determined
by applying the aforementioned extraction procedure to the initial data
set at each gridpoint of \texttt{\textbf{L}} in the region of overlap
with \texttt{\textbf{N}}. In the latter case, initial data outside the
overlap region can be set by considering the asymptotic fall-off of each
variable.

\subsubsection{Reconstruction}
\label{sec:cp_reconstruction}

An important side product of the evolution step discussed above is that
outer boundary values for \texttt{\textbf{N}} can now be computed,
although, to the best of our knowledge, this procedure has not been
implemented yet as a way to obtain outer boundary conditions. In
particular, for codes using the BSSNOK \cite{Nakamura87, Shibata95,
  Baumgarte99} or the CCZ4 \cite{Alic:2011a} formulation of the Einstein
equations, it is sufficient to provide boundary data only for $K_{i j}$,
since the interior code can calculate $\gamma_{i j}$ at the outer
boundary by integrating in time the boundary values for $K_{i j}$.

In order to compute $K_{i j}$ at an outer boundary point
of \texttt{\textbf{N}} (or any other point in the overlap
between \texttt{\textbf{N}} and ${\cal P}$), it is necessary to
reconstruct $K_{i j}$ from the multipole amplitudes and tensor spherical
harmonics. The Schwarzschild coordinate values $(r,\theta,\phi)$ of the
relevant gridpoint are first determined. Next, $(a_\times)_{_{\ell m}}$,
$(a_+)_{_{\ell m}}$, and $(h_{_{\ell m}})$ for each $(\ell,m)$ mode are
interpolated to the radial coordinate value of that point. The dependent
multipole amplitudes $(b_\times)_{_{\ell m}}$, $(b_+)_{_{\ell m}}$,
$(c_+)_{_{\ell m}}$, and $(d_+)_{_{\ell m}}$ are then computed using the
constraint equations (\ref{eq:Constraints}). Finally, the Regge--Wheeler
tensor spherical harmonics $(\hat{e}_1)_{i j}$--$(\hat{f}_4)_{i j}$ are
computed for the angular coordinates $(\theta,\phi)$ for each $(\ell,m)$
mode and the sum in Eq. (\ref{eq:kappa_expand}) is performed. This leads
to the reconstructed component of $\kappa_{i j}$ (and therefore $K_{i
j}$). A completely analogous algorithm can be used to reconstruct
$\partial_t K_{i j}$ in formulations in which this information is needed.

It is important to emphasize that this procedure allows one to compute
$K_{i j}$ at any point of \texttt{\textbf{N}} which is covered by the
perturbative region. As a result, the numerical module can reconstruct
the values of $K_{ij}$ and $\partial_t K_{i j}$ on a 2-surface of
arbitrary shape, or any collection of points outside of $\Gamma$.


\newpage
\section{Gravitational Waves in the Characteristic Approach}
\label{s-charac}

The formalism for expressing Einstein's equations as an evolution system
based on characteristic, or null-cone, coordinates is based on work
originally due to Bondi {\it et al.} \cite{Bondi1960,Bondi62} for
axisymmetry, and extended to the general case by
Sachs \cite{Sachs62}. The formalism is covered in the review by
Winicour \cite{Winicour05}, to which the reader is referred for an
in-depth discussion of its development and the associated literature.

Most work on characteristic evolution uses, or is an adpatation of, a
finite difference code that was originally developed at the University of
Pittsburgh and has become known as the PITT null code. The early work
that eventually led to the PITT code was for the case of
axisymmetry \cite{Isaacson83,Bishop90,Gomez94a}, and a general vacuum
code was developed in the
mid-1990s \cite{Bishop96,Bishop97b,Lehner98,Lehner99a,Lehner01a}. Subsequently,
the code was extended to the non-vacuum case \cite{Bishop99,Bishop05},
and code adaptations in terms of variables, coordinates and order of
accuracy have been
investigated \cite{Gomez01,Gomez03,Reisswig:2006,Reisswig:2012}.
Spectral, rather than finite difference, implementations have also been
developed, for both the axially symmetric case \cite{deOliveira:2009} and
in general \cite{Handmer:2015}. One potential difficulty, although in
practice it has not been important in characteristic extraction, is the
development of caustics during the evolution, and algorithms to handle
the problem have been proposed \cite{Stewart:1982,Corkill:1983}. There
are also approaches that use outgoing null cones but for which the
coordinates are not Bondi-Sachs \cite{Bartnik:1997,Bartnik:2000}.

Shortly after the publication of the Bondi and Bondi-Sachs metrics and
formalism, the idea of conformal compactification was introduced. This
led to the well-known asymptotic description of spacetime, and the
definitions of asymptotic flatness, past, future and spacelike infinity
($I^+,I^-,I^0$), and of past and future null infinity (${\mathcal
  J}^-,\scri$) \cite{Penrose:1963}; see
also \cite{Penrose2,Penrose:1965,Tamburino:1966} and the reviews by Adamo
{\it et al.} \cite{Adamo09} Frauendiener \cite{Frauendiener04}. The key
result is that gravitational radiation can be defined unambiguously in an
asymptotically flat spacetime only at null infinity. The waves may be expressed in
terms of the Bondi news ${\mathcal N}$ (see Eq.~(\ref{e-NB}) below), the Newman--Penrose
quantity $\psi_4$, or the wave strain $(h_+,h_\times )$.

After a characteristic code has been run using a compactified radial
coordinate as in Eq.~(\ref{e-r2x}), the metric is known at $\scri$, and
so it would seem to be straightforward to calculate the emitted
gravitational radiation. Unfortunately, this is not in general the case
because of gauge, or coordinate freedom, issues. The formulas do take a
very simple form when expressed in terms of coordinates that satisfy the
Bondi gauge condition in which the asymptotic flatness property is
obviously satisfied,
and for which conditions set at $\scri$ are propagated inwards along
radial null geodesics. However, in a numerical simulation that is not the
case: coordinate conditions are fixed on an extraction worldtube (in the
case of characteristic extraction), or perhaps on a
worldline \cite{Siebel03} or ingoing null hypersurface, and then
propagated outwards to $\scri$. The result is that the geometry at and
near $\scri$ may appear very different to one that is foliated by
spherical 2-surfaces of constant curvature. Of course, the Bondi gauge
and the general gauge are related by a coordinate transformation, and
formulas for ${\mathcal N}$ and $\psi_4$ are obtained by constructing the
transformation.

An explicit formula in the general gauge for the news was obtained
in \cite{Bishop97b} (Appendix B); and a calculation of $\psi_4$ was
reported in \cite{Babiuc:2009}, but the formula produced was so lengthy
that it was not published. These formulas have been used in the
production of most waveforms calculated by characteristic codes. An
alternative approach, in which the coordinate transformation is explicit,
rather than partially implicit, was suggested \cite{Bishop03} but has not
been further used or developed. Recently, a formula for the wave strain
$(h_+,h_\times )$, which is the quantity used in the construction
of templates for gravitational-wave data analysis, was
derived \cite{Bishop:2014}. An important special case is that of the
linearized approximation, in which deviations from the Bondi gauge are
small. The resulting formulas for ${\mathcal N}$, $\psi_4$ and $(h_+,h_\times )$,
are much simpler and so much easier to interpret than in the general
case. Further these formulas are widely used because the linearized
approximation often applies to the results of a waveform computation in a
realistic scenario.

We set the context for this section by summarizing the Einstein equations
in characteristic coordinates, and outlining the characteristic evolution
procedure. The focus of this section is formulas for gravitational waves,
and we next present the formulas in the simplest case, when the
coordinates satisfy the Bondi gauge conditions. Much of the remainder of
the section will be devoted to formulas for gravitational waves in the
general gauge, and will include a discussion of conformal
compactification. This section makes extensive use of spin-weighted
spherical harmonics and the eth formalism, which topics are discussed in
Appendix~\ref{a-sYlm}.

\subsection{The Einstein equations in Bondi-Sachs coordinates}
\label{s-nf}

We start with coordinates based upon a family of outgoing null hypersurfaces.
Let $u$ label these hypersurfaces, $\phi^{^{_A}}$ $(A=2,3)$ be angular coordinates labelling
the null rays, and $r$ be a surface area coordinate. In the resulting
$x^\alpha=(u,r,\phi^{^{_A}})$ coordinates, the metric takes the Bondi-Sachs
form
\begin{eqnarray}
 ds^2  &=&  -\left(e^{2\beta}(1 + W_c r) -r^2h_{_{AB}}U^{^{_A}}U^{^{_{_{B}}}}\right)du^2
\nonumber \\
        & &- 2e^{2\beta}dudr -2r^2 h_{_{AB}}U^{^{_{_{B}}}}dud\phi^{^{_A}} 
        +  r^2h_{_{AB}}d\phi^{^{_A}}d\phi^{^{_{_{B}}}},
\label{eq:bmet}
\end{eqnarray}
where $h^{^{_{AB}}}h_{BC}=\delta^{^{_A}}_C$ and
$\det(h_{_{AB}})=\det(q_{_{AB}})$, with $q_{_{AB}}$ a metric representing
a unit 2-sphere; $W_c$ is a normalized variable used in the code, related
to the usual Bondi-Sachs variable $V$ by $V=r+W_c r^2$. It should be
noted here that different references use various notations for what is
here denoted as $W_c$, and in particular Ref. \cite{Bishop97b} uses $W$
with $W \coloneqq r^2W_c$. As discussed in Sec.~\ref{s-dss}, we represent
$q_{_{AB}}$ by means of a complex dyad $q_{_{A}}$, then $h_{_{AB}}$ can
be represented by its dyad component $J\coloneqq
h_{_{AB}}q^{^{_A}}q^{^{_{_{B}}}}/2$. We also introduce the fields $K
\coloneqq \sqrt{1+J \bar{J}}$ and $U \coloneqq U^{^{_A}}q_{_{A}}$. The
spin-weight $s$ of a quantity is defined and discussed in
Appendix~\ref{a-swf} ; for the quantities used in the Bondi-Sachs metric
\begin{equation}
  \begin{array}{lll}
    s(W_c)=s(\beta)=0\,,\qquad & s(J)=2\,,\qquad & s(\bar{J})=-2\,, \\
    s(K)=0\,,           \qquad & s(U)=1\,,\qquad & s(\bar{U})=-1\,.
  \end{array}
\end{equation}
We would like to emphasize two matters: (a) The metric
Eq.~(\ref{eq:bmet}) applies quite generally, and does not rely on the
spacetime having any particular properties. (b) There are many different
metrics of the form Eq.~(\ref{eq:bmet}) that describe a given spacetime,
and changing from one to another is known as a gauge transformation
(about which more will be said later).

The form of the Einstein equations for the general Bondi-Sachs metric has
been known for some time, but it was only in 1997 \cite{Bishop97b} that
they were used for a numerical evolution. (See also \cite{Gomez03} for an
alternative semi-first-order form that avoids second angular derivatives
($\eth^2, \bar{\eth}^2, \bar{\eth}\eth$)). The equations are rather
lengthy, and only the hypersurface and evolution equations are given in
that paper, in an Appendix\footnote{There is a misprint in Eq. (A3) of
  the journal version of the reference, which has been corrected in the
  version on the arxiv, and also in \cite{Reisswig:2012}.}. See also
Appendix~\ref{a-algebra}. Here, in
order to make the discussion of the Einstein equations precise but
without being overwhelmed by detail, we give the equations in vacuum in
the linearized case, that is when any second-order term in the quantities
$J, \beta, U, W_c$ can be ignored. The Einstein equations are categorized into
three classes, hypersurface, evolution, and constraint. The hypersurface
equations are
\begin{eqnarray}
R_{11}: &\quad \frac{4}{r}\partial_r \beta=0\,,\\
\label{e-b}
q^{^{_A}} R_{1A}: &\quad \frac{1}{2r} \left(
4 \eth \beta - 2 r \eth \partial_r\beta + r \bar{\eth} \partial_r J
+r^3 \partial^2_r U +4 r^2 \partial_r U \right) = 0\,,\\
\label{e-rq}
h^{^{_{AB}}} R_{_{AB}}: &\quad
(4-2\eth \bar{\eth}) \beta +\frac{1}{2}(\bar{\eth}^2 J + \eth^2\bar{J})
+\frac{1}{2r^2}\partial_r(r^4\eth\bar{U}+r^4\bar{\eth}U) -2 r^2 \partial_r W_{c} -4 r W_c
=0\,.
\label{e-rw}
\end{eqnarray}
The evolution equation is
\begin{equation}
q^{^{_A}} q^{^{_{_{B}}}} R_{_{AB}}: \;\;
  -2\eth^2\beta + \partial_r(r^2 \eth U) - 2r \partial_r J
 -  r^2 \partial^2_r J 
  +2 r \partial_r\partial_u (rJ)= 0\,.
\label{e-ev}
\end{equation}
The constraint equations are \cite{Reisswig:2006}
\begin{eqnarray}
&&R_{00}: \nonumber\\  
&& \frac{1}{2r^2} \bigg( r^3 \partial^2_r W_{c}+4r^2\partial_r W_{c}+2rW_c+r\eth\bar{\eth} W_c 
  +2 \eth\bar{\eth} \beta
-4 r \partial_u\beta - r^2 \partial_u(\eth \bar{U} + \bar{\eth}U)+2 r^2 \partial_u W_{c}
 \bigg) = 0\,, \nonumber \\ \\
&&R_{01}: \nonumber\\  
&& \frac{1}{4r^2} \bigg(2r^3 \partial^2_r W_{c}+8r^2\partial_r W_{c}+4rW_c+4 \eth\bar{\eth}\beta
         -\partial_r(r^2\eth\bar{U}+r^2\bar{\eth}U)\bigg)=0\,, \nonumber
         \\ \\
&&q^{^{_A}} R_{0A}: \nonumber\\ 
&& \frac{1}{4} \bigg( 2r \eth \partial_r W_{c}+2 \eth W_c+ 2 r(4 \partial_r U
         + r \partial_r^2 U)+4 U +(\eth\bar{\eth}U-\eth^2\bar{U})
         +2 \bar{\eth}\partial_u J-2 r^2  \partial_r\partial_u U-4 \eth\partial_u\beta
         \bigg)=0\,. \nonumber \\ 
\label{e-c0A}
\end{eqnarray}
An evolution problem is normally formulated in the region of spacetime
between a timelike or null worldtube $\Gamma$ and future null infinity
(${\mathcal J}^+$), with (free) initial data $J$ given on $u=0$, and with
boundary data for $\beta,U,\partial_r U,W_c,J$ satisfying the constraints
given on $\Gamma$. (In characteristic extraction, the data satisfies the
Einstein equations inside $\Gamma$, and so the issue of ensuring that the
boundary data must satisfy the characteristic constraint equations does
not arise). The hypersurface equations are solved to find $\beta,U,W_c$,
and then the evolution equation gives $\partial_u J$ and thence $J$ on
the ``next'' null cone. See \cite{Kreiss2011,Babiuc2014} for a discussion
of the well-posedness of the problem.

\epubtkImage{}{%
\begin{figure}[htb]
 \begin{center}
   \includegraphics[width=11cm]{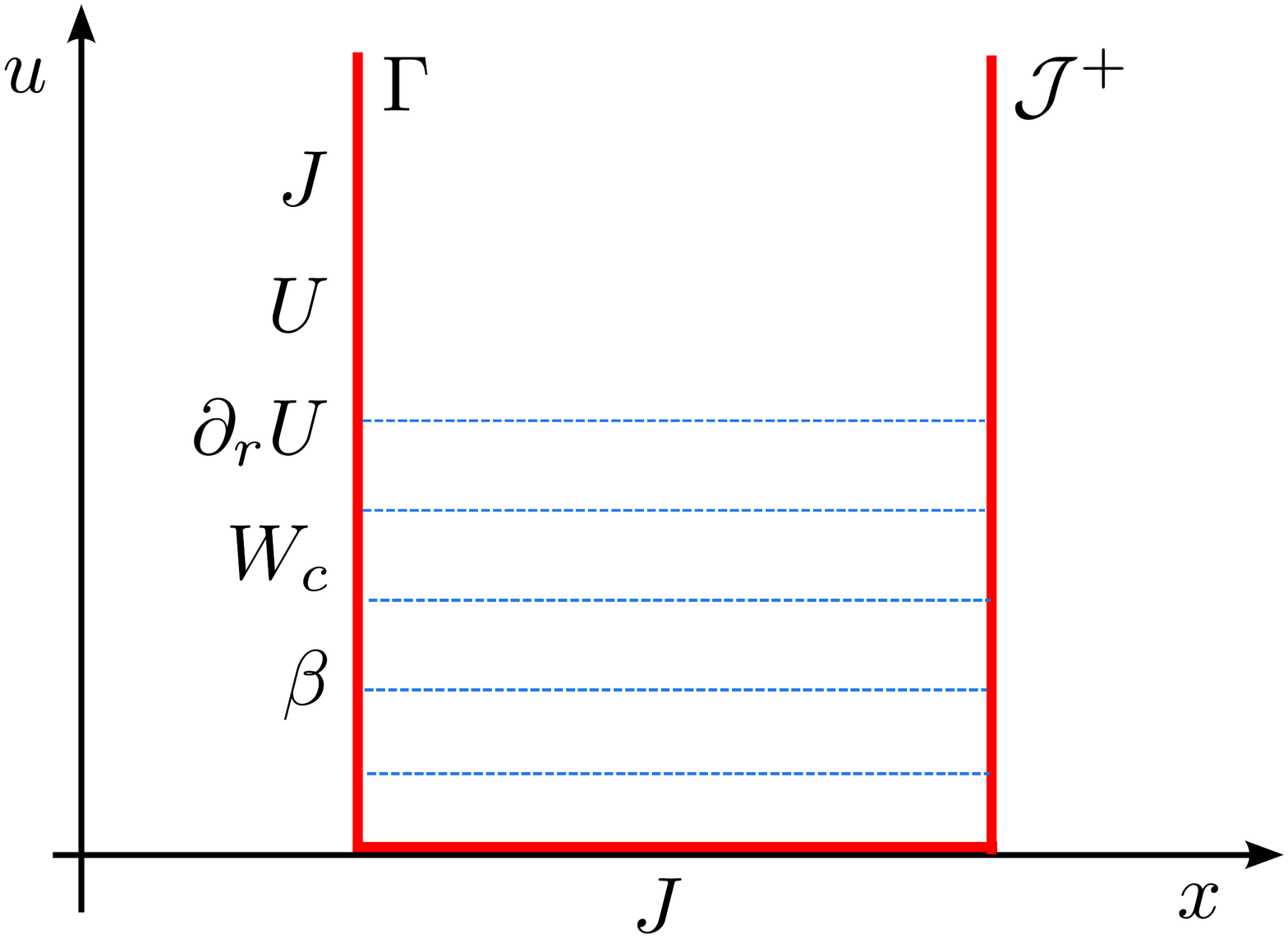}
   \caption{Schematic illustration of the boundary data required for the
     characteristic code. The data required is {{$J$ at $u=0$}} and
     {{$\beta,J, U, \partial_r U, W_c$ on the worldtube $\Gamma$}}.}
   \label{f-bscoords}
 \end{center}
\end{figure}}

We extend the computational grid to $\scri$ by
compactifying the radial coordinate $r$ by means of a transformation $r
\rightarrow x=f(r)$ where $\lim_{r \rightarrow \infty} f(r)$ is
finite. In characteristic coordinates, the Einstein equations remain
regular at $\scri$ under such a transformation. In practice, in
numerical work the compactification is usually
\begin{equation}
r \rightarrow x=\frac{r}{r+r_\Gamma}\,.
\label{e-r2x}
\end{equation}
However, for the purpose of extracting gravitational waves, it is more
convenient to express quantities as power series about $\scri$, and so we
compactify using
\begin{equation}
r\rightarrow \rho=1/r\,.
\label{er2rho}
\end{equation}
(Common practice
has been to use the notation $\ell$ for $1/r$, but since we will have
expressions involving the compactified radial coordinate and spherical
harmonics such a notation would be confusing).
Starting from the Bondi-Sachs metric Eq.~(\ref{eq:bmet}), we make the
coordinate transformation~\eqref{er2rho} to obtain
\begin{equation}
   ds^2 = \rho^{-2}\left(-\left(e^{2\beta}(\rho^2+\rho W_c)
 -h_{_{AB}}U^{^{_A}}U^{^{_{_{B}}}}\right)du^2
        +2e^{2\beta}dud\rho -2 h_{_{AB}}U^{^{_{_{B}}}}dud\phi^{^{_A}}
        + h_{_{AB}}d\phi^{^{_A}}d\phi^{^{_{_{B}}}}\right)\,.
   \label{eq:lmet}
\end{equation}
In contravariant form,
\begin{equation}
{g}^{11}=e^{-2\beta}\rho^3(\rho+ W_c)\,, \quad {g}^{1A}=\rho^2e^{-2\beta} U^{^{_A}}\,,
{g}^{10}=\rho^2e^{-2\beta}\,, \quad
{g}^{^{_{AB}}}=\rho^2h^{^{_{AB}}}\,, \quad {g}^{0A}={g}^{00}=0\,.
\label{e-bsc}
\end{equation}

Later, we will need to use the asymptotic Einstein equations, that is the
Einstein equations keeping only the leading order terms when the limit
$r\rightarrow \infty$, or equivalently $\rho\rightarrow 0$, is taken.
We write the metric variables as
$J=J_{(0)}+J_{(1)}\rho$, and similarly for $\beta, U$ and $W_c$. Each Einstein
equation is expressed as a series in $\rho$ and only leading order terms
are considered. There is considerable redundancy, and instead of 10
independent relations we find (see Appendix~\ref{a-algebra})
\begin{eqnarray}
R_{11}=0 &\rightarrow & \beta_{(1)}=0\,,
\label{e-as11} \\
q^{^{_A}} R_{1{\scriptscriptstyle A}}=0 &\rightarrow & -2\eth \beta_{(0)}
    +e^{-2\beta_{(0)}} K_{(0)} U_{(1)}+e^{-2\beta_{(0)}} J_{(0)} \bar{U}_{(1)}=0\,,
\label{e-as1A}\\
h^{^{_{AB}}}R_{_{AB}}=0 &\rightarrow & 2W_{c(0)} - \eth\bar{U}_{(0)} - \bar{\eth}U_{(0)}=0\,,
 \label{e-asAB}\\
q^{^{_A}} q^{^{_{_{B}}}} R_{_{AB}}=0 &\rightarrow & 2K_{(0)}\eth U_{(0)} + 2\partial_u J_{(0)}
+\bar{U}_{(0)}\eth J_{(0)}+U_{(0)}\bar{\eth} J_{(0)}\nonumber \\
& &+J_{(0)}\eth\bar{U}_{(0)}-J_{(0)}\bar{\eth} U_{(0)}=0\,. \label{e-asev}
\end{eqnarray}
The above are for the fully nonlinear case, with the linearized
approximation obtained by setting $K_{(0)}=e^{-2\beta_{(0)}}=1$, and
ignoring products of $J$ and $U$ terms.

\subsection{The Bondi gauge}
\label{s-Bg}

In the Bondi gauge, the form of the Bondi-Sachs metric is manifestly
asymptotically flat since it tends to Minkowskian form as $r\rightarrow
\infty$. In order to see what conditions are thus imposed, the first step
is to write the Minkowskian metric in compactified Bondi-Sachs
coordinates. Starting from the Minkowski metric in spherical coordinates
$(t,r,\phi^{^{_A}})$, we make the coordinate transformation
$(t,r)\rightarrow (u,\rho)$ where
\begin{equation}
u=t-r,\;\;\rho=\frac 1r
\label{e-M2cM}
\end{equation}
to obtain
\begin{equation}
ds^2=\rho^{-2}\left(-\rho^2 du^2+2 du\, d\rho +q_{_{AB}}d\phi^{^{_A}}\,d\phi^{^{_{_{B}}}} \right)\,.
\label{e-CompM}
\end{equation}
We use the notation $\;\tilde{ }\;$ to denote quantities in the Bondi
gauge.  The metric of Eqs.~(\ref{eq:lmet}) and (\ref{e-bsc}) still
applies, with the additional properties as ${\tilde \rho}\rightarrow 0$,
\begin{align}
\tilde{J}&=0\,,\quad\tilde{K}=1\,,\quad\tilde{\beta}=0\,,\quad\tilde{U}=0\,,\quad\tilde{W}_c=0\,,
\nonumber\\
\p_{\tilde{\rho}}\tilde{K}&=0,\;\p_{\tilde{\rho}}\tilde{\beta}=0\,,\;
\p_{\tilde{\rho}}\tilde{U}=0,\;\p_{\tilde{\rho}}\tilde{W}_{c}=0\,.
\label{e-BGC}
\end{align}
The undifferentiated conditions can be regarded as defining the Bondi
gauge, being motivated by the geometric condition that the metric
Eq.~(\ref{eq:lmet}) should take the form Eq.~(\ref{e-CompM}) in the limit
as $\rho\rightarrow 0$. The conditions on
$\p_{\tilde{\rho}}\tilde{\beta}$, $\p_{\tilde{\rho}}\tilde{U}$ and
$\p_{\tilde{\rho}}\tilde{K}$ follow from the asymptotic
Eqs.~(\ref{e-as11}), (\ref{e-as1A}), and (\ref{e-KJJb}) respectively; and
the condition on $\p_{\tilde{\rho}}\tilde{W}_{c}$ is obtained from the
asymptotic Einstein equation $h^{^{_{AB}}}R_{_{AB}}=0$ to second order in
$\rho$ and applying the Bondi gauge conditions already obtained. The null
tetrad in the Bondi gauge will be denoted by
$\tilde{\ell}^\alpha,\tilde{n}_{_{[NP]}}^\alpha, \tilde{m}^\alpha$, with
components to leading order in $\tilde{\rho}$ [obtained by applying the
  coordinate transformation~\eqref{e-M2cM} to Eq.~\eqref{e-nt}]
\begin{equation}
\tilde{\ell}^\alpha=\left(0,- \frac{\tilde{\rho}^2}{\sqrt{2}},0,0\right)\,,\qquad
\tilde{n}_{_{[NP]}}^\alpha=\left(\sqrt{2},\frac{\tilde{\rho}^2}{\sqrt{2}},0,0\right)\,,\qquad
\tilde{m}^\alpha=\left(0,0,\frac{\tilde{\rho} q^{^{_A}}}{\sqrt{2}}\right)\,.
\label{e-ntMpu}
\end{equation}

The gravitational news was defined by Bondi {\it et al.} \cite{Bondi62} and is
\begin{equation}
{\mathcal N}=\frac 12  \partial_{\tilde{u}} \partial_{\tilde{\rho}}\tilde{J}\,,
\label{e-NB}
\end{equation}
evaluated in the limit $\tilde{\rho}\rightarrow 0$, and is related to the
strain in the TT gauge by
\begin{equation}
{\mathcal N}=\frac 12 \partial_{\tilde{u}}\lim_{\tilde{r}\rightarrow\infty}
\tilde{r}\left(h_+ +i h_\times\right)
= \frac 12\partial_{\tilde{u}}H\,,
\end{equation}
where the rescaled strain $H$ is
\begin{equation}
H\coloneqq\lim_{\tilde{r}\rightarrow\infty} \tilde{r}\left(h_+ +i h_\times\right)
   =\p_{\tilde{\rho}} \tilde{J}\,,
\label{e-NHdef}
\end{equation}
which result is a straightforward consequence of the relation
$\tilde{J}=h_++ih_\times$ discussed in Appendix~\ref{s-dss}. When using
the Newman--Penrose formalism to describe gravitational waves, it is
convenient to introduce
\begin{equation}
\psi^0_4= \lim_{\tilde{r}\rightarrow \infty} \tilde{r}\psi_4
\;\;\left(=\lim_{\tilde{\rho}\rightarrow 0} \frac{\psi_4}{\tilde{\rho}}
\right)\,,
\end{equation}
since it will be important, when considering conformal compactification
(Sec.~\ref{s-cc}), to have a quantity that is defined at
$\tilde{\rho}=0$. In the Bondi gauge, as shown in
Appendix~\ref{a-algebra}, $\psi^0_4$ simplifies to
\begin{equation}
\psi^0_4= \partial^2_{\tilde{u}} \partial_{\tilde{\rho}}\bar{\tilde{J}}\,,
\label{e-psi4JB}
\end{equation}
evaluated in the limit $\tilde{\rho}\rightarrow 0$. Thus $\psi^0_4$, ${\mathcal N}$ and $H$
are related by
\begin{equation}
\psi^0_4=2 \partial_{\tilde{u}} \mathcal {\bar{N}}=
\partial^2_{\tilde{u}} \bar{H}\,.
\label{e-psi4N}
\end{equation}

\subsection{General gauge}

\epubtkImage{}{%
\begin{figure}[htb]
  \begin{center}
  \includegraphics[width=0.35\linewidth]{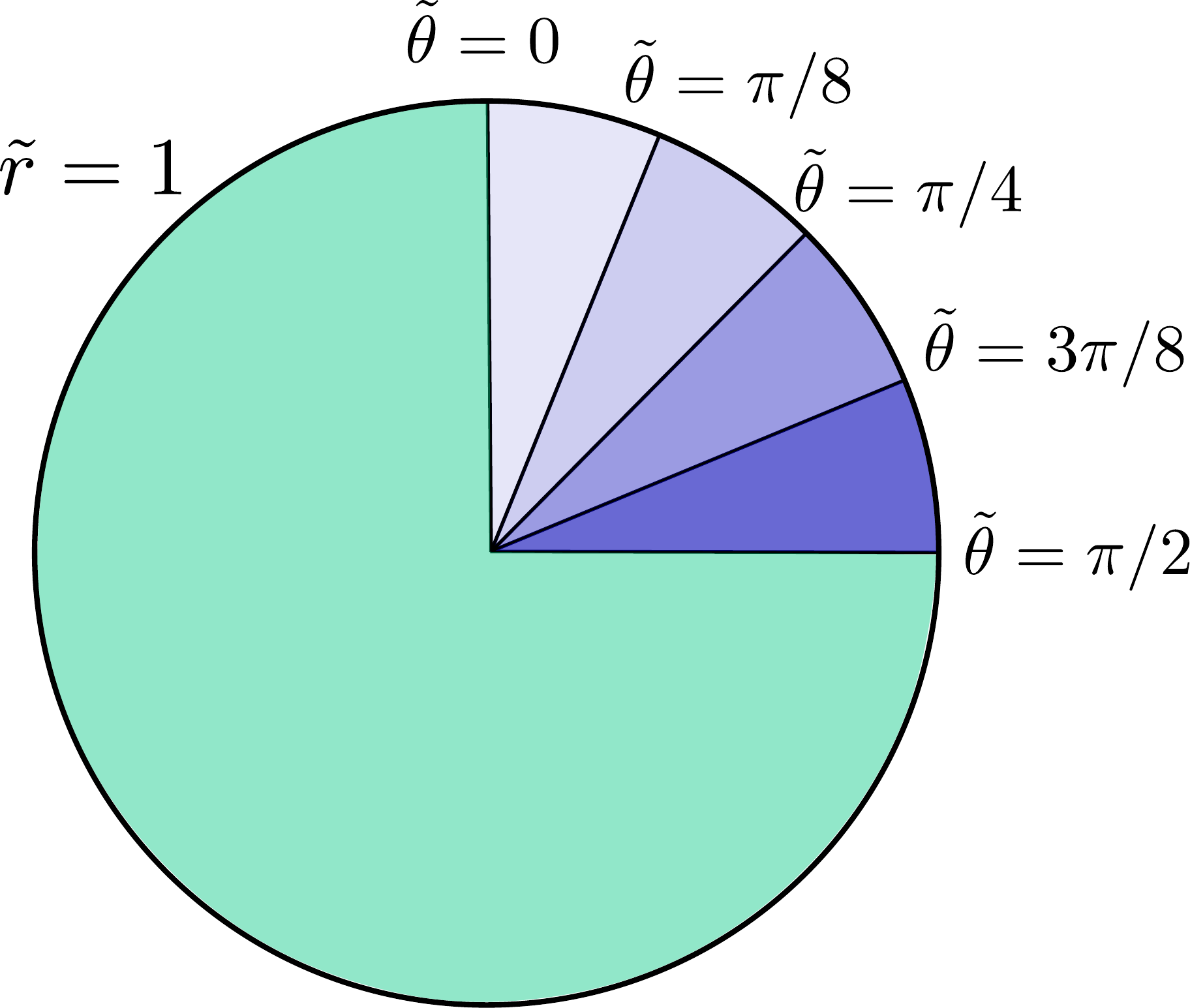}
  \includegraphics[width=0.6\linewidth]{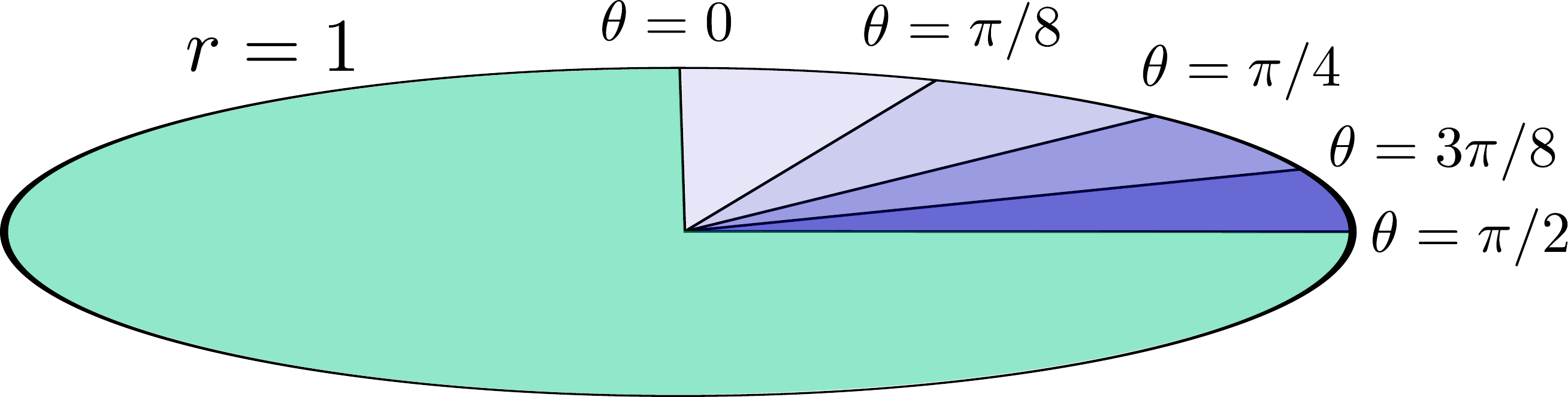}
  \end{center}
  \caption{Illustration of the relation between the Bondi and general
    gauges in Minkowski spacetime. In the Bondi gauge the unit sphere
    $\tilde{r}=1$ has constant curvature (left panel). Now consider a
    coordinate transformation $\tilde{\theta}\rightarrow\theta$ with
    $d\theta/d\tilde{\theta}>1$ near where $\tilde{\theta}$ is $0$ and
    $\pi$, and with $d\theta/d\tilde{\theta}<1$ near where
    $\tilde{\theta}$ is $\pi/2$. In these coordinates, the surface of
    constant surface area $r=1$ will be as shown in the right panel, and
    will not be a spherical surface of constant curvature.}
   \label{f-BAGG}
\end{figure}}

We construct quantities in the general gauge by means of a coordinate
transformation to the Bondi gauge, although this transformation is
largely implicit because it does not appear in many of the final
formulas.  The transformation is written as a series expansion in $\rho$
with coefficients arbitrary functions of the other coordinates. Thus it
is a general transformation, and the requirements that $g^{\alpha\beta}$
must be of Bondi-Sachs form, and that $\tilde{g}^{\alpha\beta}$ must be
in the Bondi gauge, impose conditions on the transformation
coefficients. The transformation is
\begin{equation}
u \rightarrow {\tilde u}=u+u_0+\rho A^u\,, \qquad
\rho \rightarrow {\tilde
  \rho}=\rho \omega+\rho^2 A^\rho\,, \qquad
\phi^{^{_A}} \rightarrow {\tilde
  \phi}^{^{_A}}=\phi^{^{_A}}+\phi^{^{_A}}_0+\rho A^{^{_A}}\,,
\label{e-bs2b}
\end{equation}
where the transformation coefficients
$u_0,A^u,\omega,A^\rho,\phi^{^{_A}}_0,A^{^{_A}}$ are all functions of $u$
and $\phi^{^{_A}}$ only. Conditions on the coefficients are found by
applying the tensor transformation law
\begin{equation}
{{\tilde g}}^{\alpha\beta}=\frac{\partial {\tilde x}^\alpha}{\partial x^\mu}
\frac{\partial {\tilde x}^\beta}{\partial x^\nu} { g}^{\mu\nu}\,,
\qquad \mbox{and}\qquad
{ g}_{\alpha\beta}=\frac{\partial {\tilde x}^\mu}{\partial x^\alpha}
\frac{\partial {\tilde x}^\nu}{\partial x^\beta}{\tilde g}_{\mu\nu}\,,
\label{e-gttl}
\end{equation}
for specific cases of $\alpha,\beta$, using the form of the metric in
Eqs.~(\ref{eq:lmet}) and (\ref{e-bsc}) and also applying the conditions in
Eq.~(\ref{e-BGC}) to ${\tilde g}^{\alpha\beta}$ and
${\tilde g}_{\mu\nu}$ \cite{Bishop97b,Bishop03,Bishop:2014}.The procedure is
shown in some detail for one case, with the other cases being handled in a
similar way. The actual calculations are performed by computer algebra as
discussed in Appendix~\ref{a-algebra}.

From Eqs.~(\ref{e-bsc}) and (\ref{e-BGC}), ${\tilde g}^{01}=\tilde{\rho}^2+
\mathcal{O}(\tilde{\rho}^4)$. Then using the contravariant transformation in
Eq.~(\ref{e-gttl}) with $\alpha=0,\beta=1$, we have
\begin{equation}
\tilde{\rho}^2+\mathcal{O}(\tilde{\rho}^4)=\rho^2\omega^2+\mathcal{O}({\rho}^4)=
\frac{\partial {\tilde u}}{\partial x^\mu}
\frac{\partial {\tilde \rho}}{\partial x^\nu} { g}^{\mu\nu}\,.
\end{equation}
Evaluating the right hand side to $\mathcal{O}({\rho}^2)$, the resulting
equation simplifies to give
\begin{equation}
(\partial_u+U^{^{_{_{B}}}}\partial_{_{B}})u_{0}=\omega e^{2\beta}-1\,.
\label{e-du0}
\end{equation}
The remaining conditions follow in a similar way
 \begin{eqnarray}
0+\mathcal{O} (\tilde{\rho}^4)={{\tilde g}}^{A1}\,, \qquad
 &\mbox{so that to }\mathcal{O} ({\rho}^2)\,,&\qquad (\partial_u+U^{^{_{_{B}}}}\partial_{_{B}})\phi^{^{_A}}_0=-U^{^{_A}}\,,
 \label{e-dx0}
\\
0+\mathcal{O} (\tilde{\rho}^4)={{\tilde g}}^{11}\,, \qquad
  &\mbox{so that to }\mathcal{O} ({\rho}^3)\,,&\qquad
  (\partial_u+U^{^{_{_{B}}}}\partial_{_{B}})\omega=-\omega W_c/2\,,
 \label{e-dom}
\\
 0 ={\tilde g}^{00},\qquad &\mbox{so that to }\mathcal{O}({\rho}^2)\,,&\;
 2\omega A^u=\frac{J\bar{\eth}^2u_0+\bar{J}\eth^2 u_0}{2}-K\eth u_0 \bar{\eth}u_0\,.
 \label{e-omAu}
 \end{eqnarray} 
In the next equations, $X_0=q_{_{A}} \phi_0^{^{_A}}, A=q_{_{A}}
A^{^{_A}}$; the introduction of these quantities is a convenience to
reduce the number of terms in the formulas, since $\phi_0^{^{_A}},
A^{^{_A}}$ do not transform as 2-vectors. As a result, the quantity
$\zeta=q+ip$ also appears, and the formulas are specific to stereographic
coordinates. We find
\begin{align}
&0=\tilde{q}_{_{A}} {\tilde g}^{0A}\,, 
\end{align}
so that to $\mathcal{O}({\rho}^2)$
\begin{align}
&0=2A\omega+2A_u X_0 U \bar{\zeta}  e^{-2\beta} +K\eth u_0 (2 +\bar{\eth}X_0+2X_0\bar{\zeta})
+K\bar{\eth}u_0\eth X_0 \bar{\eth}u_0 (2+\bar{\eth}X_0+2 X_0\bar{\zeta})-\bar{J}\eth u_0 \eth X_0\,,
\label{e-fA}
\end{align}
\begin{align}
&\det(q_{_{AB}})=\det(g_{_{AB}})\rho^4\,, 
\end{align}
so that at  $\rho=0$
\begin{align}
&\omega =\frac{1+q^2+p^2}{1+\tilde{q}^2+\tilde{p}^2}
      \sqrt{1+\partial_q q_{0}+\partial_p p_{0}+\partial_q q_{0}\partial_p p_{0}
-\partial_p q_{0}\partial_q p_{0}}\,,
\label{e-exom}
\end{align}
and
\begin{equation}
J=\frac{q^{^{_A}} q^{^{_{_{B}}}} g_{_{AB}}}{2} \rho^2\,,
\end{equation}
so that at $\rho=0$
\begin{equation}
\qquad J=
\frac{(1+q^2+p^2)^2}{2(1+\tilde{q}^2+\tilde{p}^2)^2\omega^2}
\eth X_0 (2+\eth\bar{X}_0+2\bar{X} \zeta)\,.
\label{e-fJ}
\end{equation}
%

Explicit expressions for the null tetrad vectors $n_{_{[NP]}}^\alpha$ and
$m^\alpha$ (but not $\ell^\alpha$) will be needed. $n_{_{[NP]}}^\alpha$
is found by applying the coordinate transformation Eq.~(\ref{e-gttl}) to
$\tilde{n}_{_{[NP]}\alpha}$ (Eq.~(\ref{e-ntMpu})) and then raising the
index, giving to leading order in $\rho$
\begin{equation}
{n}_{_{[NP]}}^\alpha=\left(\frac{e^{-2\beta}\sqrt{2}}{\omega},
  \rho \frac{e^{-2\beta}(2\partial_u\omega+\bar{U}\eth\omega+U\bar{\eth}\omega
      +2W_c\omega)}{\sqrt{2}\omega^2},
  \frac{U^{^{_A}}e^{-2\beta}\sqrt{2}}{\omega}\right)\,.
\label{e-nNPG}
\end{equation}
The calculation of an expression for $m^\alpha$ is indirect. Let
$F^{^{_A}}$ be a dyad of the angular part of the general gauge metric, so
it must satisfy Eq.~(\ref{e-qa}), then \cite{Bishop97b,Babiuc:2009},
\begin{equation}
F^{^{_A}}=\left(
\frac{q^{^{_A}}\sqrt{K+1}}{2}-\frac{\bar{q}^{^{_A}} J}{2\sqrt{(K+1)}}\right)\,,
\end{equation}
with $F^{^{_A}}$ undetermined up to an arbitrary phase factor
$e^{-i\delta(u,x^{^{_A}})}$. We then define
$m_{_{[G]}}^\alpha$
\begin{equation}
m_{_{[G]}}^\alpha= e^{-i\delta}\rho (0,0,F^{^{_A}})\,.
\label{e-mF}
\end{equation}
The suffix ${}_{_{[G]}}$ is used to distinguish the above form
from that defined in Eq.~(\ref{e-ntMpu}) since $m_{_{[G]}}^\alpha
\ne m^\alpha$. However, it will be shown later (see
Secs.~\ref{s-N}, \ref{s-psi4} and Appendix~\ref{a-algebra})
that the value of
gravitational-wave descriptors is unaffected by the use of
$m_{_{[G]}}^\alpha$ rather than $m^\alpha$ in its evaluation;
thus it is permissible, for our purposes, to approximate
$m^\alpha$ by $m_{_{[G]}}^\alpha$.
We now transform $m_{_{[G]}}^\alpha$ in Eq.~(\ref{e-mF}) to the Bondi gauge,
\begin{equation}
\tilde{m}_{_{[G]}}^\alpha=\frac{\partial \tilde{x}^\alpha}{\partial x^\beta} m_{_{[G]}}^\beta
=\left(\partial_{_{B}}u_0 e^{-i\delta}\frac{\tilde{\rho}}{\omega} F^{^{_{_{B}}}},
\partial_{_{B}}\omega e^{-i\delta}\frac{\tilde{\rho^2}}{\omega^2} F^{^{_{_{B}}}},
\partial_{_{B}} \phi^{^{_A}}_0 e^{-i\delta}\frac{\tilde{\rho}}{\omega} F^{^{_{_{B}}}}\right)\,.
\label{e-MFB}
\end{equation}
The component
$\tilde{m}^1_{_{[G]}}$ is of the same order as $\tilde{\ell}^1$, and so  $\tilde{m}_{_{[G]}}^\alpha$
and $\tilde{m}^\alpha$ are not equivalent.
It can be checked (see Appendix~\ref{a-algebra}) \cite{Bishop:2014} that the angular part
of $\tilde{m}_{_{[G]}}^\alpha$ is equivalent to $\tilde{m}^\alpha$, since
$\tilde{m}_{_{[G]}}^\alpha \tilde{m}_\alpha=0$ and
\begin{equation}
\tilde{m}_{_{[G]}}^\alpha \bar{\tilde{m}}_\alpha=\nu\,,
\label{e-Mmb}
\end{equation}
where $|\nu|=1$.

Since we actually require $\nu=1$, Eq.~(\ref{e-Mmb}) can be used to set
the phase factor $\delta$ explicitly. The result is
\begin{equation}
e^{i\delta}=\frac{F^{^{_{_{B}}}} \bar{\tilde{q}}_{_{A}}}{\omega\sqrt{2}} \p_{_{B}} \phi^{^{_A}}_0
=\frac{1+q^2+p^2}{4\omega(1+\tilde{q}^2+\tilde{p}^2)}\sqrt{\frac{2}{K+1}}
\left((K+1)(2+\eth \bar{X}_0+2\bar{X}_0\zeta)-J\bar{\eth}\bar{X}_0\right)\,.
\label{e-delta}
\end{equation}
An alternative approach \cite{Bishop97b,Babiuc:2009}, to the phase factor
$\delta$ uses the condition that $m_{_{[G]}}^\alpha$ is parallel
propagated along $\scri$ in the direction ${n}_{_{[NP]}}^\alpha$,
yielding the evolution equation
\begin{equation}
    2i(\partial_u +U^{_A}\partial_{_A})\delta = \nabla_{_A} U^{_A}
     +h_{_{AC}} \bar F^{_C} ( (\partial_u +U^{_B} \partial_{_B}) F^{_A}
             - F^{_B} \partial_{_B} U^{_A}) \,,
    \label{e-evphase}
\end{equation}
where $\nabla_{_A}$ is the covariant derivative with respect to the
angular part of the metric $h_{_{AB}}$.

\subsection{The gravitational-wave strain}

An expression for the contravariant metric $\tilde{g}^{\alpha\beta}$ in
the Bondi gauge is obtained from Eqs.~(\ref{e-bsc}) and (\ref{e-BGC}),
and each metric variable is expressed as a Taylor series about
$\tilde{\rho}=0$ (\eg $\tilde{J}=0+\tilde{\rho}
\partial_{\tilde{\rho}}\tilde{J}+{\mathcal O}(\tilde{\rho}^2)$).
Applying the coordinate transformation \eqref{e-bs2b} we find
$g^{\alpha\beta}$, then use $\tilde{\rho}=\omega\rho+A^\rho\rho^2$ to
express each component as a series in $\rho$; note that the coefficients
are constructed from terms in the Bondi gauge, \eg
$\partial_{\tilde{\rho}}\tilde{J}$. Then both sides of
\begin{equation}
J=\frac{q_{_{A}} q_{_{B}} g^{^{_{AB}}}}{2\rho^2}\,,
\end{equation}
with $g^{^{_{AB}}}$ given in Eq.~\eqref{e-bsc}, are expressed as series
in $\rho$, and the coefficients of $\rho^1$ are equated.  This leads to
an equation in which $\partial_\rho J$ depends linearly on
$\partial_{\tilde{\rho}} \tilde{J}$
($=H=\lim_{\tilde{r}\rightarrow\infty} \tilde{r}\left(h_+ +i
h_\times\right)$, the rescaled strain defined in
Eq.~(\ref{e-NHdef})) \cite{Bishop:2014},
\begin{equation}
C_1 \partial_\rho J= C_2 \partial_{\tilde{\rho}} \tilde{J} 
                    +C_3 \partial_{\tilde{\rho}}\tilde{\bar{J}}+C_4\,,
\end{equation}
which may be inverted to give
\begin{align}
H=\partial_{\tilde{\rho}} \tilde{J}&=\frac{C_1\bar{C}_2\partial_\rho J
            -C_3 \bar{C}_1 \partial_{\rho}\bar{J}
            +C_3 \bar{C}_4 -\bar{C}_2 C_4}{\bar{C}_2 C_2-\bar{C}_3 C_3}\,,
\end{align}
where the coefficients are
\begin{align}
C_1&=\frac{4\omega^2 (1+\tilde{q}^2+\tilde{p}^2)^2}{(1+q^2+p^2)^2}\,,&
C_2&=\omega (2+\eth\bar{X}_0+2\bar{X}_0\zeta)^2\,,\\
C_3&=\omega (\eth X_0)^2\,,& 
C_4&=\eth A (4+2\eth \bar{X}_0+4 
\bar{X}_0\zeta)+\eth X_0 (2\eth\bar{A}+4\bar{A}\zeta) +4\eth\omega \eth u_0\,.
\label{e-Jtrht}
\end{align}
These results are obtained using computer algebra as discussed in
Appendix~\ref{a-algebra}.
The above formula for the wave strain involves intermediate variables,
and the procedure for calculating them is to solve equations for the
variable indicated in the following order: Eq.~(\ref{e-dx0}) for
$x_0^{^{_A}}$ and thus $X_0$, Eq.~(\ref{e-exom}) for $\omega$,
Eq.~(\ref{e-du0}) for $u_0$, Eq.~(\ref{e-omAu}) for $A^u$, and
Eq.~(\ref{e-fA}) for $A$.

\subsection{Conformal compactification}
\label{s-cc}
Here we give only a brief introduction to this topic, as these matters are discussed
more fully in many standard texts and reviews, \eg \cite{Wald84,Frauendiener04}.
We have made a coordinate compactification, resulting in the metric and null
tetrad being singular at $\rho=0$, which is therefore not included in the manifold.
Thus, quantities are not evaluated at $\rho=0$, but in the limit as $\rho\rightarrow 0$.
Introducing a conformal transformation has the advantage that this technical issue is
avoided and $\scri$ at $\rho=0$ is included in the manifold; but also that the
resulting formulas for ${\mathcal N}$ and $\psi^0_4$ are simpler.
(Of course, it should be possible to use the asymptotic Einstein equations to simplify
expressions derived in physical space, but due to the complexity of the formulas this
approach has not been adopted).

We use the notation $\hat{ }$ for quantities in conformal space.
In the general gauge, the conformal metric $\hat{g}_{\alpha\beta}$ is related to the
metric Eq.~(\ref{eq:lmet}) by $g_{\alpha\beta}=\rho^{-2}\hat{g}_{\alpha\beta}$ so that
\begin{equation}
   d\hat{s}^2 = -\left(e^{2\beta}(\rho^2+\rho W_c)
 -h_{_{AB}}U^{^{_A}}U^{^{_{_{B}}}}\right)du^2
        +2e^{2\beta}dud\rho -2 h_{_{AB}}U^{^{_{_{B}}}}dud\phi^{^{_A}}
        + h_{_{AB}}d\phi^{^{_A}}d\phi^{^{_{_{B}}}}.
\end{equation}
In a similar way, the Bondi gauge conformal metric $\hat{\tilde{g}}_{\alpha\beta}$ is related
to the Bondi gauge metric $\tilde{g}_{\alpha\beta}$ by
$\tilde{g}_{\alpha\beta}=\tilde{\rho}^{-2}\hat{\tilde{g}}_{\alpha\beta}$.
Thus $\hat{g}_{\alpha\beta}$ and $\hat{\tilde{g}}_{\alpha\beta}$ are regular
at $\rho=0$ (or equivalently at $\tilde{\rho}=0$) and so in the conformal
picture $\rho=0$ is included in the manifold. The conformal metrics
$\hat{g}_{\alpha\beta}$ and $\hat{\tilde{g}}_{\alpha\beta}$ are related by
\begin{equation}
\hat{g}_{\alpha\beta}=\rho^2 g_{\alpha\beta}=\rho^2 \frac{\partial {\tilde x}^\gamma}
     {\partial x^\alpha}
\frac{\partial {\tilde x}^\delta}{\partial x^\beta}{\tilde g}_{\gamma\delta}
=\frac{\rho^2}{\tilde{\rho}^2}\frac{\partial {\tilde x}^\gamma}{\partial x^\alpha}
\frac{\partial {\tilde x}^\delta}{\partial x^\beta}{\hat{\tilde g}}_{\gamma\delta}
=\frac{1}{\omega^2}\frac{\partial {\tilde x}^\gamma}{\partial x^\alpha}
\frac{\partial {\tilde x}^\delta}{\partial x^\beta}{\hat{\tilde g}}_{\gamma\delta}
+{\mathcal O}(\rho)\,,
\label{e-conf_inv}
\end{equation}
which at $\rho=0$ is the usual tensor transformation law with an additional factor
$\omega^{-2}$. A quantity that obeys this property is said to be {\em conformally
invariant} with weight $n$ where $n$ is the power of $\omega$ in the additional factor;
thus the metric tensor is conformally invariant with weight $-2$. In practice, it is
not necessary to establish a relation of the form Eq.~(\ref{e-conf_inv}) to prove
conformal invariance. The key step is to be able to show that a tensor quantity
$T^{a\cdots}_{b\cdots}$ satisfies $\hat{T}^{a\cdots}_{b\cdots}
=\rho^{-n}T^{a\cdots}_{b\cdots}$, then conformal invariance with weight $n$ easily
follows. In Eq.~(\ref{e-conf_inv}) the error term ${\mathcal O}(\rho)$ is shown
explicitly, although it turns out to be irrelevant since the relation is evaluated
at $\rho=0$. This is generally the case, so from now on the error terms will not be
taken into account; the one exception will be in the News calculation Sec.~\ref{s-N}
which in places involves off-$\scri$ derivatives (since $\p_\rho {\mathcal O}(\rho)
={\mathcal O}(1)$).

It is important to note that not all tensor quantities are conformally
invariant, and in particular this applies to the metric connection and
thus covariant derivatives
\begin{equation}
\hat{\Gamma}^\gamma_{\alpha\beta}=\Gamma^\gamma_{\alpha\beta}
+\frac{\delta^\gamma_\alpha \partial_\beta\rho+\delta^\gamma_\beta\partial_\alpha\rho
          -\hat{g}_{\alpha\beta}\hat{g}^{\gamma\delta}\partial_\delta\rho}{\rho},
\label{e-hatGG}
\end{equation}
and to the Ricci scalar in $n$-dimensions
\begin{equation}
\hat{R}=\rho^{-2}\left[R-2(n-1)g^{ab}\nabla_a\nabla_b\ln \rho
-(n-1)(n-2)g^{ab}(\nabla_a\ln \rho)(\nabla_b\ln \rho)\right]\,.
\label{e-RS}
\end{equation}
The Weyl tensor, however, is conformally invariant,
\begin{equation}
\hat{C}^\alpha_{\beta\gamma\delta}=C^\alpha_{\beta\gamma\delta}\,,
\qquad \mbox{and}\qquad 
\hat{C}_{\alpha\beta\gamma\delta}=\rho^2 C_{\alpha\beta\gamma\delta}\,,
\end{equation}
so that the forms $C^\alpha_{\beta\gamma\delta}$ and
$C_{\alpha\beta\gamma\delta}$ are conformally invariant with weights $0$
and $-2$ respectively. The construction of the conformal null tetrad
vectors is not unique. It is necessary that orthonormality conditions
analogous to Eq.~(\ref{e-nt_ortho}) be satisfied, and it is desirable
that the component of leading order in $\rho$ should be finite but
nonzero at $\rho=0$. These conditions are achieved by defining
\begin{equation}
\hat{n}_{_{[NP]}}^\alpha=n_{_{[NP]}}^\alpha\,,
\qquad \qquad \hat{m}_{_{[G]}}^\alpha
    =\frac{m_{_{[G]}}^\alpha}{\rho}\,.
\end{equation}
Thus $\hat{n}_{_{[NP]}}^\alpha$ and $\hat{\tilde{n}}_{_{[NP]}}^a$ are
related by the usual tensor transformation law, and
\begin{equation}
\hat{\tilde{m}}_{_{[G]}}^\alpha=\frac{\tilde{m}_{_{[G]}}^\alpha}{\tilde{\rho}}=
\frac{1}{\omega\rho}\frac{\partial \tilde{x}^\alpha}{\partial x^\beta}
m_{_{[G]}}^\beta =\frac{1}{\omega}\frac{\partial
  \tilde{x}^\alpha}{\partial x^\beta} \hat{m}_{_{[G]}}^\beta\,.
\end{equation}
With these definitions, $n_{_{[NP]}}^\alpha,m_{_{[G]}}^\alpha$ are
conformally invariant with weights 0 and 1 respectively.

Considering the conformally compactified metric of the spherical
2-surface described by the angular coordinates ($\tilde{\phi}^{^{_A}}$ or
$\phi^{^{_A}}$) at $\scri$, we have
\begin{equation}
ds^2=\tilde{\rho}^{-2} d\hat{\tilde{s}}^2=
\tilde{\rho}^{-2}q_{_{AB}}d\tilde{\phi}^{^{_A}}d\tilde{\phi}^{^{_{_{B}}}}
=\rho^{-2}d\hat{s}^2=\rho^{-2}h_{_{AB}}d\phi^{^{_A}} d\phi^{^{_{_{B}}}},
\end{equation}
so that
\begin{equation}
q_{_{AB}}d\tilde{\phi}^{^{_A}}d\tilde{\phi}^{^{_{_{B}}}}
=\omega^2h_{_{AB}}d\phi^{^{_A}}d\phi^{^{_{_{B}}}},
\end{equation}
since $\omega=\tilde{\rho}/\rho$. The curvature of $\scri$ is evaluated
in two different ways, and the results are equated. The metric on the LHS
is that of a unit sphere, and therefore has Ricci scalar ${\mathcal
  R}(\tilde{x}^{^{_A}})=2$; and the metric on the RHS is evaluated using
Eq.~(\ref{e-RS})with $n=2$. Thus,
\begin{equation}
2=\frac{1}{\omega^2}\left({\mathcal R}(\phi^{^{_A}})-2h^{^{_{AB}}}\nabla_{_{A}}\nabla_{_{B}}
    \log(\omega)\right)\,,
\end{equation}
leading to
\begin{equation}
2\omega^2+2h^{^{_{AB}}}\nabla_{_{A}}\nabla_{_{B}}\log(\omega)=
2K-\bar{\eth}\eth K+\frac 12 \left(\eth^2\bar{J}+\bar{\eth}^2 J\right)
+\frac{1}{4K}\left((\bar{\eth}\bar{J})(\eth J)-(\bar{\eth}{J})(\eth \bar{J})\right)\,,
\label{e-omJK}
\end{equation}
where the relationship between ${\mathcal R}(\phi^{^{_A}})$ and $J,K$ is
derived in \cite{Gomez97}, and where
$h^{^{_{AB}}}\nabla_{_{A}}\nabla_{_{B}}\log(\omega)$ is given in terms of
the $\eth$ operator in Eq.~(B1) of \cite{Bishop97b}.  Eq.~(\ref{e-omJK})
is a nonlinear elliptic equation, and in practice is not actually solved.
However, it will be used later, when considering the linearized
approximation, since in that case it has a simple analytic solution.

\subsubsection{The news ${\mathcal N}$}
\label{s-N}
A difficulty with evaluating the gravitational radiation by means of
Eq.~(\ref{e-NB}) is that it is valid only in a specific coordinate
system, so a more useful approach is to use the definition
\cite{Penrose:1963,Bishop97b,Babiuc:2009}
\begin{equation}
{\mathcal N}= \lim_{\tilde{\rho}\rightarrow 0}\frac{\hat{\tilde{m}}^\alpha \hat{\tilde{m}}^\beta
            \hat{\tilde{\nabla}}_\alpha \hat{\tilde{\nabla}}_\beta \tilde{\rho}}
            {\tilde{\rho}}\,.
\label{e-NBp}
\end{equation}
At first sight Eqs.~(\ref{e-NB}) and (\ref{e-NBp}) do not appear to be
equivalent, but the relationship follows by expanding out the covariant
derivatives in Eq.~(\ref{e-NBp})
\begin{equation}
{\mathcal N}= \lim_{\tilde{\rho}\rightarrow
  0}\frac{\hat{\tilde{m}}^\alpha \hat{\tilde{m}}^\beta
  (\partial_\alpha\partial_\beta\tilde{\rho}-\hat{\tilde{\Gamma}}^\gamma_{\alpha\beta}
  \partial_\gamma\tilde{\rho})}{\tilde{\rho}}
=-\lim_{\tilde{\rho}\rightarrow 0}\frac{q^{^{_A}} q^{^{_{_{B}}}}
  \hat{\tilde{\Gamma}}^1_{_{AB}} } {2\tilde{\rho}}\,,
\end{equation}
then expressing the metric coefficients $\tilde{J}$ etc. as power series
in $\tilde{\rho}$ as introduced just before Eq.~(\ref{e-as11}).  Using
the Bondi gauge conditions Eq.~(\ref{e-BGC}), it quickly follows that
$-q^{^{_A}} q^{^{_{_{B}}}} \hat{\tilde{\Gamma}}^1_{(0)AB}/2 =
\partial_{\tilde{u}} \tilde{J}/2$, which is zero, and the result follows
since $-q^{^{_A}} q^{^{_{_{B}}}} \hat{\tilde{\Gamma}}^1_{(1)AB}/2 =
\partial_{\tilde{u}}\partial_{\tilde{\rho}}\tilde{J}/2$.  Computer
algebra has been used to check that replacing $\hat{\tilde {m}}^\alpha$
in Eq.~(\ref{e-NBp}) by $\hat{\tilde {m}}_{_{[G]}}^\alpha$ (with $\tilde
{m}^\alpha_{_{[G]}}$ defined in Eq.~(\ref{e-MFB})) has no
effect\footnote{This result applies only to conformal space, not to
  physical space.}.

Because covariant derivatives are not conformally invariant, transforming
Eq.~(\ref{e-NBp}) into the general gauge is a little tricky. We need to
transform to physical space, where tensor quantities with no free indices
are invariant across coordinate systems, and then to conformal space in
the general gauge. From Eq.~(\ref{e-hatGG}), and using $\tilde{\rho}=
\rho\omega$ and $\tilde{\nabla}_\gamma \tilde{\rho}=\delta^1_\gamma$,
\begin{align}
\hat{\nabla}_\alpha\hat{\nabla}_\beta\tilde{\rho}&=
\nabla_\alpha\nabla_\beta\tilde{\rho} +\frac{\hat{g}_{\alpha
    \beta}\hat{g}^{\gamma
    1}-\delta^\gamma_\alpha\delta^1_\beta-\delta^\gamma_\beta\delta^1_\alpha}{\rho}
\nabla_\gamma(\rho\omega)\,, \label{e-d2trG}\\ 
\hat{\tilde{\nabla}}_\alpha\hat{\tilde{\nabla}}_\beta\tilde{\rho}&=
\tilde{\nabla}_\alpha\tilde{\nabla}_\beta\tilde{\rho}
+\frac{\hat{\tilde{g}}_{\alpha
    b}\hat{\tilde{g}}^{11}-2\delta^1_\alpha\delta^1_\beta}{\tilde{\rho}}\,.
\label{e-d2trB}
\end{align}
Now consider $\tilde{m}_{_{[G]}}^\alpha\tilde{m}_{_{[G]}}^\beta\times$
Eq.~(\ref{e-d2trB}) $-\;m_{_{[G]}}^\alpha m_{_{[G]}}^\beta\times$
Eq.~(\ref{e-d2trG}). Using the conditions that (a) scalar quantities are
invariant in physical space so that
$\tilde{m}_{_{[G]}}^\alpha\tilde{m}_{_{[G]}}^\beta
\tilde{\nabla}_\alpha\tilde{\nabla}_\beta\tilde{\rho} - m_{_{[G]}}^\alpha
m_{_{[G]}}^\beta \nabla_\alpha\nabla_\beta\tilde{\rho}=0$, (b)
$\hat{\tilde{g}}^{11}$ is zero to ${\mathcal O} (\tilde{\rho}^2)$, (c) $
m_{_{[G]}}^\alpha\delta^1_\alpha=0$, and (d) from Eq.~(\ref{e-MFB})
\begin{equation}
\tilde{m}_{_{[G]}}^\alpha \delta^1_\alpha =e^{-i\delta}\rho^2 F^A\p_A\omega=
\rho m_{_{[G]}}^\alpha\p_\alpha\omega\,,
\end{equation}
it follows that
\begin{equation}
\tilde{m}_{_{[G]}}^\alpha\tilde{m}_{_{[G]}}^\beta
\hat{\tilde{\nabla}}_\alpha \hat{\tilde{\nabla}}_\beta\tilde{\rho}
=m_{_{[G]}}^\alpha m_{_{[G]}}^\beta
\left(\hat{\nabla}_\alpha\hat{\nabla}_\beta(\rho\omega) -\hat{g}_{\alpha
  \beta}\left(\frac{\hat{g}^{11}\omega}{\rho}
+\hat{g}^{1\gamma}\partial_\gamma\omega\right)
-\frac{2\rho\p_\alpha\omega\p_\beta\omega}{\omega}\right)\,.
\label{e-mambd2tr}
\end{equation}
Since $m_{_{[G]}}^\alpha$ and $\hat{\nabla}_\alpha\rho$ are orthogonal,
we may write $\hat{\nabla}_\alpha\hat{\nabla}_\beta(\rho\omega) =
\omega\hat{\nabla}_\alpha\hat{\nabla}_\beta\rho
+\rho\hat{\nabla}_\alpha\hat{\nabla}_\beta\omega$, so that
\begin{equation}
\tilde{m}_{_{[G]}}^\alpha\tilde{m}_{_{[G]}}^\beta \hat{\tilde{\nabla}}_\alpha
\hat{\tilde{\nabla}}_\beta\tilde{\rho}
=m_{_{[G]}}^\alpha m_{_{[G]}}^\beta \left(\omega\hat{\nabla}_\alpha\hat{\nabla}_\beta\rho
+\rho\hat{\nabla}_\alpha\hat{\nabla}_\beta\omega
-\hat{g}_{\alpha \beta}\left(\frac{\hat{g}^{11}\omega}{\rho}
+\hat{g}^{1\gamma}\partial_\gamma\omega\right)
-\frac{2\rho\p_\alpha\omega\p_\beta\omega}{\omega}\right)\,.
\label{e-mambd2trs}
\end{equation}
This expression is simplified by (a) expanding out the covariant
derivatives, (b) expressing the metric as a power series in $\rho$ and
using $m_{_{[G]}}^\alpha m_{_{[G]}}^\beta \hat{g}_{(0)\alpha\beta}=0$,
and (c) using Eq.~(\ref{e-dom}),
\begin{equation}
\tilde{m}_{_{[G]}}^\alpha\tilde{m}_{_{[G]}}^\beta \hat{\tilde{\nabla}}_\alpha
\hat{\tilde{\nabla}}_\beta\tilde{\rho}=
m_{_{[G]}}^\alpha m_{_{[G]}}^\beta\left(- \omega\hat{\Gamma}^1_{a\beta}
+\rho\partial_\alpha\partial_\beta\omega 
-\rho\partial_\gamma\omega\hat{\Gamma}^\gamma_{\alpha \beta}
-\frac{\rho\omega e^{-2\beta}W_c \partial_\rho\hat{g}_{\alpha \beta}}{2}
-\frac{2\rho\p_\alpha\omega\p_\beta\omega}{\omega}\right)\,.
\end{equation}
Finally, Eq.~(\ref{e-mF}) is used to replace $m_{_{[G]}}^\alpha$ in terms
of $F^{^{_A}}$, and the whole expression is divided by $\tilde{\rho}^3$,
yielding~ \cite{Bishop97b,Babiuc:2009}
\begin{align}
{\mathcal N}&= 
\lim_{\tilde{\rho}\rightarrow 0}
\frac{\hat{\tilde{m}}^\alpha \hat{\tilde{m}}^\beta
            \hat{\tilde{\nabla}}_\alpha \hat{\tilde{\nabla}}_\beta \tilde{\rho}}
            {\tilde{\rho}}
= \lim_{\tilde{\rho}\rightarrow 0}
\frac{\hat{\tilde{m}}_{_{[G]}}^\alpha \hat{\tilde{m}}_{_{[G]}}^\beta
            \hat{\tilde{\nabla}}_\alpha \hat{\tilde{\nabla}}_\beta \tilde{\rho}}
            {\tilde{\rho}}\nonumber \\
&= \frac{e^{-2i\delta}}{\omega^2}\bigg[
-\lim_{\rho\rightarrow 0}\frac{F^{^{_A}}F^{^{_{_{B}}}} \hat{\Gamma}^1_{_{AB}}}{\rho}\nonumber \\
&+F^{^{_A}}F^{^{_{_{B}}}}\left(\frac{\partial_{_A}\partial_{_B}\omega}{\omega}
             -\frac{\hat{\Gamma}^{\gamma}_{_{AB}}\partial_\gamma\omega}{\omega}
             -\frac{\partial_\rho\hat{g}_{_{AB}}e^{-2\beta}W_c}{2}
             -\frac{2\partial_{_A}\omega\partial_{_B}\omega}{\omega^2}
\right)
\bigg].
\label{e-NG}
\end{align}
The limit is evaluated by expressing each metric coefficient as a power
series in $\rho$, \eg $J=J_{(0)}+\rho J_{(1)}$, and then writing
$F^{^{_A}}F^{^{_{_{B}}}} \hat{\Gamma}^1_{_{AB}}= F^{^{_A}}F^{^{_{_{B}}}}
\hat{\Gamma}^1_{_{AB}(0)}+ \rho F^{^{_A}}F^{^{_{_{B}}}}
\hat{\Gamma}^1_{_{AB}(1)}$. Direct evaluation combined with use of the
asymptotic Einstein Eq.~\eqref{e-asev} shows that
$F^{^{_A}}F^{^{_{_{B}}}} \hat{\Gamma}^1_{_{AB}(0)}=0$ (see
Appendix~\ref{a-algebra}), so that the limit evaluates to
$F^{^{_A}}F^{^{_{_{B}}}} \hat{\Gamma}^1_{_{AB}(1)}$. Further evaluation
of Eq.~(\ref{e-NG}) into computational $\eth$ form is handled by computer
algebra, as discussed in Appendix~\ref{a-algebra}.

The attentive reader may have noticed that the derivation above used
$\tilde{\rho}=\rho\omega$ rather than $\tilde{\rho}=\rho(\omega+\rho
A^\rho)$, so that $\p_\rho\omega$ should not be taken as $0$ but as
$A^\rho$. However, the corrections that would be introduced remain
${\mathcal O}(\rho)$ since (a) $\hat{m}_{_{[G]}}^1=0$, (b) in
Eq.~(\ref{e-mambd2trs}) the term $\hat{g}^{11}A^\rho$ contained in
$\hat{g}^{1\gamma}\p_\gamma\omega$ is ${\mathcal O}(\rho) A^\rho$, and
(c) in Eq.~(\ref{e-NG}) the term $F^A F^B \hat{\Gamma}^1_{_{AB}}A^\rho$
contained in $F^A F^B \hat{\Gamma}^\gamma_{_{AB}}\p_\gamma\omega$ is also
${\mathcal O}(\rho) A^\rho$.

\subsubsection{The Newman--Penrose quantity $\psi^0_4$}
\label{s-psi4}
The Newman--Penrose quantity $\psi_4$, and its re-scaled version
$\psi_4^0$, were introduced in section~\ref{s-NP}, and defined there for
the case of physical space.  Because the Weyl tensor is conformally
invariant, it is straightforward to transform the earlier definition into
one in the conformal gauge. Thus, in the conformal Bondi gauge,
\begin{equation}
\psi^0_4=\lim_{\tilde{\rho}=0} \frac{\hat{\tilde{C}}_{\alpha\beta\mu\nu}
  \hat{\tilde{n}}_{_{[NP]}}^\alpha \bar{\hat{\tilde{m}}}^\beta
  \hat{\tilde{n}}_{_{[NP]}}^\mu
  \bar{\hat{\tilde{m}}}^\nu}{\tilde{\rho}}\,,
\label{e-psi4B}
\end{equation}
and again, as in the case of the news ${\mathcal N}$, the limiting
process means that the metric variables need to be expressed as power
series in $\tilde{\rho}$. Calculating the Weyl tensor is discussed in
Appendix~\ref{a-algebra}, and the result is
$\psi^0_4=\p^2_{\tilde{u}}\p_{\tilde{\rho}}\bar{\tilde{J}}$ as given in
Eq.~(\ref{e-psi4JB}). The Appendix also checks that replacing
$\hat{\tilde {m}}^\alpha$ in Eq.~(\ref{e-psi4B}) by $\hat{\tilde
  {m}}^\alpha_{_{[G]}}$ (with $\tilde {m}^\alpha_{_{[G]}}$ defined in
Eq.~(\ref{e-MFB})) does not affect the result for $\psi^0_4$.

In this case, transforming Eq.~(\ref{e-psi4B}) to the conformal general
gauge is straightforward, since the tensor quantities are conformally
invariant and the net weight is 0. The result is
\begin{equation}
\psi^0_4=\frac{1}{\omega}\lim_{\rho=0} \frac{\hat{C}_{\alpha\beta\mu\nu}\hat{n}_{_{[NP]}}^\alpha
 \bar{\hat{m}}_{_{[G]}}^\beta \hat{n}_{_{[NP]}}^\mu \bar{\hat{m}}_{_{[G]}}^\nu}{\rho}\,,
\label{e-psi4G}
\end{equation}
where $\hat{m}_{_{[G]}}^\alpha=e^{-i\delta}(0,0,F^A)$, and is further
evaluated, by directly calculating the Weyl tensor, in
Appendix~\ref{a-algebra} \cite{Babiuc:2009} (but note that this reference
uses a different approach to the evaluation of $\psi^0_4$).

\subsection{Linearized case}

In the linearized case the Bondi-Sachs metric variables $\beta,J,U,W_c$
and the coordinate transformation variables
$u_0,A^u,(\omega-1),A^\rho,x^{^{_A}}_0,A^{^{_A}}$ are regarded as
small. Algebraically, the approximation is implemented by introducing a
parameter $\epsilon=$ max$(|\beta|,|J|,|U|,|W_c|)$ in a neighbourhood of
$\scri$. Then, the metric variables are re-written as
$\beta\rightarrow\epsilon\beta$ etc., and quantities such as ${\mathcal
  N},\psi^0_4$ are expressed as Taylor series in $\epsilon$ with terms
${\mathcal O}(\epsilon^2)$ ignored, leading to considerable
simplifications. It is common practice to assume that the error in the
approximation is about $\epsilon^2$. While computational results do not
contradict this assumption, a word of caution is needed: no work on
establishing a formal error bound for this problem has been reported.

Equations~(\ref{e-du0}) to (\ref{e-dom}), and Eq.~(\ref{e-fA}), simplify
to
\begin{equation}
\partial_u u_0=(\omega-1)+2\beta\,,\qquad
\partial_u \phi^{^{_A}}_0=-U^{^{_A}}\,,\qquad
\partial_u \omega=-W_c/2\,,
\;\; A=-\eth u_0\,.
\label{e-d_du_x0}
\end{equation}
It will also be useful to note the linearized form of Eq.~\eqref{e-fJ},
\begin{equation}
J=\eth X_0\,.
\label{e-JXlin}
\end{equation}
In the linearized case, Eq.~(\ref{e-omJK}) takes the form
\cite{Bishop-2005b}
\begin{equation}
2+4 (\omega-1) +2\bar{\eth}\eth \omega=2+\frac 12
\left(\eth^2\bar{J}+\bar{\eth}^2 J\right)\,,
\end{equation}
which is solved by decomposing $\omega$ and $J$ into spherical harmonic
components
\begin{equation}
\omega=1+\sum_{\ell\ge 2,|m|\le\ell}Y^{\ell\,m}\omega_{\ell\,m},
\;\;J=\sum_{\ell\ge 2,|m|\le\ell}\,{}_2Y^{\ell\,m}J_{\ell\,m},
\end{equation}
leading to
\begin{equation}
\omega_{\ell\,m}(4-2\Lambda)=-\Re(J_{\ell\,m})\Lambda(2-\Lambda)
        \sqrt{\frac{1}{(\ell+2)\Lambda(\ell-1)}}\,,
\end{equation}
[recall that $\Lambda=\ell(\ell+1)$] so that
\begin{equation}
\omega_{\ell\,m}=-\frac{\Lambda}{2}\sqrt{\frac{1}{(\ell+2)\Lambda(\ell-1)}}\Re(J_{\ell\,m})\,.
\label{e-omlm}
\end{equation}

Evaluating Eq.~\eqref{e-NG} for the news, and using Eq.~\eqref{e-JXlin},
is discussed in Appendix~\ref{a-algebra}. The result is
\cite{Bishop-2005b}
\begin{equation}
{\mathcal N}=\frac{1}{2\rho}\left(\partial_u J+\eth U\right)
+\frac 12 \left(\eth^2 \omega +\partial_u\partial_\rho J+\partial_\rho\eth U\right)\,.
\label{e-Nlinraw}
\end{equation}
Now from the linearized asymptotic Einstein equations, $\partial_u J+\eth
U=0$ and $\partial_\rho U-2\eth \beta=0$, so we get
\begin{equation}
{\mathcal N}=\frac 12 \left(\eth^2 \omega +\partial_u\partial_\rho J+2\eth^2\beta\right)\,.
\end{equation}
The result is more convenient on decomposition into spherical harmonics,
${\mathcal N}=\sum\,{}_2Y^{\ell\,m}{\mathcal N}_{\ell\,m}$,
\begin{equation}
{\mathcal N}_{\ell\,m}=-\frac{\Lambda\Re(J_{\ell\,m})}{4}
        +\frac{\partial_u\partial_\rho J_{\ell\,m}}{2}
        +\sqrt{(\ell+2)\Lambda(\ell-1)}\beta_{\ell\,m}.
\label{e-linNlm}
\end{equation}

In the linearized case, the evaluation of $\psi^0_4$ is straightforward,
because the Weyl tensor is a first-order term so the null tetrad vectors
need be correct only to zeroth order. As discussed in
Appendix~\ref{a-algebra}, we find~ \cite{Babiuc:2009}
\begin{equation}
\psi^0_4=\lim_{\tilde{\rho}\rightarrow 0}\left[
\frac{\partial_u\bar{\eth}\bar{U}+\partial^2_u\bar{J}}{\tilde{\rho}}
      +\frac{\rho}{\tilde{\rho}}
           \left(-\bar{\eth}\bar{U} +\partial_u\partial_\rho\bar{\eth}\bar{U}
          - \partial_u\bar{J} +\partial^2_u\partial_\rho\bar{J}
      -\frac 12 \bar{\eth}^2W_c \right)\right]\,.
\label{e-psi4linraw}
\end{equation}
It would appear that $\psi^0_4$ is singular, but applying the asymptotic
Einstein equation Eq.~(\ref{e-asev}) we see that these terms cancel;
further, to linear order the deviation of $\omega$ from unity is
ignorable, so that
\begin{equation}
\psi^0_4=\left(- \bar{\eth}\bar{U} +\partial_u\partial_\rho\bar{\eth}\bar{U}
          -\partial_u\bar{J} +\partial^2_u\partial_\rho\bar{J}
          -\frac 12 \bar{\eth}^2W_c \right)\,.
\label{e-psi4a}
\end{equation}
Eq.~(\ref{e-psi4N}) stated a relationship between $\psi^0_4$ and the news
${\mathcal N}$ which should remain true in this general linearized
gauge. In order to see this, we modify Eq.~(\ref{e-psi4a}) by applying
Eq.~(\ref{e-d_du_x0}), Eq.~(\ref{e-as1A}) and Eq.~(\ref{e-asev}) to the
terms $W_c$, $\partial_\rho\bar{U}$ and $\partial_u\bar{J}$,
respectively, yielding
\begin{equation}
\psi^0_4=2\partial_u\bar{\eth}^2\beta+ \bar{\eth}^2 \partial_u\omega
     +\partial^2_u\partial_\rho\bar{J}\,,
\end{equation}
from which it is clear that $\psi^0_4=2\partial_u\bar{\mathcal N}$.

In the linearized approximation, the wave strain Eq.~(\ref{e-Jtrht})
simplifies to \cite{Bishop:2014}
\begin{equation}
H=\partial_\rho J-\eth A\,,
\label{e-JtrhtLin}
\end{equation}
and using Eq.~(\ref{e-d_du_x0}) to replace $A$,
\begin{equation}
H=\partial_\rho J+\eth^2 u_0\,.
\label{e-Jtrhtl}
\end{equation}
An expression for $u_0$ is obtained using the first relationship in
Eq.~(\ref{e-d_du_x0}), $\p_u u_0=(\omega-1)+2\beta$, which is integrated
to give $u_0$. It is clear that $u_0$ is subject to the gauge freedom
$u_0\rightarrow u_0^\prime = u_0 + u_{_G}$, provided that $\partial_u
u_{_G}=0$ so that $u_{_G}=u_{_G}(x^{_A})$. Thus the wave strain $H$ is
subject to the gauge freedom $H\rightarrow H^\prime = H + H_{_G}$, where
$H_{_G}=\eth^2 u_{_G} (x^{_A})$. Decomposing $H$ into spherical
harmonics, $H=\sum\,{}_2Y^{\ell\,m}H_{\ell\,m}$, it follows that
\begin{equation}
H_{\ell\,m}(u)=\p_\rho J_{\ell\,m}(u)+\sqrt{(\ell+2)\Lambda(\ell-1)}
\int^u \omega_{\ell\,m}(u^\prime)+2\beta_{\ell\,m}(u^\prime) du^\prime\,,
\label{e-Jtrhtllm}
\end{equation}
with $\omega_{\ell\,m}$ given by Eq.~\eqref{e-omlm}. The gauge freedom
now appears as a constant of integration for each spherical harmonic mode
in Eq.~\eqref{e-Jtrhtllm}. This freedom needs to be fixed by a gauge
condition. Normally the spacetime is initially dynamic but tends to a
final state that is static, for example the Kerr geometry.  In such a
case, we impose the condition $H_{\ell\,m}(u)\rightarrow 0$ as
$u\rightarrow \infty$.  The same gauge freedom would occur if the wave
strain $H$ were obtained by time integration of the news ${\mathcal N}$,
in this case appearing as an arbitrary ``constant'' of integration
$f(\tilde{x}^{_A})$.

\newpage
\section{Numerical Implementations of the Characteristic Approach}
\label{s-CE}

The idea of combining the ``3+1'' and characteristic approaches to
extract the gravitational-wave signal from a numerical simulation was
introduced in the 1990s \cite{Bishop92,Bishop93}. However, this early
work focused on using the combination for the whole spacetime and was
called Cauchy-characteristic matching (CCM). Subsequently, implementation
difficulties with CCM led to the development of something less ambitious
known as characteristic extraction (CE). Under certain conditions (which
in practice are achievable), CE is just as accurate as CCM. The advantage
of CCM, if it can be implemented, is that it has the potential to make a
significant contribution to overall code efficiency \cite{Bishop96}.  An
outline of what is meant by CCM and CE, and the differences between the
two approaches, is illustrated in Fig.~\ref{f-ccem} and described in the
caption.

\epubtkImage{}{%
\begin{figure}[htb]
  \centering
   \includegraphics[width=9cm]{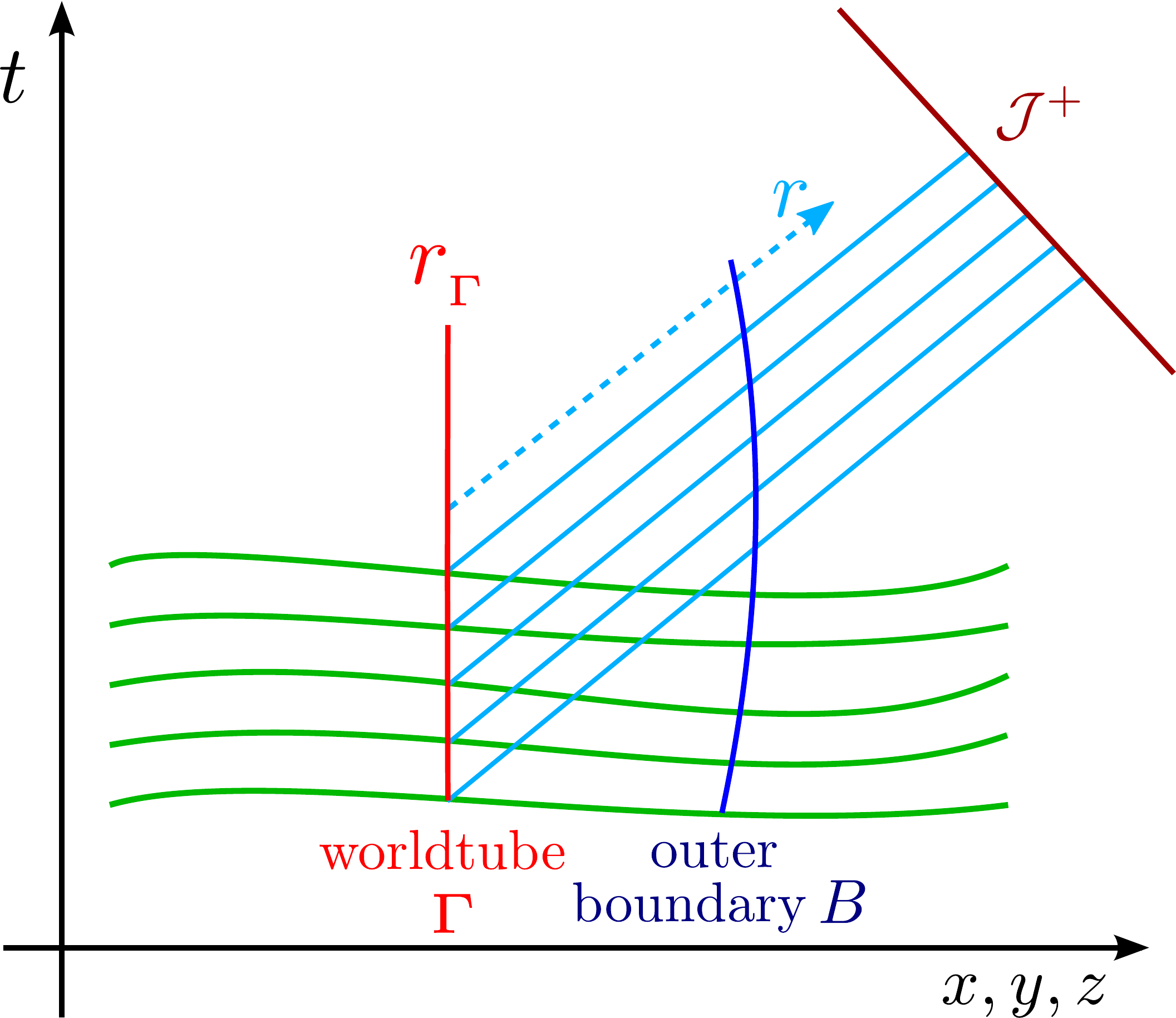}
   \caption{Schematic illustration of Cauchy-characteristic matching (CCM)
      and characteristic extraction(CE). In both cases there is a Cauchy
      evolution {{green slices}}, and a characteristic evolution {{light
      blue slices}} between $r_\Gamma$ and $\scri$. The difference is
      that in CCM the outer boundary of the Cauchy evolution is at the
      {{worldtube $r_\Gamma$}} with boundary data supplied by the
      characteristic evolution; and in CE the outer boundary of the
      Cauchy evolution is as shown in {{dark blue}} and is subject to a
      boundary condition that excludes incoming gravitational
      waves.}  \label{f-ccem}
\end{figure}}

As first steps towards CCM in relativity, it was implemented for the
model problem of a nonlinear scalar wave
equation \cite{Bishop-etal-1996:Cauchy-characteristic-matching,Bishop97}
without any symmetries, and for the Einstein equations with a scalar
field under the condition of spherical
symmetry \cite{Gomez96,Lehner2000}. There has been a series of papers on
CCM under axial symmetry \cite{Clarke:1994,Clarke:1995,dInverno-Vickers-1997,dInverno-Dubal-Sarkies-2000,dInverno:96,Dubal:1995,Dubal1998}.
A detailed algorithm for CCM in relativity in the general case was
presented in \cite{Bishop98a}. The stable implementation of matching is
quite a challenge, and this goal has not yet been
achieved \cite{Szilagyi00a,Szilagyi00}; although a stable implementation
without symmetry has been reported with the Einstein equations linearized
and using harmonic ``3+1''
coordinates \cite{Szilagyi02a,Szilagyi02b}. The issue of progress towards
CCM is much more fully discussed in the review by
Winicour \cite{Winicour05}.

As a consequence of the difficulties with a stable implementation of CCM,
in the 2000s attention shifted to the issue of developing CE, for which
stability is not expected to be an issue. Further, although CCM has the
advantage of high computational efficiency \cite{Bishop96}, it was
realized that CE can be as accurate as CCM, provided the outer boundary
of the Cauchy evolution is sufficiently far from the worldtube $\Gamma$
that the two are not causally related,
as indicated in Fig.~\ref{fig:causal_bc}. The implementation of CE for a
test problem was described in 2005 \cite{Babiuc:2005pg}. Subsequently,
codes have been developed that yield useful results for the astrophysical
problem of the inspiral and merger of two black
holes \cite{Reisswig:2009us,Reisswig:2010,Reisswig:2010a,Babiuc:2011,Babiuc:2011b}. Work that uses, rather than develops, characteristic
extraction includes \cite{Ott:2011,Reisswig:2011a,Reisswig2012b}. There is
an alternative approach~\cite{Helfer2010} that yields the emitted energy, momentum, and angular momentum, although it has not been implemented
numerically.

\epubtkImage{}{%
\begin{figure}[htb]
  \begin{center}
    \includegraphics[width=14cm]{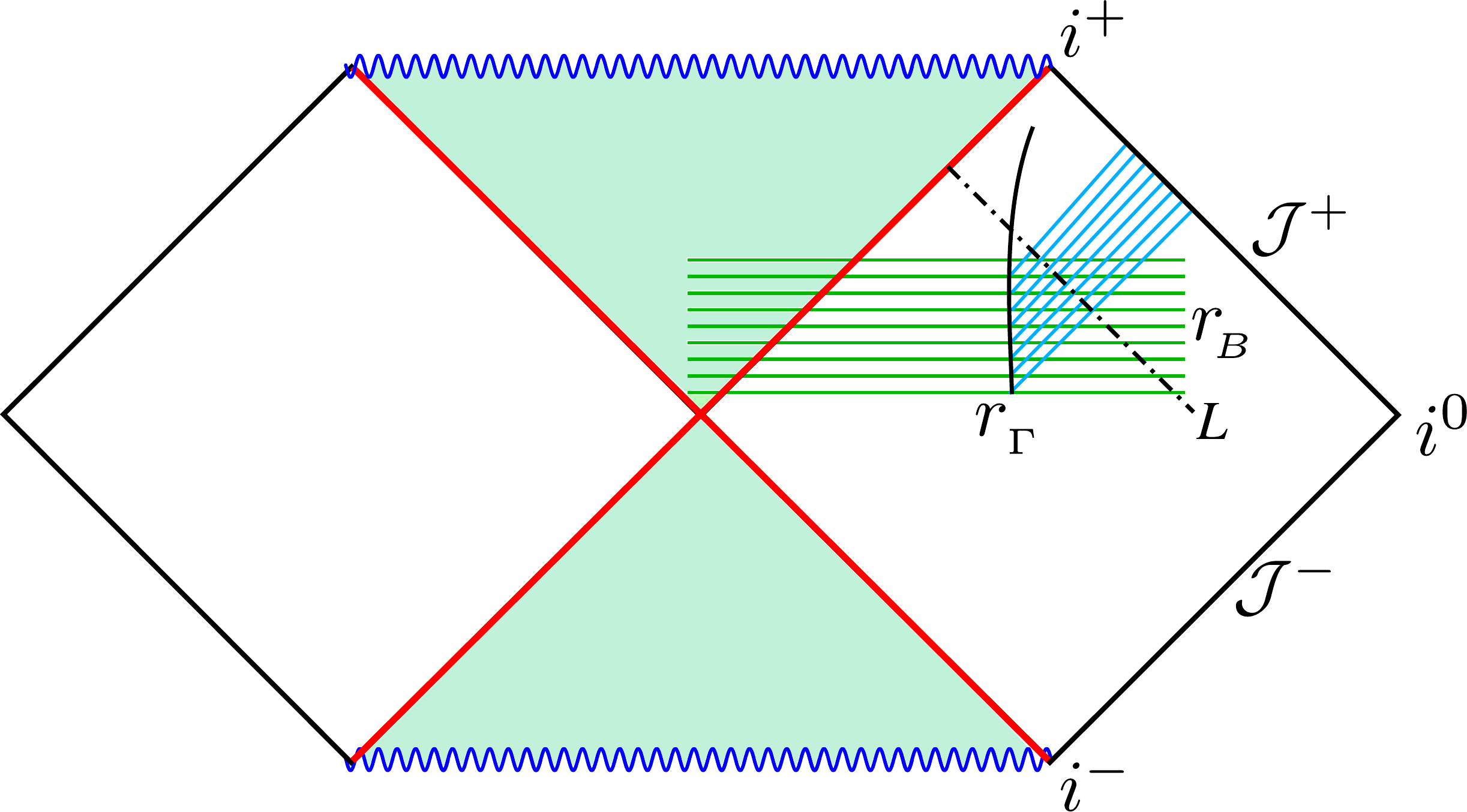}
  \end{center}
\caption{Portions of the Kruskal diagram that are determined
  numerically. The green horizontal lines indicate the region of
  spacetime that is determined by the Cauchy evolution and which has
  finite spatial extent with artificial outer boundary at $r_{_B}$. The
  light blue diagonal lines indicate the region that is determined by
  characteristic evolution, and which starts off from a worldtube
  $\Gamma$ located at $r_{_\Gamma}$ using boundary data from the Cauchy
  evolution. The future Cauchy horizon of the Cauchy initial data is
  indicated by the dotted diagonal line $L$ parallel to $\scri$. As long
  as the worldtube $\Gamma$ is located within the future Cauchy horizon,
  the numerically evolved subset of the spacetime is consistently
  determined. }
   \label{fig:causal_bc}
\end{figure}}

The key feature of characteristic extraction is just a coordinate
transformation, from Cauchy to characteristic coordinates in a
neighbourhood of the worldtube $\Gamma$. However, it is a more
complicated procedure than it might appear, because the Bondi-Sachs
radial coordinate $r$ is a surface area coordinate, so it cannot be
expressed explicitly in terms of the ``3+1'' coordinates. This
complicates the matter in two ways
\begin{enumerate}
\item The coordinate transformation has to be made in two steps, firstly
  to a null coordinate system in which the radial coordinate is an affine
  parameter on the outgoing null radial geodesics, and secondly to
  Bondi-Sachs coordinates.

\item In general $\Gamma$ is not a worldtube of constant $r$, so setting
  data at the innermost radial grid point of the Bondi-Sachs system
  requires special care.
\end{enumerate}

Implementations of characteristic extraction mainly
follow \cite{Bishop98a}, but do differ in certain aspects. Further, the
field clearly needs to develop since most implementations are second-order
accurate. This is a significant limitation since codes with higher order
accuracy have existed for some time on the Cauchy side; and recently the
first fourth order characterictic code has been
reported \cite{Reisswig:2012}, as well as a spectral characteristic
code \cite{Handmer:2015}. An important recent development is the
implementation of spectral characteristic extraction~\cite{Handmer:2015b,
Handmer:2016}, so that the whole computation is spectrally convergent.
Below we outline the characteristic extraction
procedure including some of the variations currently in use.

\subsection{Worldtube boundary data}

Characteristic extraction is conceptually a post-processing procedure, as
opposed to CCM in which the ``3+1'' and characteristic codes must run in
step with each other. The transformation to Bondi-Sachs coordinates will
require of the ``3+1'' data the full four-metric and all its first
derivatives on the extraction worldtube $\Gamma$. Thus the first issue is
to consider precisely what data the ``3+1'' code should dump to file, for
subsequent processing by the characteristic extraction code. The
simplistic solution of just dumping everything is not practical, because
the data set is too large, and this is even the case should the data dump
be restricted to those grid-points that are close to $\Gamma$. Some form
of data compaction is required. On a given timeslice the extraction
surface is spherical, so the natural compaction procedure is
decomposition in terms of spherical harmonics. It turns out that this
procedure also has some beneficial side-effects
\begin{enumerate}
\item It filters out high frequency noise.
\item It greatly simplifies the process of interpolation onto a regular angular
grid.
\end{enumerate}

Assuming that the ``3+1'' spacelike coordinates are approximately
Cartesian $x^i_{\scriptscriptstyle [C]}=(x,y,z)$, the extraction
worldtube $\Gamma$ is defined by
\begin{equation}
R^2=x^2+y^2+z^2,
\end{equation}
for some fixed radius $R$. Of course the above is a simple
coordinate-specific, rather than a geometric, definition; but in practice
the definition has worked well and $\Gamma$ has not exhibited any
pathologies such as becoming non-convex. As indicated above, the
extraction code will need the full four-metric and its first-derivatives,
and it is a matter of choice as to whether conversion from the ``3+1''
variables (such as lapse, shift and three-metric) to four-metric is performed in
the ``3+1'' code or in the extraction routine; for simplicity, this
discussion will be on the basis that the conversion is performed in the
``3+1'' code. The conversion formulas are given in
Eq.~\eqref{eq:ADM}. The time derivatives of the four-metric could be found
by finite differencing, but the results are likely to be less noisy if
they can be expressed in terms of other variables in the ``3+1'' code --
\eg the ``1+log'' slicing condition and the hyperbolic
$\tilde{\Gamma}$-driver condition \cite{Pollney2011b}, if being used,
would mean that time derivatives of the lapse and shift are known
directly, and the time derivative of the three-metric may be obtainable from
the extrinsic curvature. For the spatial derivatives of the four-metric, it
is sufficient to calculate and write to file only the radial derivative,
since $\p_x$, $\p_y$, $\p_z$ can later be reconstructed from the radial
derivative and angular derivatives of the spherical harmonics which are
known analytically. The radial derivative in terms of the Cartesian
derivatives is
\begin{equation}
\p_R = \frac{1}{R} \left(x\p_x+y\p_y+z\p_z \right)\,.
\end{equation}
Having calculated the above variables and derivatives at ``3+1'' grid
points in a neighbourhood of $\Gamma$, they must each be interpolated
onto points on the coordinate sphere $\Gamma$ using (at least)
fourth-order interpolation. Then each quantity $A$, whether scalar,
vector or tensor, is decomposed as
\begin{equation}
A_{\ell\,m} = \int_{S^2} d\Omega\, \bar{Y}^{\ell\,m} A(\Omega)\,,
\end{equation}
and the $A_{\ell\,m}$ are written to file. The decomposition is performed
for $\ell\le\ell_{\max}$, and in practice $\ell_{\max}\approx 8$.

A variation of the above procedure was introduced
by \cite{Babiuc:2011}. Instead of calculating the radial derivative of a
quantity $A$, the idea is to decompose $A$ into a product of spherical
harmonics and Chebyshev polynomials in $r$. More precisely, we consider a
``thick'' worldtube $R_1<R<R_2$ and the idea is to seek coefficients
$A_{k,\ell,m}$ such that we may write
\begin{equation}
A=\sum_{k,\ell,m}A_{k\ell\, m}U^k(\tau(R))Y^{\ell\,m}\,,
\label{e-UkY}
\end{equation}
where $U^k$ is a Chebyshev polynomial of the second kind, and
\begin{equation}
\tau(R)=\frac{2R-R_1-R_2}{R_2-R_1}\,,
\end{equation}
so that, within the thick worldtube, the argument of $U^k$ has the
required range of -1 to 1. The coefficients $A_{k,\ell,m}$ are then
determined by a least squares fit to the data at each ``3+1'' grid point
within the thick worldtube. The decomposition is carried out for $k\le
k_{\max}$, and in practice \cite{Babiuc:2011} takes $k_{\max}=6$. Thus,
this procedure involves writing three times as much data to file compared
to that of calculating the radial derivative, but is probably more
accurate and has the flexibility of being able to reconstruct data,
including radial derivatives, at any point within the thick
worldtube. Ref. \cite{Babiuc:2011} also introduced the option of
calculating time derivatives via a Fourier transform process, so being
able to filter out high frequency noise.

\subsection{Reconstruction from spectral modes}

The variables are reconstructed via
\begin{equation}
A = \sum_{\ell\, m} A_{\ell\,m} Y^{\ell\,m}\,
\end{equation}
in the case of decomposition only into angular modes, or via Eq.~(\ref{e-UkY}) in
the case of decomposition into both angular and radial modes. The radial derivatives
at the extraction worldtube $R=R_\Gamma$ are obtained either directly, or by
analytic differentiation in the case that the reconstruction is in terms of
Chebyshev polynomials. We then need to obtain the Cartesian derivatives in terms of
radial and angular derivatives, and by the chain rule
\begin{equation}
\p_i A =\sum_{\ell\,m}\left(Y^{\ell\,m}\p_i R\p_{_R} A_{\ell\,m}
+ A_{\ell\,m}\p_i\phi^{2}\p_{\phi^2} Y^{\ell\,m}
+ A_{\ell\,m}\p_i\phi^{3}\p_{\phi^3} Y^{\ell\,m}
\right)\,,
\end{equation}
where $\phi^{_A}=\phi^2,\phi^3$ are the angular coordinates. The
angular derivatives of the $Y^{\ell\,m}$ may be re-expresed in terms of spin-weighted
spherical harmonics, and $\p_i \phi^{_A}$ expressed explicitly in terms of the Cartesian
coordinates. The details depend on the specific angular coordinates being used, and in the
common case of stereographic coordinates the formulas are \cite{Reisswig:2010a}
\begin{equation}
\p_i A = \sum_{\ell\,m}\left(\frac{A_{\ell\,m}\sqrt{\ell(\ell+1)}}{1+q^2+p^2}
  \left[{}_1Y^{\ell\,m}(\p_i q-i\p_i p)- {}_{-1}Y^{\ell\,m}(\p_i q+i \p_ip) \right]
  + \p_i R \p_{_R} A_{\ell\,m}Y^{\ell\,m}\right)\,,
\label{eq:radial-deriv}
\end{equation}
where
\begin{align}
\p_i R &= \p_i\sqrt{x^2+y^2+z^2} = \frac{(x,y,z)}{R}\,, \\
\p_i q &= \p_i\left(\frac{x}{\sqrt{x^2+y^2+z^2}\pm z} \right) = \frac{1}{(R\pm z)^2}
\left(R\pm z-x^2/R,\, -xy/R,\, -xz/R\mp x \right)\,,\\
\p_i p &= \p_i\left(\frac{\pm y}{\sqrt{x^2+y^2+z^2}\pm z} \right) = \frac{1}{(R\pm z)^2}\left(\mp xy,\, \pm R+z\mp y^2/R,\, -y\mp yz/R \right)\,,
\end{align}
where the upper sign is valid for the north patch and the lower sign is valid for the
south patch.

\subsection{Transformation to null affine coordinates}

In this section we construct the coordinate transformation from the
Cartesian like ``3+1'' coordinates to a null coordinate system in which
the radial coordinate is an affine parameter rather than the Bondi-Sachs
surface area coordinate. As already mentioned, we need this intermediate
step because the surface area coordinate cannot be expressed as a function
of only the ``3+1'' coordinates, but would also need terms involving
the three-metric $\gamma_{ij}$. The procedure is illustrated schematically in
Fig.~\ref{f-cce-idea}.

\epubtkImage{}{%
\begin{figure}[htb]
\begin{center}
\includegraphics[scale=0.4]{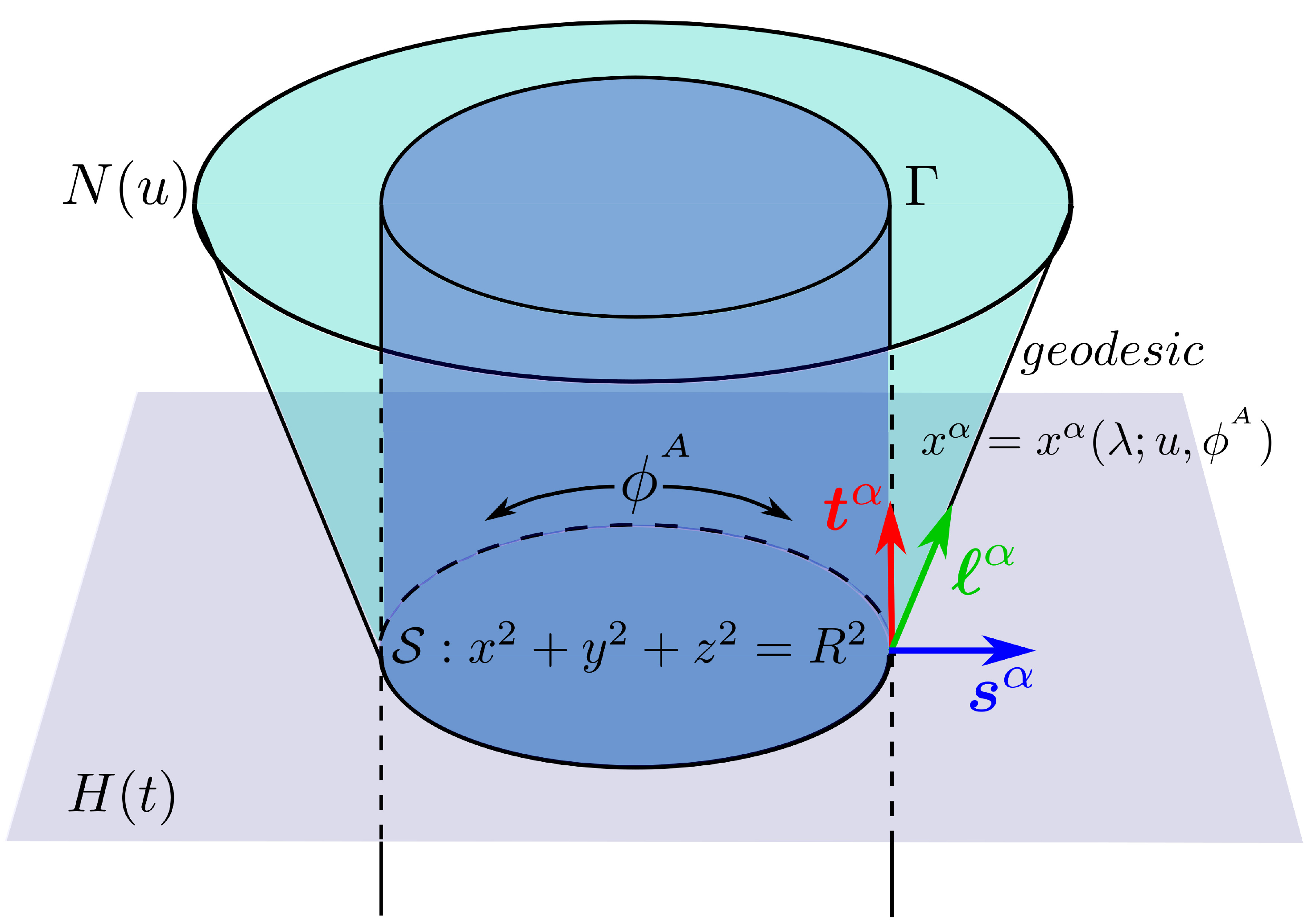}
\end{center}
\caption{Schematic illustration of the (first stage) construction of
  characteristic coordinates and metric.}
\label{f-cce-idea}
\end{figure}}

On the extraction worldtube $\Gamma$ we can simply define the coordinate
transformation, but off $\Gamma$ it will need to be calculated. The null
affine coordinates are $x_{\scriptscriptstyle
  [N]}^\alpha=(u,\lambda,q,p)$, and the relation to the ``3+1''
coordinates $x_{\scriptscriptstyle [C]}^\alpha=(t,x,y,z)$ on $\Gamma$ is
defined to be
\begin{equation}
u=t,\;\lambda=0\,,
\label{e-utl}
\end{equation}
with $q, p$ given by Eq.~(\ref{e-C2S}) for $r=R_\Gamma$.

Although the above is given in terms of stereographic angular coordinates
$(q,p)$, rather than general angular coordinates $\phi^{^{_A}}$, the
formulas that follow will not be specific to stereographic coordinates.

The unit normal $n^{\mu}$ to the hypersurface $\Sigma_{t}$ is determined
from the lapse and shift as stated in Eq.~(\ref{eq:eulerian_observer}).
Let $s^{\alpha}=(s^{i},0)$ be the outward pointing unit normal to the
section $S_{t}$ of the worldtube at time $t^{n}$. By construction,
$s^{i}$ lies in the slice $\Sigma_{t}$, and is given by Eq.~\eqref{e-sxyz}.
The generators $\ell^{\alpha}$ of the outgoing null cone through
$S_{t}$ are given on the worldtube by
\begin{equation}
   \ell^{\alpha} = \frac{n^{\alpha} + s^{\alpha}} 
   {\alpha - \gamma_{ij} \beta^{i} s^{j}}\,,
   \label{e-ell}
\end{equation}
which is normalized so that $\ell^{\alpha} t_{\alpha} = -1$, where
$t^{\alpha} = (1,0,0,0)$ is the Cauchy evolution vector.

We may now build the coordinate transformation between the ``3+1''
Cartesian coordinates $x^{\alpha}$ and the (null) affine coordinates
${y}^{\alpha}$. As already discussed, we need this in a neighborhood of
the worldtube, not just on the worldtube. Along each outgoing null
geodesic emerging from $S_{t}$, angular and time coordinates are defined
by setting their values to be constant along the rays, and equal to their
values on the worldtube.  Geometrically, the idea is that we define
$(u,q,p)$ to be constant on each null geodesic generator. Algebraically,
given $(u,\lambda,q,p)$, the ``3+1'' coordinates are given by
\begin{equation}
   x_{\scriptscriptstyle [C]}^{\alpha} = x_{\scriptscriptstyle [C]}^{(0)}{}^{\alpha}
+ \ell^{\alpha} \lambda   + O(\lambda^{2})\,,
   \label{eq:geodesic}
\end{equation}
where $x_{\scriptscriptstyle [C]}^{(0)}{}^{\alpha}$ is given by
Eq.~(\ref{e-utl}), and where $\ell^{\alpha}$ is given by
Eq.~(\ref{e-ell}).  This expression determines $x_{\scriptscriptstyle
  [C]}^{\alpha}(u,\lambda,q,p)$ to $O(\lambda^{2})$.  Consequently, the
calculation of any quantity off-$\Gamma$ is restricted to be be
second-order accurate. If higher order is required, we would need to take
into account how the geodesic generators, \ie the $\ell^{\alpha}$, vary
off-$\Gamma$, which would mean using information provided by the geodesic
equation.

Then the metric $g_{{\scriptscriptstyle [N]}\alpha\beta}$ in null affine
coordinates $x_{\scriptscriptstyle [N]}^\alpha=(u,\lambda,q,p)$ is
expressed in terms of the ``3+1'' metric as
\begin{equation}
  {g}_{{\scriptscriptstyle [N]}{\alpha}{\beta}}=
   \frac{\partial x_{\scriptscriptstyle [C]}^{\mu}}  {\partial x_{\scriptscriptstyle [N]}^{\alpha}}
   \frac{\partial x_{\scriptscriptstyle [C]}^{\nu}}  {\partial x_{\scriptscriptstyle [N]}^{\beta}}
   g_{\mu\nu}\,.
\end{equation}
The Jacobian of the coordinate transformation is now expressed as a
series expansion in the parameter $\lambda$. We do not need the
$\p_\lambda x_{\scriptscriptstyle [C]}^{\mu}$ because the coordinate
$\lambda$ is an affine parameter of the null geodesics, which fixes the
$g_{{\scriptscriptstyle [N]}\lambda{\mu}}$:
\begin{equation}
  {g}_{{\scriptscriptstyle [N]}\lambda\lambda}=
  {g}_{{\scriptscriptstyle [N]}\lambda {\scriptscriptstyle A}} = 0\,, 
  \qquad {g}_{{\scriptscriptstyle [N]}\lambda {u}} = -1\,,
  \label{eq:knowneta}
\end{equation}
with the numerical value of the last condition a consequence of the
normalization condition $t^{\alpha}\ell_{\alpha}=-1$.  The relevant part
of the coordinate transformation is then
\begin{equation}
  \p_\xi x_{\scriptscriptstyle [C]}^{\mu} \coloneqq
   \frac{\partial x_{\scriptscriptstyle [C]}^{\mu}}  {\partial x_{\scriptscriptstyle [N]}^{\xi}} =
   \p_\xi x_{\scriptscriptstyle [C]}^{(0)}{}^{\mu} + \p_\xi x_{\scriptscriptstyle [C]}^{(1)}{}^{\mu} \lambda
   + O(\lambda^{2})\,, 
   \quad \p_\xi x_{\scriptscriptstyle [C]}^{(1)}{}^{\mu} =\p_\xi \ell^{(0)}{}^{\mu}\,,
   \quad {\rm for} \quad {\xi} = (u,q,p)\,.
	\label{eq:jacob}
\end{equation}

The order ${\mathcal O}(\lambda^0)$ part of the Jacobian is evaluated by
analytic differentation of Eq.~(\ref{e-utl}). From Eq.(\ref{eq:jacob}),
the ${\mathcal O}(\lambda^1)$ part of the Jacobian is obtained from
$\p_\xi \ell^{(0)}{}^{\mu}$ with ${\xi} = (u,q,p)$; since
$\ell^{(0)}{}^{\mu}$ is known on the worldtube analytically in the
angular directions from the spherical harmonic decomposition, and
analytically or on a regular $(q,p,u)$ grid in the time direction,
$\p_\xi \ell^{(0)}{}^{\mu}$ can easily be found by analytic
differentiation or finite differencing.

\subsection{Null affine metric}
\label{sec:nullmetric}

The $\lambda$-derivative of the Cauchy $4$-metric at the worldtube can be
expressed as
\begin{equation}
  (\p_\lambda g_{\alpha\beta})_{|\Gamma} = \p_\mu g^{(0)}_{\alpha\beta} 
  \ell^{(0)}{}^{\mu}\,,
   \label{eq:g1}
\end{equation}
so that the null affine metric takes the form
\begin{equation}
   {g}_{{\scriptscriptstyle [N]}{\alpha}{\beta}} =
   {g}^{(0)}_{{\scriptscriptstyle [N]}{\alpha}{\beta}}
   + \p_\lambda {g}_{{\scriptscriptstyle [N]}{\alpha}{\beta}} \lambda + O(\lambda^{2})\,.
\end{equation}
In the above Eq., the ${}^{(0)}$ coefficients are
\begin{align}
   {g}^{(0)}_{{\scriptscriptstyle [N]}{u}{u}} & =  g_{tt}{}_{|\Gamma}\,,
   \nonumber \\
   {g}^{(0)}_{{\scriptscriptstyle [N]}{u}{\scriptscriptstyle A}} & = 
    \left(\p_{_{A}}x_{\scriptscriptstyle [C]}^{i} g_{it}{}\right)_{|\Gamma}\,,
   \nonumber \\
   {g}^{(0)}_{\scriptscriptstyle [N]AB} & =  \left(\p_{_{A}}x_{\scriptscriptstyle [C]}^{i}
     \p_{_{B}}x_{\scriptscriptstyle [C]}^{j} g_{ij}{}\right)_{|\Gamma}\,, 
   \label{eq:eta0}
\end{align}
and the $\lambda$ derivative coefficients are
\begin{align}
   \p_\lambda{g}_{{\scriptscriptstyle [N]}{u}{u}} & =  \left[\p_\lambda g_{tt} 
   + 2\, \p_u \ell^{\mu} g_{\mu t}\right]_{|\Gamma} + O(\lambda)\,,
   \nonumber \\
   \p_\lambda{g}_{{\scriptscriptstyle [N]}{u} {\scriptscriptstyle A}} & =  
        \left[\p_{_A} x_{\scriptscriptstyle [C]}^{k} \left(
      \p_u \ell^{\mu} g_{k\mu} + \p_\lambda g_{kt} \right)
      +\p_{_A} \ell^{k} g_{kt} + \p_{_A}\ell^{t} g_{tt} \right]_{|\Gamma} 
      + O(\lambda)\,, 
   \nonumber \\
   \p_\lambda{g}_{\scriptscriptstyle [N]AB} & =  
      \left[ \p_{_A} x_{\scriptscriptstyle [C]}^{k} \p_{_B}x_{\scriptscriptstyle [C]}^{l}
   \p_\lambda g_{kl}
   + \left( \p_{_A} \ell^{\mu} \p_{_B} x_{\scriptscriptstyle [C]}^{l} 
      + \p_{_B}\ell^{\mu}\p_{_A} x_{\scriptscriptstyle [C]}^{l} \right) 
     g_{\mu l} \right]_{|\Gamma}
   + O(\lambda)\,,
   \label{eq:eta1}
\end{align}
and where the $\lambda$-derivatives of the Cauchy metric are evaluated as
\begin{equation}
\p_\lambda g_{\alpha\beta}=\ell^\gamma\p_\gamma g_{\alpha\beta}\,.
\end{equation}

The contravariant null affine metric, ${g}_{\scriptscriptstyle
  [N]}^{\alpha\beta}$, is also expressed as an expansion in $\lambda$,
\begin{equation}
   {g}_{\scriptscriptstyle [N]}^{{\mu}{\nu}} = g_{\scriptscriptstyle [N]}^{(0)}{}^{{\mu}{\nu}}
   + \p_\lambda {g}^{{\mu}{\nu}}_{\scriptscriptstyle [N]} \lambda
   + O(\lambda^{2}) \,.
\end{equation}
The coefficients are obtained from the conditions
\begin{equation}
   {g}_{\scriptscriptstyle [N]}^{(0)}{}^{{\mu}{\alpha}} 
   {g}^{(0)}_{{\scriptscriptstyle [N]}{\alpha}{\nu}} = \delta^{{\mu}}_{{\nu}}\,, \qquad
   \p_\lambda{g}^{{\mu}{\nu}}_{\scriptscriptstyle [N]}  = - {g}^{{\mu}{\alpha}} \,
   {g}_{\scriptscriptstyle [N]}^{{\beta}{\nu}} \, 
      \p_\lambda{g}_{{\scriptscriptstyle [N]}{\alpha}{\beta}} \,,
\end{equation}
as well as the requirement that certain components are fixed (which
follows from Eq.~(\ref{eq:knowneta}))
\begin{equation}
   {g}_{\scriptscriptstyle [N]}^{\lambda {u}} = -1\,, \qquad
   {g}_{\scriptscriptstyle [N]}^{ u {\scriptscriptstyle A}} = 
       {g}_{\scriptscriptstyle [N]}^{{u}{u}} = 0 \,.
   \label{eq:tildeup}
\end{equation}
Thus the contravariant null affine metric and its $\lambda$ derivative are
\begin{align}
   {g}_{\scriptscriptstyle [N]}^{_{AB}} {g}_{\scriptscriptstyle [N] BC} & =  
         \delta^{_{A}}_{\, \,_{C}}\,, 
   \nonumber \\
   {g}_{\scriptscriptstyle [N]}^{\lambda{\scriptscriptstyle A}} & = 
        {g}_{\scriptscriptstyle [N]}^{_{AB}} g_{{\scriptscriptstyle [N]}u{\scriptscriptstyle B}}\,,
   \nonumber \\
   {g}_{\scriptscriptstyle [N]}^{\lambda\lambda} & =  - g_{{\scriptscriptstyle [N]}{u}{u}}
   + {g}_{\scriptscriptstyle [N]}^{\lambda {\scriptscriptstyle A}}
         g_{{\scriptscriptstyle [N]}u{\scriptscriptstyle A}}\,, \nonumber \\
   \p_\lambda g^{_{AB}}_{\scriptscriptstyle [N]} & = 
      - {g}_{\scriptscriptstyle [N]}^{_{AC}}
        {g}_{\scriptscriptstyle [N]}^{_{BD}} 
   \p_\lambda g_{\scriptscriptstyle [N]CD}\,,
   \nonumber \\
  \p_\lambda g^{\lambda {\scriptscriptstyle A}}_{\scriptscriptstyle [N]} & =  
       {g}_{\scriptscriptstyle [N]}^{_{AB}}
   \p_\lambda\left( g_{{\scriptscriptstyle [N]}{u}{\scriptscriptstyle B}} - {g}_{\scriptscriptstyle [N]}^{\lambda {\scriptscriptstyle C}} 
  \p_\lambda g_{\scriptscriptstyle [N]CB} \right)\,,
   \nonumber \\
  \p_\lambda g^{\lambda\lambda}_{\scriptscriptstyle [N]} & =  
      - \p_\lambda g_{{\scriptscriptstyle [N]}{u}{u}} 
   + 2\, {g}_{\scriptscriptstyle [N]}^{\lambda {A}}\p_\lambda  
         g_{{\scriptscriptstyle [N]}{u}{\scriptscriptstyle A}} 
   - {g}_{\scriptscriptstyle [N]}^{\lambda {\scriptscriptstyle A}} 
           {g}_{\scriptscriptstyle [N]}^{\lambda 
                {\scriptscriptstyle B}} 
   \p_\lambda g_{\scriptscriptstyle [N]AB} \,.
\end{align}

\subsection{Metric in Bondi-Sachs coordinates}
\label{sec:bondimetric}

We are now able to construct the surface area coordinate
$r(u,\lambda,\phi^{_A})$
\begin{equation}
   r  =  \left( \frac{\det({g}_{\scriptscriptstyle [N]AB})} {\det(q_{_{AB}})} \right)
   ^{{1}/{4}}\,.
   \label{eq:r}
\end{equation}
In order to make the coordinate transformation $x_{\scriptscriptstyle
  [N]}^{\alpha}=(u,\lambda,\phi^{_A}) \rightarrow x_{\scriptscriptstyle
  [B]}^{\alpha}=(u,r,\phi^{_A})$, we need expressions for $\p_\lambda r$,
$\p_{_A} r$ and $\p_u r$. From Eq.~(\ref{eq:r}),
\begin{align}
   \p_\lambda r &= \frac{r}{4} {g}_{\scriptscriptstyle [N]}^{_{AB}} 
       \p_\lambda g_{\scriptscriptstyle [N]AB}\,,  \\
   \p_{_C} r &= \frac{r}{4} \left({g}_{\scriptscriptstyle [N]}^{_{AB}} \p_{_C}
        g_{\scriptscriptstyle [N]AB} 
   - q^{^{_{AB}}}\p_{_C} q_{_{AB}} \right)\,,
   \label{eq:rl}
\end{align}
where
\begin{equation}
   \p_{_C} g_{\scriptscriptstyle [N]AB}  =  
   \left(\p_{_A}\p_{_C} x_{\scriptscriptstyle [C]}^{i}\, \p_{_B}x_{\scriptscriptstyle [C]}^{j}
        + \p_{_A}x_{\scriptscriptstyle [C]}^{i}\,  
         \p_{_B}\p_{_C}x_{\scriptscriptstyle [C]}^{j} \right) g_{ij}
   + \p_{_A}x_{\scriptscriptstyle [C]}^{i}\,\p_{_B} x_{\scriptscriptstyle [C]}^{j}\, 
        \p_{_C}x_{\scriptscriptstyle [C]}^{k}\,\p_k g_{ij} \,,
\end{equation}
in which the $\p_{_A}\p_{_C}x_{\scriptscriptstyle [C]}^{i}$ are evaluated
analytically in terms of $\phi^{_A}$. 
An expression for $\p_u r$ will be required later but only on the worldtube
$\Gamma$, so that Eq.~(\ref{eq:eta0}) may be used when simplifying $\p_u$
applied to  Eq.~(\ref{eq:r}); further on $\Gamma$, $\p_u=\p_t$, and by
construction $\p_{_A}x_{\scriptscriptstyle [C]}^{i}$ is independent of time.
Thus,
\begin{align}
   \p_u r =&
\frac{r}{4}\, {g}_{\scriptscriptstyle [N]}^{\scriptscriptstyle AB}
\p_u  {g}_{\scriptscriptstyle [N] AB} \nonumber \\
=&\frac{r}{4}\, {g}_{\scriptscriptstyle [N]}^{\scriptscriptstyle AB}
\p_u\left(\p_{_A}x_{\scriptscriptstyle [C]}^{i}\, 
        \p_{_B}x_{\scriptscriptstyle [C]}^{j}\, g_{ij} \right)
        \nonumber \\
=&     \frac{r}{4}\, {g}_{\scriptscriptstyle [N]}^{\scriptscriptstyle AB} 
        \p_{_A}x_{\scriptscriptstyle [C]}^{i}\, 
        \p_{_B}x_{\scriptscriptstyle [C]}^{j}\,\p_t g_{ij}\,.
\end{align}

The metric $g_{\scriptscriptstyle [B]}^{\alpha\beta}$ in Bondi-Sachs coordinates
is obtained from the coordinate transformation
\begin{equation}
   g_{\scriptscriptstyle [B]}^{\alpha\beta} = 
   \frac{\partial x_{\scriptscriptstyle [B]}^{\alpha}} {\partial x_{\scriptscriptstyle [N]}^{\mu}}
   \frac{\partial x_{\scriptscriptstyle [B]}^{\beta}}  {\partial x_{\scriptscriptstyle [N]}^{\nu}}
   g_{\scriptscriptstyle [N]}^{{\mu}{\nu}}\,.
   \label{eq:etaup}
\end{equation}
Note that the spherical part of the metric is unchanged, \ie
$g_{\scriptscriptstyle [B]}^{\scriptscriptstyle AB}=g_{\scriptscriptstyle
  [N]}^{\scriptscriptstyle AB}$, and only the components
$g_{\scriptscriptstyle [B]}^{11}$, $g_{\scriptscriptstyle
  [B]}^{1{\scriptscriptstyle A}}$ and $g_{\scriptscriptstyle [B]}^{01}$
on $\Gamma$ need to be determined. From Eq.~(\ref{eq:tildeup}),
\begin{align}
   g_{\scriptscriptstyle [B]}^{11} & =  \p_\alpha \, \p_\beta r\,
g_{\scriptscriptstyle [N]}^{{\alpha}{\beta}} 
   = \left(
\p_\lambda r\right)
^2 g_{\scriptscriptstyle [N]}^{11} 
   + 2\, \p_\lambda r\,
\left(
\p_{_A}r\,g_{\scriptscriptstyle [N]}^{1 {\scriptscriptstyle A}} - \p_u r
\right)
   +\p_{_A} r\,\p_{_B}r\, g_{\scriptscriptstyle [N]}^{_{AB}}\,,
   \nonumber \\
   g_{\scriptscriptstyle [B]}^{1{\scriptscriptstyle A}} & = \p_\alpha r\, 
      g_{\scriptscriptstyle [N]}^{{\alpha} {\scriptscriptstyle A}} 
   =\p_\lambda r\, g_{\scriptscriptstyle [N]}^{1 {\scriptscriptstyle A}} + 
     \p_{_{B}}r\, g_{\scriptscriptstyle [N]}^{_{AB}}\,,
   \nonumber \\
   g_{\scriptscriptstyle [B]}^{01} & =  \p_{{\alpha}}r\,
g_{\scriptscriptstyle [N]}^{0{\alpha}} = - \p_{\lambda}r\,.
   \label{eq:bondim}
\end{align}

As discussed in Sec.~\ref{s-nf}, the characteristic Einstein equations
are not formulated directly in terms of the metric components, but in
terms of quantities derived from the metric, specifically $J,\beta,U$ and
$W_c$. Explicitly, the relations between these quantities and the
contravariant metric components are
\begin{equation}
J=-\frac{q_{_{A}} q_{_{B}} g_{\scriptscriptstyle [B]}^{^{_{AB}}}}{2
  r^2}\,, \qquad \beta=-\frac 12 \log(g_{\scriptscriptstyle [B]}^{01})\,,
\qquad U=\frac{g_{\scriptscriptstyle [B]}^{1A}}{g_{\scriptscriptstyle
    [B]}^{01}}q_{_{A}}\,, \qquad W_c=-\frac{g_{\scriptscriptstyle
    [B]}^{11}+g_{\scriptscriptstyle [B]}^{01}} {g_{\scriptscriptstyle
    [B]}^{01}r}\,.
\label{e-JbU}
\end{equation}

\subsection{Starting up the null code at the worldtube}
\label{sec:nullbdry}

As already mentioned, a difficulty faced is that Eq.~(\ref{e-JbU}) gives
metric quantities on the worldtube $\Gamma$, which is not in general a
hypersurface at a constant value of the $r$-coordinate. The original
method for tackling the problem makes use of a Taylor series in $\lambda$
\cite{Bishop98a}, and has been implemented in
\cite{Szilagyi00a,Szilagyi00,Babiuc:2005pg,Reisswig:2009us,Reisswig:2010a}.
Recently, a method that uses a special integration algorithm between the
worldtube and the first characteristic grid-point, has been proposed and
tested \cite{Babiuc:2011,Babiuc:2011b}. Both approaches are outlined
below.

\subsubsection{Taylor series method}

The Taylor series method is based on writing, for some quantity $A$,
\begin{equation}
A(u,\lambda,q,p)=A(u,0,q,p)+\lambda \p_{\lambda}A(u,0,q,p) +{\mathcal
  O}(\lambda^2)\,,
\end{equation}
where $A$ represents $J,\beta,U$ and $W_c$. $A$ needs to be written as a
function of null affine coordinates, so that $\p_{\lambda}A$ can be
evaluated. Also, using Eqs.~(\ref{eq:r}) and (\ref{eq:rl}) evaluated on
the worldtube, we need to find the value of $\lambda$ at which
$r(\lambda)=r_i$, where $r_i$ is an $r$-grid-point near the worldtube;
this needs to be done for each grid-point in both the angular and time
domains. The derivation of the Taylor expansions is straightforward, with
second $\lambda$-derivatives eliminated using the Einstein equations
 \cite{Bishop98a}. The results are
\begin{equation}
   \p_\lambda J  =  -\frac{1}{2\,r^{2}} q_{_A} q_{_B} \p_\lambda 
      g^{^{_{AB}}}_{\scriptscriptstyle [N]}
   - 2\,\frac{\p_\lambda r}{r} J\,,
   \label{eq:jl}
\end{equation}
\begin{equation}
   \p_\lambda \beta = \frac{r}{8}\, \p_\lambda r
   \left(\p_\lambda J \p_\lambda\bar{J}- 
   \frac{1}{1 + J \bar{J}} 
   \left(\bar{J}\p_\lambda J +\p_\lambda\bar{J} J\right)^2  \right)\,.
   \label{eq:nbetal}
\end{equation}
\begin{equation}
  \p_\lambda U = - \left( \p_\lambda g^{1{\scriptscriptstyle A}}_{\scriptscriptstyle [N]}
                 + \frac{\p_\lambda \p_{_B} r}{\p_\lambda r} g_{\scriptscriptstyle [N]}^{_{AB}}
                 + \frac{\p_{_B}r}{\p_\lambda r} \p_\lambda g^{_{AB}}_{\scriptscriptstyle [N]} 
                    \right) q_{_{A}}
         + 2 \, \p_\lambda \beta \left( U 
                     + g_{\scriptscriptstyle [N]}^{1{\scriptscriptstyle A}} q_{_{A}} \right)\,,
\end{equation}
\begin{align}
   \p_\lambda W_{c}  = &  - \frac{\p_\lambda r}{r} 
   \left(
         \left( \frac{\p_\lambda r}{r} + 2\, \p_\lambda\beta \right)
         g_{\scriptscriptstyle [N]}^{11}
         - \p_\lambda g^{11}_{\scriptscriptstyle [N]} 
         - \frac{1}{r}
   \right)
   + \frac{2}{r} 
     \left( \frac{\p_\lambda r \p_u r}{r} - \p_\lambda \p_u r \right)
   \nonumber \\
     & + \frac{2}{r} 
         \left( 
               \p_\lambda\p_{_A}r
               - \frac{\p_\lambda r\p_{_A}r}{r} 
         \right)
         g_{\scriptscriptstyle [N]}^{1{\scriptscriptstyle A}}
   + 2\frac{\p_{_A}r}{r}\,\p_\lambda g^{1{\scriptscriptstyle A}}_{\scriptscriptstyle [N]}
   \nonumber \\
     & +  \frac{\p_{_B}r}{r\,\p_\lambda r} 
     \left(
           2 \, \p_\lambda \p_{_A}r g_{\scriptscriptstyle [N]}^{_{AB}}
         + 2 \, \p_\lambda\beta \p_{_A}r
         + \p_{_A}r \p_\lambda g^{_{AB}}_{\scriptscriptstyle [N]} 
     \right)
   - \frac{\p_{_A}r \, \p_{_B}r}{r^2}\, g_{\scriptscriptstyle [N]}^{_{AB}}\,.
\end{align}

\subsubsection{Special evolution routine between the worldtube and the first
radial grid-point}

In this approach, on a null cone say $u=u_n$, we need only the values of
the Bondi-Sachs metric variables at the angular grid-points on the
worldtube. We also suppose that the value of $J$ is known at all
grid-points of the Bondi-Sachs coordinate system on the given null cone,
either as initial data or from evolution from the previous null cone. A
mask is set to identify those radial grid-points for which $x_i-x_\Gamma
< \Delta x$, and these points will be called ``B points''. The special
algorithm is concerned with setting data at the points $i=B+1$, called
``B+1 points''. The first hypersurface equation in the hierarchy is the
one for $\beta$, and is the simplest one to hadle. The algorithm is
\begin{equation}
\beta_{{\scriptscriptstyle B}+1}=\beta_\Gamma+\Delta_r \frac{r_{{\scriptscriptstyle B}+1}
    +r_\Gamma}{16\Delta_r^2}\left(
(J_{{\scriptscriptstyle B}+1}-J_\Gamma)(\bar{J}_{{\scriptscriptstyle B}+1}-\bar{J}_\Gamma)
     -(K_{{\scriptscriptstyle B}+1}-K_\Gamma)^2
\right)\,,
\end{equation}
where $\Delta_r=r_{{\scriptscriptstyle B}+1}-r_{_\Gamma}$. The local truncation error associated
with this algorithm is ${\mathcal O}(\Delta_r^3)$. The remaining
hypersurface equations involve angular derivatives, which cannot be
evaluated on the worldtube because it is not, in general, a hypersurface
of constant $r$. Consequently, the right hand sides of these equations
are evaluated at the B+1 points rather than at the points mid-way between
$r_{{\scriptscriptstyle B}+1}$ and $r_{\scriptscriptstyle\Gamma}$. Schematically, the
hypersurface equations are
of the form $(r^n A)_{,r}=f$, and the algorithm is
\begin{equation}
A_{{\scriptscriptstyle B}+1}=\frac{r_{\scriptscriptstyle\Gamma}^n A_{\scriptscriptstyle\Gamma} 
    +\Delta_r f_{{\scriptscriptstyle B}+1}}{r_{{\scriptscriptstyle B}+1}^n}\,.
\end{equation}
The result is that the local truncation error for these equations is
reduced to ${\mathcal O}(\Delta_r^2)$. Even so, one start-up step with
error ${\mathcal O}(\Delta_r^2)$ is consistent with the global error of
${\mathcal O}(\Delta_x^2)$.

Since the value of $r$ varies on the worldtube, it may happen that the
angular neighbour of a B+1 point is a B point. Thus, the code must also
set data for the metric variables at the B points, even though much of
this data will not be needed.

\subsection{Initial data}
\label{s-char_id}
The above discussion has shown how data should be set at, or on a
neighbourhood of, the inner worldtube $\Gamma$, but in order to run a
characteristic code data for $J$ is also required on an initial null cone
$u=$ constant. Earlier work has adopted the simplistic but unphysical
approach of just setting $J=0$, assuming that the error so introduced
would quickly be eliminated from the system. Refs. \cite{Babiuc:2011,Bishop:2011}
investigated the matter. It was found that the error due to simplistic
initial data is usually small, but it can take a surprisingly long time,
up to $800M$, until saturation by other effects occurs. In terms of
observations by a gravitational-wave detector, the effect of the error in
search templates is not relevant. However, if a signal is detected, the
effect would be relevant for accurate parameter estimation at large SNR (signal to
noise ratio), but no quantitative estimates have been given.

Two methods for setting physically realistic initial data for a
characteristic evolution have been proposed and tested. In \cite{Bishop:2011} the
initial data is set by means of fitting the boundary data to a general
form of a linearized solution to the vacuum Einstein equations.
On the other hand, \cite{Babiuc:2011} sets the initial data by means of
the simple condition
\begin{equation}
J=J|_\Gamma\frac {r_\Gamma} r\,,
\end{equation}
as in this case there should be no incoming radiation since the
Newman--Penrose quantity $\psi_0=0$.

\subsection{Implementation summary}
\label{s-characIS}
The issues summarized here are: (1) setting up a characteristic code that starts
from the output of a ``3+1'' code; (2) estimating the gravitational waves from metric data
in a compactified domain output by a characteristic code; (3) estimating
quantities derived from the gravitational waves, \ie the energy, momentum and angular momentum.

\subsubsection{Setting worldtube boundary data for the characteristic code}
\label{s-CEIS}

The coding of characteristic extraction is a complex process, and is not
simply a matter of implementing a few of the formulas derived earlier in
this section. Below, we outline the key steps that are required. The
reader is also referred to Appendix~\ref{a-codes} for information about
computer code that implements characteristic extraction.\begin{enumerate}

\item Within the ``3+1'' code, write a routine that uses
  Eq.~(\ref{e-UkY}) to perform a spectral decomposition of the
  three-metric, lapse and shift, and outputs the data to file.

\item In a front-end to the characteristic code, write a routine that
  reads the data from the file created in the previous step, and
  reconstructs the four-metric and its first derivatives at the angular
  grid-points of the extraction worldtube.

\item Construct the generators $\ell^\alpha$ of the outgoing null cone
  using Eq.~(\ref{e-ell}), and then the Jacobian $\partial
  x_{\scriptscriptstyle [C]}^{\mu}/\partial x_{\scriptscriptstyle
    [N]}^{\alpha}$ as a series expansion in the affine paramenter
  $\lambda$, for each angular grid-point on the worldtube.

\item As described in section \ref{sec:nullmetric}, construct the null
  affine metric $g_{{\scriptscriptstyle [N]}\alpha\beta}$ and its first
  $\lambda$-derivative at the angular grid-points of the extraction
  worldtube; then construct the contravariant forms
  $g_{\scriptscriptstyle [N]}^{\alpha\beta}$ and $\p_\lambda
  g_{\scriptscriptstyle [N]}^{\alpha\beta}$.

\item From Eq.~(\ref{eq:r}), determine the surface area coordinate $r$
  and its first derivatives at the angular grid-points of the extraction
  worldtube.

\item Construct the Jacobian $\partial x_{\scriptscriptstyle
  [B]}^{\mu}/\partial x_{\scriptscriptstyle [N]}^{\alpha}$, and thus the
  Bondi-Sachs metric $g_{\scriptscriptstyle [N]}^{\alpha\beta}$ and then
  the metric coefficients $\beta,J,U,W_c$ at the angular grid-points of
  the extraction worldtube.

\item Implement either of the special start-up procedures described in
  section \ref{sec:nullbdry}.

\item The construction of a characteristic code is not described in this
  review, but see Appendix~\ref{a-codes} for information about the
  availability of such codes.
\end{enumerate}

\subsubsection{Estimation of gravitational waves}

In Sec.~\ref{s-charac}, many formulas used $\rho$ ($=1/r$) as the radial
coordinate, but that is unlikely to apply in practice. In the case that
the radial code coordinate is $x$ given by Eq.~\eqref{e-r2x}, the
relation between $\p_x$ and $\p_\rho$ at $\scri$ is
\begin{equation}
\p_\rho= -\frac{ \p_x}{r_\Gamma}\,.
\end{equation}

If all the metric coefficients near $\scri$ are small (which, in
practice, is often but not always the case), then the linearized formulas
apply, and:
\begin{itemize}

\item $\psi^0_4$ is evaluated using Eq.~\eqref{e-psi4a}.

\item The news ${\mathcal N}$ is evaluated, decomposed into spherical
  harmonics, using Eq.~\eqref{e-linNlm}.

\item The strain $H$ is evaluated, decomposed into spherical harmonics,
  using Eq.~\eqref{e-Jtrhtllm}.

\end{itemize}

In the general (nonlinear) case, it is first necessary to evaluate the
coordinate transformation functions $\phi_0^{_A}(u,x^{_A})$,
$\omega(u,x^{_A})$ and $u_0(u,x^{_A})$. The reason for doing so is that
it is then possible to determine, at each $(u,\phi^{_A})$ grid point, the
corresponding values of the Bondi gauge coordinates
$(\tilde{u},\tilde{\phi}^{_A})$. Thus the gravitational-wave quantities
can be expressed as functions of the (physically meaningful) Bondi gauge
coordinates, rather than as functions of the code coordinates. This issue
did not arise in the linearized case because it is a second-order effect
and thus ignorable. The procedure for evaluating these functions is:

\begin{itemize}

\item $\phi_0^{_A}(u,x^{_A})$. Solve the evolution problem
  Eq.~\eqref{e-dx0} with initial data $\phi_0^{_A}(0,x^{_A})=0$. This
  initial condition assumes that the initial data for $J$ has been set
  with $J=0$ at $\scri$.

\item $\omega(u,x^{_A})$. Either solve the evolution problem
  Eq.~\eqref{e-dom} with initial data $\omega(0,x^{_A})=1$, or evaluate
  the explicit formula Eq.~\eqref{e-exom}.

\item $u_0(u,x^{_A})$. Solve the evolution problem Eq.~\eqref{e-du0}. In
  this case, there is a gauge freedom to set the initial data
  $u_0(0,x^{_A})$ arbitrarily.
\end{itemize}

In the cases of ${\mathcal N}$ and $\psi^0_4$, the phase factor
$\delta(u,x^{_A})$ also needs to be evaluated. This can be done either
explicitly, Eq.~\eqref{e-delta}, or by solving the evolution problem
Eq.~\eqref{e-evphase} with initial data $\delta(0,x^{_A})=0$. Then:

\begin{itemize}

\item $\psi^0_4$ is evaluated using Eq.~\eqref{e-psi4G}.

\item The news ${\mathcal N}$ is evaluated using Eq.~\eqref{e-NG}.

\item The strain $H$ is evaluated using Eq.~\eqref{e-Jtrht}.

\end{itemize}

\subsubsection{Energy, momentum and angular momentum in the waves}

The formulas for the energy, momentum and angular momentum have already
been given in terms of $\psi_4$ in Sec.~\ref{s-psi4EMA}, and these
formulas are directly applicable here on substituting $\psi_4$ by
$\psi^0_4/r$. The resulting formulas involve one or two time integrals of
$\psi^0_4$, and it is useful to note that here all such integration can
be avoided by using
\begin{equation}
\int_{-\infty}^t \int_{-\infty}^{t^\prime}\psi^0_4 dt^\prime dt^{\prime\prime}
=\bar{H}\,,\qquad\qquad\qquad
\int_{-\infty}^t \psi^0_4 dt^\prime = 2\bar{\mathcal N}\,.
\end{equation}

\newpage
\section{A Comparison Among Different Methods}
\label{s-comp}
This review has described the following methods for extracting the
gravitational-wave signal from a numerical simulation
\begin{itemize}
\item The quadrupole formula, including various modifications, leading to
  the wave strain $(h_+,h_\times)$;
\item $\psi_4$ (fixed radius) and $\psi_4$ (extrapolation), leading to
  the Newman-Penrose quantity $\psi_4$;
\item Gauge-invariant metric perturbations, leading to the wave strain
  $(h_+,h_\times)$;
\item Characteristic extraction, leading to the wave strain
  $(h_+,h_\times)$, the gravitational news ${\mathcal{N}}$, or the
  Newman-Penrose quantity $\psi_4$.
\end{itemize}
There are a number of factors that need to be taken into account in
deciding the appropriate method for a particular simulation. In outline,
these factors are:
\begin{itemize}
\item {\bf Physical problem motivating the simulation}. The most
  appropriate method for extracting gravitational waves is affected by
  how the result is to be used. It may be that only moderate accuracy is
  required, as would be the case for waveform template construction for
  use in searches in detector data; on the other hand, high accuracy
  would be needed for parameter estimation of an event in detector data
  at large SNR. Further, the purpose of the simulation may be not to
  determine a waveform, but to find the emitted momentum of the radiation
  and thus the recoil velocity of the remnant.
\item {\bf Domain and accuracy of the simulation}. The domain of the
  simulation may restrict the extraction methods that can be used. All
  methods, except that using the quadrupole formula, require the
  existence of a worldtube, well removed from the domain boundary, on
  which the metric is Minkowskian (or Schwarzschild) plus a small correction.
  As discussed
  in Sec.~\ref{s-psi4Extrap}, extrapolation methods need these worldtubes
  over an extended region. Further, the accuracy of the simulation in a
  neighbourhood of the extraction process clearly limits the accuracy
  that can be expected from any gravitational-wave extraction method.
\item {\bf Ease of implementation of the various extraction methods}. All
  the methods described in this review are well understood and have been
  applied in different contexts and by different groups. Nevertheless,
  the implementation of a new gravitational-wave extraction tool will
  always require some effort, depending on the method, for coding,
  testing and verification.
\item {\bf Accuracy of the various extraction methods}. Theoretical
  estimates of the expected accuracy of each method are known, but
  precise data on actual performance is more limited because suitable
  exact solutions are not available. In a simulation of a realistic
  astrophysical scenario, at least part of the evolution is highly
  nonlinear, and the emitted gravitational waves are oscillatory and of
  varying amplitude and frequency. On the other hand, exact solutions are
  known in the linearized case with constant amplitude and frequency, or
  in the general case under unphysical conditions (planar or cylindrical
  symmetry, or non-vacuum). One exception is the Robinson-Trautman
  solution \cite{Robinson:1962zz}, but in that case the gravitational
  waves are not oscillatory and instead decay exponentially.

Thus, in an astrophysical application, the accuracy of a computed
waveform is estimated by repeating the simulation using a different
method; then the difference between the two waveforms is an estimate of
the error, provided that it is in line with the theoretical error
estimates. In some work, the purpose of comparing results of different
methods is not method testing, but rather to provide validation of the
gravitational-wave signal prediction. The only method that is, in
principle, free of any systematic error is characteristic extraction, but
the method was not available for general purpose use until the early
2010s.
It should also be noted that there remains some uncertainty about factors
that could influence the reliability of a computed waveform~\cite{Boyle2016}.
\end{itemize}

\subsection{Comparisons of the accuracy of extraction methods}
\label{s-CE-WS}

A number of computational tests have been reported, in which the accuracy
of various extraction methods is compared. Such tests are, of course,
specific to a particular physical scenario, and to the choice of ``3+1''
evolution code, initial data, gauge conditions, etc. Some of the tests
reported are now outlined, together with the results that were
obtained. While it is natural to want to generalize these results, a word
of caution is needed since the testing that has been undertaken is quite
limited. Thus any generalization should be regarded as providing only a
guide to which there may well be exceptions.
\begin{itemize}

\item Ref. \cite{Nagar:2005fz} investigates various modifications of the
  standard quadrupole formula in comparison to results obtained using
  gauge-invariant metric perturbations for the case of oscillating
  accretion tori. Good results are obtained when back-scattering is
  negligible, otherwise noticeable differences in amplitude occur.

\item Ref. \cite{Balakrishna:2006ru} computes $\psi_4$ (fixed radius) and
  gauge-invariant metric perturbations for gravitational waves from boson
  star perturbations, but detailed comparisons between the two methods
  were not made.

\item Ref. \cite{Pollney:2007ss} compares gauge-invariant metric
  perturbations to $\psi_4$ (fixed radius) extraction for the recoil
  resulting from a binary black hole merger. It was found that results
  for the recoil velocities are consistent between the two extraction
  methods.

\item Refs. \cite{Shibata:2003ga,Baiotti:2008nf} compare gauge-invariant
  metric perturbations, modified quadrupole formula and $\psi_4$ (fixed
  radius) extraction for a perturbed neutron star. While the results are
  generally consistent, each method experienced some drawback. The
  gauge-invariant method has a spurious initial junk component that gets
  larger as the worldtube radius is increased. In $\psi_4$ extraction,
  fixing the constants of integration that arise in obtaining the wave
  strain can be a delicate issue, although such problems did not arise in
  this case. The generalized quadrupole formula led to good predictions
  of the phase, but to noticeable error in the signal amplitude.

\item Refs. \cite{Reisswig:2009us,Reisswig:2010a} compared $\psi^0_4$
  from characteristic extraction and from $\psi_4$-extrapolation for
  Binary Black Hole (BBH) inspiral and merger in spinning and
  non-spinning equal mass cases. The ``3+1'' evolution was performed
  using a finite difference BSSNOK code \cite{Pollney2011b}.  A
  comparison was also made in Ref. \cite{Babiuc:2011} for the equal mass,
  non-spinning case. Recently, a more detailed investigation of the same
  problem and covering a somewhat wider range of BBH parameter space, was
  undertaken \cite{Taylor:2013} using SpEC for ``3+1''
  evolution \cite{Szilagyi:2009qz}.

These results lead to two main conclusions. (1) The improved accuracy of
characteristic extraction is not necessary in the context of constructing
waveform templates to be used for event searches in detector data. (2)
Characteristic extraction does provide improved accuracy over methods
that extract at only one radius. The $\psi_4$-extrapolation method
performs better, and there are results for $\psi^0_4$ that are equivalent
to characteristic extraction in the sense that the difference between the
two methods is less than an estimate of other errors. However, that does
not apply to all modes, particularly the slowly varying $m=0$ ``memory''
modes.

\item A study of gravitational-wave extraction methods in the case of
  stellar core collapse \cite{Reisswig:2011a} compared characteristic
  extraction, $\psi_4$ extraction (fixed radius), gauge-invariant metric
  perturbations, and the quadrupole formula. In these scenarios, the
  quadrupole formula performed surprisingly well, and gave results for
  the phase equivalent to those obtained by characteristic extraction,
  with a small under-estimate of the amplitude. However, quadrupole
  formula methods fail if a black hole forms and the region inside the
  horizon is excised from the spacetime. The gauge-invariant metric
  perturbation method gave the poorest results, with spurious high
  frequency components introduced to the signal. In characteristic
  extraction and $\psi_4$ extraction the waveform was obtained via a
  double time integration, and the signal was cleaned up using Fourier
  methods to remove spurious low frequency components.

\item It is only very recently \cite{Bishop:2014} that a method was
  developed in characteristic extraction to obtain the wave strain
  directly instead of via integration of ${\mathcal N}$ or
  $\psi^0_4$. That work also compared the accuracy of the waveform
  obtained to that found from integration of $\psi_4^0$ using
  $\psi_4$-extrapolation, in two cases -- a binary black hole merger, and
  a stellar core collapse simulation. When comparing the wave strain
  from characteristic extraction to that found by time integration of
  $\psi^0_4$, good agreement was found for the dominant (2,2) mode, but
  there were differences for $\ell\ge 4$.
\end{itemize}

\newpage

\section*{Acknowledgements}
\label{sec:acknowledgements}

NTB thanks the National Research Foundation, South Africa, for financial
support, and the Max Planck Institute for Gravitational Physics (Albert
Einstein Institute), the Institute for Theoretical Physics, Frankfurt,
and the Inter-University Centre for Astronomy and Astrophysics, India,
for hospitality while this article was being completed. The authors thank
Alessandro Nagar and Christian Reisswig for comments on the
article. Partial support comes from ``NewCompStar'', COST Action MP1304,
from the LOEWE-Program in HIC for FAIR, from the European Union's Horizon
2020 Research and Innovation Programme under grant agreement No. 671698
(call FETHPC-1-2014, project ExaHyPE), and from the ERC Synergy Grant
``BlackHoleCam - Imaging the Event Horizon of Black Holes'' (Grant
610058).

\newpage

\appendix
\newpage
\section{Notation}
\label{a-not}
\subsection{Latin symbols}
\begin{tabular}{ll}
$A_{\mu\nu}, A_+,A_\times$	& Wave amplitude tensor and coefficients\\
$A,A^A,A^u,A^\rho$	& Coefficients in transformation between Bondi and general gauges \\
$a$			& Term in Schwarzschild metric \\
$b$			& Term in Schwarzschild metric \\
$C_{abcd}$		& Weyl tensor \\
$d\Omega$		& Surface area element of a sphere \\
$E$			& Energy \\
$E^{\ell m}_{_A}$	& Vector spherical harmonic \\
$\boldsymbol{e}_{+},\boldsymbol{e}_{\times},\boldsymbol{e}_{_{\rm L}},\boldsymbol{e}_{_{\rm R}}$
			& Polarization tensors \\
$\bs{e}_x$		& Unit vector in the direction of the coordinate $x$ \\
$F^{_A}$		& Dyad of general Bondi-Sachs gauge metric \\
${\mathcal F}$		& Fourier transform operator \\
$G_{ab}$		& Einstein tensor \\
$G$			& Quantity in Cauchy-perturbative matching \\
$g_{\alpha\beta}$	& four-metric \\
$g{\!\!\!\!\; \raisebox{-0.1ex}{$^{^0}$}}_{\mu\nu}$   & Background metric \\
$H$			& Re-scaled wave strain,
			\ie $\lim_{\tilde{r}\rightarrow \infty} r(h_++ih_\times)$ \\
$H_0,H_1$		& Quantity in Cauchy-perturbative matching \\
$h_{_{AB}}$	        & Angular part of metric \\
$h_{\mu\nu}$		& Metric perturbation \\
$\htt_{\mu\nu}$		& Metric perturbation in TT gauge \\
$\bar{h}_{\mu\nu}$	& Trace-free metric perturbation \\
$h^{(0)}_{_A}, h$	& Quantity in Cauchy-perturbative matching \\
$\Itf_{jk}$		& Trace-less mass quadrupole \\
$\scri$			& Future null infinity \\
$J$			& Bondi-Sachs metric variable \\
$J_i$			& Angular momentum \\
$K$			& Bondi-Sachs metric variable \\
$K$			& Quantity in Cauchy-perturbative matching \\
$K_{ij}$		& Extrinsic curvature \\
$k_{_A}$		& Quantity in Cauchy-perturbative matching \\
${\cal L}$		& Lie derivative operator \\
$L^{\ell m}, L^{\ell m}_{_A}$	& Quantities in Cauchy-perturbative matching \\
$\ell^\alpha$		& Newman--Penrose tetrad vector, tangent to outgoing null geodesic \\
$m^\alpha$		& Newman--Penrose tetrad vector \\
$m_{_{[G]}}^\alpha$	& Approximation to $m^\alpha$ \\
${\mathcal N}$		& Gravitational news \\
$N^\mu_\nu$		& Time projection operator \\
$n_{_{[NP]}}^\alpha$	& Newman--Penrose tetrad vector \\
$n^\alpha$		& Unit normal to $\Sigma_t$ \\ 
$P_i$			& three-momentum \\
$p$			& Stereographic coordinate \\
$Q^{(0)}$		& Quantity in Cauchy-perturbative matching \\
$q$			& Stereographic coordinate \\
$q^{^{_A}}$		& Complex dyad vector \\
$q_{_{AB}}$	        & Unit sphere metric \\
\end{tabular}
\newpage
\begin{tabular}{ll}
$R$			& ``3+1'' radial coordinate, $R=\sqrt{x^2+y^2+z^2}$ \\
$R, R_{ab}, R_{abcd}$	& Ricci scalar, Ricci tensor, Riemann tensor \\
${\mathcal R}$		& Intrinsic 2-curvature \\
$r$			& Bondi-Sachs radial coordinate \\
$r$			& Radial coordinate in Cauchy perturbative approach \\
$r_*$			& ``Tortoise'' radial coordinate \\
$S^{(0)}$		& Quantity in Cauchy-perturbative matching \\
$S_t$			& $\Gamma \cap \Sigma_t$ \\
$S^{\ell m}_c,S^{\ell m}_{cd}$		& Vector spherical harmonic, tensor spherical harmonic \\
$s$			& Radial like coordinate \\
$s^i$			& Unit normal to $S_t$ \\
$T_{ab}$		& Stress-energy tensor \\
$T^{\ell m}, T^{\ell m}_{_A}$	& Quantities in Cauchy-perturbative matching \\
$t_{ab}$		& Stress-energy tensor of gravitational waves (averaged) \\
$t$			& ``3 + 1'' time coordinate \\
$U$			& Bondi-Sachs metric variable \\
$U^k(x)$		& Chebyshev polynomial of the second kind \\
$u$			& Bondi-Sachs time coordinate \\
$u^\mu$			& Fluid four-velocity \\
$V_\ell^{(0)}$		& Quantity in Cauchy-perturbative matching \\
$W$			& Lorentz factor \\
$W^{\ell m}$		& Quantities in Cauchy-perturbative matching \\
$W_c$			& Bondi-Sachs metric variable \\
$x^\alpha\,, x^i$	& Coordinates \\
$x_{\scriptscriptstyle [B]}^\alpha$	& Bondi-Sachs coordinates \\
$x_{\scriptscriptstyle [C]}^\alpha$	& Minkowski-like coordinates \\
$x_{\scriptscriptstyle [N]}^\alpha$	& Null affine coordinates \\
$X^{\ell m}$		& Quantities in Cauchy-perturbative matching \\
$Y^{\ell\,m},{}_sY^{\ell\,m}$	& Spherical harmonic, spin-weighted version \\
$Z^{\ell\,m}{}_sZ^{\ell\,m}$	& ``Real'' spherical harmonic, spin-weighted version\\
$Z^{\ell\,m}_{_{CD}}$	& Tensor spherical harmonic \\
    \end{tabular}

\subsection{Greek symbols}
\begin{tabular}{ll}
$\alpha$		& Lapse \\
$\beta$			& Bondi-Sachs metric variable \\
$\beta^i$		& Shift \\
$\Gamma$		& Extraction worldtube \\
$\gamma_{ij}$		& three-metric\\
$\delta$		& Phase factor \\
$\eta_{\mu\nu}$		& Metric of special relativity \\
$\theta$		& Spherical polar coordinate \\
$\kappa_\alpha$		& Wave propagation vector \\
$\kappa_1,\kappa_2$	& Quantities in Cauchy-perturbative matching \\
$\Lambda$		& $\ell(\ell + 1)$ \\
$\lambda$		& Affine parameter \\
$\xi^\alpha$		& Vector defining gauge transformation \\
$\xi$			& $(u,q,p)$ \\
$\rho$			& Compactified radial coordinate \\
$\Sigma_t$		& Timelike slice \\
$\Phi^{(o)},\Phi^{(e)}$	& Quantities in Cauchy-perturbative matching \\
$\phi^{^{_A}}$		& Angular coordinates \\
$\phi$			& Spherical polar coordinate \\
$\Upsilon$		& $- q^{\scriptscriptstyle A}\bar{q}^{\scriptscriptstyle B}
  \nabla_{\scriptscriptstyle A} q_{\scriptscriptstyle B}/2$, factor in definition of $\eth$ \\
$\chi$			& Quantity in Cauchy-perturbative matching \\
$\Psi^{(o)},\Psi^{(e)}$	& Quantities in Cauchy-perturbative matching \\
$\psi_0 \cdots \psi_4$	& Newman--Penrose quantities \\
$\psi_4^0$              & Re-scaled $\psi_4$, \ie $\lim_{r\rightarrow \infty} r\psi_4$ \\
$\Omega_\mu$		& $\nabla_\mu t$ \\
$\omega$		& Relation between the radial coordinate in the general and Bondi gauges \\
$\omega$		& Wave frequency \\
    \end{tabular}

\subsection{Operators}
\begin{tabular}{ll}
$\nabla_\alpha$		& Covariant derivative operator \\
$\p_\alpha$		& Partial derivative operator \\
$\Box$			& Wave operator \\
$\eth$			& Spin-weighted angular derivative operator \\
$\hat{ }$		& Quantity in conformally compactified gauge \\
$\hat{ }$		& Index in another coordinate system \\
$\tilde{ }$		& Quantity in Bondi gauge \\
$\tilde{ }$		& Fourier transformed quantity \\
$\bar{ }$		& Complex conjugate (except see above for $\bar{h}_{\mu\nu}$) \\
    \end{tabular}

\subsection{Indexing conventions}
\begin{tabular}{ll}
${}_{\alpha,\beta,\cdots}$	&$(0,1,2,3)$ Spacetime indices \\
${}_{i,j,\cdots}$		&$(1,2,3)$ Spacelike indices\\
${}_{\scriptscriptstyle A,B, \cdots}$	&$(2,3)$ Angular indices \\
${}_{a,b,\cdots}$		& $(0,1)$ Non-angular indices \\
${}_{\ell m}$			& Spherical harmonic indices \\
\end{tabular}

\newpage

\section{Spin-weighed spherical harmonics and the $\eth$ operator}
\label{a-sYlm}

A convenient way to represent vector and tensor quantities over
the sphere, including their angular derivatives, is to use spin-weighted
quantities and the $\eth$ operator. The
formalism was introduced by Newman and collaborators in the
1960s \cite{Newman-Penrose-1966,Goldberg:1967}, and has been described in
text books such as \cite{Penrose84,Stewart:1990uf}. Even so, the theory
is not well known and there are variations in notation and conventions
(which topic is discussed further in Sec.~\ref{s-convention}),
so the theory will be presented here in some detail based on the
conventions of \cite{Gomez97}. We will describe the theory, both in
general terms and with specific reference to the coordinates commonly
used, \ie spherical polar and stereographic. Further, the spin-weighted
spherical harmonic functions will be introduced, as well as the vector
and tensor spherical harmonics \cite{Newman06}. All of these can be used
as basis functions on the sphere.

\subsection{The complex dyad}
\label{s-qA}
Let $\phi^{^{_A}}$, $q_{_{AB}}$ and $q^{^{_{AB}}}$ ($2\le A,B \le3$) be coordinates and
the associated metric of a unit sphere. For example, in standard
spherical polars, $\phi^{^{_A}}=(\theta,\phi)$ and
\begin{equation}
q_{_{AB}}=\left(
\begin{array}{cc}
1 & 0 \\
0 & \sin^2\theta
\end{array}
\right)\,.
\end{equation}
The first step is to introduce a complex dyad $q^{^{_A}}$. Geometrically,
the dyad is a 2-vector that can be written as
$q^{^{_A}}=\Re(q^{^{_A}})+i\Im(q^{^{_A}})$ where $\Re(q^{^{_A}})$,
$\Im(q^{^{_A}})$ are real and orthonormal. In other words, the real and
imaginary parts of $q^{^{_A}}$ are unit vectors at right-angles to each
other. From this definition, it is straightforward to verify the
following properties
\begin{equation}
q^{^{_A}} q_{_{A}}=0,\; q^{^{_A}} \bar{q}_{_{A}}=2,\;
q_{_{AB}}=\frac{1}{2}\left(q_{_{A}} \bar{q}_{_{B}}+\bar{q}_{_{A}} q_{_{B}}\right)\,.
\label{e-qa}
\end{equation}
Clearly, the dyad $q^{^{_A}}$ is not unique being arbitary up to a
rotation and/or reflection, so that $p^{^{_A}}=e^{i\gamma}q^{^{_A}}$ and
$p^{^{_A}}=\bar{q}^{^{_A}}$ are also dyads. Even so, it is convenient to
write the dyad in a way that is natural to the coordinates being used,
which for a diagonal metric means that the real part of the dyad should
be aligned to the $\phi^2$ direction, and the imaginary part to the
$\phi^3$ direction. Thus, the dyad usually used in spherical polar
coordinates is
\begin{equation}
q^{^{_A}}=\left(1,\frac{i}{\sin\theta}\right)\,, \mbox{ and }
q_{_{A}}=(1,i\sin\theta)\,.
\end{equation}

The possibility of different parities, \ie of a reflection, introduces an
additional complication, which can be avoided by convention: All the
coordinate systems used in this review are right-handed, and we will not
consider dyads that are related by complex conjugation. This is
equivalent to embedding the sphere into Euclidean space with coordinates
$(r,\phi^2,\phi^3)$ that are always \textit{right-handed}, \ie the vector
product $\Re(\bs{q}) \times \Im(\bs{q})$ points in the positive
$r$-direction.

The dyad $q^{^{_A}}$ resembles the angular components of the tetrad vector
$m^\alpha$ given in Eq.~(\ref{e-nt}) when evaluated on a unit sphere, but
there is a difference of a factor of $\sqrt{2}$. This is an example of
different conventions used by different authors, which matter is
discussed further in Sec.~\ref{s-convention}.

\subsection{Spin-weighted fields}
\label{a-swf}
Having defined the dyad $q^{^{_A}}$, we are now in a position to construct
spin-weighted fields from vector and tensor fields. In the simplest
case, suppose that $\mu(\phi^2,\phi^3)$ is a scalar field. Then $\mu$
is also a spin-weighted field with spin-weight $s=0$. Given a vector
field $v_{_{A}}(\phi^2,\phi^3)$, we define
\begin{equation}
V=q^{^{_A}} v_{_{A}}
\end{equation}
to be a field with spin-weight $s=1$. Since $V$ is a complex quantity, it
contains two independent fields and thus uniquely represents the two
components of $v_{_{A}}$. We can also define the quantity $\bar{V}=\bar{q}^{^{_A}}
v_{_{A}}$ with spin-weight $s=-1$, but of course it is not independent of
$V$. From a second-rank tensor $t_{_{AB}}$ we can construct two independent
fields
\begin{equation}
T=q^{^{_A}} q^{^{_{_{B}}}} t_{_{AB}},\; \tau=q^{^{_A}} \bar{q}^{^{_{_{B}}}} t_{_{AB}},
\end{equation}
with spin-weights $s=2$ and $s=0$ respectively. Together $T$ and $\tau$
have 4 independent fields to represent the 4 components of $t_{_{AB}}$. In
general, given a tensor field $w_{A\cdots D}$, the quantity
\begin{equation}
W=w_{_{A \cdots BC\cdots D}}q^{^{_A}} \cdots q^{^{_{_{B}}}} \bar{q}^{^{_C}} \cdots \bar{q}^{^{_D}},
\label{e-Ww}
\end{equation}
with $m$ factors $q$ and $n$ factors $\bar{q}$ is defined to be a quantity with
spin-weight $s=m-n$.

Spin-weighted quantities may also be defined in terms of their transformation
properties. However, here, we use the definition above and later derive the
transformation rule Eq.~(\ref{e-eig-H}).

\subsection{Differentiation and the $\eth$ operator}
We would now like to define derivative operators $\eth$ and $\bar{\eth}$
that act on spin-weighted fields, and that are consistent with covariant
differentiation. This means that if $W$ is defined as in
Eq.~(\ref{e-Ww}), then we would like
\begin{equation}
\eth W=\nabla_{_{_E}}w_{_{A \cdots BC\cdots D}}q^{^{_A}} \cdots q^{^{_{_{B}}}} \bar{q}^{^{_C}} \cdots \bar{q}^{^{_D}} q^{^{_E}},\;
\bar{\eth} W =\nabla_{_{_E}}w_{_{A \cdots BC\cdots D}}
\cdots q^{^{_{_{B}}}} \bar{q}^{^{_C}} \cdots \bar{q}^{^{_D}} \bar{q}^{^{_E}}.
\label{e-ethd}
\end{equation}
We see immediately that, if $W$ has spin-weight $s$, then $\eth W$ has
spin-weight $s+1$ and $\bar{\eth} W$ has spin-weight $s-1$. We
achieve the desired effect by the definition
\begin{equation}
\eth W=q^{^{_A}} \partial_{_{A}} W + s \Upsilon W,\;\;
\bar{\eth} W=\bar{q}^{^{_A}} \partial_{_{A}} W - s \bar{\Upsilon} W
\label{e-etha}
\end{equation}
where
\begin{equation}
\Upsilon=-\frac{1}{2} q^{^{_A}}\bar{q}^{^{_{_{B}}}} \nabla_{_{_A}}q_{B}.
\label{e-Gamma}
\end{equation}
(Normally, the quantity denoted here by $\Upsilon$ is given the notation
$\Gamma$; but we use the notation $\Upsilon$ because  $\Gamma$ is used to
represent the extraction worldtube). We demonstrate the above for a
spin-weight $1$ field $V=q^{^{_A}} v_{_{A}}$. Starting from the definition
Eq.~(\ref{e-ethd}) and the ansatz Eq.~(\ref{e-etha}) with $\Upsilon$ to be
determined, we obtain
\begin{equation}
q^{^{_A}}\partial_{_{_A}}(q^{^{_{_{B}}}} v_{_{B}}) +\Upsilon q^{^{_C}} v_{_{C}}
=q^{^{_A}} q^{^{_{_{B}}}} \partial_{_{_B}}v_{_{A}}
-q^{^{_A}} q^{^{_{_{B}}}} v_{_{C}} \Gamma^{^{_C}}_{_{AB}},
\end{equation}
and thus
\begin{equation}
v_{_{C}}\left( q^{^{_C}}\Upsilon +q^{^{_A}}\left(\partial_{_{_A}}q^{^{_C}}
+q^{^{_{_{B}}}}\Gamma^{^{_C}}_{_{AB}}\right)\right)
=0
\end{equation}
after renaming some dummy indices. This must be true for all $v_{_{C}}$, and
the bracketed term is just a covariant deriavtive, so we have
\begin{equation}
q^{^{_C}}\Upsilon +q^{^{_A}} \nabla_{_{_A}}q^{^{_C}}=0\,.
\end{equation}
We lower the free superscript ${}^{^{_C}}$ and then contract with $\bar{q}^{^{_C}}$
to obtain the desired result.

It should be noted that, in general, $\eth$ and $\bar{\eth}$ do not commute. The commutator is
\begin{equation}
(\bar{\eth}\eth - \eth\bar{\eth})W= 2sW,
\label{e-ethbeth}
\end{equation}
so that the operators commute only in the case of a quantity with spin-weight $s=0$.

\subsection{Coordinate transformation of spin-weighted quantities:
Rotation factors $\exp(i\gamma)$}

Spin-weighted quantities are defined in a way that they have no free
tensorial indices so it would appear that they are scalars, but this is
misleading because different dyads are used in the different coordinate
systems. Suppose that we have two coordinate systems $S_{(q)}$ and
$S_{(p)}$ with natural dyads $\bs{q}$ and $\bs{p}$ respectively. Each dyad  has
components in each of the coordinate systems, and so we define $q^{^{_A}}_{(q)},
q^{^{_A}}_{(p)}$ to mean the components of $\bs{q}$ in $S_{(q)},S_{(p)}$
respectively; and similarly for 
$p^{^{_A}}_{(q)}$ and $p^{^{_A}}_{(p)}$. Assuming that $S_{(q)}$ and $S_{(p)}$
have the same parity, then their dyads are related by a rotation and, as discussed
just after Eq.~(\ref{e-qa}),
\begin{equation}
p^{^{_A}}_{(p)}=\exp(i\gamma)q^{^{_A}}_{(p)}\,.
\label{e-eiq}
\end{equation}

Suppose that $\bs{v}$ is a vector and that $V_{(q)},V_{(p)}$ are the corresponding
spin-weighted quantities with respect to the dyads $\bs{q},\bs{p}$ respectively.
Thus
\begin{eqnarray}
V_{(p)}=p^{^{_A}}_{(p)}
v_{(p){\scriptscriptstyle A}}=\exp(i\gamma)q^{^{_A}}_{(p)}v_{(p){\scriptscriptstyle A}}=
\exp(i\gamma)q^{^{_A}}_{(q)}v_{(q){\scriptscriptstyle A}}=\exp(i\gamma)V_{(q)}\,.
\end{eqnarray}
Generalizing to the case where $V$ is defined with $m$ factors $q^A$ and $m-s$
factors $\bar{q}^A$ so that the spin-weight is $s$, we find
\begin{align}
V_{(p)} &= V_{(q)} \exp(i m \gamma)\exp(-i(m-s)\gamma)\nonumber \\
&=V_{(q)} \exp(i s\gamma)\,.
\label{e-eig-H}
\end{align}

\subsection{Specific coordinate systems}
It is very convenient to define coordinate systems on the unit sphere in
terms of a coordinate transformation from a Cartesian system.  The
Cartesian coordinates are denoted by $x^a_{_{[C]}}=(x,y,z)$, and the
spherical coordinates by $x^a=(r,\phi^2,\phi^3)$.  We use computer
algebra to construct the Jacobian of the transformation bewteen spherical
and Cartesian coordinates, and then to find the components of the dyad
with respect to the Cartesian system. We use the notation $Q^i$ to denote
the components of a dyad with respect to Cartesian coordinates. The
computer algebra also evaluates the quantity $\Upsilon$ used in the
definition of the $\eth$ operator Eq.~(\ref{e-Gamma}), see
Appendix~\ref{a-algebra}.

\subsubsection{Spherical polar coordinates}
\label{s-polar}
We use coordinates $(r,\theta,\phi)$ for standard spherical polar
coordinates. Only one coordinate patch is required, but the coordinate
system is singular at the poles $\theta=0$ and $\theta=\pi$. The relation
to Cartesian coordinates is
\begin{equation}
x=r\sin\theta\cos\phi\,,\qquad
y=r\sin\theta\sin\phi\,,\qquad
z=r\cos\theta
\label{e-C2sp}
\end{equation}
and the inverse transformation is
\begin{equation}
r=\sqrt{x^2+y^2+z^2}\,,\qquad
\theta=\arccos\left(\frac{z}{\sqrt{x^2+y^2+z^2}}\right)\,,\qquad
\phi=\arctan\left(\frac{y}{x}\right)\,.
\label{e-C2spi}
\end{equation}
Transforming the Cartesian metric to $(r,\theta,\phi)$ coordinates,
we find, as is well known,
\begin{equation}
ds^2=dr^2+r^2(d\theta^2+\sin^2\theta d\phi^2),
\end{equation}
which is normally represented by the dyad
\begin{equation}
q^{^{_A}}=\left(1,\frac{i}{\sin\theta}\right)\,, 
\qquad \mbox{or}\qquad
q_{_{A}}=(1,i\sin\theta)\,.
\label{e-qsp}
\end{equation}
The components of the dyad with respect to the Cartesian coordinates are
\begin{equation}
Q^i=\left(\cos\theta \cos\phi - i\sin\phi,
          \cos\theta \sin\phi + i\cos\phi, -\sin\theta \right)\,,
\end{equation}
and the quantity $\Upsilon$ is
\begin{equation}
\Upsilon=- \cot\theta.
\end{equation}

\subsubsection{Stereographic coordinates}
\label{sec:stereo}

\label{s-stereo}
\epubtkImage{}{%
\begin{figure}[htb]
  \begin{center}
    \includegraphics[width=110mm,clip,trim=0 0 0 0]{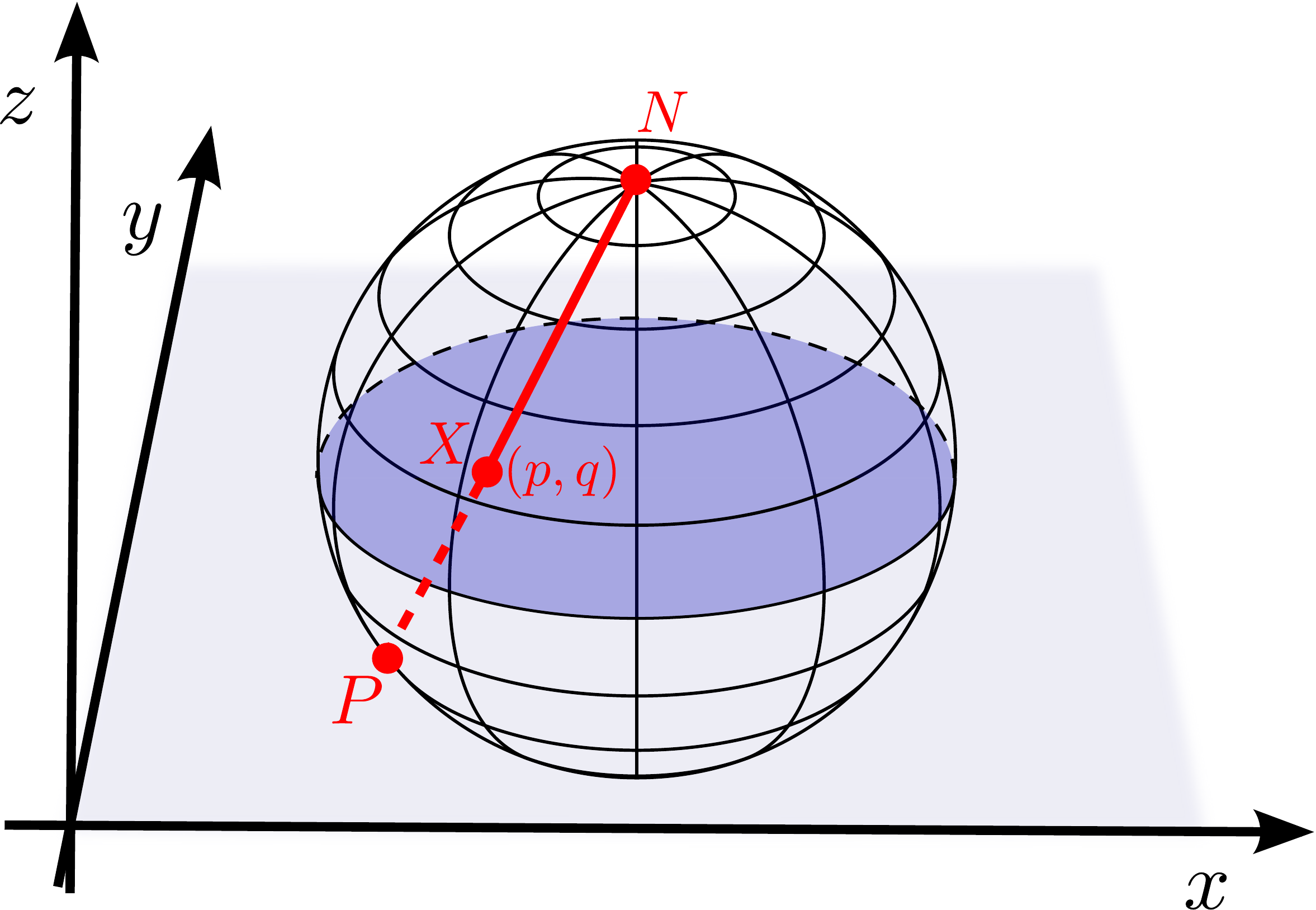}
  \end{center}
  \caption{Illustration of stereographic coordinates. Consider a unit
    sphere with centre at the origin of Cartesian $(x,y,z)$
    coordinates. Construct a straight line from the north pole $N$ of the
    sphere (at $x=y=0,z=1$) to a point $P$ on the sphere, and let the
    line meet the plane $z=0$ at $X$. Let the \emph{Cartesian}
    coordinates of $X$ be $(x,y,z)=(q_{_{[S]}},p_{_{[S]}},0)$, then the
    southern patch \emph{stereographic} coordinates of $P$ are
    $(q_{_{[S]}},p_{_{[S]}})$. The northern patch stereographic
    coordinates are constructed in a similar way, with the straight line
    in this case starting from the south pole.}
   \label{f-stereo}
\end{figure}}

In stereographic coordinates, the sphere is described by means of two
patches, called North and South, with local coordinates
$x^a_{_{[N]}}=(r,q,p)$ and $x^a_{_{[S]}}=(r,q,p)$ defined on each patch. Where
it is necessary to distinguish between $(q,p)$ on the North and South
patches, we will use the suffix ${}_{_{[N]}}$ or ${}_{_{[S]}}$, but otherwise
the suffix will be omitted. The relation to Cartesian coordinates is
\begin{eqnarray}
\mbox{North}&\!\!\!\!:&\qquad
x=\frac{2qr}{1+q^2+p^2}\,,\quad y=\frac{2pr}{1+q^2+p^2}\,,\quad
z=\frac{r(1-q^2-p^2)}{1+q^2+p^2}\,, \nonumber \\
\mbox{South}&\!\!\!\!:&\qquad
x=\frac{2qr}{1+q^2+p^2}\,,\quad y=\frac{-2pr}{1+q^2+p^2}\,,\quad
z=-\frac{r(1-q^2-p^2)}{1+q^2+p^2}\,.
\label{e-C2N}
\end{eqnarray}
The inverse transformation is
\begin{eqnarray}
\mbox{North}&\!\!\!\!:&\;\;
r=\sqrt{x^2+y^2+z^2}\,,\quad q=\frac{x}{\sqrt{x^2+y^2+z^2}+z}\,,\quad
p=\frac{y}{\sqrt{x^2+y^2+z^2}+z}\,, \nonumber \\
\mbox{South}&\!\!\!\!:&\;\;
r=\sqrt{x^2+y^2+z^2}\,,\quad q=\frac{x}{\sqrt{x^2+y^2+z^2}-z}\,,\quad
p=-\frac{y}{\sqrt{x^2+y^2+z^2}-z}\,.
\label{e-C2S}
\end{eqnarray}
It is then straightforward to construct the Jacobian and to transform the
Cartesian metric into the metric in $(r,q,p)$ coordinates. We find, on
both patches,
\begin{equation}
ds^2=dr^2+\frac{4r^2}{(1+q^2+p^2)^2}\left(dq^2+dp^2\right)\,,
\end{equation}
which is normally represented by the dyad
\begin{equation}
q^{^{_A}}=\frac{1+q^2+p^2}{2}(1,i)\,, 
\qquad \mbox{or}\qquad
q_{_{A}}=\frac{2}{1+q^2+p^2}(1,i)\,.
\end{equation}
The relationship between $(r_{_{[N]}},q_{_{[N]}},p_{_{[N]}})$ and
$(r_{_{[S]}},q_{_{[S]}},p_{_{[S]}})$ is found by going via the Cartesian
coordinates. We apply Eq.~(\ref{e-C2N}) to find values for $(x,y,z)$, and
then apply Eq.~(\ref{e-C2S}) to find the corresponding values for
$(r_{_{[S]}},q_{_{[S]}},p_{_{[S]}})$. The result is
\begin{equation}
r_{_{[S]}}=r_{_{[N]}}\,,\qquad
q_{_{[S]}}=\frac{q_{_{[N]}}}{q_{_{[N]}}^2+p_{_{[N]}}^2}\,,\qquad
p_{_{[S]}}=-\frac{p_{_{[N]}}}{q_{_{[N]}}^2+p_{_{[N]}}^2}\,,
\end{equation}
which may be expressed more compactly and informatively as
\begin{equation}
q_{_{[S]}}+i p_{_{[S]}}=\frac{1}{q_{_{[N]}}+i p_{_{[N]}}}\,,.
\label{e-qS2N}
\end{equation}
The components of the dyads with respect to the Cartesian coordinates
are
\begin{equation}
Q^i_{_{[N]}}=\left(\frac{1 - q^2 + p^2 - 2 i q p}{1+q^2+p^2}\,,\;
\frac{i + i q^2 -i p^2 - 2 q p}{1+q^2+p^2}\,,\;
-\frac{2(q + ip)}{1+q^2+p^2}\right)\,,
\end{equation}
\begin{equation}
Q^i_{_{[S]}}=\left(\frac{1 - q^2 + p^2 - 2 i q p}{1+q^2+p^2}\,,\;
-\frac{i + i q^2 -i p^2 - 2 q p}{1+q^2+p^2}\,\;
\frac{2(q + ip)}{1+q^2+p^2}\right)\,.
\end{equation}
The rotation factor between the dyads on the North and South patches, as
defined in Eq.~(\ref{e-eiq}), is
\begin{equation}
\exp(i\gamma)=\frac{Q^i_{_{[S]}}\bar{Q}^j_{_{[N]}}\delta_{ij}}{2}
=-\frac{q_{_{[N]}}-ip_{_{[N]}}}{q_{_{[N]}}+ip_{_{[N]}}}\,,
\end{equation}
where Eq~(\ref{e-qS2N}) has been used. The quantity $\Upsilon$ is
\begin{equation}
\Upsilon=q + i p\,.
\end{equation}

\subsection{The spin-weighted spherical harmonics ${}_s Y^{\ell\,m}$}
\label{s-sYlm}

The standard, \ie spin-weight zero, spherical harmonics are given in
Cartesian coordinates by
\begin{equation}
Y^{\ell\, 0}=\sqrt{\frac{2\ell +1}{4\pi}}2^{-\ell}
\sum_{k=0}^{\left\lfloor\ell/2\right\rfloor}
(-1)^k \frac{(2\ell -2k)!}{k!(\ell-k)!(\ell-2k)!}
\left(\frac{z}{r}\right)^{\ell-2k}\,,
\end{equation}
and
\begin{eqnarray}
\left( \begin{array}{c} Y^{\ell\,m} \\
                        Y^{\ell\, -m} \end{array} \right)
&=&\left( \begin{array}{c} (-1)^m(A_m+iB_m) \\
                          (A_m-iB_m) \end{array} \right)
\sqrt{\frac{2\ell +1}{4\pi}}2^{-\ell}\sqrt{\frac{(\ell-m)!}{(\ell+m)!}} \nonumber \\
&\times &
\sum_{k=0}^{\left\lfloor(\ell-m)/2\right\rfloor} 
\frac{(-1)^k(2\ell -2k)!}{k!(\ell-k)!(\ell-2k-m)!}
\frac{z^{\ell-2k-m}}{r^{\ell-2k}}\,,
\end{eqnarray}
where in the above formula $m>0$, and where
\begin{eqnarray}
A_m(x,y)&=&\sum_{k=0}^m\frac{m!}{k!(m-k)!}x^ky^{m-k}\cos\left(\frac{\pi(m-k)}{2}\right)\,,
\nonumber \\
B_m(x,y)&=&\sum_{k=0}^m\frac{m!}{k!(m-k)!}x^ky^{m-k}\sin\left(\frac{\pi(m-k)}{2}\right)\,.
\end{eqnarray}
The symbol $\lfloor \ \rfloor$ means truncation to an integer; for
example, $\lfloor 4/2 \rfloor=\lfloor 5/2 \rfloor=2$. The (spin-weight
zero) spherical harmonics in angular coordinates are then obtained by
simply substituting the appropriate coordinate transformation
$x^a_{_{[C]}}=x^a_{_{[C]}}(x^a_{_{[S]}})$ into the above formulas for
$Y^{\ell\,m}$, and it will be found that $r$ cancels out of the result,
leaving a formula in terms only of the angular coordinates. The constant
factors in the definitions are chosen so that the $Y^{\ell\,m}$ satisfy
the orthonormality condition
\begin{equation}
\int_{S^2}Y^{\ell\,m} \bar{Y}^{\ell^\prime m^\prime}d\Omega = \delta^{\ell \ell^\prime}
\delta^{m m^\prime}\,,
\label{e-orthoY}
\end{equation}
where integration is over the unit sphere, and where $d\Omega
=\sqrt{\det(q_{_{AB}})} d\phi^2 d\phi^3$ -- \eg in spherical polars
$d\Omega=\sin\theta d\theta d\phi$.

The spin-weighted spherical harmonics are found by repeated application of the
operators $\eth$ or $\bar{\eth}$:
\begin{eqnarray}
{}_sY^{\ell\,m} &=& \sqrt{\frac{(\ell-s)!}{(\ell+s)!}}\eth^s
Y^{\ell\,m}\,, \quad s>0, \nonumber \\ 
{}_{s}Y^{\ell\,m} &=& (-1)^s
\sqrt{\frac{(\ell+s)!}{(\ell-s)!}}  \bar{\eth}^{-s} Y^{\ell\,m}\,, \quad
s<0\,.
\label{e-sYlm}
\end{eqnarray}
The ${}_sY^{\ell\,m}$ are defined only in the cases that $|s|\le \ell$
and $|m|\le \ell$, and in the spin-weight zero case it is usual to omit
the ${}_s$, \ie ${}_0Y^{\ell\,m}=Y^{\ell\,m}$. The ${}_sY^{\ell\,m}$
satisfy the same orthonormality condition as the $Y^{\ell\,m}$ in
Eq.~(\ref{e-orthoY}) and this is the origin of the square root factors in
Eq.~(\ref{e-sYlm}). However orthonormality does not fix the phase, and
this raises issues that are pursued in Sec.~\ref{s-convention}. The
${}_sY^{\ell\,m}$ constitute a large number of different cases, and are
obtained from computer algebra scripts, see Appendix~\ref{a-algebra}.

The $Y^{\ell m}$ satisfy the property
\begin{equation}
\eth\bar{\eth}Y^{\ell m}=\bar{\eth}\eth Y^{\ell m}=-\Lambda Y^{\ell m},\qquad
\mbox{where }\Lambda=\ell(\ell + 1)\,,
\label{e-L2Y}
\end{equation}
so that $\eth\bar{\eth}$ is the Laplacian operator on the sphere. Using Eqs.~
\eqref{e-L2Y} and \eqref{e-ethbeth}, it follows that
\begin{equation}
\eth^2\bar{\eth}^2Y^{\ell m}=\bar{\eth}^2\eth^2 Y^{\ell
  m}=(\Lambda^2-2\Lambda) Y^{\ell m}\,.
\label{e-L22Y}
\end{equation}

\subsection{Spin-weighted representation of deviations from spherical symmetry}
\label{s-dss}

As already discussed, any quantity that can be regarded as a vector or
tensor on the 2-sphere can be given a spin-weighted representation, and
this includes quantities that describe the angular part of the metric in
a curved spacetime. Suppose that a general metric has angular part
$ds^2=r^2 h_{_{AB}}dx^{^{_A}} dx^{^{_{_{B}}}}$ \cite{Bishop96}; for the
applications considered here we restrict attention to the case that $r$
is a surface area coordinate (see \cite{Gomez97} for the general case),
so that $\det(h_{_{AB}})=\det(q_{_{AB}})$ for some unit sphere metric
$q_{_{AB}}$. Then we define
\begin{equation}
J=\frac 12 q^{^{_A}} q^{^{_{_{B}}}} h_{_{AB}}=-\frac 12 q_{_{A}} q_{_{B}}
h^{^{_{AB}}}\,,
\qquad \mbox{ and } \qquad
K=\frac 12 q^{^{_A}} \bar{q}^{^{_{_{B}}}} h_{_{AB}}=\frac 12 q_{_{A}} \bar{q}_{_{B}} h^{^{_{AB}}}\,,
\end{equation}
with the spherically symmetric case characterized by $J=0$. From the
determinant condition, it follows that
\begin{equation}
K^2=1+J\bar{J}\,.
\label{e-KJJb}
\end{equation}
The metric can be written interms of $J$ and $K$. We find in spherical
polars
\begin{eqnarray}
h_{_{AB}}&=&\left(
\begin{array}{cc}
\displaystyle
\frac 12 (J+\bar{J}+2K) & 
\displaystyle
\frac i2 (\bar{J}-J)\sin\theta \\ \\
\displaystyle
\frac i2 (\bar{J}-J)\sin\theta  & 
\displaystyle
\frac 12 (2K-J-\bar{J})\sin^2\theta
\end{array}
\right)\,,\nonumber \\
h^{^{_{AB}}}&=&\left(
\begin{array}{cc}
\displaystyle
\frac 12 (2K-J-\bar{J}) & 
\displaystyle
\frac {i}{2\sin\theta}(J-\bar{J}) \\ \\
\displaystyle
\frac {i}{2\sin\theta}(J-\bar{J}) &
\displaystyle
\frac {1}{2 \sin^2\theta}( 2K+J+\bar{J})
\end{array}
\right)\,,
\label{e-hJK}
\end{eqnarray}
and in stereographic coordinates \cite{Bishop97b}
\begin{eqnarray}
h_{_{AB}}&=&\frac{2}{(1+q^2+p^2)^2}\left(
\begin{array}{cc}
J+\bar{J}+2K & i(\bar{J}-J) \\ \\
i(\bar{J}-J) & 2K-J-\bar{J}
\end{array}
\right)\,,\nonumber \\
h^{^{_{AB}}}&=&\frac{(1+q^2+p^2)^2}{8}\left(
\begin{array}{cc}
2K-J-\bar{J} & i(J-\bar{J}) \\ \\
i(J-\bar{J}) & 2K+J+\bar{J}
\end{array}
\right)\,.
\end{eqnarray}

The quantity $J$ is simply related to the strain in planar coordinates,
$J=h_++ih_\times$.  To see this, suppose that a plane with Cartesian-like
coordinates $(x,y)$ is tangent to the unit sphere at a given point
$(\theta_0,\phi_0)$ (The argument is simpler when using specific, rather
than general, angular coordinates). In a neighbourhood of
$(\theta_0,\phi_0)$, the coordinate transformation is
$x=\theta-\theta_0,y=(\phi-\phi_0)/\sin\theta_0$. Now, $h_+,h_\times$ are
weak field quantities, and in this case $|J|\ll 1$. Further,
$J+\bar{J}=2\Re(J)$ and $i(\bar{J}-J)=2\Im(J)$, and thus
Eq.~(\ref{e-hJK}) may be simplified to
\begin{eqnarray}
h_{_{AB}}&=&\left(
\begin{array}{cc}
\displaystyle
1+\Re{J} & 
\displaystyle
\Im(J)\sin\theta \\ \\
\displaystyle
\Im(J)\sin\theta  & 
\displaystyle
1-\Re(J)\sin^2\theta
\end{array}
\right)\,.
\end{eqnarray}
Then transforming from $(\theta,\phi)$ to $(x,y)$ coordinates, and in a
neighbourhood of $(x,y)=(0,0)$, gives
\begin{eqnarray}
h_{_{AB}}&=&\left(
\begin{array}{cc}
\displaystyle
1+\Re{J} & 
\displaystyle
\Im(J) \\ \\
\displaystyle
\Im(J)  & 
\displaystyle
1-\Re(J)
\end{array}
\right)\,,
\end{eqnarray}
which is the expected form in the TT gauge.

\subsection{$Z^{\ell\,m}$, the ``real'' $Y^{\ell\,m}$}
Since metric quantities are real, a decomposition in terms of the
$Y^{\ell\,m}$ may introduce mode mixing between $\pm m$ modes. This can
be avoided by making use of the formalism described in
\cite{Zlochower03,Bishop-2005b}, and using basis functions which, in the
spin-weight 0 case, are purely real; following \cite{Bishop-2005b}, these
are denoted as ${}_sZ^{\ell\,m}$.
\begin{eqnarray}
{}_s Z^{\ell\,m} &=& \frac{1}{\sqrt{2}} \left[{}_s Y^{\ell\,m}
   +(-1)^m {}_s Y^{\ell\, -m}\right] \quad \mbox{ for } m>0\,, \nonumber \\
{}_s Z^{\ell\,m} &=& \frac{i}{\sqrt{2}} \left[(-1)^m{}_s Y^{\ell\,m} 
   -{}_s Y^{\ell\, -m} \right] \quad \mbox{ for }  m<0\,, \nonumber \\
{}_s Z^{\ell\, 0} &=& {}_s Y^{\ell\, 0}\,.
\end{eqnarray}
The ${}_sZ^{\ell\,m}$ obey orthonormal properties similar to those of the
$Y^{\ell\,m}$ in Eq.~(\ref{e-orthoY}), and have a relationship with
$\eth^s Z^{\ell\,m}$ similar to that for the ${}_sY^{\ell\,m}$ in
Eq.~(\ref{e-sYlm}).

\subsection{Vector and tensor spherical harmonics}
\label{app:vtsh}

Particularly within the context of gauge invariant perturbation theory,
it is common practice to use vector and tensor spherical harmonics rather
than spin-weighted spherical harmonics \cite{Nagar05}. The vector
spherical harmonics are defined, in the even parity case
\begin{equation}
E^{\ell\,m}_{_{A}}= \nabla_{_{A}}Y^{\ell\,m}\,,
\end{equation}
and in the odd parity case
\begin{equation}
S^{\ell\,m}_{_{C}}=\epsilon_{_{CD}} q^{^{_{DE}}} \nabla_{_{E}}Y^{\ell\,m}\,,
\end{equation}
where $\epsilon_{_{CD}}$ is the Levi-Civita completely antisymmetric tensor
on the 2-sphere; for example, in spherical polars
$\epsilon_{\theta\theta}=\epsilon_{\phi\phi}=0,
\epsilon_{\theta\phi}=-\epsilon_{\phi\theta}=\sin\theta$. The tensor
spherical harmonics are defined, in the even parity case
\begin{equation}
Z^{\ell\,m}_{_{CD}}= \nabla_{_{C}}\nabla_{_{D}} Y^{\ell\,m}+\frac 12 \ell(\ell+1)
      q_{_{CD}}Y^{\ell\,m}\,,
\end{equation}
and in the odd parity case
\begin{equation}
S^{\ell\,m}_{_{CD}}=\frac 12 \left(\nabla_{_{D}} S^{\ell\,m}_{_{C}}
     +\nabla_{_{C}}S^{\ell\,m}_{_{D}}\right)\,.
\end{equation}

The vector and tensor spherical harmonics are related to the $\eth$
operator and thereby to the spin-weighted spherical harmonics
\begin{equation}
q^{^{_A}} E^{\ell\,m}_{_{A}}=\eth Y^{\ell\,m}=\sqrt{(\ell+1)\ell}\,{}_1 Y^{\ell\,m}\,,
\end{equation}
\begin{equation}
q^{^{_C}} S^{\ell\,m}_{_{C}} =-i\eth Y^{\ell\,m}=-i\sqrt{\ell(\ell+1)}\,{}_1 Y^{\ell\,m}\,,
\end{equation}
\begin{equation}
q^{^{_C}} q^{^{_D}} Z^{\ell\,m}_{_{CD}}= \eth^2 Y^{\ell\,m} = \sqrt{(\ell+2)(\ell+1)\ell(\ell-1)}
     \,{}_2 Y^{\ell\,m}\,,
\end{equation}
\begin{equation}
S^{\ell\,m}_{_{CD}} q^{^{_C}} q^{^{_D}}=-i\eth^2 Y^{\ell\,m}
=-i\sqrt{(\ell+2)(\ell+1)\ell(\ell-1)}\,{}_2 Y^{\ell\,m}\,.
\end{equation}

\subsection{Regge--Wheeler harmonics}
\label{app:rwh}

We report below the explicit expressions of the Regge--Wheeler harmonics,
$(\hat e_1)_{i j},\cdots,(\hat f_4)_{i j}$, which have been introduced in
section \ref{subsec:angdecomp} when discussing the numerical
implementation of the Cauchy-perturbative method. In particular, the
tensor spherical harmonics $(\hat{e}_{1})_{ij}$ and $(\hat{e}_{2})_{ij}$
in have the rather lengthy but otherwise straightforward expressions
\begin{equation}
\left( \hat{e}_{1}\right)_{ij}=\left(
\begin{tabular}{ccc}
$0$ & $\displaystyle -\frac{1}{\sin \theta }\partial_{\phi }Y_{\ell m}$ & 
	$\sin \theta \partial_{\theta }Y_{\ell m}$ \\ \\
$\displaystyle -\frac{1}{\sin \theta }\partial_{\phi }Y_{\ell m}$ & 
	$0$ & $0$ \\ \\

$\sin \theta \partial_{\theta }Y_{\ell m}$ & $0$ & $0$%
\end{tabular}
\right) \,,
\end{equation}
and
\begin{equation}
\label{e2ij}
\left( \hat{e}_{2}\right)_{ij}=\left( 
\begin{tabular}{ccc}
$0$ & $0$ & $0$ \\ \\
$0$ & $\displaystyle \frac{1}{\sin \theta }
	\left( \partial_{\theta \phi }^{2}-\cot \theta
	\partial_{\phi }\right) Y_{\ell m}$ 
	& $\displaystyle\frac{1}{2}\left( 
	\displaystyle\frac{1}{\sin^{2}\theta }
	\partial_{\phi }^{2}-\cos \theta \partial_{\theta }-
	\sin\theta \partial_{\theta }^{2}\right) Y_{\ell m}$ \\ \\
	$0$ & $\displaystyle\frac{1}{2}\left(
	\displaystyle\frac{1}{\sin^{2}\theta }\partial_{\phi}^{2} - 
	\cos \theta \partial_{\theta }-\sin \theta \partial_{\theta }^{2}%
	\right) Y_{\ell m}$ & 
	$-\left( \sin \theta \partial_{\theta \phi }^{2}-
	\cos\theta \partial_{\phi }\right) Y_{\ell m}$%
\end{tabular}
\right) \ .
\end{equation}

Similarly, the tensor spherical harmonics $(\hat{f}_{1})_{ij} -
(\hat{f}_{4})_{ij}$ which enter in the decomposition of even-parity
perturbations have the form
\begin{equation}
\left( \hat{f}_{1}\right)_{ij}=\left( 
\begin{tabular}{ccc}
$0$ & $\partial_{\theta }Y_{\ell m}$ & $\partial_{\phi }Y_{\ell m}$ \\ \\
$\partial_{\theta }Y_{\ell m}$ & $0$ & $0$ \\ \\
	$\partial_{\phi }Y_{\ell m}$ & $0$ & $0$%
\end{tabular}
\right)\,,
\end{equation}
\begin{equation}
\left( \hat{f}_{2}\right)_{ij}=\left( 
\begin{tabular}{ccc}
$Y_{\ell m}$ & $0$ & $0$ \\ \\
$0$ & $0$ & $0$ \\ \\
$0$ & $0$ & $0$%
\end{tabular}
\right) \,,
\end{equation}
\begin{equation}
\left( \hat{f}_{3}\right)_{ij}=\left( 
\begin{tabular}{ccc}
$0$ & $0$ & $0$ \\ \\
$0$ & $Y_{\ell m}$ & $0$ \\ \\ 
$0$ & $0$ & sin$^{2}\theta Y_{\ell m}$%
\end{tabular}
\right) \,,
\end{equation}
and
\begin{equation}
\left( \hat{f}_{4}\right)_{ij}=\left( 
\begin{tabular}{ccc}
$0$ & $0$ & $0$ \\ \\
	$0$ & $\partial_{\theta }^{2}Y_{\ell m}$ & 
	$\left( \partial_{\theta \phi}^{2}-\cot \theta 
	\partial_{\phi }\right) Y_{\ell m}$ \\ \\
$0$ & $\left( \partial_{\theta \phi }^{2}-\cot \theta 
	\partial_{\phi }\right) Y_{\ell m}$ & $\left( \partial_{\phi }^{2}
	-\sin \theta \cos \theta \partial_{\theta }\right) Y_{\ell m}$%
\end{tabular}
\right) . 
\end{equation}

\subsection{Issues of convention in the definitions of spin-weighted 
quantities}
\label{s-convention}

The reader needs to be aware that different authors use different
conventions in the definitions of quantities discussed in this section,
and that it is therefore inadvisable to use expressions from different
sources without first carefully checking the conventions used.

Fortunately, the definitions of the $Y^{\ell\,m}$ do seem to be standard
throughout the mathematical-physics community. However, there are
differences in the definition of the complex dyad, here denoted by
$q^{^{_A}}$ and normalized so that $q^{^{_A}} \bar{q}_{_{A}}=2$; much
other work uses $m^{^{_A}}$ with normalization $m^{^{_A}}
\bar{m}_{_{A}}=1$ so that $q^{^{_A}}=\sqrt{2}m^{^{_A}}$. Clearly the dyad
definition -- $q^{^{_A}}$ or $m^{^{_A}}$ -- affects the definition of
spin-weighted quantities. The $\eth$ operator is defined so that for a
spin-weight 0 scalar $V$, $\eth\bar{\eth}V=\nabla^2 V$ where $\nabla^2$
is the Laplacian operator on the unit sphere, and the various definitions
of $\eth$ all have the same magnitude. However, there can be a variation
in sign. For example, the definition used in \cite{Alcubierre:2008} is
$-1$ times that used here. This implies that the definition of
${}_sY^{\ell\,m}$ is $-1$ times that used here for any \textit{odd} $s$,
positive or negative. However, for \textit{even} $s$ definitions of
${}_sY^{\ell\,m}$ are consistent.

The definitions of the \textit{even} vector and tensor spherical
harmonics seem to be consistent, although the notation can vary. However,
there are sign differences in the definition of the \textit{odd} vector
and tensor spherical harmonics, for example \cite{Alcubierre:2008,
  Baumgarte2010a}

\newpage

\section{Computer codes and scripts}
\label{a-codesscripts}

\subsection{Computer algebra}
\label{a-algebra}

This Appendix describes the computer algebra (Maple) files used to derive
a number of equations in the main text. The files are available at [TO BE
  SPECIFIED BY JOURNAL EDITOR]. The maple script files are named
name.map with output in name.out. In some cases the main script files
call auxilliary scripts as detailed below.

The files \textbf{gamma.map}, \textbf{gamma.out}, \textbf{R.map} and
\textbf{contraint.map} are used to derive the vacuum nonlinear Einstein
equations for the Bondi-Sachs metric in Sec.~\ref{s-nf}. The linearized
Einstein equations were given as Eqs.~(\ref{e-b}) to (\ref{e-c0A}). The
file \textbf{gamma.map} calculates the Christoffel symbols, and then the
script \textbf{R.map} reads gamma.out and calculates the hypersurface and
evolution equations, confirming the formulas given in
\cite{Bishop97b}. The files \textbf{J.map}, \textbf{k.map},
\textbf{U.map} and \textbf{W.map} are auxilliary scripts used by
\textbf{R.map}. The script \textbf{constraint.map} uses
\textbf{gamma.out} and \textbf{k.map}, and evaluates $R_{00},R_{01}$ and
$q^AR_{0A}$. The asymptotic Einstein Eqs.~(\ref{e-as11}) to
(\ref{e-asev}), as well as the condition
$\partial_{\tilde{\rho}}\tilde{W}_c=0$ in Eq.~(\ref{e-BGC}), are derived
in the script \textbf{asympt.map}, and using the auxilliary files {\bf
  gamma-asympt.map, gamma-asympt.out}.

The script \textbf{C\_trans.map} uses \textbf{compactified.map} and
derives Eqs.~(\ref{e-du0}) to (\ref{e-fA}) and (\ref{e-nNPG}). The script
{\bf J\_om\_Jrho\_delta.map} uses \textbf{compactified.map} and derives
Eqs.~(\ref{e-delta}), (\ref{e-Jtrht}) and (\ref{e-JtrhtLin}). The script
\textbf{JK.map} also uses \textbf{compactified.map} and derives
Eqs.~(\ref{e-exom}) and (\ref{e-fJ}), and checks that $|\nu|=1$ in
Eq.~(\ref{e-Mmb}) and that $\tilde{m}_0^\alpha \tilde{m}_\alpha = 0$. The
script \textbf{NewsBondi.map} uses \textbf{conformal.map} and evaluates
the news ${\mathcal N}$ in the Bondi gauge, confirming that
Eq.~(\ref{e-NBp}) reduces to
$\partial_{\tilde{u}}\partial_{\tilde{\rho}}\tilde{J}/2$ independently of
whether $\hat{\tilde {m}}^\alpha_{(0)}$ or $\hat{\tilde {m}}^\alpha$ is
used in Eq.~(\ref{e-NBp}). The script \textbf{checkFA\_FB\_GamAB1\_0.map}
uses \textbf{gamma-asympt.map} and shows that $F^{^{_A}}F^{^{_{_{B}}}}
\hat{\Gamma}^1_{_{AB}(0)}=0$, as discussed just after
Eq.~(\ref{e-NG}). The script \textbf{NewsGen.map} uses
\textbf{conformal.map} for the further evaluation of Eq.~(\ref{e-NG}) to
obtain an expression for ${\mathcal N}$ in the general case. The script
\textbf{psi4Bondi.map} uses \textbf{weyl\_asympt.map} to evaluate
Eq.~(\ref{e-psi4B}) confirming that, in the Bondi gauge,
$\psi^0_4=\p^2_{\tilde{u}}\p_{\tilde{\rho}}\bar{\tilde{J}}$,
independently of whether $\hat{\tilde {m}}^\alpha_{(0)}$ or $\hat{\tilde
  {m}}^\alpha$ is used. The script \textbf{psi4Gen.map} again uses
\textbf{weyl\_asympt.map} to reduce, in the general gauge,
Eq.~(\ref{e-psi4G}) to computational $\eth$ form. In the linearized case,
\textbf{NewsLin.map} uses \textbf{news.map} and \textbf{conformal.map} to
derive Eq.~(\ref{e-Nlinraw}) for ${\mathcal N}$, and
\textbf{psi4\_lin.map} uses \textbf{weyl\_asympt.map} to derive
Eq.~(\ref{e-psi4linraw}) for $\psi^0_4$.

The files \textbf{polars.map} and \textbf{stereo.map} are used in
Appendix~\ref{a-sYlm}. They specify the coordinate transformation between
spherical and Cartesian coordinates, as well as the metric, dyad and
$\Upsilon$ (used in the evaluation of the operator $\eth$) in the
spherical coordinates. Each file passes these quantities to the procedure
``C2P'' in the file \textbf{procs.map}, which checks that all the
relations given in Eqs.~(\ref{e-qa}) and (\ref{e-Gamma}) are satisfied;
this procedure also calculates and outputs the dyad transformed into
Cartesian coordinates, \ie $Q^a$. The procedure ``sYlm'' in
\textbf{procs.map} requires as input the various quantities defined in
the coordinate-specific files, together with values for $s,\ell$ and $m$;
it then calculates ${}_sY^{\ell\,m}$ in the appropriate coordinate patch
using the equations given in Sec.~\ref{s-sYlm}. The file
\textbf{sYlm.map} reads each of the driver files in turn, and for each
coordinate patch calculates all the ${}_sY^{\ell\,m}$ for $\ell\le
\ell_{\mbox{max}}$, $|s|\le \min(\ell_{\mbox{max}},s_{\mbox{max}})$, with
default values $\ell_{\mbox{max}}=3,s_{\mbox{max}}=2$. The output is
written to {\bf polars-sYlm.out, stereoNorth-sYlm.out} and
\textbf{stereoSouth-sYlm.out}.

\subsection{Numerical codes}
\label{a-codes}

The Einstein toolkit (\url{http://einsteintoolkit.org}) contains a wide
range of numerical relativity codes using the Cactus code framework
(\url{http://cactuscode.org}). In particular, it contains the following
thorns relevant to gravitational-wave extraction: ``\texttt{Extract}''
which implements the Cauchy-perturbative method, ``\texttt{WeylScal4}''
which implements $\psi_4$ extraction. The process of characteristic
extraction is started in ``\texttt{NullSHRExtract}'' which constructs the
worldtube boundary data, and then the various thorns listed under
``\texttt{PITTNullCode}'' are used for the characteristic evolution and
the determination of gravitational-wave descriptors at $\scri$.

\newpage

\bibliography{refs}

\end{document}